# Studies of braided non-Abelian anyons using anyonic tensor networks

By

**Babatunde Moses Ayeni**

A thesis submitted to Macquarie University
for the degree of Doctor of Philosophy
Department of Physics
July 2017

MACQUARIE
University
SYDNEY · AUSTRALIA

Except where acknowledged in the customary manner, the material presented in this thesis is, to the best of my knowledge, original and has not been submitted in whole or part for a degree in any university.

Babatunde Moses Ayeni

*To my parents, Mary Olufunke and Emmanuel Oluwadamisi Ayeni*

# Acknowledgements

As it is with any successful journey, I have safely got to my "destination"— the end of my doctoral research programme, but definitely not the end of my research life. My voyage has been quite an adventure and full of wonderful experiences. Overall, I give God thanks for giving me the ability to complete this programme.

As it might be expected, a lot of people contributed to my success. I will start by thanking my supervisors: A/Prof. Gavin Brennen and Dr. Robert Pfeifer. I am grateful to Gavin for accepting to supervise my PhD at a point when I was at a crossroad of indecision (or confusion); he took me under his wings, and steered me in the right direction. For this I am eternally grateful. Thank you for your motivations, encouragements, and reassurances during my moments of doubts. I also thank Robert Pfeifer for accepting to be my co-supervisor when he moved to Macquarie University from Perimeter Institute. He made tremendous contributions to the success of my research. Alongside, I also thank Sukhwinder Singh for enlightening discussions on various concepts relating to my research on non-Abelian anyons, tensor networks, and physics in general, and for giving me his anyonic-TEBD MATLAB codes which formed the basis of my initial interaction with non-Abelian symmetries in tensor networks. In addition, I thank Prof. Jason Twamley and Dr. James Cresser, for their efforts, supervision, and understanding during the time spent as my initial supervisors at Macquarie. I will also like to acknowledge the hard work of the entire staff of the Department of Physics and Astronomy, and those of the ARC Centre of Excellence in Engineered Quantum Systems (EQUS), to whom the success of the department and the centre should be partly credited.

I am very grateful to some of my friends in Sydney and around the world for being my good companions. I want to specially acknowledge the following fine people: Hossein Dinani for being the buddy that makes things happen for us (his buddies), and manages to make everyone happy; Andreas Tabacchini; Alex Busse; Nora Tischler; Dr. Ben Okwuofu-Thomas and family, Andre Lim, Suhki, and others. I also want remember my all-time cheerful friend, Safriat Adunni Yusuff, who sadly passed on to glory around May 2016. She was the freest and yet morally sound person I have ever met. You left a vacuum in my heart, Safriat. I really miss you!

I would not have been able to complete my doctorate programme without the support of my family. They shared in my pains, worries, doubts, and all that. For this, I say thank you for your contagious and unfading love. I appreciate my most caring mummy, Mary Ayeni, for her unparalleled love and efforts in ensuring my total well-being. I thank my very principled dad, Emmanuel Ayeni, for his love and support in every way. You both are appreciated and highly valued. I thank my siblings: David, Joshua, Peace, and Deborah, for their cares and prayers. I thank my sister, Aunty Funmi, and my nephew Boluwade. God richly bless you all.

My doctoral research was funded by Macquarie University through the "international Macquarie University Research Excellence Scholarship" (iMQRES) scheme, and also by the Centre of Excellence in Engineered Quantum Systems (EQuS), Project No. CE110001013, through the Australian Research Council (ARC). I am very grateful to everyone whose money has contributed to this scholarship. In return, I hope this research has advanced the scope of human knowledge. In addition, I thank the National eResearch Collaboration Tools and Resources (Nectar), supported by the Australian government, for providing me the computing resources that I used to run my long and computationally intensive simulations, which as a whole, ran for several months.

Science never ends, at least not until we unravel all the mysteries of nature, which in my opinion, is still quite far. So, as it is being said, “as one journey ends, so another begins.” Onwards still!

# List of Publications

1. Babatunde M. Ayeni, Sukhwinder Singh, Robert N. C. Pfeifer, and Gavin K. Brennen, *Simulation of braiding anyons using matrix product states*, Physical Review B, **93**, 165128 (2016).

2. Babatunde M. Ayeni, Robert N. C. Pfeifer, and Gavin K. Brennen, *Phase transitions in a braided non-Abelian anyonic system*, (In preparation, 2017).

3. Babatunde M. Ayeni, James Cresser, and Jason Twamley, *Using a biased quantum random walk as a quantum lumped element router*, Physical Review A, **90**, 012339 (2014).

## 0.1 Organisation of Papers in thesis

Here is how the papers listed above are arranged in this thesis: Paper 1 forms the basis of Chapters 4 and 5. Paper 2, which is still in preparation, forms the basis of Chapter 6. Finally, Paper 3 forms the basis of Chapter 7.

# Abstract

The content of this thesis can be broadly summarised into two categories: first, I constructed modified numerical algorithms based on tensor networks to simulate systems of anyons in low dimensions, and second, I used those methods to study the topological phases the anyons form when they braid around one another.

Anyons are point-like particles which are neither bosons nor fermions. All point-like particles in our three dimensional world are either bosons or fermions and have an exchange factor of $+1$ and $-1$ respectively, when a pair of those particles exchange positions. Anyons on the other hand represent new possibilities that are capable of existing only in two dimensions, and have non-trivial exchange factors, which can either be a complex number or even a matrix acting on a space of degenerate states. These unusual particles have some surprising properties. For example, states with anyonic excitations are known be resilient against local perturbations, which simply means that, you may perturb their local environments, and they would still retain their quantum properties—a property exclusive only to anyons. This property has motivated scientists to sometimes call them *topological charges*. Anyons have motivated many recent developments in science and technology. I mention two of them. Firstly, just as bosons give rise to unique phases of matter such as Bose-Einstein condensates, and fermions allow for degenerate Fermi gases, interacting anyons can form new phases of matter, and it is still a subject of intense research, as we have as yet little understanding of these new forms of matter. Secondly, the topological nature of anyons, i.e. their fault tolerance, make them to be among the leading prospective candidate particles to realise a quantum computer. This inspired a different field of research called *topological quantum computation*. As important as anyons tend to be, in reality, collective systems of anyons are very challenging to study analytically and numerically, due to the exponential growth of the state space of such systems with the number of particles. It is in fact this exponential growth that makes anyons to be usable for quantum computation.

In the past, common numerical techniques used to study quantum many body systems included exact diagonalisation (ED) and Monte Carlo (MC) methods. Tensor network techniques are a more modern approach which have lots of advantages. I will list only two of them: 1) they are able to simulate very large system sizes and even infinite-sized systems which are not possible with ED, and 2) they can treat systems of fermions and anyons which are problematic for MC methods.

In the first phase of this thesis, I extended the anyonic tensor network algorithms, by incorporating U(1) symmetry to give a modified ansatz Anyon$\times$U(1) tensor networks, which are capable of simulating anyonic systems at any rational filling fraction. By testing the ansatz with the *time evolving block decimation algorithm*, I benchmarked these methods by using it to confirm previously known results for simple models, including a one-dimensional chain of itinerant and localised anyons, and a two-leg ladder where anyons hop between sites and use available vacancies to exchange positions with other anyons. I performed test simulations with four different particle types, namely, Fibonacci and Ising anyons, which are non-Abelian anyons, and spinless fermions and hardcore bosons, which can be treated as “simple types” of anyons. I compared the results with known exact results obtained using other methods, and got very good performances with relative error of around $\epsilon < 10^{-4}$ in the ground state energy.

After being satisfied with the performance of the numerical algorithms, I then proceeded to the second phase where I used the numerical methods to study some models of non-Abelian anyons that naturally allows for exchange of anyons. I proposed a lattice model of anyons, which I dubbed *anyonic Hubbard model*, which is a pair of coupled chains of anyons (or simply called *anyonic ladder*). Each site of the ladder can either host a single anyonic charge, or it can be empty. The anyons are able to move around, interact with one another, and exchange positions with other anyons, where vacancies exist. Exchange of anyons is a non-trivial process which may influence the formation of different kinds of new phases of matter. I studied this model using the two prominent species of anyons: Fibonacci and Ising anyons, and made a number of interesting discoveries about their phase diagrams. I identified new phases of matter arising from the interaction of these anyons and their exchange “braid” statistics.

# Contents

# List of Figures

# List of Tables

# 1 Introduction

Particles in our three-dimensional (3D) world are classified into either bosons or fermions. Fermions are the particles that form matter, while bosons are the force carriers and mediate interactions between matter. Bosons were discovered by Bose and Einstein in 1924 [3, 4], and they obey Bose-Einstein statistics, while fermions were discovered by Fermi and Dirac in 1926 [5, 6], and they obey Fermi-Dirac statistics. Bosons and fermions are responsible for some of the most interesting phenomena discovered in human history. For example, a Bose gas (i.e. a gas of bosonic particles) forms a Bose-Einstein condensate at very low temperatures, which is a phenomenon where all the bosons occupy the same single quantum state, and form a large macroscopic fluid which exhibits unusual behaviour such as superfluidity (i.e. flow without resistance) and macroscopic interference. Fermions on the other hand are responsible for phenomena such as conductivity, magnetism, superconductivity, etc., without which today's technology would be impossible. These observed phenomena represent different phases of matter which are described by *Landau theory of phase transitions* where we have a definable local order parameter such as number and phase fluctuations, average particle density, local magnetisation, and so on. The formation of different phases is dictated by various competing terms between the particles of a system, including the interaction between neighboring particles, hopping of particles from one site to another, geometric frustrations of the particles, dimensionality and topology of the system, and particle statistics [7–9]. The importance of these terms have since been known and appreciated. What was initially less emphasized was the effect of topology on the physical system. Topology is the branch of pure mathematics that studies the properties of a system that are preserved through deformations of geometric objects, where deformations include any or all of, stretching and twisting, but not tearing. For example, a circle, an ellipse, or any other non-self-intersecting closed line, are all topologically equivalent. Arguably, the concept of topology first entered into physics with the discovery of Aharonov-Bohm effect [10] by Aharonov and Bohm, where if a charge $q$ winds around a magnetic flux $\Phi$, the wavefunction picks up a phase factor of $e^{inq\Phi}$, where $n$ is the number of times the charge winds around that flux. This phase factor does not depend on the nature of the path, or the rate at which the charge traverses the path, but on the number of circulation $n$. Topology was formally and rigorously introduced into physics with the seminal work of Witten [11], which connected the field of topological quantum field theories with knot theory, conformal field theories, etc.

Bosons and fermions have an exchange symmetry. The wavefunction of a collection of bosons or fermions acquires a phase factor of $+1$ or $-1$ respectively, when a pair of them are exchanged. Theoretical curiosity into other possibilities in lower dimensions (particularly in two-dimensions) revealed that there are other particles that break the dichotomy of classification of particles into either bosons or fermions in three-dimensions [12]. These particles have exotic exchange statistics, where if a pair of them are exchanged, their wavefunction can acquire any arbitrary (complex) phase factor $e^{i\theta}$, or even a unitary matrix $\hat{U}$ for a degenerate ground space manifold. Because they could pick *any* phase factor when a pair is exchange, they were named *anyons* [13], and in analogy with Aharonov-Bohm effect, this exchange does not depend on the path or the rate of traversing the path, but on the number of circulation (or *winding number*). Thus systems of anyons are *topological*. Anyons are divided into two classes: Abelian anyons with a phase exchange factor, and non-Abelian anyons with a matrix exchange factor. All these discoveries created many new different research directions in physics and technology, including the discovery of new phases of matter, the possibility of building a fault-tolerant quantum computer using anyons [14], and the development of aesthetically beautiful and conceptually powerful tools for modern theoretical physics. Theoretical proposals of systems where anyons may be found include fractional quantum Hall systems and two-dimensional spin liquids [15–28], one dimensional nanowires [29–32], and ultra-cold atoms in optical lattices [33]. Experimentally, the recent evidence for Majorana edge modes (i.e. Ising anyons) might be closing the gap between theory and practical realisation of anyons [32, 34].

The study of phase transitions is an important problem in science. When the parameters of a physical system change, there can be a transition from one phase into another. For example, changing the external temperature and pressure of water can change it from ice to liquid, or to gas. These phase transitions are understood in terms of breaking the symmetry of the system, a concept put forward by Landau. Symmetries are described by group theory, and phase transitions are determined using a local order parameter, for example, the average particle density $n$, or magnetisation $m$. But it was soon found out that Landau symmetry breaking was not enough, i.e. it was discovered that there are some phase transitions which are characterised by non-local order such as degeneracy that depends on topology and rigid transformation rules between ground state sectors [35]. The low-energy excitations of these new phases of matter are anyons, and they have a rich structure of combination (fusion) and exchange (braiding) statistics. This has motivated several studies on topologically-ordered systems, i.e. systems which host anyons.

Many-body systems of anyons are generally intractable except for a small number of exactly soluble systems, which in part, has hindered the theoretical progress made in understanding topological phases. Anyons are described by the powerful mathematical framework of unitary braided tensor categories (UBTC), and each anyon model comes with its own topological data such as F-matrix, R-matrix, topological spin, and so on. Numerically, they are also hard to simulate, due to the exponential growth in the Hilbert space dimension with the number of particles in the system. A pure quantum state $|\psi\rangle$ on a lattice $\mathcal{L}$ with $L$ sites, each having a $d$ dimensional state space, is written generally as,

$$|\psi\rangle = \sum_{i_1,\dots,i_L} c_{i_1\dots i_L} |i_1 \dots i_L\rangle \,. \tag{1.1}$$

where the sum contains $d^L$ number of coefficients $c_{i_1\dots i_L}$, and hence, is exponentially large in the system size $L$. Working directly in this representation is not feasible in numerical simulations, except for very small system sizes. Alternatively, one can use other approximate schemes like Monte Carlo (MC) methods, but MC methods suffers from the “sign problem” for fermions, and anyons [36]. Tensor networks offer a more robust alternative, being able to represent the ground state and/or the low-energy subspace of a quantum system with low entanglement. They can handle large system sizes, and possibly, infinite-size systems if the system is translation invariant. They can also simulate all kinds of particles, including frustrated magnets, fermions, Abelian and non-Abelian anyons. Examples of popular tensor networks include matrix product states (MPS), projected entangled-pair states (PEPS), multi-scale renormalisation ansatz (MERA). Examples of algorithms used in manipulating the tensor networks are density matrix renormalisation group (DMRG), time-evolving block decimation (TEBD), variational Monte Carlo methods, etc. These ansatz states and algorithms are reviewed in [37, 38]. The use of tensor network to simulate anyons was pioneered independently by Pfeifer et al. [39] and by König and Bilgin [40]. Their works opened the pathway into using existing tensor network approaches to study systems of anyons, including the more challenging non-Abelian anyons. After that, a few number of other anyonic tensor network states and algorithms have been put forward, including anyonic MPS and anyonic TEBD algorithm [41], Anyon×U(1) MPS [42], and anyonic DMRG algorithm [43].

## 1.1 Overview

In this thesis, I use anyonic tensor networks to study the physics of systems of anyons where the particles braid around each other. These systems can be at any arbitrary filling fraction, achieved either by fixing the global particle density, which is a U(1) symmetry in addition to the anyonic symmetry of the tensor network, or alternatively by introducing a chemical potential term into the Hamiltonian, thereby bypassing the need for U(1) symmetry, but while still retaining the anyonic symmetry.

This thesis is arranged as follows:

In Chapter Two, I review the general concepts of the tensor network states and algorithms mentioned above. I introduce the necessary diagrammatic conventions that I adopt throughout this thesis. In particular, I give a quite extensive review of matrix product states (MPS) and TEBD, which form the basis of the ansatz and algorithms I develop later in the thesis.

In Chapter Three, I review the algebraic theory of anyons, and recall the diagrammatic notations which persist throughout this thesis. The diagrammatic formalism is conceptually powerful and convenient for our purpose.

In Chapter Four I present my first original work, in which I show how U(1) symmetry in terms of particle number can be combined with anyonic symmetry to construct a modified ansatz: Anyonic×U(1) MPS, which is an ansatz that exploits both the anyonic (quantum) symmetry of anyonic systems and a possible U(1) symmetry of the system. I show how this MPS can be used to encode the ground state of an arbitrary anyonic system, including systems in the thermodynamic limit.

In Chapter Five I present a practical implementation of anyonic-TEBD algorithm. Including non-Abelian symmetries into tensor networks is a non-trivial task, and can be daunting at first. As interacting non-Abelian anyons are challenging, so are the tensor networks that encode them. This chapter may therefore benefit anyone wanting to simulate anyons using tensor networks. In addition, in this chapter, I test our Anyonic×U(1) MPS with the anyonic TEBD algorithm. I will apply it to a number of test models with known results. I later present a few new original results.

In Chapter Six I study a new model of itinerant and interacting non-Abelian anyons on the ladder, which I dub the *anyonic Hubbard model*. Particles in these model have braiding degrees of freedom, and also interact with a Heisenberg-like interaction. I present the ground state phase diagrams of Fibonacci and Ising anyons, and show the influence of anyonic braid statistics on the system. We find evidence of a braiding induced phase transition.

Chapter Seven is the first work on my PhD studies which focused on out of equilibrium dynamics of a single quantum particle via the quantum walk formalism. This is in contrast to the equilibrium dynamics of the previous chapters. In this work I derived the "probability current" of a single walker, and study the steady state of the system. Using the anyonic MPS, the methods developed in this chapter can be applied to study out-of-equilibrium dynamics of a small number of anyons.

I end this thesis with a summary and outlook in Chapter Eight.

# 2

# Tensor Network States and Algorithms

## 2.1 Introduction

The study of quantum many-body systems is both exciting and challenging for many reasons. On one hand, it is exciting because of the new surprising phenomenoma constantly being discovered. Some of the most recent developments include, the discovery and characterisation of new exotic phases of quantum matter and quantum phase transitions [44], the role of entanglement in these phases, the possible realisation of a quantum computer [45], and many more. On the other hand, theoretical progress in these research areas is hindered, generally by a lack of analytical tools to study these systems. Except for a small number of exactly solvable systems, solutions to most of the many-body systems, either rely on perturbation theory or exact diagonalisation, which both give results that are far from ideal, and can even be sometimes misleading. For instance, the method of perturbation theory only works at the weakly interacting regime, not the strongly interacting regime—where most of the interesting physics lies. The instance at which the method breaks down might not be exactly clear. On the contrary, exact diagonalisation gives exact results for all interaction strengths, but can only be applied to small system sizes. Therefore, apart from the quest of discovering new exciting physical phenomena, there is also a search for new efficient numerical techniques to probe many-body quantum systems in the thermodynamic limit.

One central problem of interest in modern theoretical condensed matter physics is to understand the low-energy properties of quantum lattice systems. Let $\mathcal{L}$ be a lattice of $L$ sites, with a tensor product Hilbert space $\mathcal{H} = \mathcal{H}_{\text{site}}^{\otimes L}$, where $\mathcal{H}_{\text{site}}$ is a $d$-dimensional Hilbert space at each site. A general quantum state defined on this lattice $\mathcal{L}$ is

$$|\psi\rangle = \sum_{i_1,\ldots,i_L} c_{i_1\ldots i_L} |i_1 \ldots i_L\rangle, \tag{2.1}$$

where $i_k \in \{1,\ldots,d\}$ is the index label for the orthonormal basis vectors $\{|1\rangle\,|2\rangle\ldots|d\rangle\}$ of the Hilbert space $\mathcal{H}_k$ of each site $k$. The dimension of the tensor product space $\mathcal{H}$ is $d^L$, which is exponential in the system size $L$, and therefore implies that, we need an exponential number of parameters $c_{i_1\ldots i_L}$ to specify the state $|\psi\rangle$. An immediate consequence of this is that, large quantum systems cannot be studied using exact diagonalisation (ED), as already mentioned.

However, other conventional techniques like Monte Carlo (MC) methods can handle large, but finite systems. Both methods (i.e. ED and MC methods) depend on finite-size extrapolation to get results, such as the ground state, in the thermodynamic limit. Another drawback is that, MC methods suffers from the so-called sign problem [36] for certain problems, including frustrated magnets, interacting fermions [46] and systems of anyons.

Tensor networks are relatively new approaches, which have been successful in studying large system sizes, and even infinite-sized systems if translational invariance of the system is exploited. They are a generalisation of a precursory successful method: the density matrix renormalisation group (DMRG) [47–50], which is the most favourable method for simulating one-dimensional quantum systems. A tensor network is a network of tensors which are either connected—when the tensors share indices, or disjointed—when they do not share indices. As it can be imagined, it will be possible to construct many such tensor networks. The mathematical framework of these constructions culminates in what is widely referred to as *tensor network theory*. The usefulness of tensor networks is that they can be used to encode the amplitudes $c_{i_1 \dots i_L}$ occurring in Eq. (2.1) efficiently, for states with a limited amount of entanglement, i.e. states which obey the “area law” of entanglement entropy [51]. The ground (or low-energy) state properties of the system can be computed using some known *tensor network algorithms*.

Some examples of tensor network are: matrix product states [52–54]—which are ansatz states for algorithms such as DMRG and time-evolving block decimation (TEBD) [55, 56]—, tree tensor network [57–64], projected entangled-pair states (PEPS), which is a generalisation of the MPS ansatz to two and higher dimensions [65–72].

The ansatz states and algorithms, either constructed or used in this thesis, to study systems of anyons, are based on MPS and TEBD. I will, therefore, review the conventional MPS and TEBD in this chapter.

My plan for the rest of this chapter is as follows: In Section 2.2, I review the notion of tensors and their manipulations. I introduce the diagrammatic notations which are used to conveniently represent tensors and their operations. In Section 2.3, I review the matrix product states for finite systems, and its extension to infinite systems in Section 2.4. In Section 2.5, I review the conventional time-evolving block decimation algorithm, which can used in imaginary-time to compute the ground state of finite and infinite systems, and also in real-time to compute the time evolution of any quantum state that has limited entanglement.

## 2.2 Tensors and their diagrammatic notations

In this section, I review the notion of the mathematical objects called *tensors*, and how they are manipulated, and also introduce their conventional diagrammatic notation. My brief review follows closely Section II of Ref. [73]. The reader should consult this paper for a more detailed review.

In the context of this thesis, a tensor $T$ is a multidimensional array of complex numbers $T_{i_1 i_2 \dots i_n} \in \mathbb{C}$, where these numbers are indexed by the labels $i_1$, $i_2$,..., and $i_n$. Note that we make no distinction between covariant and contravariant indices of a tensor. Each $i_k$ takes values from an index set, e.g. $i_k \in \{1, 2, 3, \dots, d\}$, which can be concisely written as $i_k \in [1 : d]$, where the ordering of the numbers is understood. For simplicity, the tensor $T$ can be assumed to be an object constructed over a tensor product space $(\mathbb{C}^d)^{\otimes n}$, and $i_k$ indexes the orthonormal basis of a $d$-dimensional vector space $\mathbb{C}^d$ for each site $k$, but this need not always be so. The number of index labels on a tensor determines its *order*. Here, the order of the tensor $T$ is $n$, (or equivalently, that $T$ is an order-$n$ tensor). For example, when $n = 0$, $T$ is a scalar (or an order-0 tensor), when $n = 1$, $T$ is a vector (or an order-1 tensor), when $n = 2$, $T$ is a matrix (or an order-2 tensor), and so on. When the order $n$ of a tensor is much smaller than the lattice size $L$, e.g. $n = 1, 2, 3, \ldots$, i.e. $n \ll L$, then the tensor will be referred to as a low-order tensor. It is convenient to have a graphical denotation for a tensor. A tensor $T$ with $n$ indices will be denoted diagrammatically by a shape (e.g. a circle, square, diamond, etc.) and its $n$ indices by $n$ number of lines (similarly called "legs") protruding from it. Examples are shown in Fig. 2.1. These tensors can be operationally manipulated. We consider some examples of the manipulations needed for the MPS, some among which are: (i) multiplication of tensors (ii) factorisation of a tensor, (iii) permutation of the elements of a tensor (iv) Reshaping of a tensor, to either reduce or increase its order, and finally the (v) trace of a tensor. All these operations generalise operations performed on matrices.

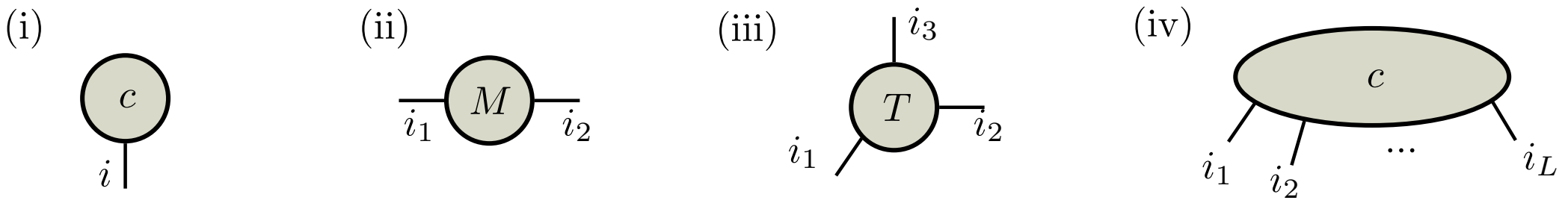

Figure 2.1: Diagrammatic notation for tensors. Fig. (i) represents a vector $c$, with its label "$c$" inscribed inside a circle, and has a single line emanating from the circle which enumerates the orthonormal basis of the vector. As mentioned in the text, $i$ takes values $i \in \{1, 2, \ldots, d\}$. Fig. (ii) represents a matrix $M$ with its rows and columns (in matrix representation) denoted by two lines labelled by $i_1$ and $i_2$. Fig. (iii) is an order-3 tensor $T$. Fig. (iv) represents the tensor $c_{i_1 i_2 \ldots i_L}$ defining the state $|\psi\rangle$ in Eq. (2.1). We have not imposed any preferred orientation of the legs, so the diagrams can be rotated at will, so long as legs are not crossed.

The multiplication of two tensors $S$ and $T$, for example is

$$R_{ipq} = \sum_{j,k} S_{ijk} T_{pjqk} \, , \tag{2.2}$$

where the sum is over two (repeated) indices $j$ and $k$ of tensors $S$ and $T$, and is formally referred to as "contraction over indices $j$ and $k$". When there is no shared index between the tensors, the elements are multiplied directly as,

$$R_{ijkprqs} = S_{ijk} T_{prqs}, \tag{2.3}$$

which is equivalent to the conventional notion of "tensor product" of the two tensors. The order $n_R$ of the new tensor $R$ can be computed from the relation $n_R = n_S + n_T - 2n_{(S,T)}$, where $n_S$ ($n_T$) is the order of tensor $S$ ($T$), and $n_{(S,T)}$ is the number of shared indices between tensors $S$ and $T$. For example, in Eq. (2.2), $n_S = 3$, $n_T = 4$, and $n_{(S,T)} = 2$, therefore, $n_R = 3$, whereas $n_{(S,T)} = 0$ in Eq. (2.3), and hence, $n_R = 7$. The multiplication of tensors generalises the multiplication of two matrices (or vectors). Examples of different contractions are shown in Fig.2.2.

The structure of a tensor $T$ can be changed by two elementary single-tensor operations: *reshape* and *permutation* of the indices of the tensor $T$. A tensor $T$ can be permuted by changing the positions of its elements, and swapping the positions of the index labels as,

$$(T')_{acb} = \text{Permute}(T_{abc}), \tag{2.4}$$

where $T'$ is the new tensor obtained after permutation. Also, a tensor $T$ can be reshaped by "fusing" or "splitting" its indices. Fusion of the indices of a tensor $T$ is,

$$(T')_{ad} = \text{Reshape}(T_{abc}), \tag{2.5}$$

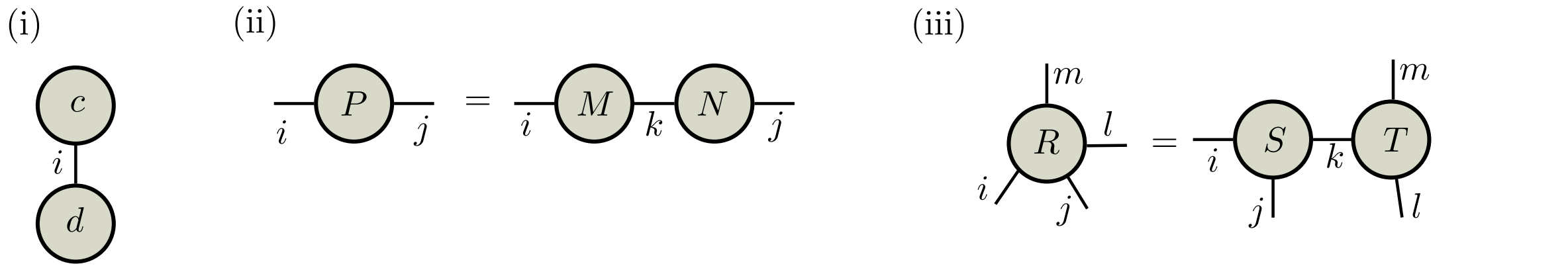

Figure 2.2: Some examples of the multiplication of two tensors. (i) The inner product of vectors $c$ and $d$, which gives a number. In components form $\langle d|c\rangle = \sum_i c_i d_i^*$. (ii) The multiplication of two matrices $M$ and $N$ to give a new matrix $P$. In components form, $P_{ij} = \sum_k M_{ik} N_{kj}$. (iii) Tensor multiplication—a generalisation of matrix multiplication. This represents the contraction of two tensors $S$ and $T$ to give a new tensor $R$. In components form, $R_{ijlm} = \sum_k S_{ijk} T_{klm}$, which is a contraction over a single shared index $k$.

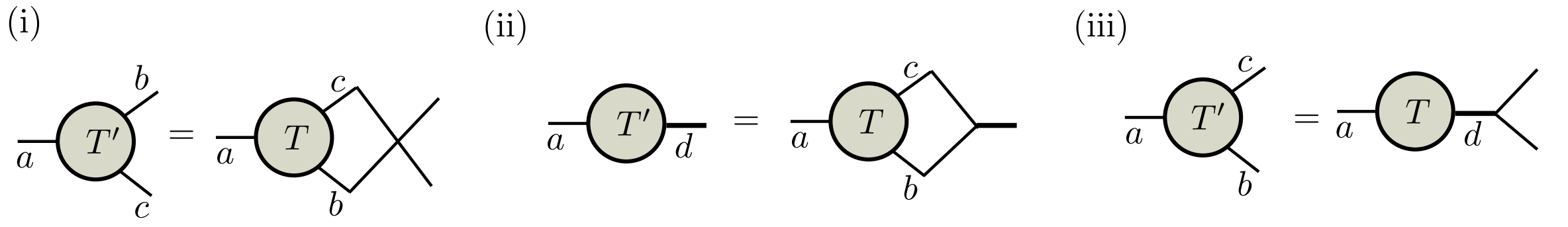

Figure 2.3: Permuting and reshaping a tensor $T$ to give a new tensor $T'$. (i) Permuting the indices $b$ and $c$, represented by the crossing. (ii) Fusing the indices $b$ and $c$ of tensor $T$ into a new index $d$ of tensor $T'$. The "fattened" line of tensor $T'$ diagrammatically represent the fusion operation. (iii) Splitting the index $d$ of tensor $T$ into the indices $b$ and $c$ of tensor $T'$, diagrammatically represented by the $90^o$ clockwise rotated "Y" shape.

where the "new" index $d$ on tensor $T'$ is obtained by mapping the pair of indices $(b, c)$ uniquely to $d$. Similarly, the index of a tensor $T$ can be split into two indices as,

$$(T')_{abc} = \text{Reshape}(T_{ad}), \tag{2.6}$$

where the indices $(b, c)$ is obtained by mapping index $d$ uniquely to the indices $b$ and $c$. The diagrammatic representation of the permutation, splitting and fusion of the indices of a tensor are shown in Fig. 2.3

A matrix $M$ can be decomposed into a product of two (or more) matrices in several different ways. Some of the most popular (and useful) decompositions in tensor networks are, singular value decompositions (SVD), QR-decompositions, eigenvalue decompositions, among many others. For instance the SVD of a matrix $M$ is,

$$M_{ij} = \sum_k U_{ik} \lambda_k V_{kj}, \tag{2.7}$$

where the matrix $U$ and $V$ are unitary, i.e. $U^\dagger U = UU^\dagger = \mathbb{I}$ and $V^\dagger V = VV^\dagger = \mathbb{I}$, and the diagonal matrix $\lambda$ contains the singular values of the matrix $M$, which are real positive numbers, $\lambda_k \in \mathbb{R}_+$. On the other hand, the eigenvalue decomposition of a diagonalisable square matrix $M$ is,

$$M_{ij} = \sum_k D_{ik} \lambda_k (D^{-1})_{kj}, \tag{2.8}$$

where $D$ is the invertible matrix of the eigenvectors of matrix $M$, $D^{-1}$ is the inverse of $D$, and $\lambda$ is a diagonal matrix of real numbers, $\lambda_k \in \mathbb{R}$. The decomposition of matrices generalise to tensors. A higher order tensor $T$ can be decomposed into any number of factors. First, the tensor is converted into a matrix by using a combination of *reshape* and *permutation* of the indices of the tensor. The resulting matrix is decomposed into its matrix factors, and finally, undo any reshape and permutation of the indices.

The trace $\text{Tr}(T)$ of a tensor $T$ is obtained by summing over the repeated indices of the tensor. For example, the new tensor $T'$

$$(T')_{ij} = \sum_k T_{ijkk}, \tag{2.9}$$

is obtained by summing over the repeated index $k$ of tensor $T$. This is represented diagrammatically as

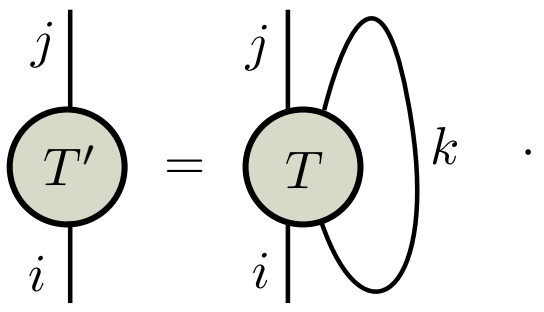

(2.10)

This example can be generalised to tensors of arbitrary order.

## 2.3 Matrix Product States

Having reviewed the notion of tensors and their diagrammatic notations, we are now ready to introduce the matrix product state (MPS), which is arguably the simplest tensor network state for (non-critical) one-dimensional quantum systems. A pure state $|\psi\rangle$ generally written as,

$$|\psi\rangle = \sum_{i_1,\ldots,i_L} c_{i_1\ldots i_L} |i_1 \ldots i_L\rangle, \tag{2.11}$$

has $d^L$ number of coefficients $c_{i_1\ldots i_L}$, which is exponentially large in the system size $L$. Working directly in this representation is not feasible in numerical simulations, except for very small system sizes.

The matrix product state ansatz provides an efficient parametrisation of the amplitudes $c_{i_1\ldots i_L}$ for quantum states with limited amount of entanglement. There are two different, but equivalent, expressions for MPS. The first one is given in terms of matrices as,

$$c_{i_1\ldots i_L} = \text{Tr}\left(A^{[1]i_1} A^{[2]i_2} \ldots A^{[L]i_L}\right), \tag{2.12}$$

where $i_k \in \{1, \ldots, d\}$ indexes the orthonormal basis vector of the Hilbert space $\mathcal{H}_k$ of site $k$, and $A^{[k]i_k}$ is a matrix for each local basis $|i_k\rangle$. The MPS given is for systems with periodic boundary condition (PBC). For an MPS with open boundary condition (OBC), $A^{[1]i_1}$ and $A^{[L]i_L}$ are row and column vectors respectively, and no trace is needed, as the multiplication of all the matrices gives a number. This MPS can be conveniently represented in the graphical notations introduced in Sec. 2.2 as

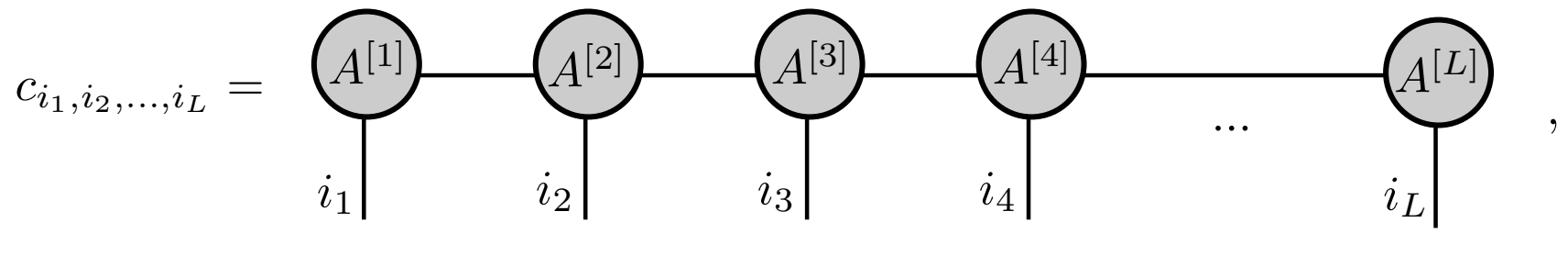

(2.13)

where the labelled vertical lines are called *physical legs* and enumerates over the basis states of the site Hilbert space, and the connecting lines are called *bonds*. The connection between two matrices implies contraction (or matrix multiplication in this case). The MPS as given is called *left canonical* if the set of matrices $A^{[k]i_k}$ on each site $k$ satisfy the left-normalisation condition

$$\mathbb{I} = \sum_{i_k} \left(A^{[k]i_k}\right)^{\dagger} A^{[k]i_k}. \tag{2.14}$$

This is represented graphically as

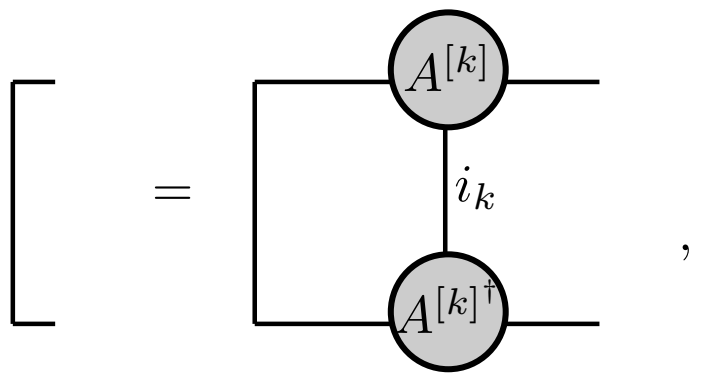

(2.15)

where the left hand diagram represents the identity operator.

If the MPS is *right canonical*, rather than using the notation $A$, the matrices are often denoted using the letter '$B$' in the literature (e.g. see Ref. [74]). The set of matrices $B^{[k]i_k}$ on each site $k$ satisfy the right-normalisation condition,

$$\mathbb{I} = \sum_{i_k} B^{[k]i_k} \left(B^{[k]i_k}\right)^{\dagger}, \tag{2.16}$$

which is also graphically represented as

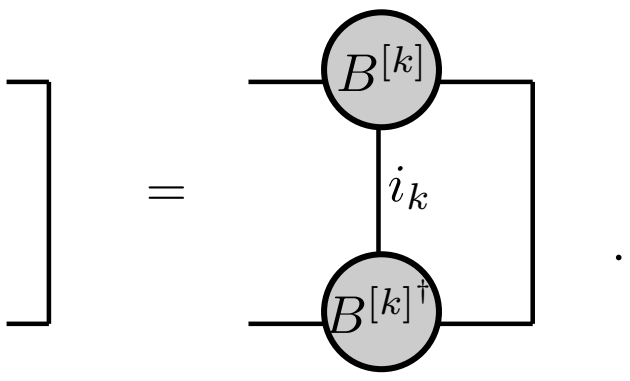

(2.17)

The second frequently used definition of MPS, for OBC, is expressed in terms of order-3 and order-2 tensors as,

$$c_{i_1 \dots i_L} = \sum_{\alpha_1, \dots, \alpha_{L-1}} \left(\Gamma^{[1]i_1}_{\alpha_1} \lambda^{[1]}_{\alpha_1} \Gamma^{[2]i_2}_{\alpha_1 \alpha_2} \lambda^{[2]}_{\alpha_2} \dots \lambda^{[L-1]}_{\alpha_{L-1}} \Gamma^{[L]i_L}_{\alpha_{L-1}}\right), \tag{2.18}$$

where $\Gamma^{[k]}$ is an order-3 tensor for each site $k$, except for the boundary where it becomes a matrix, and $\lambda^{[k]}$ is a diagonal matrix containing the Schmidt coefficients of the state. I will favour the use of this second expression throughout this thesis.

In the convenient graphical language, this MPS is represented as,

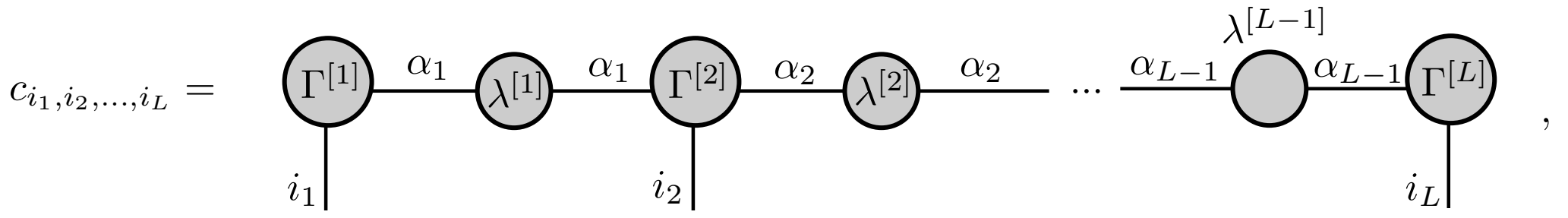

(2.19)

where, as before, the labelled vertical lines on the $\Gamma$-tensors are the physical legs, and the connecting labelled lines between the $\Gamma$-tensors and the $\lambda$-matrices are the bonds. The label $i_k$ enumerates the basis on each site $k$, while $\alpha_k$ enumerates the Schmidt basis—described below in Sec. 2.3.1—on each bond $k$. Let the dimension of the Hilbert space on each bond $k$ be $\chi_k$, referred to as *bond dimension*. For simplicity, the bond dimension can be restricted to a constant value, $\chi$, on each bond. The number of parameters in the MPS representation is now therefore of order $O(Ld\chi^2)$, which is linear in the system size $L$ for a constant bond dimension $\chi$, and not exponential as in the representation in Eq. (2.11). In principle, $\chi \leq d^{L/2}$, so the number of parameters in the MPS can also be exponentially large[1], but this quantity depends on the amount of entanglement in the state—which may be regulated. A state with logarithmic entanglement is one where the $\chi$ scales polynomially with the system size $L$, i.e. $\chi = \text{Poly}(L)$ rather than scaling exponentially.

The MPS representation was given without any mathematical justification. I will now give a short derivation of the MPS based on successive Schmidt decomposition [45] of the lattice $\mathcal{L}$ into a bipartite system $\mathcal{A}$ and $\mathcal{B}$.

### 2.3.1 Derivation of MPS

We start with the state $|\psi\rangle \in \mathcal{H}$ in Eq. (2.11),

$$|\psi\rangle = \sum_{i_1,\ldots,i_L} c_{i_1\ldots i_L} |i_1 \ldots i_L\rangle . \tag{2.20}$$

The order-$L$ tensor $c_{i_1\ldots i_L}$ can be decomposed into a network of tensors of lower orders using techniques from linear algebra, such as singular value decomposition (SVD) or QR decomposition. Here, we will use SVD to achieve our desired decomposition as follows: reshape the tensor $c_{i_1 i_2\ldots i_L}$ into a matrix $c_{i_1(i_2\ldots i_L)}$, where the indices $i_2, i_3, \ldots, i_L$ are grouped together. Decompose the matrix using SVD,

$$c_{i_1\ldots i_L} \equiv c_{i_1(i_2\ldots i_L)} = \sum_{\alpha_1} \Gamma^{[1]i_1}_{\alpha_1} \lambda^{[1]}_{\alpha_1} V^{[1]}_{\alpha_1(i_2\ldots i_L)}. \tag{2.21}$$

Note that the diagonal matrix $\lambda$ obtained from SVD is purely real with entries $\lambda_\alpha \geq 0$. We recombine the $\lambda^{[1]}$ and $V^{[1]}$ and reshape the resulting tensor into $c_{(\alpha_1 i_2)(i_3\ldots i_L)}$, where the bond index $\alpha_1$ and the physical index $i_2$ are grouped together.

$$c_{i_1\ldots i_L} = \sum_{\alpha_1} \Gamma^{[1]i_1}_{\alpha_1} c_{(\alpha_1 i_2)(i_3\ldots i_L)}. \tag{2.22}$$

Now decompose $c_{(\alpha_1 i_2)(i_3\ldots i_L)}$ using SVD as $c_{(\alpha_1 i_2)(i_3\ldots i_L)} = \sum_{\alpha_2} U_{(\alpha_1 i_2)\alpha_2} \lambda^{[2]}_{\alpha_2} V_{\alpha_2(i_3\ldots i_L)}$. Substitute this back into Eq. (2.22) and restore back $\lambda^{[1]}$ from Eq. (2.21) into the equation by letting $U_{(\alpha_1 i_2)\alpha_2} = \sum_{\alpha_1} \lambda^{[1]}_{\alpha_1} \Gamma^{[2]i_2}_{\alpha_1\alpha_2}$. This gives

$$c_{i_1\ldots i_L} = \sum_{\alpha_1,\alpha_2} \Gamma^{[1]i_1}_{\alpha_1} \lambda^{[1]}_{\alpha_1} \Gamma^{[2]i_2}_{\alpha_1\alpha_2} \lambda^{[2]}_{\alpha_2} V_{\alpha_2(i_3\ldots i_L)}. \tag{2.23}$$

Continue this iteration until the last site $L$ to give the desired expression,

$$c_{i_1\ldots i_L} = \sum_{\alpha_1,\ldots,\alpha_{L-1}} \left( \Gamma^{[1]i_1}_{\alpha_1} \lambda^{[1]}_{\alpha_1} \Gamma^{[2]i_2}_{\alpha_1\alpha_2} \lambda^{[2]}_{\alpha_2} \ldots \lambda^{[L-1]}_{\alpha_{L-1}} \Gamma^{[L]i_L}_{\alpha_{L-1}} \right), \tag{2.24}$$

which is the MPS representation given in Eq. (2.18) and graphically in Eq. (2.19). In summary, at each iteration, we reshaped the tensor into a matrix and do a singular value decomposition. The number of singular values at each iteration is then the minimum of either the rows or columns of the matrix. When the lattice $\mathcal{L}$ of $L$ sites divides into two equal halves with $L/2$ sites, the resulting matrix will be $d^{[L/2]} \times d^{[L/2]}$. Therefore, the number of singular values (i.e. bond dimension) $\chi$ satisfies $\chi \leq d^{[L/2]}$. Therefore, I have shown that the MPS is just an alternative re-expression of the same state $|\psi\rangle$, and is in principle exponential in system size $L$. The advantage of using the MPS representation (and indeed other TN states) rather than the primitive representation in Eq. (2.20), is that the amount of entanglement $\chi$ can be easily controlled in a numerical simulation, and allows us to approximately target the right subspace of the huge Hilbert space. In order words, we approximate the target state with fewer parameters.

[1]The MPS representation is just an alternative parametrisation of the state $|\psi\rangle$ and has as many parameters as the state $|\psi\rangle$ itself.

We noted above that the two different representations for MPS are equivalent. Depending on how we grouped adjacent $\Gamma$ and $\lambda$ tensors of Eq. 2.19 together, we can recover the representation Eq. 2.13 or a similar representation with the $B$ matrices depending on the "direction" of the singular value decompositions. The decomposition which starts from the left and proceeds rightward yields a left-canonical MPS, while a decomposition in the reverse direction yields a right-canonical one. If we contract the tensors $\lambda^{[i-1]}$ and $\Gamma^{[i]}$ in Eq. 2.19 we get the $A^{[i]}$ set of matrices in Eq. 2.13, and hence the two representations are shown to be equivalent.

## 2.4 Infinite Matrix Product State

The MPS representation of the state of a finite quantum system can be extended to an infinite-sized system at thermodynamic limit by exploiting the translational invariance of the system. The infinite MPS (iMPS) [56] can be derived from the finite MPS by following a very simple argument. Let us assume the boundary of the finite-MPS (shown in Eq. (2.19)) extends to infinity, such that there are no more boundary effects. We impose translational invariance on the state $|\psi\rangle$ of the system. Translate the MPS $|\psi\rangle$ by one site to give the state $|\psi'\rangle$, $|\psi'\rangle = \hat{T}|\psi\rangle$, where $\hat{T}$ is the translation operator. Translational invariance of the MPS demands that the tensors,

$$\Gamma^{[k]} = \Gamma^{[k+1]}, \qquad \lambda^{[k]} = \lambda^{[k+1]}, \tag{2.25}$$

for all $k \in \mathbb{Z}$. Therefore, a translation invariant MPS is made up of a set of two site-independent tensors $\{\Gamma, \lambda\}$ which are infinitely repeated to form the iMPS.

In general, the iMPS can be extended to represent the state of systems that are $k$-sites periodic, by demanding translational invariance after $k$ sites. A single block $\mathcal{B}$ of the iMPS will, therefore, be made up of a set of $k$ number of tensors, i.e. $\mathbb{B} = \{\Gamma^{[1]}, \lambda^{[1]}, \Gamma^{[2]}, \lambda^{[2]}, \ldots, \Gamma^{[k]}, \lambda^{[k]}\}$, arranged into a chain within the block, and the blocks are infinitely repeated to form the iMPS. Invariance of the iMPS which is $k$ sites periodic implies that,

$$\Gamma^{[l]} = \Gamma^{[l+nk]}, \qquad \lambda^{[l]} = \lambda^{[l+nk]}, \tag{2.26}$$

where $k$ is the period of the iMPS, and $n \in \mathbb{Z}$ is an integer. Optimisation of the iMPS involves optimising only the tensors making up a single block. We shall use this idea later in Chapter 4 to construct Anyon×U(1) MPS ansatz for simulating infinite anyonic systems at fixed global particle density.

## 2.5 Time-Evolving Block Decimation Algorithm

The time evolving block decimation algorithm (TEBD), developed by Vidal [55], is an algorithm which can be both used to compute the time evolution of state $|\psi(t)\rangle$, written as an MPS, and also the ground state $|\mathrm{GS}\rangle$ of a Hamiltonian $\hat{H}$ in imaginary-time. The TEBD method is based on the time evolution of a quantum state.

Let $\hat{H}$ be the Hamiltonian of a system, with an eigenstate state $|\psi_0\rangle$ (presumably an initial state at time $t = 0$). The time evolution of this state is,

$$|\psi(t)\rangle = \hat{G}(t)\,|\psi_0\rangle, \tag{2.27}$$

where $\hat{G}(t) = e^{-i\hat{H}t}$ is the unitary evolution operator[2] (i.e. the unitary gate) of a time-independent Hamiltonian $\hat{H}$. In general, the Hamiltonian could be time dependent, in which case the unitary evolution is a path ordered integral of the exponential of $H(t)$. For simplicity, we will focus on time-independent Hamiltonians. Typically, most Hamiltonians of quantum many-body systems are local, i.e. written as a sum of terms acting over a few nearest neighbouring sites at a time. For the sake of simplicity, let the Hamiltonian $\hat{H}$ be,

$$\hat{H} = \sum_{k\in\mathbb{Z}} \hat{h}^{[k,k+1]}, \tag{2.28}$$

where $h^{[k,k+1]}$ is a nearest-neighbour local Hamiltonian on two sites. The term $\hat{h}^{[k,k+1]}$ is used as a shorthand notation for $\otimes_{i<k} \mathbb{I}_i \otimes \hat{h}^{[k,k+1]} \otimes_{j>k+1} \mathbb{I}_j$. We have seen that, using tensor network manipulations, the state $|\psi\rangle$ in an Hilbert space $\mathcal{H}$ can be rewritten as a tensor network (e.g. as an MPS) of low-order tensors local to each physical site $i$ . In MPS representation, the unitary gate $\hat{G}(t)$ cannot be efficiently applied to the MPS; it also needs to be decomposed into a network of local low-order tensors. The exponential of the Hamiltonian does not directly factor into a product of exponentials of local Hamiltonians, as the terms in the Hamiltonian are in general non-commuting. Using the Suzuki-Trotter decomposition [75, 76], that decomposition can be achieved approximately.

The Suzuki-Trotter decomposition approximates the exponential of a sum of non-commuting operators into a product of exponentials of each operator. For any arbitrary non-commuting operators $\hat{A}$ and $\hat{B}$,

$$e^{(\hat{A}+\hat{B})t} \approx (e^{\hat{A}t/n} e^{\hat{B}t/n})^n + O(1/n), \tag{2.29}$$

for all $t \in \mathbb{R}$. Let the $\delta t = t/n$ in the above formula. The accuracy of this approximate decomposition relies on the "smallness" of $\delta t$ or "largeness" of the number of partitions $n$.

We can now apply the Suzuki-Trotter formula to decompose the unitary gate $\hat{G}(t)$,

$$\hat{G}(t) = e^{-it\hat{H}} = e^{-it\sum_k \hat{h}^{[k,k+1]}}, \tag{2.30}$$

into a product of exponentials of local Hamiltonian terms. We will present the "even-odd" decomposition in Ref. [77]. Let the Hamiltonian $\hat{H} = \sum_{k\in\mathbb{Z}} \hat{h}^{[k,k+1]}$ be rewritten as a sum of terms on even and odd sites, $\hat{H} = \hat{H}_e + \hat{H}_o$, where $\hat{H}_e = \sum_{k\in\mathbb{Z}} \hat{h}^{[2k,2k+1]}$ is the sum of commuting terms on even sites, and $\hat{H}_o = \sum_{k\in\mathbb{Z}} \hat{h}^{[2k-1,2k]}$ is the sum of commuting terms on odd sites. The unitary gate $\hat{G}(t)$, therefore decomposes as,

$$\hat{G}(t) = (e^{-i\delta t\hat{H}_e} e^{-i\delta t\hat{H}_o})^n + O(1/n). \tag{2.31}$$

[2] In this thesis, I will always use $\hat{G}(t)$ to represent the evolution operator, which is referred to as the "unitary gate" in quantum information theory, as opposed to the using $\hat{U}(t)$ as in conventional quantum theory. This is because, the ansatz states and algorithms used or constructed in this thesis are based on tensor networks, which are methods coming from quantum information theory.

Let $\hat{G}_e(\delta t) = e^{-i\delta t \hat{H}_e}$ and $\hat{G}_o(\delta t) = e^{-i\delta t \hat{H}_o}$. Therefore, $\hat{G}(t) = \prod^n \hat{G}(\delta t)$, where $\hat{G}(\delta t) = \hat{G}_e(\delta t)\hat{G}_o(\delta t)$. The terms $\hat{G}_e(\delta t)$ and $\hat{G}_o(\delta t)$ decompose exactly into products as,

$$\hat{G}_e(\delta t) = \prod_{k\in\mathbb{Z}} e^{-i\delta t \hat{h}^{[2k,2k+1]}}, \tag{2.32}$$

and

$$\hat{G}_o(\delta t) = \prod_{k\in\mathbb{Z}} e^{-i\delta t \hat{h}^{[2k-1,2k]}}, \tag{2.33}$$

since the terms on disjoint sites always commute. The time-evolved state $|\psi(t)\rangle$ is therefore computed from $|\psi_t\rangle = \hat{G}(t)\,|\psi_0\rangle$, where one step of the update at time $t$ is $|\psi_{t+\delta t}\rangle = \hat{G}(\delta t)\,|\psi_t\rangle$. This can be represented graphically as,

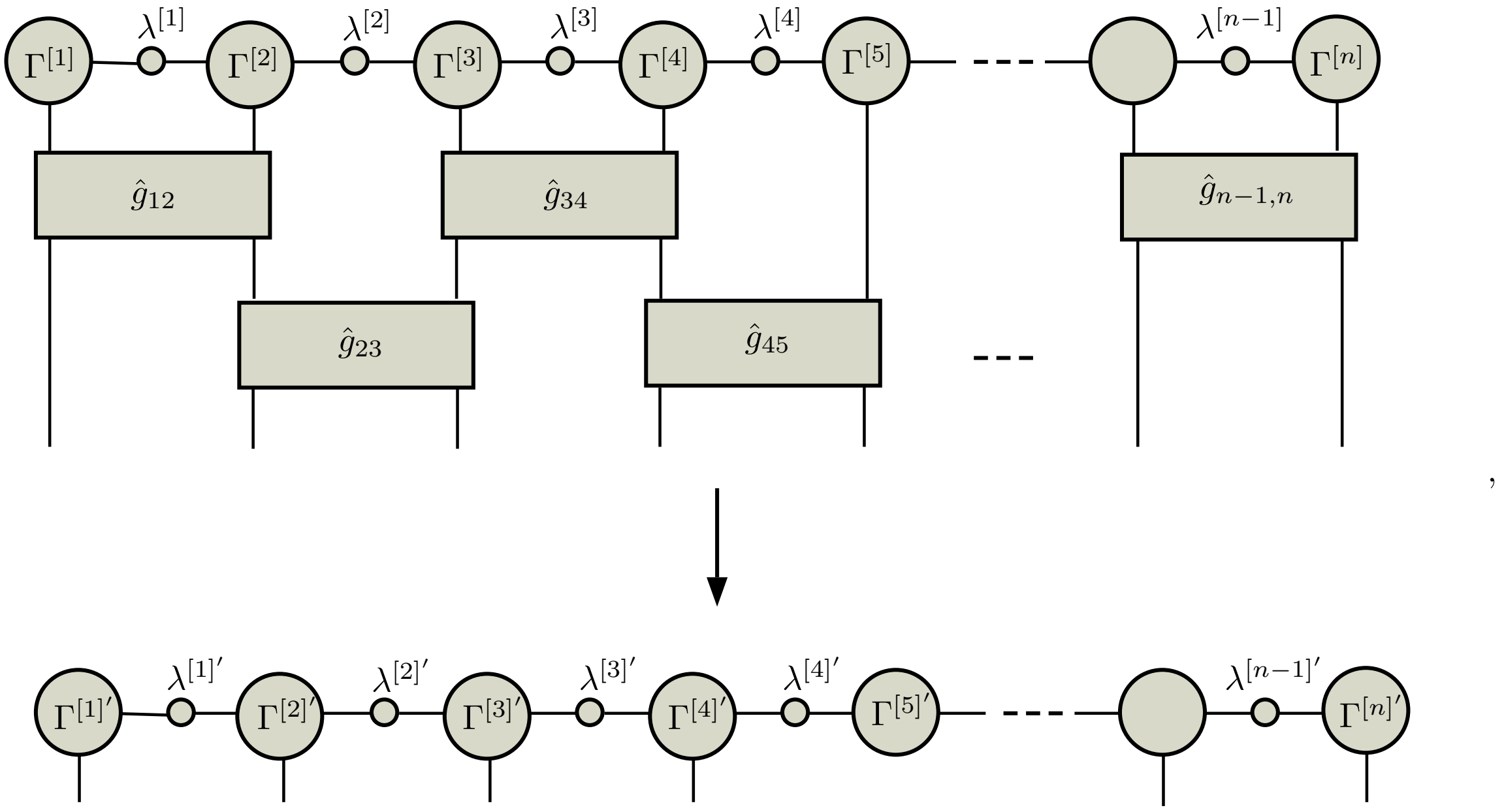

, (2.34)

where $\hat{g}_{k,k+1} = e^{-it\hat{h}^{[k,k+1]}}$ is the unitary gate on either even or odd sites. The tensors $\{\Gamma^{[k]}, \lambda^{[k]}\}$ are updated to $\{\Gamma^{[k]'}, \lambda^{[k]'}\}$ for each site $k$. It can be seen that the contraction of this network depends recursively on knowing how to contract the two sites MPS [56] explained in Fig. 2.4.

The TEBD algorithm can be used to compute the ground state $|\mathrm{GS}\rangle$ of a Hamiltonian $\hat{H}$ in imaginary-time by setting $t = i\tau$, where $i$ is the complex number. To see this, we consider the "evolution" of a general quantum state in imaginary time $\tau$. Let $\{|i\rangle\}$ be the set of eigenstates of a Hamiltonian $\hat{H}$. We let $|i = 0\rangle$ be the ground state (GS) denoted by $|\mathrm{GS}\rangle$ and having a corresponding GS energy $E_0$, and all other $|i\rangle$ are excited states with the corresponding energies $E_i$, where $E_i > E_0$. Since the Hamiltonian $\hat{H}$ is Hermitian, its eigenvalues (i.e. energies) are real and its eigenstates are orthogonal, and therefore span the Hilbert space. Below, we show that for long enough time, i.e. $\tau \to \infty$, an arbitrary initial state $|\psi\rangle$ gets projected to the ground state $|\mathrm{GS}\rangle$ of the Hamiltonian $\hat{H}$, as long as that initial state is non-orthogonal to the ground state, i.e. the ground state should have a support on the initial state.

A general state $|\psi\rangle$ can be expanded in terms of the orthonormal basis states $\{|i\rangle\}$ as,

$$|\psi\rangle = \sum_i c_i\,|i\rangle, \tag{2.35}$$

where the complex numbers $c_i$ are the amplitudes of the state, and can be represented as an MPS, as already seen in Sections 2.3 and 2.4. As the imaginary-time gate $\hat{G}(\tau)$ is not unitary, the state has to be re-normalised after the gate acts on it. We apply the imaginary-time projector $\hat{G}(\tau) = e^{-\tau \hat{H}}$ to $|\psi\rangle$ as,

$$|\psi'\rangle = \hat{G}(\tau)\,|\psi\rangle = e^{-\tau \hat{H}}\left(\sum_i c_i\,|i\rangle\right), \tag{2.36}$$

$$= \sum_i c_i e^{-\tau E_i}\,|i\rangle, \tag{2.37}$$

$$= c_0 e^{-\tau E_0}\left[|0\rangle + \sum_{i\neq 0}\left(\frac{c_i}{c_0}\right) e^{-\tau(E_i - E_0)}\,|i\rangle\right], \tag{2.38}$$

where the amplitude $c_0$ of the ground state is not zero. In the limit $\tau \to \infty$,

$$|\psi'\rangle \to |0\rangle, \tag{2.39}$$

after normalisation, where we have assumed that the degeneracy of the ground state is 1. For a higher degeneracy, the evolution will project onto one of the states in the ground space manifold. The excited states decay exponentially for a gapped spectrum, where the energy gap $E_i - E_0 > 0$, for states above the ground state. The convergence to GS is determined by the gap of the GS above the excited states. The evolution converges exponentially to the ground state when the time $\tau > (E_i - E_0)$. A system for which the gap of its spectrum goes to zero in the thermodynamic limit is called critical, and is usually the hardest to simulate with TEBD, though in most cases, it is still often possible to get a reliable approximation of ground state properties, such as average energy density, entropy, etc.

## 2.6 Computing quantities from iMPS

In this section, I review how to compute some physical quantities of interest from the infinite MPS.

### 2.6.1 Expectation value of an observable

We illustrate how to compute the expectation value of local operators and correlation functions, using the example of two operators $\hat{O}_i$ and $\hat{O}_j$, acting on two sites $i$ and $j$. Let $\hat{O} : \mathcal{H} \to \mathcal{H}$ be an observable of states in the Hilbert space $\mathcal{H}$, where for instance, $\hat{O} = \hat{O}_i \otimes \mathbb{I}_{i+1} \otimes \ldots \otimes \mathbb{I}_{j-1} \otimes \hat{O}_j$, though this can be generalised to other arbitrary operators. The expectation value $\langle\psi|\,\hat{O}\,|\psi\rangle$ can be computed from the iMPS, and is given diagrammatically as,

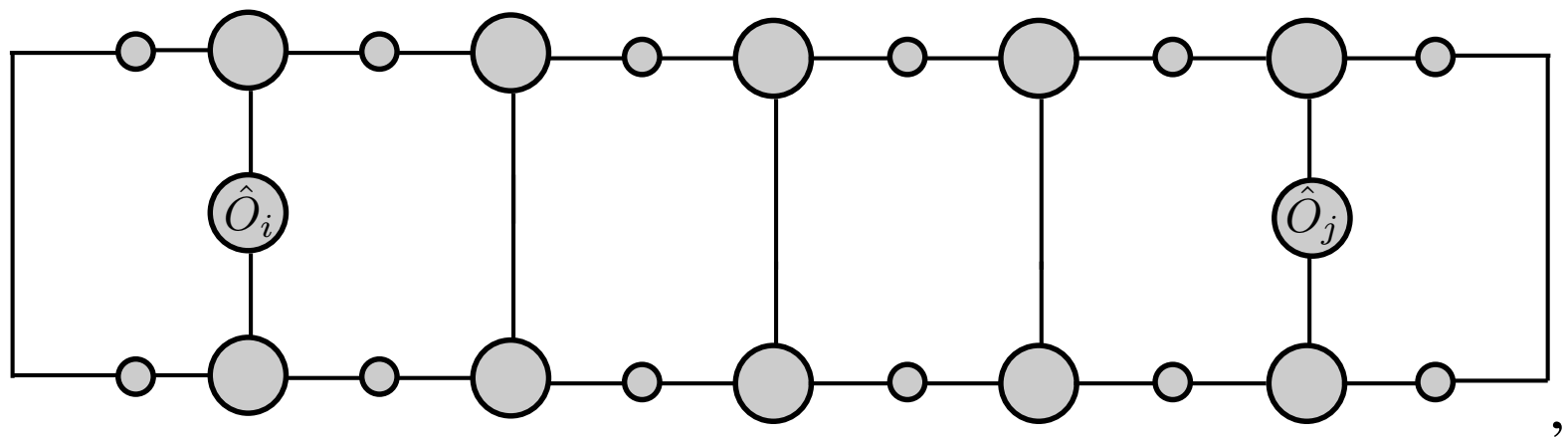

, (2.40)

where the tensor labels and index labels have been suppressed. This is a simple network which can be contracted proceeding from left to right and top to bottom.

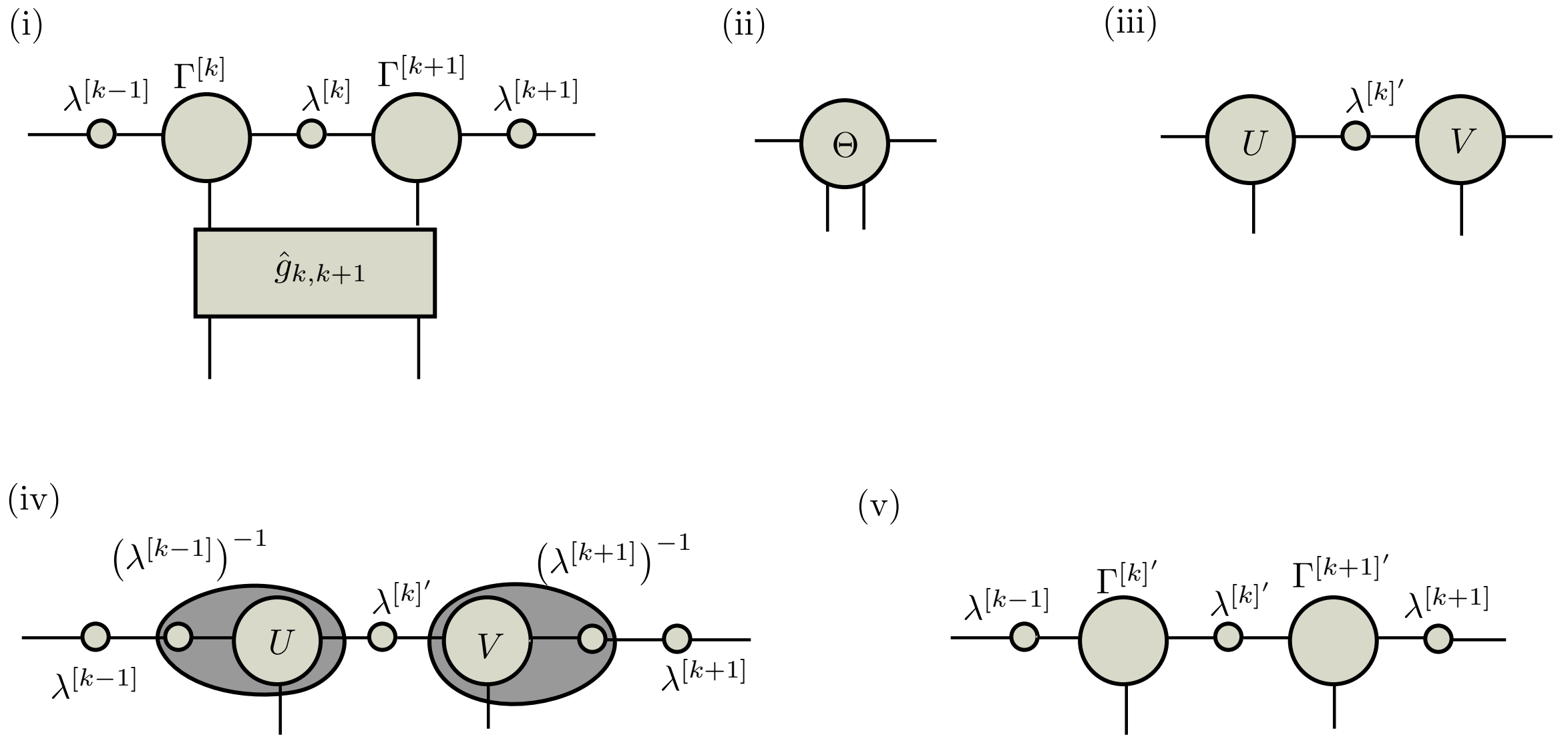

Figure 2.4: Update of the two sites of the iMPS with the unitary gate $\hat{g}_{k,k+1}$ on sites $k$ and $k+1$. (i) Contract the tensor network to give the order-4 tensor $\Theta$ in Fig. (ii). Reshape $\Theta$ into a matrix and decompose using SVD into its matrix factors $U\lambda^{[k]'}V$. Reshape the matrices $U$ and $V$ back into tensors as in Fig. (iii), which can be followed by an optional truncation of the bond dimension, based on the number of Schmidt values to be kept. Restore the canonical form of the tensor network by introducing identity matrices $\mathbb{I} = \lambda^{[k-1]}\left(\lambda^{[k-1]}\right)^{-1}$ and $\mathbb{I} = \left(\lambda^{[k+1]}\right)^{-1}\lambda^{[k+1]}$ to the left and right hand sides as shown in Fig.(iv). The shaded part are then contracted to produce the updated $\Gamma^{[k]'}$ and $\Gamma^{[k+1]'}$ shown in Fig. (v). The index labels on the legs and bonds have been suppressed for clarity.

One may wonder if this tensor network can be efficiently contracted, since in principle, contraction of arbitrary tensors can become hard to achieve computationally. I now show that this contraction is computational easy to achieve. To this end, let us define the following tensors,

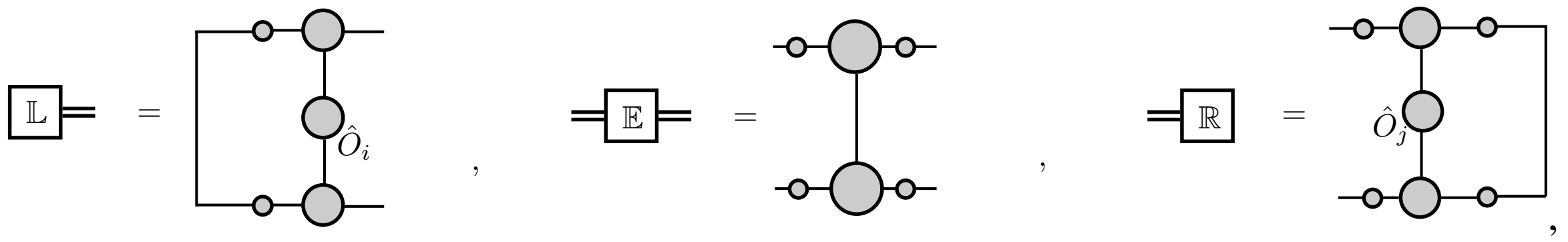

(2.41)

where $\mathbb{L}$ is an order-one tensor from the left boundary, and has $O(\chi^2)$ elements, $\mathbb{R}$ is an order-one tensor from the right boundary, also with $O(\chi^2)$ elements, and $\mathbb{E}$ is the transfer operator on each site which has $O(\chi^4)$ elements. The two close lines on the objects imply grouping together of indices (or fusion). All these objects can be efficiently computed, requiring only a small number of operations of order $O(d)$. The contraction diagram Eq. 2.40 for correlation function then becomes,

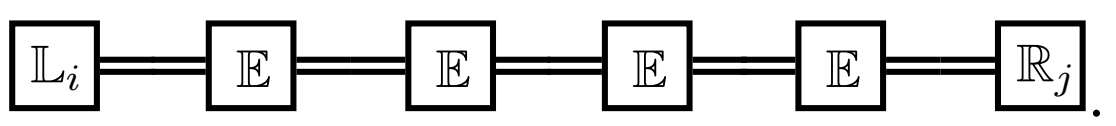

(2.42)

Let us consider the leftmost and rightmost objects as row and column vectors respectively, each with $\chi^2$ elements. The object $\mathbb{E}$ as a matrix with $\chi^2 \times \chi^2$. The diagram is then just a sequence of vector-matrix multiplications. Proceeding from the left, and multiplying a vector and the adjacent matrix pairwisely requires $O(\chi^4)$ basic operations. The total cost of the whole contraction is therefore $O(|i-j|\chi^4)$, which is polynomial in $\chi$, and hence efficient. For $|i-j| \gg 1$, one can compute the highest weight eigenvector and eigenvalue of the transfer operator to compute correlation functions by taking the eigenvalue to the power $|i-j|$.

In fact, the cost of contracting the diagram Eq. 2.40 can be further reduced by a factor of $\chi$ if, rather than forming transfer operators, we contract the tensors appearing in Eq. 2.40 in a pairwise manner.

### 2.6.2 Entropy

The Von-Neumann entropy $S$ of a state described by a density matrix $\hat{\rho}$ is,

$$S = -\mathrm{Tr}(\hat{\rho}\,\log\hat{\rho}). \tag{2.43}$$

The density matrix $\hat{\rho}$ can be written in diagonal form as,

$$\hat{\rho} = \sum_i P_i \left|i\right\rangle\left\langle i\right|, \tag{2.44}$$

where $\{P_i\}$ are the eigenvalues of the density matrix, and $\{|i\rangle\}$ are the corresponding eigenvectors of $\hat{\rho}$. Then, the entropy $S$ can be computed from,

$$S = -\sum_i P_i \log P_i. \tag{2.45}$$

This quantity can be computed directly from the MPS representation Eq. (2.19). For an MPS of pure state, which is a split into a bipartition with one subsystem consisting of the sites to the left including site $i$ and the other subsystem comprising the rest of the sites, the entropy of either subsystem is

$$S = -\sum_i \lambda_i \log\lambda_i, \tag{2.46}$$

where $\lambda_i$ are the Schmidt coefficients of the reduced density matrix of any part of the bipartite system. If this quantity is computed from the iMPS, the computed entropy $S$ is the entropy of entanglement of half infinite chain.

# 3

# Theory of Anyons

## 3.1 Introduction

In this chapter, I briefly review the theory of anyons which sets the theoretical framework that will be combined with tensor networks, reviewed in Chapter 2, to construct an anyonic tensor network ansatz state and algorithm in Chapters 4 and 5.

As point-like particles, anyons exist in two-dimensional systems. An anyon model is defined by specifying certain algebraic data, namely, a particle set $\mathcal{A}$ for the type of particles in the theory, the fusion rules of those particles, braid matrix $R$, F-matrix $F$, and quantum dimensions of each particle type.

The remaining discussion in this chapter is split as follows: in Section 3.2, I discuss the concept of fusion algebra for a set of anyons. In Section 3.3, I discuss the idea of a Hilbert space for anyonic systems in analogy to familiar concepts from the linear algebra in (basic) quantum mechanics, and in Section 3.4, I introduce the powerful diagrammatic representations frequently employed to describe states and operators of systems of anyons. In Section 6.5, I end with a conclusion.

## 3.2 Fusion Algebra

An anyon model has a set $\mathcal{A} = \{a, b, c, \ldots, k\}$ of finite number of charges with labels $a, b, c$ and so on. These charges are interchangeably called anyonic or topological charges, particle types, charge labels, superselection sectors, or some other name. The set satisfies a commutative and associative fusion algebra,

$$a \times b = \sum_c N_{ab}^c \, c, \tag{3.1}$$

where $N_{ab}^c$ is the multiplicity tensor that specifies the number of copies of charge $c$ obtained from fusing charges $a$ and $b$. The associativity constraints at the level of fusion of charges means,

$$a \times b \times c = (a \times b) \times c = a \times (b \times c) \equiv d, \tag{3.2}$$

which means the fusion of three anyons having charges $a, b, c$ will give the same charge outcome $d$ irrespective of the order of fusion; starting from the left, we can fuse $a$ and $b$ first, then fuse the outcomes with $c$ to give $d$, or starting from the right, fuse charges $b$ and $c$ first, then fuse the outcomes with $a$ to give $d$. There is a trivial particle in the set $\mathcal{A}$, referred to as the vacuum charge $\mathbb{I}$, that fuses trivially with other charges in the set,

$$\mathbb{I} \times a = a, \tag{3.3}$$

for all $a \in \mathcal{A}$. Also, for each $a$, there is a corresponding conjugate charge $\bar{a}$ or "antiparticle," that fuses with $a$ as,

$$a \times \bar{a} = \mathbb{I} + \cdots, \tag{3.4}$$

which means that, fusion of a charge and its conjugate can either annihilate to vacuum ($\mathbb{I}$) or produce a new charge "$\cdots$". Anyons are broadly classified into two categories based on their fusion rules: Abelian and non-Abelian. An anyon model is referred to as non-Abelian if for some pairs of charges $a$ and $b$, $\sum_c N_{ab}^c > 1$, otherwise they are Abelian.

Each charge $a \in \mathcal{A}$ has a *quantum dimension* $d_a$, which is loosely the dimension of the "Hilbert space" associated with each anyon type. The quantum dimension of the charges similarly obey the relation,

$$d_a d_b = \sum_c N_{ab}^c d_c. \tag{3.5}$$

If this equation is re-written as, $\sum_c (N_a)_b^c d_c = d_a d_b$, as a set of linear equations for each of charge $b \in \mathcal{A}$. Then $(N_a)\mathbf{d} = d_a \mathbf{d}$, where $\mathbf{d}$ is a vector of quantum dimensions of all the charges. The largest eigenvalue obtained by solving the eigenvalue equation is the quantum dimension $d_a$ of charge $a$. The total quantum dimension is defined as $\mathcal{D} = \sqrt{\sum_a d_a^2}$.

In the rest of this chapter, I will restrict to only multiplicity free models with $N_{ab}^c = 0, 1$, which are some of the most important anyons models in experiment, including Majorana fermions (or Ising anyons) and Fibonacci anyons.

## 3.3 Vector spaces: Bra-Ket notation

To a large extent, anyon models can be described in terms of the tools of conventional quantum mechanics, and is hence amenable to the concepts of linear algebra, such as vector spaces, basis vectors, and operators that act between these spaces.

When two anyonic particles having charges $a$ and $b$ combine, the composite object has a total charge $c$ from a set of possible outcomes determined by the fusion rules of the anyon model. To each unique fusion process, where the charges $a$ and $b$ fuse to $c$, i.e. $a \times b \to c$, there is assigned a fusion Hilbert space $V_{ab}^c$ whose dimension is $\dim V_{ab}^c = N_{ab}^c$. A dual process is where charges $a$ and $b$ split from charge $c$ having a corresponding splitting space $V_c^{ab}$. The fusion space can be regarded as the dual of the splitting space, $V_{ab}^c = \left(V_c^{ab}\right)^*$ (or vice versa). I will refer to these space as the *elementary vector spaces*—since, as will be shown below, they form the building blocks of "larger" vector spaces of more than two anyons.

Let $|ab; c\rangle$ be an orthonormal basis of the space $V_c^{ab}$, and $\langle ab; c| \in V_{ab}^c$ as the conjugate basis in the dual space. Inner product is defined on the vector space as

$$\langle ab; c' | ab; c\rangle = \delta_{cc'}. \tag{3.6}$$

Since the charges are conserved, the splitting space $V^{ab}$ of two anyons with charges $a$ and $b$ is $V^{ab} = \oplus_c V_c^{ab}$, which is a direct sum of the vector spaces associated with each total charge outcome $c$. The identity operator on $V^{ab}$ is

$$\mathbb{I} = \sum_c |ab; c\rangle \langle ab; c| . \tag{3.7}$$

Apart from fusion, another physically relevant operation is *braiding*. Two particles with charges $a$ and $b$ can be exchanged in space, a process which is technically referred to as braiding when exchange of particles applies to two dimensional systems. Exchange does not change the total charge outcome $c$ resulting from the fusion of the anyonic charges $a$ and $b$. Let the vector space of the swapped particles be $V_c^{ba}$, which is isomorphic to $V_c^{ab}$. Let $\hat{R}$ be the braid operator with the mapping

$$\hat{R} : \bigoplus_c V_c^{ab} \to \bigoplus_c V_c^{ba}. \tag{3.8}$$

If we choose orthonormal bases $\{|ab; c\rangle\}$ and $\{|ba; c\rangle\}$ for the spaces, then $\hat{R}$ is a unitary matrix which can be expressed as

$$\hat{R} = \sum_c R_c^{ab} |ba; c\rangle \langle ab; c| . \tag{3.9}$$

These ideas can be extended to more than two anyons. Three particles with charges $a, b, c$ fusing to yield a total charge $d$ has an associated vector space $V_d^{abc}$. Since fusion is associative, it means there is more than one way to decompose this vector space into products of vector spaces of pair of charges. For the specific case of three anyons,

$$V_d^{abc} \cong \bigoplus_e V_e^{ab} \otimes V_d^{ec} \cong \bigoplus_f V_d^{af} \otimes V_f^{bc}. \tag{3.10}$$

The corresponding bases with respect to the two alternative decompositions may be written respectively as

$$|ab; e, ec; d\rangle \equiv |ab; e\rangle \otimes |ec; d\rangle , \tag{3.11}$$

$$|af; d, bc; f\rangle \equiv |af; d\rangle \otimes |bc; f\rangle , \tag{3.12}$$

which should be related by a unitary transformation

$$|ab; e, ec; d\rangle = \sum_f \left(F_d^{abc}\right)_e^f |af; d, bc; f\rangle , \tag{3.13}$$

where the matrix $F_d^{abc}$ is called the F-matrix.

We can generalise to a system of $n$ anyons having charges $a_1, a_2, \ldots, a_n$ having total charge $c$. If we assume that the charges are fusing in the "standard convention," that is, pairwise from the left to the right, the vector space $V_c^{a_1 a_2 \ldots a_n}$ can be decomposed as

$$V_c^{a_1 a_2 \ldots a_n} \cong \bigoplus_{b_1, b_2, \ldots, b_{n-2}} V_{b_1}^{a_1 a_2} \otimes V_{b_2}^{b_1 a_3} \otimes \ldots \otimes V_c^{b_{n-2} a_n}. \tag{3.14}$$

It should be noted that the space $V_c^{a_1 a_2 \ldots a_n}$ does not have a natural decomposition in terms of tensor product of spaces associated with the subsystems, but rather in terms of direct sums of vector spaces of the intermediate charges derived from successive fusion process. This renders the vector space $V_c^{a_1 a_2 \ldots a_n}$ to be *nonlocal*, and is therefore referred to as a *topological*

*Hilbert space*. Let the basis of this space be $|a_1a_2; b_1, b_1a_3; b_2, \ldots, b_{n-2}a_n; c\rangle$, which, in line with the decomposition of the vector space, can be written as a product of basis of pairs of charges

$$|a_1a_2; b_1, b_1a_3; b_2, \ldots, b_{n-2}a_n; c\rangle = |a_1a_2; b_1\rangle\, |b_1a_3; b_2\rangle \ldots |b_{n-2}a_n; c\rangle, \tag{3.15}$$

where there is an implied tensor product between adjacent ket vectors $|\bullet\rangle$.

As it was previously done for the case of two anyons, we can also introduce the inner product, and the identity operator on the space $V_c^{a_1\ldots a_n}$ of a system of $n$ anyons. The inner product is

$$\begin{aligned}\langle a_1'a_2'; b_1', b_1'a_3'; b_2', \ldots, b_{n-2}'a_n'; c'|\, a_1a_2; b_1, b_1a_3; b_2, \ldots, b_{n-2}a_n; c\rangle \\ = \delta_{a_1'a_1} \ldots \delta_{a_n'a_n}\delta_{b_1'b_1} \ldots \delta_{b_{n-2}'b_{n-2}}\delta_{c'c},\end{aligned} \tag{3.16}$$

and the identity operator is

$$\mathbb{I} = \sum_{b_1,\ldots,b_{n-2},c} |a_1a_2; b_1, \ldots, b_{n-2}a_n; c\rangle\, \langle a_1a_2; b_1, \ldots, b_{n-2}a_n; c| \,. \tag{3.17}$$

As in conventional many-body quantum physics, we proceed to specify a quantum state of a system of $n$ anyons, and the operators that act on them. A quantum state of an anyonic system can be written as

$$|\psi\rangle = \sum_{b_1,\ldots,b_{n-2}} \psi_{b_1\ldots b_{n-2}}\, |a_1a_2; b_1, b_1a_3; b_2, \ldots, b_{n-2}a_n; \mathbb{I}\rangle, \tag{3.18}$$

where I have assumed that the charges $(a_1, a_2, \ldots, a_n)$ have a single fixed value, and hence does not appear in the sum in $|\psi\rangle$. Note that the total outcome charge $c$ has been set to the vacuum charge $\mathbb{I}$, to correspond to a physically valid system where anyonic excitations are created in pairs (i.e. particle and antiparticle ) from the vacuum. As is customary, we can normalise the state $|\psi\rangle$ by,

$$1 = \sum_{b_1,\ldots,b_{n-2}} |\psi_{b_1\ldots b_{n-2}}|^2, \tag{3.19}$$

since the basis are orthonormal. We can also construct an operator $\hat{Q}$ that act on a subset (or all) of the anyons in the system. The operator is constructed based on the physical operation to be done on the anyons. For example, in a typical condensed matter physics setting, a Hamiltonian $\hat{H}$ can be defined for how pairs of anyons in the system interact. A local Hamiltonian is written as $\hat{H} = \sum_i h^{[i,\hat{i}+1]}$, where $h^{[i,i+1]}$ is the operator acting on a pair of anyons on sites $i$ and $i+1$ having charges $a_i$ and $a_{i+1}$, respectively. We may then be interested in studying properties of the ground state $|\mathrm{GS}\rangle$ such like ground state energy, correlation functions of some operators, and the amount of entanglement in the state.

To recapitulate, I have shown that physical and mathematical operations can be formulated in the language of linear algebra as done in basic quantum mechanics using Dirac "bra-ket" notation. Starting from below, I will adopt a different fashion, where I employ the use of the already well established graphical notations, where anyonic charges, basis vectors, state vectors, and operators are represented using diagrams. Operations performed on anyons then correspond to certain manipulations of those diagrams. All these culminate in what is widely referred to as graphical calculus.

## 3.4 Graphical Calculus for anyons

It is convenient to employ diagrammatic representation [1, 78] for the basis elements of the elementary vector spaces. Each anyon of charge $a$ will be represented by an oriented line with a label $a$, which is also equivalent to a line with an opposite orientation and the conjugate charge $\bar{a}$,

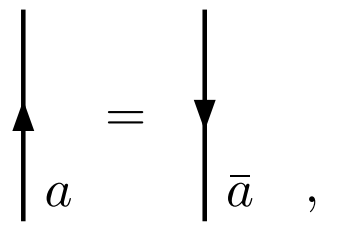

, (3.20)

the upward orientation can be interpreted as the direction of increasing "time". The splitting and fusion basis will be represented, respectively, by a trivalent vertex as,

$|a, b; c\rangle = \left(\frac{d_c}{d_a d_b}\right)^{1/4}$ 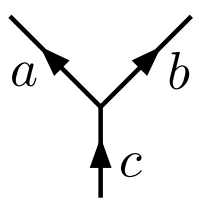

; $\langle c; a, b| = \left(\frac{d_c}{d_a d_b}\right)^{1/4}$ 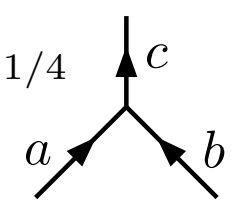

, (3.21)

where the *vertex normalisation factor* $\left(\frac{d_c}{d_a d_b}\right)^{1/4}$ ensures that the diagrams are in diagrammatic isotopy convention [1]. These diagrams can also be interpreted as representing the history of the particles. Let me briefly state that working in the mentioned diagrammatic isotopy convention, where the vertex in each diagram is normalised as prescribed above, is not strictly necessary to describe states and operators of anyons. As I have shown above in Sec. 3.3, basis states and operators of anyons can be formulated entirely in terms of bra-ket notation, which reduces to basic vector and matrix computations. One could in fact associate those basis vectors with diagrams (as will be shown in this section) but without any pre-factor, and then proceed to build the vector and matrix representations of the basis of states and operators. The need for vertex normalisation factor becomes important if we introduce diagrammatic manipulations that associate a numerical weight to loops appearing in diagrams, and also bends a charge line in a direction opposite to its original orientation. In order words, diagrams are invariant with respect to any topological manipulation that distort and change the orientation of charge lines. I adopt this normalisation convention since it proves useful in tensor network manipulations where one may bend the "legs" of a tensor in order to bring the tensor into a form that is convenient to use.

Inner product $\langle c'; a, b|\, a, b; c\rangle$ involves stacking the diagrams of the splitting and fusion basis, from which we get the relation,

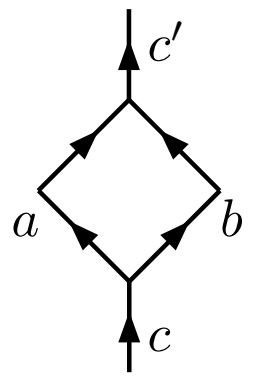

$= \delta_{c,c'}\sqrt{\frac{d_a d_b}{d_c}}$ 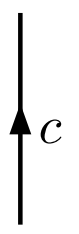

, (3.22)

which diagrammatically encodes conservation of the anyonic charge. Some important deductions from this relation are: ($i$) the "no tadpole" rule, which forbids diagrams such as this:

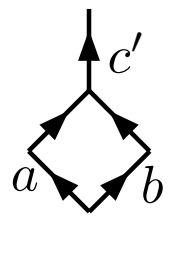

, (3.23)

where charges $a$ and $b$ are created from vacuum, but $c'$ in not the vacuum charge $\mathbb{I}$. This diagram by itself evaluates to zero, and if it occurs within any other diagram, it sets that diagram to zero. (*ii*) The diagrammatic representation of the quantum dimension $d_a$ of charge $a$ is,

$d_a =$ 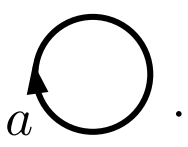 (3.24)

The identity operator $\mathbb{I}_{ab}$ over the vector space of two anyons $a$ and $b$ is,

$$\mathbb{I}_{ab} = \sum_c |a, b; c\rangle \langle c; a, b| , \tag{3.25}$$

which is represented diagrammatically as,

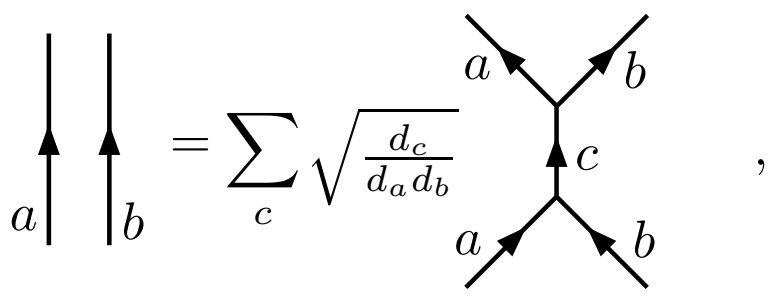

(3.26)

where the charge $c$ of the fusion and splitting trees has been matched on the trunk for charge conservation.

If $V_d^{abc}$ is the splitting space of three anyons labelled $a$, $b$, and $c$, with $d$ as the total charge outcome, then it has two natural isomorphic decompositions in terms of the splitting spaces of pairs of particles as

$$V_d^{abc} \cong \bigoplus_e V_e^{ab} \otimes V_d^{ec} \cong \bigoplus_f V_d^{af} \otimes V_f^{bc} . \tag{3.27}$$

The basis in these two isomorphic spaces are shown below in Eq. (3.28) (*a*) and (*b*) respectively:

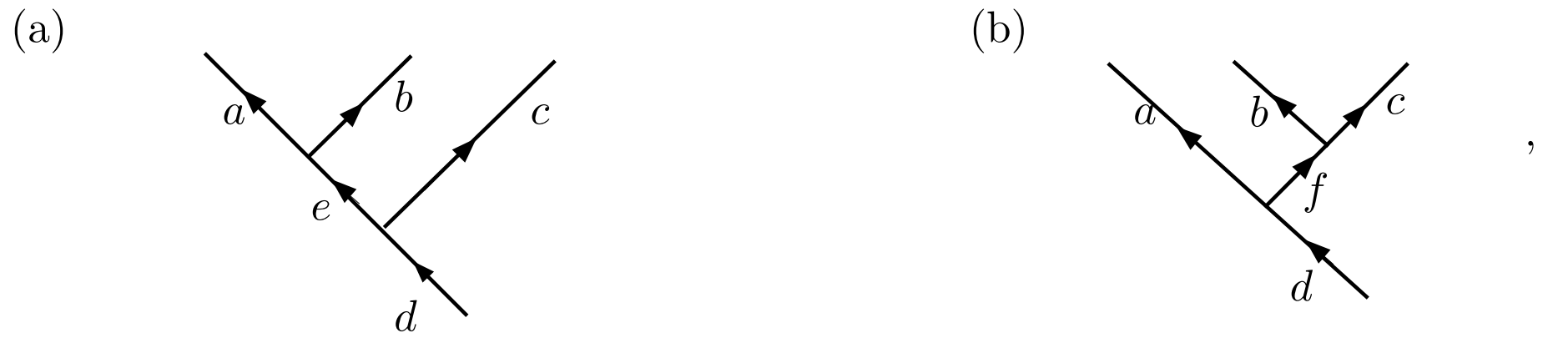

(3.28)

which are two isomorphic bases in the space $V_d^{abc}$, and they are related by a unitary transformation, called "F-move," which is given as:

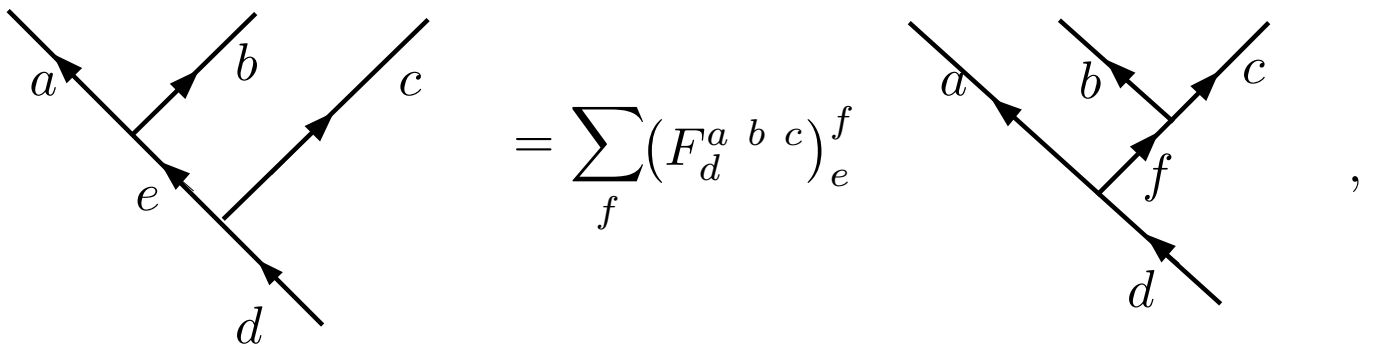

(3.29)

where the $F_d^{abc}$ is the F-matrix that encodes the coefficients of the transformation. The F-matrix implements associativity at the level of vector spaces, and has to satisfy the Pentagon equation for a valid non-Abelian anyon model. [79] We shall only be concerned with *unitary theory* in this thesis. An anyon model is said to be unitary if the F-matrix is unitary, i.e. $F^\dagger F = \mathbb{I}$.

There is an alternative F-move, which is also quite useful. This F-move involves two incoming and two outgoing anyons, which can be referred to as "2-2" F-move, and represented

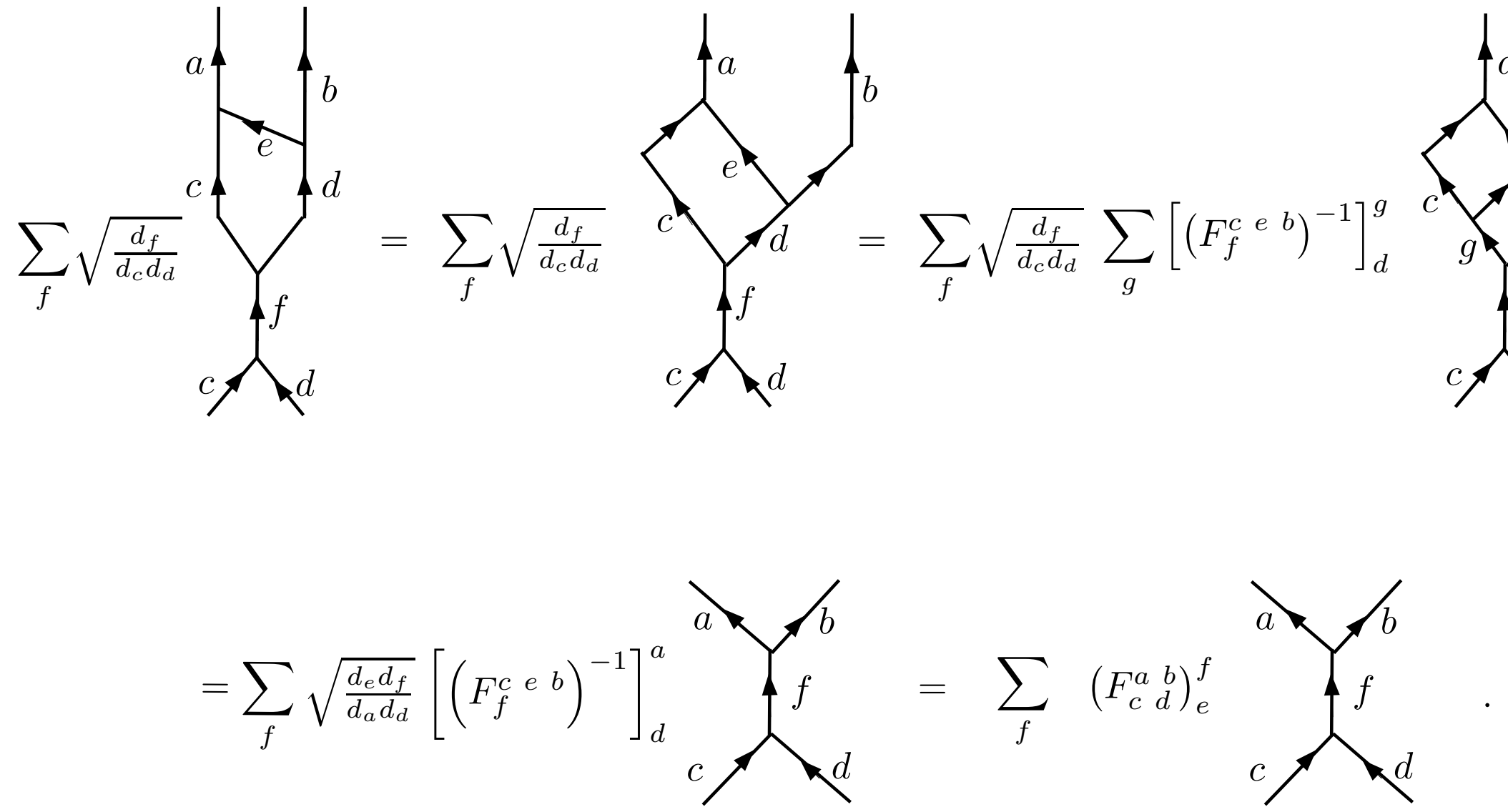

Figure 3.1: Derivation of the "2-2" F-move. Starting from the left-hand side, insert the identity operator Eq. (3.26) to the bottom of the diagram (i.e. Eq. (3.30)). Distort the diagram, which is still topologically equivalent to the preceding diagram. Then apply the inverse of the F-move Eq. (3.29). Finally, remove the loop using Eq. (3.22) to give the desired diagrammatic form.

diagrammatically as,

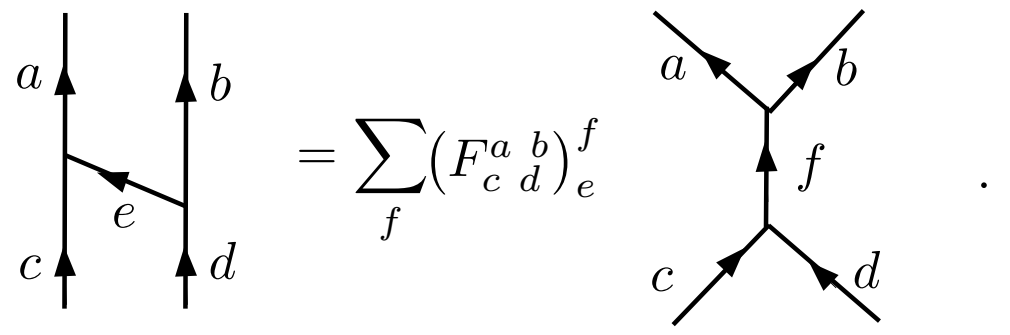

(3.30)

This particular F-move relates tunnelling of charges $e$ between two sites to the combined charge $f$ of the two sites. The expression for the F-matrix $\left(F^{a\,b}_{c\,d}\right)$ can be derived from manipulations of this diagrammatic equation, by applying Eqs. 3.26, 3.29 and 3.22, as shown in Fig. 3.1. The expression for $\left(F^{a\,b}_{c\,d}\right)^f_e$ can be read-off to be,

$$\left(F^{a\,b}_{c\,d}\right)^f_e = \sqrt{\frac{d_e d_f}{d_a d_d}} \left[\left(F^{c\,e\,b}_f\right)^d_a\right]^*, \tag{3.31}$$

for unitary theories.

If the tunnelling charge $e$ is right-directed in the "2-2" F-move as in,

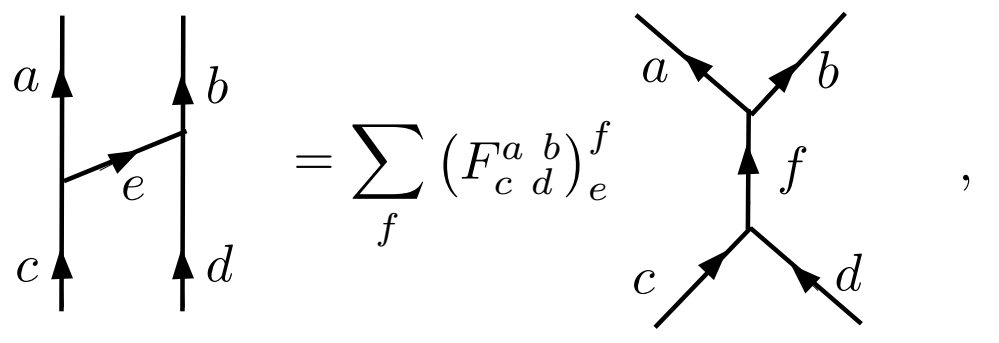

(3.32)

the expression for $\left(F^{a\,b}_{c\,d}\right)^f_e$ can be similarly derived, and is given as,

$$\left(F^{a\,b}_{c\,d}\right)^f_e = \sqrt{\frac{d_e d_f}{d_c d_b}} \left(F^{a\,e\,d}_f\right)^b_c. \tag{3.33}$$

We will now use these two "2-2" F-move expressions to derive factors when vertically bending a charge line.

One of the advantages of using diagrammatic representations is that we can perform (and visualise) topological manipulations such as bending charge lines up, down, or sideways, which correspond to certain transformations applied to the basis states of some anyons. Some definitions are in order. A vertical bend is defined as a manipulation which reverses the orientation of a charge line. [See Fig. (3.20) for an example of a directed charge line]. If in the case of an upward-going line, the manipulation reverses the orientation of the line to become downward-going, thereby forming something like a "cap" shape, or vice versa, i.e. an originally downward-going line becomes an upward-going line, thereby forming something like a "cup" shape, then that manipulation will be called a "vertical bend," which is nontrivial as will be shown below. A "horizontal bend" is defined as a manipulation which, irrespective of direction, bends a line sideways, and does not reverse the orientation of the line. Horizontal bends of an anyon charge line is trivial, meaning the state does not acquire any factor whatsoever. However, a vertical bend around the vertex of a splitting and fusion diagram acquires a factor, which can be computed from the "2-2" F-move equations Eqs. 3.30 and 3.32, and given as,

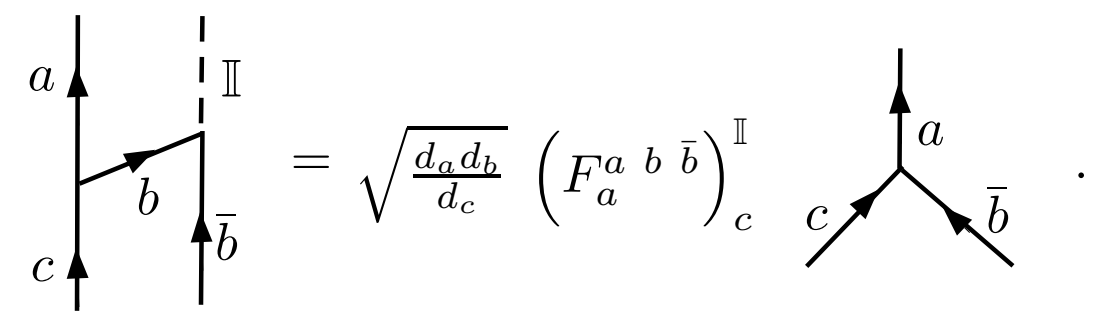

(3.34)

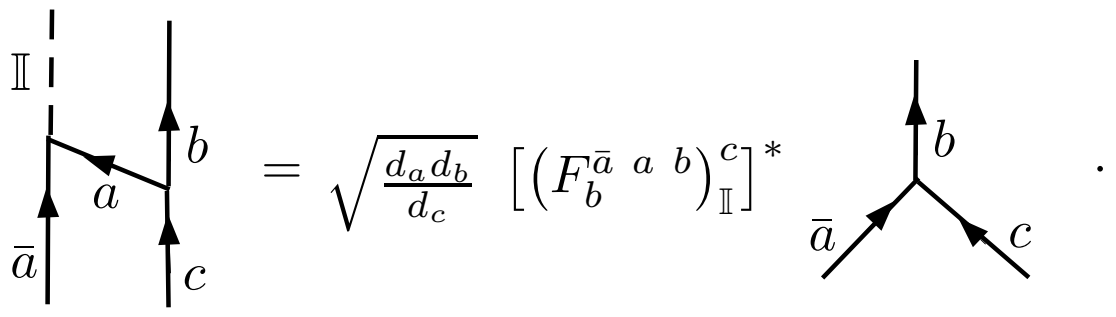

(3.35)

In the first equation, the outgoing leg with charge $b$ bends down to become an incoming leg with the conjugate charge $\bar{b}$. The same reasoning applies to the second equation. The bend factors are the pre-factors in front of the diagrams. In addition, a "zig-zag" bend can be straightened,

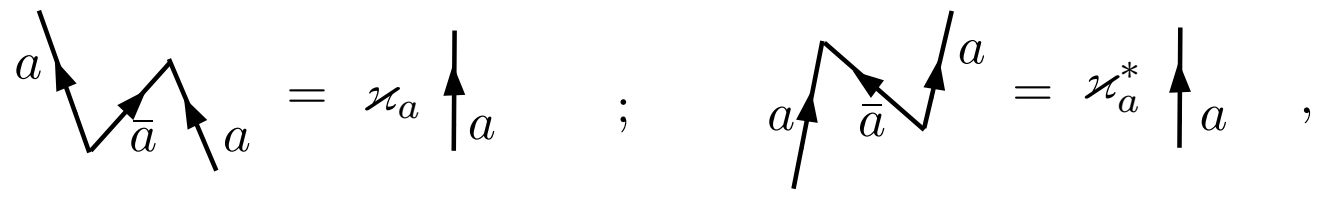

(3.36)

where $\varkappa_a = d_a \left(F_a^{a\bar{a}a}\right)^{\mathbb{I}}_{\mathbb{I}}$ is a phase factor. These equations show that bends introduce, at worst, a phase factor when working in diagrammatic isotopy. The phase factor can be made equal to 1 for non self-dual charges by a choice of gauge but for self-dual charges, this factor is $+1$ or $-1$ and is known as the gauge invariant Frobenius-Schur indicator.

Anyons on two sites with charges $a$ and $b$ can be swapped, which is referred to as a braiding of the anyons. This is a generalisation of permutation of particles in three-dimensions. Let $R_{ab} : V^{ab} \to V^{ba}$ be the braid operator, which is a unitary representation of the braid group. This is represented diagrammatically as,

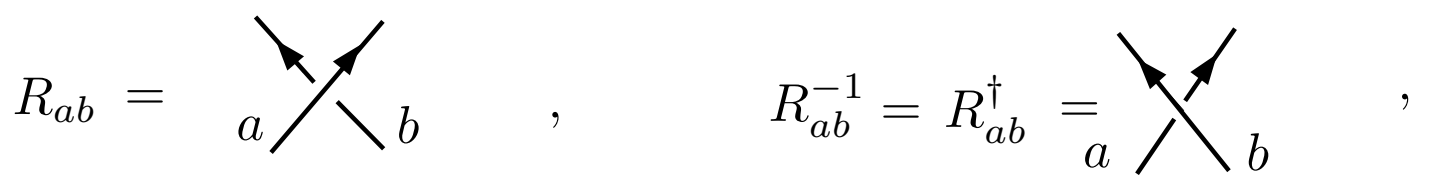

(3.37)

where the braid operator $R_{ab}$ is an "over-crossing" for when anyon $a$ and $b$ are exchanged. By convention $R_{ab}$ is taken for a counter-clockwise exchange in two dimensions. The $R_{ab}^{-1}$ is for the inverse process where anyon $a$ goes around anyon $b$ in a clockwise manner. The diagram is referred to as an "under-crossing." In terms of the splitting and fusion basis states, the braid operator $R_{ab}$ is represented as,

$$R_{ab} = \sum_c \sqrt{\frac{d_c}{d_a d_b}} R_c^{ab}$$ 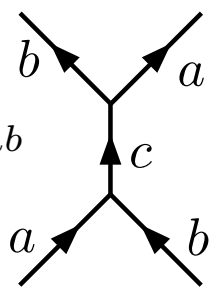

, (3.38)

where $R_c^{ab}$ is the braid factor when charges $a$ and $b$ fuse to $c$. The braid matrix has to satisfy the hexagon equation, which ensures consistency of braiding with fusion of anyons.

With all the necessary ingredients introduced, we are now ready to specify how to diagrammatically construct a pure state and operators of a many-body anyonic system. Let $V_{\mathbb{I}}^{a_1,a_2,\ldots,a_n}$ be a topological Hilbert space of $n$ anyons with charges $a_1, a_2, \ldots, a_n$ fusing into an overall vacuum charge $\mathbb{I}$. A pure state in this space is written as as superposition of valid labelled diagrams as,

$$|\psi\rangle = \sum_{b_1,\ldots,b_{n-2}} \psi_{b_1\ldots b_{n-2}} \left(\frac{1}{d_{a_1}\cdots d_{a_n}}\right)^{1/4}$$ 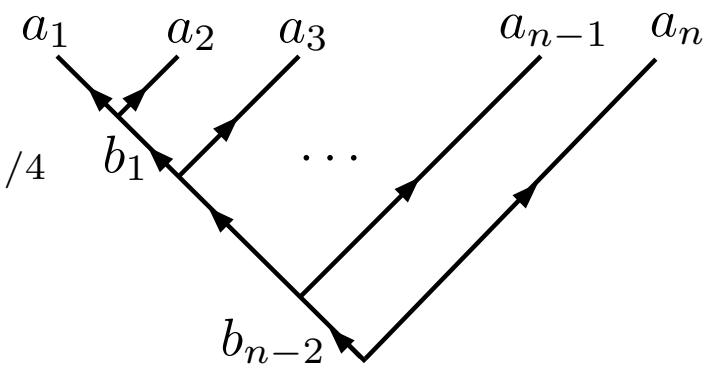

, (3.39)

where $a_1, a_2, \ldots, a_n$ are the labels of the anyonic charges on sites, with each taking a fixed value, and $\mathbf{b} = (b_1, \ldots, b_{n-2})$ is a list of in-between fusion outcomes. The labelled diagram with the vertex normalisation factor is taken as a whole to be orthonormal. If we demand that the state $|\psi\rangle$ is normalised, then $\sum_{\mathbf{b}} |\psi_{\mathbf{b}}|^2 = 1$.

An operator $\hat{O}$ acting on an anyonic system, where $\hat{O} \in V_{a_1',a_2',\ldots,a_m'}^{a_1,a_2,\ldots,a_n}$, in general, takes $m$ anyons to $n$ anyons. This is represented diagrammatically as,

$$\hat{O} = \sum_{\substack{b_1,\ldots,b_{n-2} \\ b_1',\ldots,b_{m-2}' \; c}} O_{b_1'\ldots b_{m-2}'}^{b_1\ldots b_{n-2}\; c}$$ 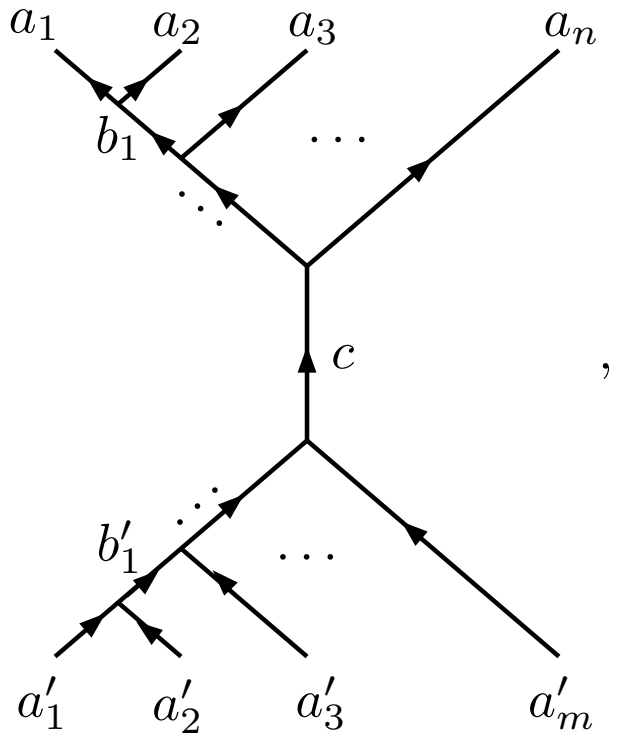

, (3.40)

where charge $c$ on the root of the splitting and fusion tree has been matched on the trunk so as to ensure charge conservation of the operator. The vertex normalisation factors of the labelled diagram have been absorbed into the coefficients defining the operator. Succinctly, this operator can be written as a direct sum $\hat{O} = \oplus_c \hat{O}_c$ of the operators $\hat{O}_c$ acting in each charge sector $c$.

## 3.5 Conclusion

In conclusion, the diagrammatic techniques used in manipulating anyonic states and operators provide powerful algorithms which can be combined with the techniques of tensor networks (that was reviewed in Chapter 2) to construct *anyonic tensor networks* ansatz states. I will show this for the specific case of MPS in Chapter 4. This ansatz together with the TEBD (or DMRG) algorithm can be used to simulate anyonic systems of both finite and infinite extent, and at arbitrary filling fractions either by tuning a chemical potential term in the Hamiltonian or exploiting the U(1) symmetry of the model.

4

# Anyonic Matrix Product States with U(1) Symmetry

## 4.1 Introduction

Anyons are point-like (quasi)particles which exist only in two-dimensional systems and have richer exchange statistics than bosons or fermions. One of the main interests in anyons is in their application to implementing fault-tolerant (topological) quantum computation [14, 22, 80]. Anyons have also garnered a substantial theoretical interest since they are proposed to exist in systems as diverse as fractional quantum Hall systems and two-dimensional spin liquids [15–28], one dimensional nanowires [29–32], and ultra-cold atoms in optical lattices [33]. Recent experiments showing evidence for Majorana edge modes (i.e. Ising anyons) in nanowires [32] might bring us closer to working with anyons in the laboratory, with far-reaching scientific and technological applications.

One dimensional chains of static $SU(2)_k$ anyons with a local antiferromagnetic Heisenberg-like interaction have been studied extensively since, for example, they are critical and realize all minimal models of conformal field theories (CFTs) [81]. It is also natural to ask whether interesting states and phases appear in anyon models where the anyons are allowed to hop on a lattice and braid around one another. Braiding pairs of anyons generally transforms the anyonic state in a non-trivial way, in contrast with bosons and fermions which merely pick up a factor of $\pm 1$. For anyons, braiding is a topological interaction, with the meaning that the interaction is independent of the distance between the anyons and arises only from the inherent anyonic statistics. In Refs. [82, 83] the authors report on some phases that appear in lattice models of itinerant anyons, where the anyons—coupled by a Heisenberg interaction—are located on the sites of a lattice, with vacancies which allow for anyons to hop between sites but without braiding around one another. In Ref. [84] the authors study the real time dynamics of a single anyon moving between the sites of a ladder lattice with static anyons pinned to the plaquettes of the ladder, which serves as a model of coherent noise in topological quantum memories, and uncovers a signature that distinguishes abelian anyons from non-Abelian anyons based on their transport properties. Noise models for medium sized topological memories based on real time stochastic dynamics of braiding Ising models anyons [85], Fibonacci anyons [86], and quantum double model anyons [87] have also been studied. In this chapter, I construct an ansatz to simulate ground states of 1D and quasi-1D models of itinerant anyons at fixed particle density, which may or may not involve braiding, and possibly include a Heisenberg interaction. I benchmark the ansatz using different physical models whose ground state energy and entanglement entropy are given in chapter 5.

Large anyonic systems, like generic quantum many-body systems, are hard to simulate on a classical computer due to the exponential growth in the dimension of the state space with the number of particles. Until recently, numerical studies of anyons have primarily used exact diagonalization [2, 81, 82, 88–90] which limits analysis to small system sizes and relies on finite-size scaling to extract properties in the thermodynamic limit. A more successful approach uses tensor networks (TNs) which describes quantum many body states using a network of low-order tensors which can be contracted together to compute relevant quantities such as ground state energy, correlations, subsystem entropy, etc. One of the simplest tensor networks is the matrix product state (MPS) which forms the basis of highly successful algorithms, namely, the density matrix renormalization group (DMRG) [47, 48, 74] and the time-evolving block decimation (TEBD) [77, 91, 92] to simulate the ground state and dynamics of 1D and quasi-1D quantum many-body systems. Exploiting translation invariance in TN states has allowed the study of systems directly at thermodynamic limit, circumventing the limitation on size encountered in exact diagonalization [56].

Owing to their success for spin systems, tensor network algorithms have recently been adapted to simulate quantum many-body systems of anyons [39, 40, 93]. In particular, anyonic versions of the Matrix Product States (MPS), and of the TEBD and DMRG algorithms have been proposed and tested with a high degree of accuracy for anyonic chains [41, 43]. Tensor network algorithms are adapted to anyons by explicitly hardwiring the constraints implied by the fusion rules of the anyon model into the tensor network ansatz. This provides two important advantages. First, an anyonic TN representing a many-body anyonic state contains fewer complex coefficients than a non-symmetric TN description of the same state that does not explicitly encode the anyonic symmetry, thus providing for computational speedup. Secondly, using an anyonic TN as an ansatz in numerical simulations guarantees that one remains in the physically relevant sector of the Hilbert state, namely, one with the desired total anyonic charge, and thus avoiding leakage into states that are not allowed by the physics of the system, due to numerical errors.

In this chapter and the next, I describe how to simulate the ground state of a system of itinerant anyons by means of the anyonic TEBD algorithm that additionally incorporates a U(1) symmetry corresponding to conservation of particle number density. Our construction of the combined Anyon $\times$ U(1) symmetric MPS is the first to allow for simulating these systems with an arbitrary, *specified* rational particle number density (or filling fraction), and gives direct access to Hilbert space sectors enumerated by anyonic charge and particle number density. Our MPS ansatz also allows us to simulate bosons, fermions, and anyons using the same algorithm, since bosons and fermions can be treated as simple types of anyons.

In this chapter, I develop the Anyon $\times$ U(1) symmetric MPS formalism. The Anyon $\times$ U(1) symmetric MPS combines the recently proposed anyonic MPS [41, 43] with the implementation of a U(1) symmetry in the MPS [73, 94], and as such our presentation contains some review of both elements separately, which serves both as a reminder of important concepts and also introduces useful terminologies that persist throughout the thesis. The structure of this chapter is as follows: in Sec. 4.2 I review the basics of anyonic tensors—which constitute objects making up any anyonic tensor network. In Sec. 4.3, I review the anyonic MPS. In Sec. 4.4, I review the implementation of a U(1) symmetry in the MPS corresponding to conservation of particle number density, in particular, showing how it can be achieved as an instance of the anyonic MPS and how an arbitrary filling fraction is realized at the level of the ansatz. In Sec. 4.5 I construct the combined MPS ansatz that incorporates both the anyonic symmetry and the U(1) symmetry. Then, I present test models and benchmarking results in Chapter 5.

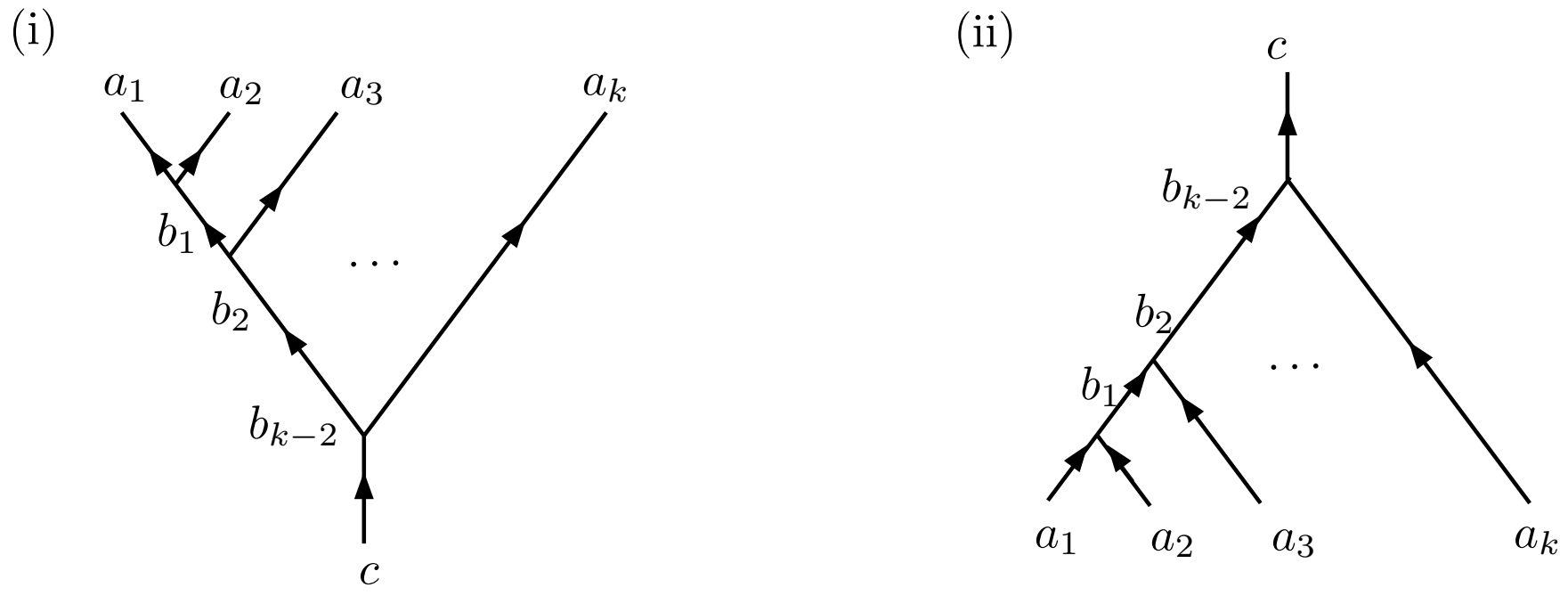

Figure 4.1: (i) Splitting tree and (ii) fusion tree, defining the "ket' and "bra" bases respectively for a total number $k$ of anyonic charges $(a_1, a_2, \ldots, a_k)$ on the leaves, and $(b_1, b_2, \ldots, b_{k-2})$ as the fusion products on the links of the trees. The charge $c$ on the trunk can, in principle, take all possible charge values permissible by the anyon model. If however, the fusion tree defines the basis of a pure quantum state, the charge $c$ can only be the vacuum charge $\mathbb{I}$.

## 4.2 Basics of Anyonic Tensor Networks

Using tensor networks to simulate quantum systems involves choosing a network pattern of connected tensors along with a choice of an algorithm that optimizes the representation of the many body state [37]. In what follows, I review the basic objects common to most anyonic tensor networks (TN), and then later use these objects to construct the anyonic MPS, and our modified Anyonic-U(1) MPS ansatz.

### 4.2.1 Components of anyonic tensor networks

In this section, I use the theory of anyons that was briefly reviewed in Chapter 3, to construct the components of an anyonic tensor network.

The basis of the Hilbert space of anyons is described by a labelled directed fusion tree (see Fig.4.1) where the charge $c$ on any incoming edge at a vertex is determined from the charges $a$ and $b$ of the two outgoing edges around the same vertex, according to the fusion rules of the anyon theory

$$a \times b \to \sum_c N_{ab}^c \, c, \tag{4.1}$$

which implies that charges $a$ and $b$ are allowed to fuse to, possibly, several different charges $c$.

The labeled fusion/splitting tree in Fig.4.1 contains many charge labels, and can be extremely verbose when dealing with large anyonic systems. While explicit labelling of fusion trees is possible in principle, it is not very practical for anyonic tensor network simulations. A better alternative is to enumerate the labelled fusion trees having a particular charge $c$ at the trunk of the tree. To this end, let $c$ be the total charge at the trunk of the fusion tree and introduce a new index $\mu_c$ that enumerates each unique labelled fusion tree in increasing numerical order, $\mu_c = 1, 2, \cdots, \nu_c$. Here $\mu_c$ is called the degeneracy index, and $\nu_c$ is the degeneracy of the charge sector $c$. The term "degeneracy" in symmetric TN does not refer to the "degeneracy of energy levels" as used in many-body physics, but to the number of configuration states forming a basis in a particular symmetry sector. All the fusion trees are therefore concisely labelled by the multi-index $\gamma = (c, \mu_c)$, with $c$ as the total charge label and $\mu_c$ as its degeneracy index.

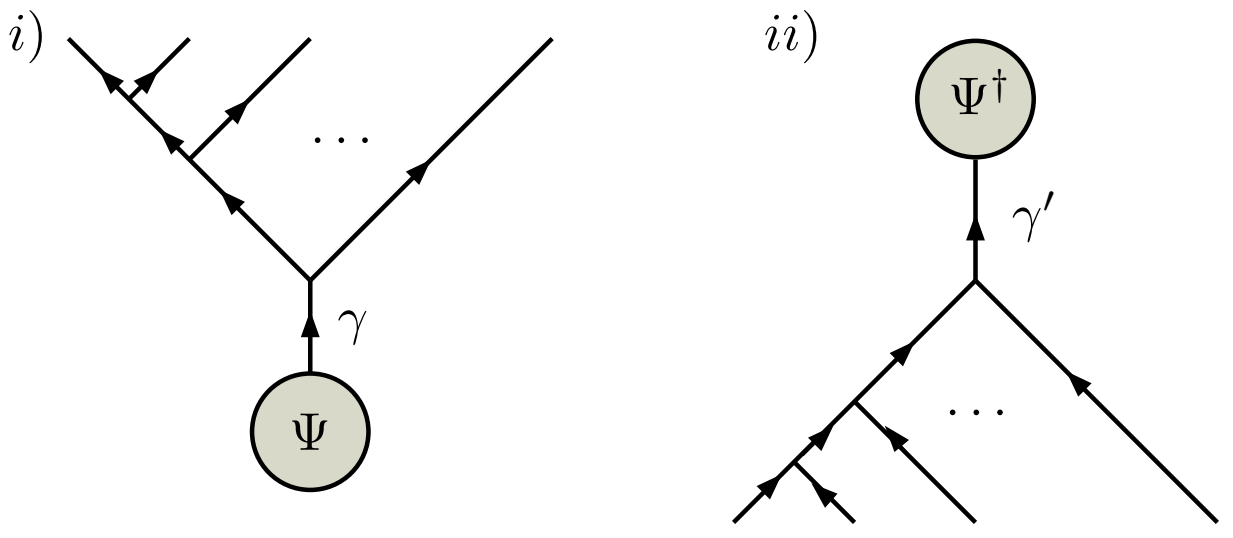

Figure 4.2: (i) Anyonic state vector and (ii) its Hermitian conjugate.

The basic objects in any tensor network include vectors (or one-index tensors), matrices (or two-index tensors), and more generally, $n$-index tensors. I now examine how the anyonic equivalent of these objects are created. The (orthonormal) basis of these objects are indexed by the greek index $\{\gamma = (c, \mu_c)\}$, where $c$ is the charge label of the sector, and $\mu_c = 1, 2, \ldots, \nu_c$.

### Anyonic state vector

An anyonic quantum state $|\Psi\rangle$ can be written as a weighted superposition of all labellings of a fusion tree having a total vacuum charge $\mathbb{I}$.[1] More compactly, in the multi-index notation, the quantum state can be written as

$$|\Psi\rangle = \sum_{\gamma} \Psi^{\gamma} |\gamma\rangle, \tag{4.2}$$

where $\gamma = (\mathbb{I}, \mu_{\mathbb{I}})$ is an index enumerating all the valid fusion trees. If all the enumerated fusion trees are associated with normalized anyonic diagrams, which is referred to as the *implicit normalization* scheme, then the state amplitudes $\Psi^{\gamma}$ can be arranged as a column vector in the standard basis. Following the diagrammatic notations employed for anyonic tensors in Ref. [39], I depict the anyonic quantum state by a filled circle with a central leg enumerating all the multi-indexed bases, and an unlabeled tree structure, as shown in Fig.4.2(i). It should be noted that if topological manipulations were to be performed on the fusion tree, such as vertically bending a line opposite to its orientation, it is preferable to use diagrammatic isotopy convention. In such a case, we adopt the prescription given in Ref. [1], where the fusion diagrams are weighted with certain pre-factors of quantum dimensions of the anyonic charges on the fusion tree. For the fusion basis shown in Fig.4.1, the pre-factor would be $\left(\frac{d_c}{d_{a_1} d_{a_2} \cdots d_{a_k}}\right)^{1/4}$. These normalization factors are then also absorbed into the amplitudes defining the anyonic state vector. This is referred to as the *explicit normalization* scheme. During topological manipulations, all the charges labelling a particular fusion tree can be recovered from the fusion lookup tables and used in computing the necessary data associated to that operation. I work exclusively in the explicit normalization scheme, where each vertex is normalized according to diagrammatic isotopic convention. The differences between working in *implicit* and *explicit* normalization scheme, collectively called *mixed normalization*, are treated in the recent anyonic DMRG paper [43].

[1] A system of anyons which does not fuse to the total vacuum charge $\mathbb{I}$ can be considered to be within a larger system with a nontrivial boundary charge that enforces a complete annihilation of the anyonic charges to the vacuum charge.

The Hermitian conjugate of the state $|\Psi\rangle$, written as

$$\langle\Psi| = \sum_{\gamma} \Psi_{\gamma}^{\dagger} \langle\gamma| , \tag{4.3}$$

is represented diagrammatically as in Fig.4.2(ii) where the unlabeled fusion tree is reflected vertically and all its arrows are reversed. The coefficients of the vector are also complex-conjugated.

### Anyonic Matrix Operator

In conventional quantum theory, an operator $\hat{O} : \mathbb{V} \to \mathbb{V}'$ is written in the bra-ket notation as,

$$\hat{O} = \sum_{j',j} O_{j',j} |j'\rangle \langle j| . \tag{4.4}$$

where the indices $j$ and $j'$ enumerates basis in $\mathbb{V}$ and $\mathbb{V}'$. Example of such operators include Hamiltonians, density matrices, projectors, etc.

In a similar vein, an anyonic operator acting on a set of anyonic charges with total charge $c$ does not change the total charge. The operator $\hat{O}_c : \mathbb{V}_c^{a_1,a_2,\cdots a_k} \to \mathbb{V}_c^{a'_1,a'_2,\cdots a'_{k'}}$ takes states of anyons $a_1, a_2, \cdots, a_k$ to states of anyons $a'_1, a'_2, \cdots, a'_{k'}$ without changing the conserved total charge $c$. As such, the operator $\hat{O} = \bigoplus_c \hat{O}_c$ can be constructed as a block-diagonal matrix with each block indexed by the conserved anyonic charge $c$. Each block matrix $\hat{O}_c$ is constructed by enumerating (as in Fig.4.1) all the fusion tree bases fusing to that charge. As such the charge-conserving matrix is indexed by the multi-index $\gamma = (c, \mu_c)$ for fusion trees and $\gamma' = (c, \mu'_c)$ for splitting trees. The anyonic operator can therefore be written as

$$\hat{O}_c = \sum_{\gamma',\gamma} \hat{O}_{\gamma',\gamma} |\gamma'\rangle \langle\gamma| , \tag{4.5}$$

where $\gamma = (c, \mu_c)$ and $\gamma' = (c, \mu'_c)$ implying charge conservation. The matrix elements will depend on the particular physics of the system. The anyonic matrix operator is represented diagrammatically by Fig. 4.3(i), where the multi-indices $\gamma = (c, \mu_c)$ and $\gamma' = (c, \mu'_c)$ enumerate all the fusion and splitting trees. The vertex normalization factors of the fusion/splitting trees are absorbed into the matrix operator.

### Anyonic order-3 tensor

The anyonic matrix operator can be extended to an order-3 tensor where the tensor elements are indexed by three multi-indices $\alpha$, $\beta$ and $\gamma$. An anyonic order-3 tensor $T_{\gamma}^{\alpha,\beta}$ may be represented in the manner shown in Fig. 4.3(ii) where the leaves on each branch of the tree are enumerated and assigned a multi-index notation $\alpha = (a, \mu_a)$, $\beta = (b, \mu_b)$ and $\gamma = (c, \mu_c)$. All the vertices on the leaves fulfill the fusion rules during enumeration of the basis, and the implicit vertex contained within the grey circle also obeys the fusion rules of the anyon model.

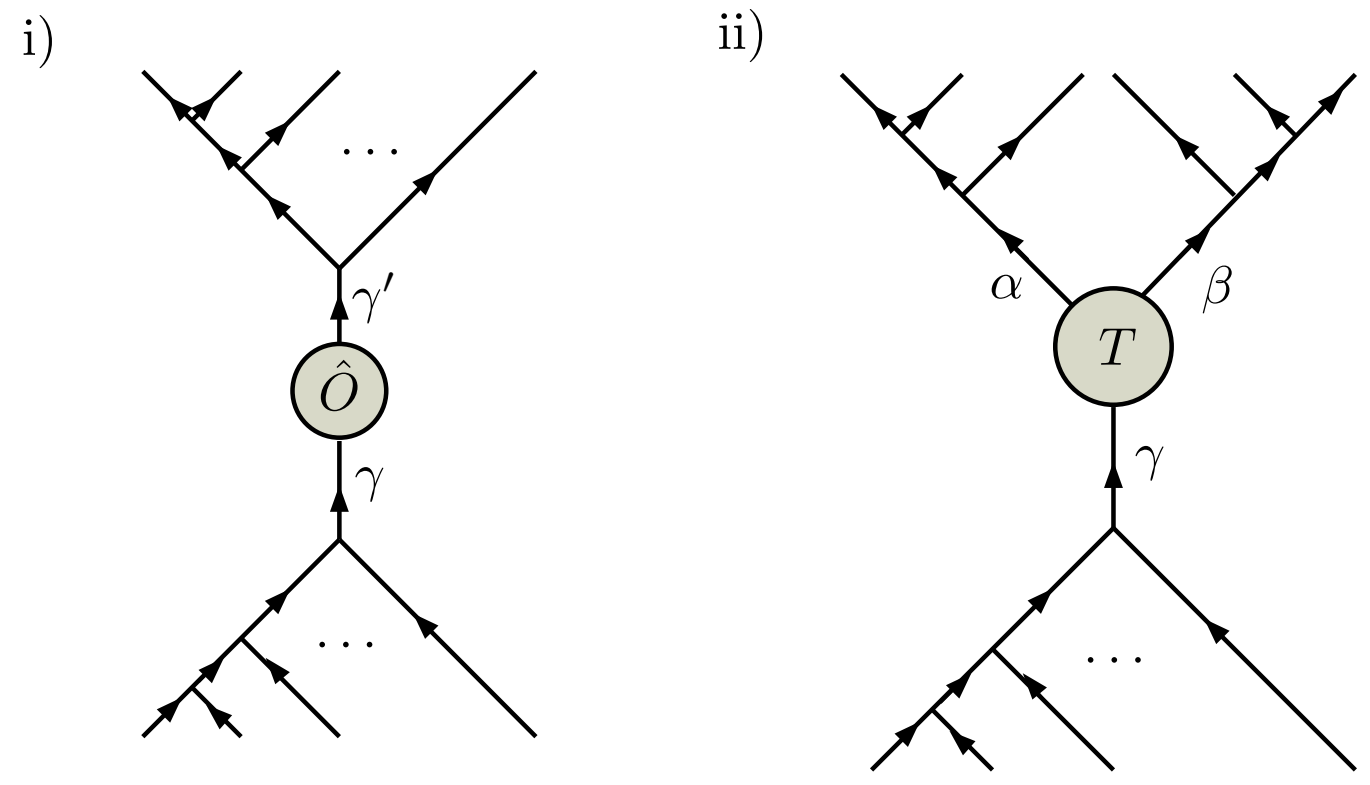

Figure 4.3: (i) The anyonic matrix operator $\hat{O}$ and (ii) the anyonic order-3 tensor $T$. The tree is assumed to have been normalized according to diagrammatic isotopy convention.

The tensor $T_{\gamma}^{\alpha,\beta}$ is indexed by $\gamma = (c, \mu_c)$, $\alpha = (a, \mu_a)$ and $\beta = (b, \mu_b)$, where the charge triplet $(a, b, c)$, obtained from each of the subtrees have to be compatible i.e. $a \times b \to c$. The explicit form of the tensor $T$ is

$$T = \bigoplus_{a,b,c} N_{ab}^{c} \sum_{\mu_a,\mu_b,\mu_c=1}^{\nu_a,\nu_b,\nu_c} \left(T_c^{a,b}\right)_{\mu_c}^{\mu_a,\mu_b} |\mu_a\mu_b\rangle \langle\mu_c| \,. \tag{4.6}$$

The direct sum implies that tensor $T$ is composed blockwise from tensors indexed by the charges of the subtrees, with each block then being indexed by the degeneracy index of the compatible fusion trees.

There are more objects that could be implemented to manipulate anyonic tensor networks [39, 40, 43], but as MPS is a trivalent tensor network, the objects introduced above are sufficient to construct the matrix product state and the algorithms that act on it.

## 4.3 Anyonic Matrix Product States

Using the anyonic tensors reviewed above, the simplest tensor network for anyonic systems can be constructed, namely the anyonic matrix product states (MPS). The anyonic MPS constructed in Refs. [41] and [43], is a network which consists of order-3 (trivalent) tensors and order-2 tensors—which encodes the Schmidt values of the state.

One convenient form of the conventional MPS ansatz is that given by Vidal [77], which is an array of two-index and three-index tensors forming a linear network of tensors. For a finite lattice with open boundary condition, the tensors on the boundary of the MPS (i.e. the first and last sites) are two-index tensors while the “bulk” of the network consists both of two-index tensors (Schmidt vectors) and three-index tensors for each of the other $(n-2)$ sites.

Analogously, the MPS was adapted to anyons by Singh et. al. in Ref. [41], using the basic anyonic tensors (two-index and three-index anyonic tensors) after the pattern of the conventional MPS. Each three-index tensor is indexed by both the charge and the degeneracy of the anyons making up each site. The charges on the trivalent vertex of the tensor are compatible in accordance with the fusion rules of the anyon model. The Schmidt vectors, which are two-index tensors, are charge-conserving diagonal matrices. The basis labelling $\alpha_i = (a_i, \mu_{a_i})$ for each site of the anyonic lattice is given by the set of charges $a_i$ and the degeneracies $\mu_{a_i}$ of each charge. The labels $\mu_{a_i}$ take fixed value of 1 if there is only one configuration for each possible charge label at each site, e.g. if the possible physical states are merely the presence or the absence of a charge.

Formally, for a lattice $\mathcal{L}$ of $L$ sites with anyonic charges $\alpha_1 = (a_1, \mu_{a_1}), \alpha_2 = (a_2, \mu_{a_2}), \ldots, \alpha_L = (a_L, \mu_{a_L})$, the anyonic MPS encoding the ground state $\Psi_{\mathrm{GS}}$ is given diagrammatically as

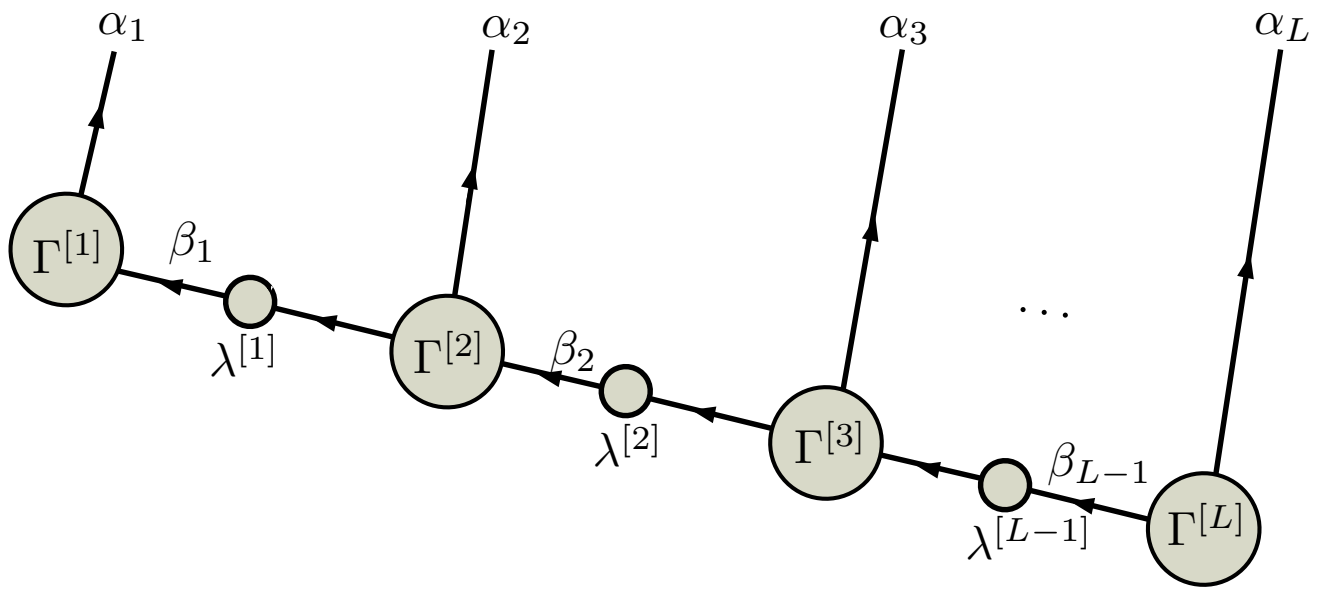

(4.7)

where the multi-indices $\beta_{i+1} = (b_{i+1}, \mu_{b_{i+1}})$ on the bonds are obtained by an iterative fusion of the multi-indices $\beta_i = (b_i, \mu_{b_i})$ and $\alpha_{i+1} = (a_{i+1}, \mu_{a_{i+1}})$,

$$\beta_i \times \alpha_{i+1} \rightarrow \beta_{i+1}, \tag{4.8}$$

where, as before, the charge $b_{i+1}$,

$$b_{i+1} = \sum_{b_i, a_{i+1}} N^{b_{i+1}}_{b_i a_{i+1}} (b_i \times a_{i+1}), \tag{4.9}$$

and the total degeneracy $\nu_{b_{i+1}}$ of the charge $b_{i+1}$ is determined by

$$\nu_{b_{i+1}} = \sum_{b_i, a_{i+1}} N^{b_{i+1}}_{b_i a_{i+1}} \nu_{b_i} \nu_{a_{i+1}}. \tag{4.10}$$

It should be noted that this anyonic MPS has been drawn with site indices going upwards, to make apparent the visual similarity with anyonic fusion tree diagrams, but it is essentially the same ansatz as given in Ref. [41]. Due to the iterative fusion process down the tree of the anyonic MPS, the dimensions of the tensors $\{\Gamma^{[i]}\}$ required to construct an arbitrary state will vary, but in practice an upper bound is imposed on the bond dimension $\chi$ ahead of time. The bound chosen usually depends on the amount of entanglement and correlations needed to faithfully represent the state of the system (and on computational resources available). As such, anyonic MPS provides a systematic way of handling anyonic systems, specifying both the basis (i.e. the fusion tree) and encoding the amplitudes of the state in the tensors.

As a proof-of-principle example, this anyonic-MPS ansatz has been used to simulate, together with the anyonic-TEBD algorithm, a chain of interacting non-Abelian anyons (e.g. Fibonacci and Ising anyons) coupled by a Heisenberg interaction. The charge multi-index $\alpha_i$ on each site $i$ of the leaves of the anyonic-MPS is set (in the case of Fibonacci anyons) to $\alpha_i = (\tau, 1)$, where $\tau$ is the Fibonacci anyon charge, and the number 1 is the degeneracy of the $\tau$ charge on site $i$ (i.e. the number of different configurations on the site consistent with a total charge of $\tau$). The anyonic MPS is, however, a general ansatz capable of dealing with systems with any quantum group symmetry, and hence, can be adapted to work with other symmetries, Abelian or non-Abelian. For instance, by replacing the anyonic charges with particle number charges, the anyonic-MPS can serve as a U(1)-MPS [73], which can be used to simulate physical systems having a global particle number $N$ on a finite lattice $\mathcal{L}$.

On an infinite lattice with translation invariance of the Hamiltonian, if the U(1) charge is identified with particle number then the U(1)-MPS is primitively a zero-density ansatz [i.e. one favouring a mean U(1) charge per site of 0], and cannot directly be used to simulate an infinite lattice with a finite non-zero particle density. In the next section I show how to tune the U(1)-MPS to simulate an infinite lattice system at non-zero density, and in Section 4.5 I propose a modified ansatz, the Anyonic-U(1) MPS, that conserves both particle density and anyonic charge symmetry, and which can be used to simulate anyonic systems (including braiding of anyons) at a specified rational filling fraction.

## 4.4 U(1)-MPS and Particle Density Conservation

In the last section we alluded to the fact that the anyonic MPS can serve as a U(1)-MPS by replacing the anyonic charge labels with the particle number charge labels. Specifically, let us consider a lattice $\mathcal{L}$ of $L$ sites, where each site can accommodate a finite number of particles, $n = 0, 1, 2, \ldots, d-1$. The positive integers $n$ can be regarded as the irreps of the $U(1)$ symmetry, which can intuitively be understood as: $n = 0$ is the absence of a particle, $n = 1$ is the presence of one particle, $n = 2$ is the presence of two particles, and so on. The total number of particles $N$ on the lattice of $L$ sites is $N = \sum_{i=1}^{L} n_i$, with a particle density of $\nu = N/L$.

The Hilbert space of the lattice, $\mathbb{V}^{\mathcal{L}} = \bigotimes_{i=1}^{L} \mathbb{V}^{(i)}$, can be alternative written as, $\mathbb{V}^{\mathcal{L}} = \bigoplus_{n=0}^{N} \mathbb{V}_n$, a direct sum over subspaces with fixed numbers of particles $n$. Utilizing this alternative structure a particle-number conserving Hamiltonian $\hat{H}$ can be directly diagonalized in the $\mathbb{V}_n$ subspace, offering savings on the computational cost. The U(1)-MPS ansatz for $N$ particles on an $L$-site lattice can be derived from the anyonic MPS by fixing the particle number $N$ and degeneracy $\nu_N = 1$ at the "right end" of the last tensor, and charge 0 (i.e. zero) on the "left end" of the first tensor. The on-site multi-indices of the "bulk" $(L-2)$ tensors carry $\alpha_i = (n_i, \mu_i)$, where $n_i$ is the U(1) charge on site $i$, and $\mu_i$ enumerates the degeneracy of that charge, for all $i \in \mathcal{L}$. The MPS bonds also carry charge and degeneracy indices, but unlike systems of anyons where degeneracy comes from the fusion rules of the anyon model, degeneracy in U(1)-symmetric lattice models comes from the number of combinatorial arrangements of the charges on the lattice.

Therefore, with a properly constructed ansatz and an optimization algorithm like TEBD or DMRG [49], one can compute the ground state of a local U(1)-symmetric Hamiltonian on a finite lattice. If this finite U(1)-MPS is naively extended to simulate an infinite lattice model, the ansatz would correspond to a zero-density ansatz because of the finite size of the bond dimension $\chi$ and the assumption that the U(1) charge labels exhibited on this bond are finite. In the next subsections I give a heuristic proof of this statement, and I then propose a technique which can be employed to tune the U(1)-MPS away from being a zero-density ansatz, to any desired non-zero particle density.

### 4.4.1 Zero-density U(1)-MPS

Restricted to a finite bond dimension $\chi$ carrying finite U(1) charges, the U(1)-MPS with integer charge labels on an infinite lattice is a zero-density MPS ansatz. Consider a section of the infinite MPS in Fig.4.4 with the charge-degeneracy indices on physical sites

$$\gamma_a = (n_1, \mu_{n_1}), \gamma_b = (n_2, \mu_{n_2}), \gamma_c = (n_3, \mu_{n_3}),$$

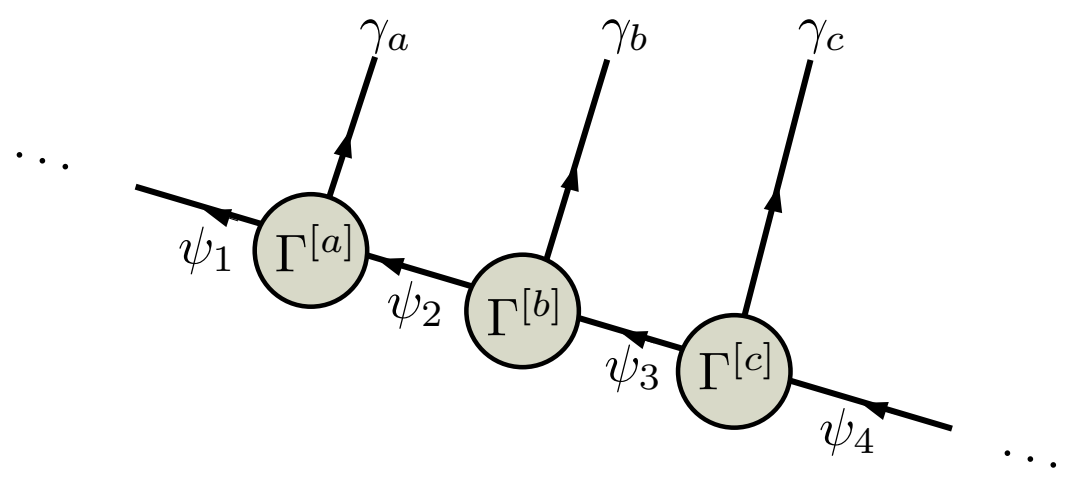

Figure 4.4: An example of an infinite MPS with a block made up of three tensors $\Gamma^{[a]}$, $\Gamma^{[b]}$ and $\Gamma^{[c]}$. By translational invariance of the Hamiltonian, an infinite MPS corresponds to an infinite repetition of the block and hence optimization to ground state is performed only on the tensors within a single block.

and on the links

$$\psi_1 = (m_1, \mu_{m_1}), \psi_2 = (m_2, \mu_{m_2}), \psi_3 = (m_3, \mu_{m_3}),$$
$$\psi_4 = (m_4, \mu_{m_4}).$$

The on-site charges $n_i$ are set to take positive integer charges corresponding to particle number (e.g. hardcore boson has $n_i \in \{0, 1\}$). The charges $m_i$ on the links take only a finite number of charges with degeneracy index $\mu_{m_i} = 1, 2, \cdots, \nu_{m_i}$. The charges and degeneracies on the bond are constrained by the finite bond dimension $\chi$ and given as $\chi = \sum_{m_i} \nu_{m_i}$, where $m_i$ labels the charge on the link $i$. For any realistic computer simulation, the charge labels on the MPS bonds are all finite. Assume we cut the infinite lattice into two partitions. There exists a finite amount of charge $k$ on the link of the left partition, corresponding to a finite number of particles, and the density on the left half-chain is therefore $\nu = k/\infty \to 0$ and therefore the infinite U(1) MPS is a zero-density ansatz. However it is possible to remedy this and have a nonzero density U(1) MPS by shifting the on-site charges so that a U(1) charge of zero corresponds to the desired filling fraction. I present this transformation below.

### 4.4.2 Non-zero density U(1)-MPS

By employing translation invariance, an infinite U(1)-symmetric MPS consists of a block of repeated U(1)-symmetric tensors, albeit such an ansatz is zero-density and will yield a ground state of an empty lattice as seen above. However, by transforming the on-site charges of the MPS, we can make a U(1) charge of zero to correspond to the desired density.

For simplicity and without loss of generality we consider hardcore particles, with charge labels $n \in \{0, 1\}$ on each site of the U(1)-MPS lattice. Let the desired density on the infinite lattice be $\nu = p/q$, which can be interpreted as having an average of $p$ particles on every $q$ sites, with $q > p$. Using the additive (abelian) fusion rules of U(1) charges, a U(1)-MPS with $p$ particles corresponds to having $p$ sites with charge $n = 1$ and the remaining $q - p$ sites with holes $n = 0$. In an infinitely increasing block, the number of particles $p$ increases infinitely, but by "subtracting off" the $p$ number of particles, we can re-center the relevant subspace to be labeled by the charge 0, which is retained in a practical simulation. Formally, by using the transformation,

$$n' = q\left(n - \frac{p}{q}\right) = qn - p, \quad p \leq q, \tag{4.11}$$

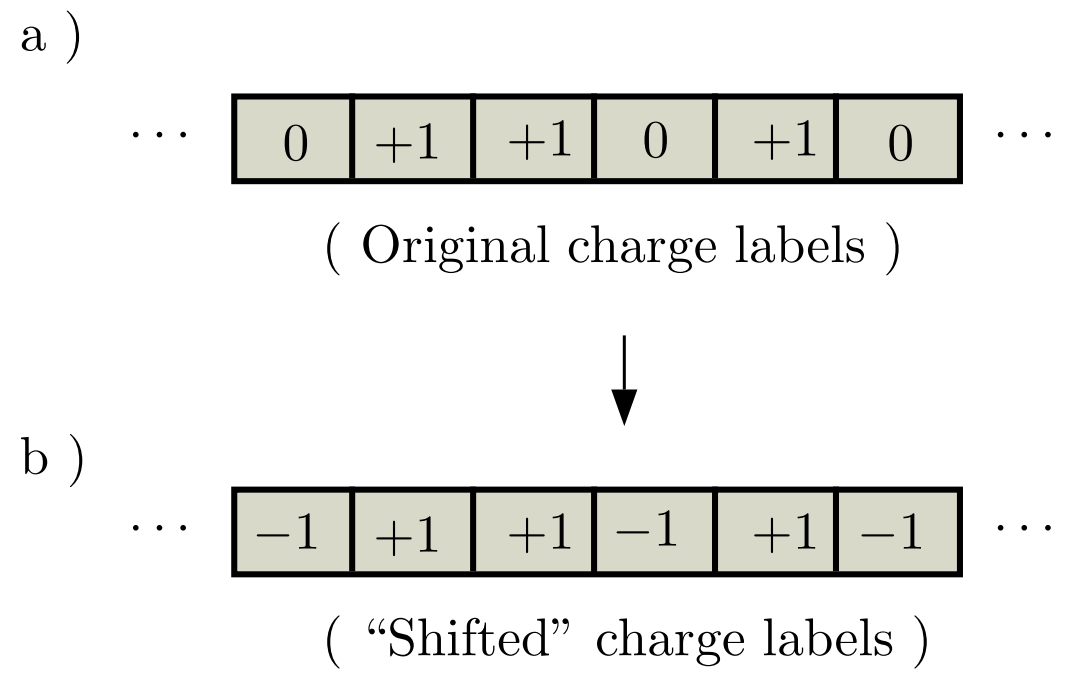

Figure 4.5: Schematic representation of an infinite lattice with a) a typical half-filled configuration with one particle on every two sites, and b) a "shifted" version of a) but with average of zero particle on every two sites.

the on-site charges transform as,

$$\begin{aligned} n = 0 \quad &\rightarrow \quad n' = -p \\ n = 1 \quad &\rightarrow \quad n' = q - p \end{aligned}$$

where multiplication by $q$ in Eq. (4.11) is purely for convenience and ensures that the $n'$ charges, like the $n$ charges, are integer. In essence, before this transform, the desired filling fraction in the MPS would correspond to having $p$ occurrences of charge 1 and $q - p$ holes 0, summing to a total charge of $p$. But after the transform, the desired filling corresponds to having $p$ occurrences of particles with charge $q - p$, and $q - p$ holes with charge $-p$, which sums to a total charge of zero. The charge distribution on any link on a U(1)-symmetric infinite MPS is centered on the zero charge sector, which now corresponds to a particle density of $p/q$. Thus it becomes possible to tune the U(1)-MPS to the desired filling fraction without using tensors with more than three legs.[2] An example of how this transform applies to the half-filling is presented as an example below.

### Example: Half-filled MPS ansatz for hardcore bosons

Consider a particular configuration of an infinite lattice at half filling, where there is on average, one particle on every two sites as shown in Fig. 4.5(a). Each box represents a site and the charge on the site is indicated inside the box. There is on average one particle for every two sites, and assuming that this average density is maintained, this will correspond to half filling on the infinite lattice. This is of course not the only way to achieve half filling, but the example will suffice to illustrate how to achieve a half-filled U(1)-MPS.

With only nonnegative charges on each site, i.e. $n \in \{0, 1\}$, the charges on the links of the MPS—which are derived by fusion of all charges leading to that link—are also all nonnegative. However the implementation requirement that the charge indices be finite (and the finite size of the bond dimension) places an upper bound on the set of charges on the links which are retained after truncation of the Hilbert space of the link. Hence the dominant larger-$N$ states in the infinite lattice are truncated. However, by using $n' = 2n - 1$, the on-site charges are re-defined as $0 \to -1$ and $1 \to 1$, to give the "shifted" configuration in Fig.4.5(b), for which the dominant states now inhabit the zero particle sector. Nearby charge sectors such as $\pm 1$ on the bonds represent small fluctuations in filling fraction relative to a baseline of $\nu = 1/2$. I emphasize that the complex amplitudes of the state are not changed, only that their index is relabelled.

[2]Alternative approaches for identifying a U(1) charge of zero with the desired filling fraction, for example by inserting ancillary indices which remove $p$ U(1) charges every $q$th tensor, either add extra tensors to the network or require tensors with more than three legs, both of which are undesireable as they increase the complexity of the network.

## 4.5 Anyon × U(1)-symmetric MPS

### 4.5.1 Composite charges and fusion rules

In the last Section, I reviewed the U(1) MPS and explained how to achieve an arbitrary rational filling fraction on an infinite lattice. In this Section, I investigate how anyonic systems at arbitrary filling fraction can be simulated using an ansatz that conserves both the anyonic (quantum group) symmetry and the U(1) symmetry.

We first recognize that the two symmetry groups are described by particle spectra with differing fusion rules. Similar to creating a new group from product of two groups, we introduce the Cartesian product of the anyonic charge spectrum $\mathcal{A} = \{a, b, c, \cdots, d\}$ and the U(1) charge spectrum, which will be designated as $\mathcal{U} = \{n, m, \cdots, z\}$ as a set of integer charges, where $n$ is an integer label in $\mathbb{Z} = \{-\infty, \ldots, 0, \ldots, +\infty\}$. The product of the two particle spectra is defined as $\mathcal{A} \times \mathcal{U} = \{(a, n) \mid a \in \mathcal{A}, n \in \mathcal{U}\}$, where the label $(a, n)$ is referred to as the composite charge, and we can write it concisely as $\tilde{a} = (a, n)$. When we define a physical system and, hence also construct the MPS, the charges on physical sites and on links of the MPS are taken from the set $\mathcal{A} \times \mathcal{U}$.

The "new" fusion rules for the composite charges are derived from the known fusion rules of the individual charges. If two charges having the labels $\tilde{a}_1$ and $\tilde{a}_2$ are fused, we let the the charge of the outcome charge be $\tilde{a}_12$, where $\tilde{a}_1 \times \tilde{a}_2 \to \tilde{a}_12$. Writing this out explicitly,

$$
\begin{aligned}
(a_1, n_1) \times (a_2, n_2) &= (a_1 \times a_2, n_1 \times n_2) \\
&= \sum_{a_{12}} \left( N^{a_{12}}_{a_1 a_2} a_{12}, n_1 + n_2 \right),
\end{aligned}
\tag{4.12}
$$

where as aforementioned $n_1 \times n_2$ has a unique outcome $(n_1 + n_2)$ with an additive fusion rule, while the nonabelian anyons have, in general, more than one fusion outcome, and hence the need for the summation $\sum_{a_{12}}$ over all possible charge outcomes $a_{12}$. The vacuum charge of the composite charge spectrum $\mathcal{A} \times \mathcal{U}$ is $(\mathbb{I}, 0)$, where $\mathbb{I}$ is the anyonic vacuum charge and 0 is the U(1) vacuum or the absence of any particle. In this thesis, I impose a hardcore constraint on the systems I study to prevent a site from hosting more than one particle. In other words, there is either a single nontrivial anyonic particle on a site or the site is vacant. The presence of a single nontrivial anyonic charge is represented by $(a, 1)$, where $a \neq \mathbb{I}$ and the U(1) charge 1 imposes a hardcore constraint of a single charge on the site. The use of the U(1) charge allows the counting of the anyonic charges fusing into a particular fusion channel irrespective of the outcome anyonic charge. A simple example is shown in Fig.4.6.

The anyonic MPS ansatz and the U(1) symmetry discussed in previous Sections can be used together to realize an Anyon × U(1)-symmetric MPS ansatz with the desired particle density. The minor modification needed in the new ansatz involves using the composite charges along with the composite fusion rules. To have an ansatz for a particular anyonic filling fraction, the method of shifting the U(1) charges can be employed. This only amounts to a shift in the U(1) charge labels, while the labels on the anyonic fusion space are not altered. The diagrammatic representation of tensors with the new symmetry group and the MPS ansatz constructed using them are the same as given in Section 4.2 and will not be reproduced again.

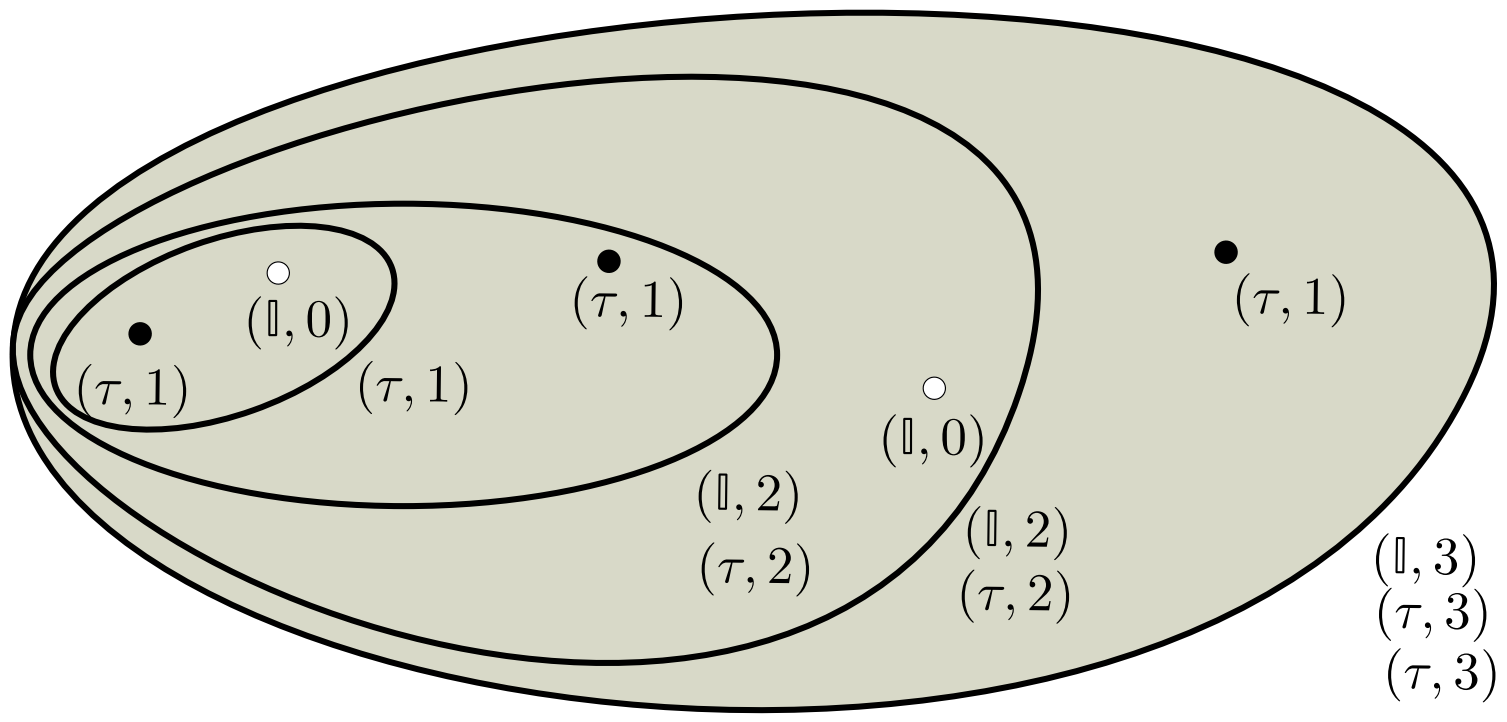

Figure 4.6: Fusion of composite charges situated on a manifold which supports either a single nontrivial anyonic Fibonacci charge $(\tau, 1)$ or a vacuum charge $(\mathbb{I}, 0)$ at each point. We impose a linear ordering where fusion proceeds from left to right. The set of charge outcomes is $\{(\mathbb{I}, 3), (\tau, 3), (\tau, 3)\}$. With the knowledge of the number of Fibonacci anyons on the manifold, the U(1) charge '3' indicates that there are 3 Fibonacci anyons fusing into both the $\mathbb{I}$ sector and the $\tau$ sector. As there are two outcomes with label $(\tau, 3)$, we introduce a degeneracy index $\mu_{(\tau,3)} = 1, 2$ to enumerate these outcomes. This simple example illustrates how composite charges fuse in symmetric tensor networks admitting both anyonic and U(1) symmetries. The U(1) helps "record" the number of nontrivial charges fusing into a particular channel. For instance, the outcome charge $(\mathbb{I}, 3)$—"heavy vacuum" is not the same as $(\mathbb{I}, 0)$—the ordinary vacuum charge.

### 4.5.2 Manipulations of Anyon $\times$ U(1) tensors

Topological manipulations such as F-moves, R-moves, fusion, splitting, elimination of loops, and vertical bends which apply to anyonic fusion trees can also be applied to anyonic tensors. Below, I present how this works in the case of tensors with Anyon$\times$U(1) symmetry. These are the typical machineries needed to manipulate (and contract) tensors during the optimisation of an anyonic MPS.

**F-moves**

The first topological manipulation required is that of changing the fusion order of the composite charges represented by the fusion tree. Let the basis fusion tree where fusion of charges proceeds from left to right be referred to as the standard basis. If instead a different fusion ordering is chosen, such as fusion from right to left, the charge outcomes are still the same, a fact guaranteed by the constraint of associativity. Formally, this associativity constraint corresponds to the Pentagon Equations, as given in e.g. [1]. The corresponding operation is the F-move, which transforms from one fusion basis to another one and is given

diagrammatically as

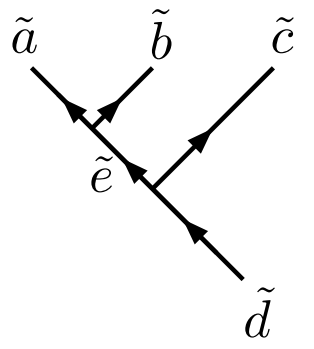

$$= \sum_{\tilde f} \left(F^{\tilde a \tilde b \tilde c}_{\tilde d}\right)^{\tilde f}_{\tilde e}$$

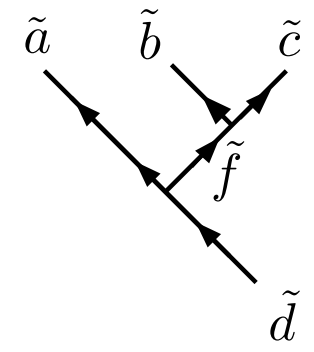

, (4.13)

where the coefficient $\left(F^{\tilde a \tilde b \tilde c}_{\tilde d}\right)^{\tilde f}_{\tilde e}$ decomposes into its anyonic and U(1) counterparts as,

$$\left(F^{\tilde a \tilde b \tilde c}_{\tilde d}\right)^{\tilde f}_{\tilde e} = \left(F^{abc}_{d}\right)^{f}_{e} \; \left(F^{n_a n_b n_c}_{n_d}\right)^{n_f}_{n_e}. \tag{4.14}$$

The factor $\left(F^{abc}_{d}\right)^{f}_{e}$ is given by the F coefficients of the anyon model while the U(1) factor is given by $\left(F^{n_a n_b n_c}_{n_d}\right)^{n_f}_{n_e} = N^{n_e}_{n_a n_b} N^{n_d}_{n_e n_c} N^{n_f}_{n_b n_c} N^{n_d}_{n_a n_f}$ which equals one if the charges are compatible or zero otherwise. F-moves may also be applied to any other pair of contiguous vertices appearing within a larger diagram.

It was noted in Ref. [95] that a symmetric tensor decomposes into a linear superposition of the degeneracy tensor and its spin network for systems with nontrivial symmetries such as SU(2). This result generalises to any quantum symmetries including anyonic symmetries. Therefore any section of the anyonic MPS can be decomposed into its degeneracy tensor and anyonic fusion network as

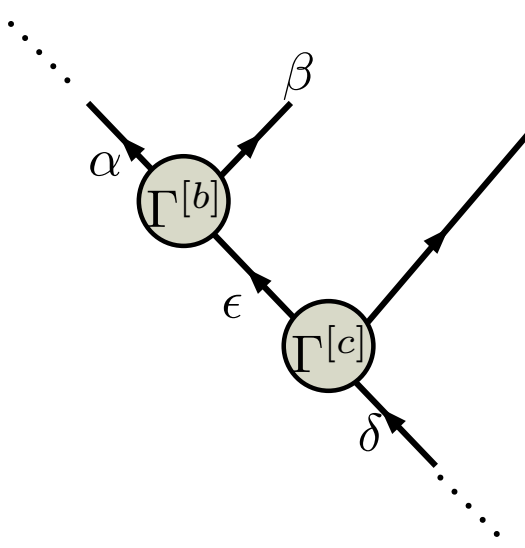

$$=$$

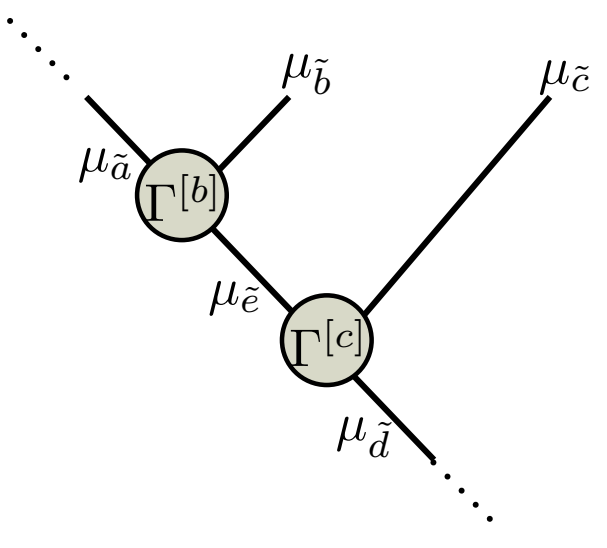

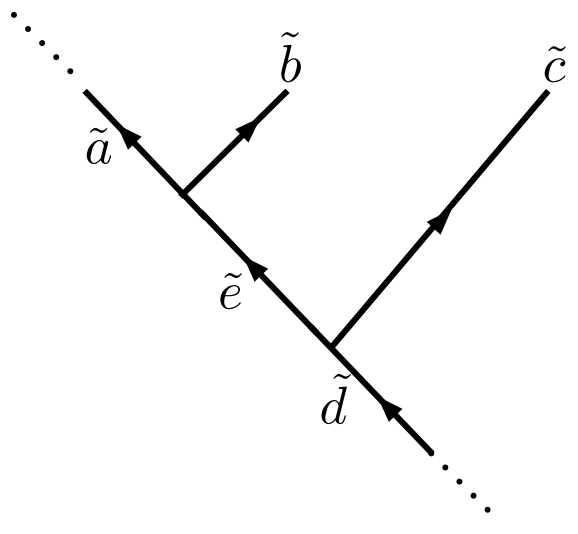

. (4.15)

The F-move is then applied on the anyonic diagram and the resulting F-factors are absorbed into the tensor resulting from contraction of the degeneracy tensor network. As shown, this process is valid for any portion of the diagram where the F-move operation can be applied.

### R-moves

Anyons have very rich particle exchange statistics which are neither bosonic nor fermionic. The exchange factors are encoded in the R-matrix, which is a matrix representation of the braid (or R-) move. The braid operation for composite anyonic charges is

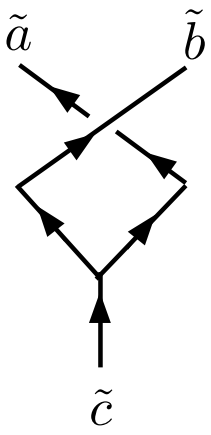

$$= R^{\tilde a \tilde b}_{\tilde c}$$

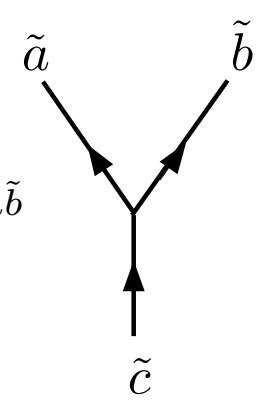

, (4.16)

where the factor $R^{\tilde a \tilde b}_{\tilde c}$ decomposes as

$$R^{\tilde a \tilde b}_{\tilde c} = R^{ab}_{c} R^{n_a n_b}_{n_c}, \tag{4.17}$$

and $R^{n_a n_b}_{n_c} = 1$ if $n_a + n_b = n_c$. The factors $R^{ab}_c$ are given by the specific anyon model.

The braid operation can be applied to a trivalent tensor to swap the two outgoing legs around the vertex while leaving the incoming leg untouched, similar to the braid diagram above. Algebraically, let $\mathcal{R}$ be the braid operator that acts on an anyonic tensor. Applied on $T^{\beta\alpha}_{\gamma}$, we have

$$T^{\alpha\beta}_{\gamma} = \mathcal{R}\left(T^{\beta\alpha}_{\gamma}\right). \tag{4.18}$$

Diagrammatically, this is represented as

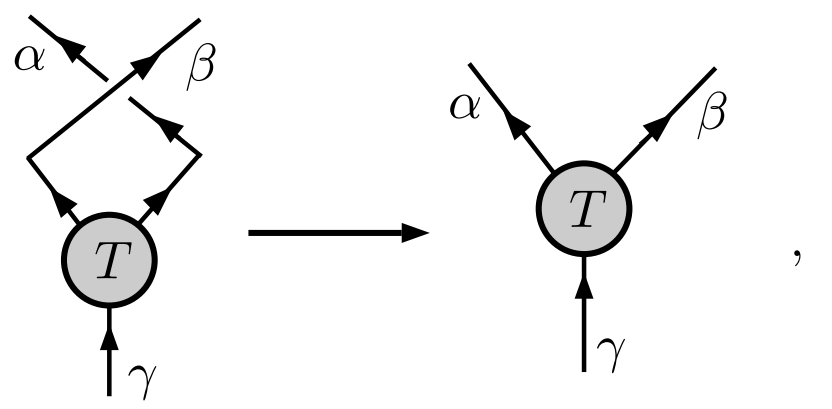

(4.19)

where the crossing in the diagram represent the braiding operation. The detail of the braiding process in shown in Fig. 4.7

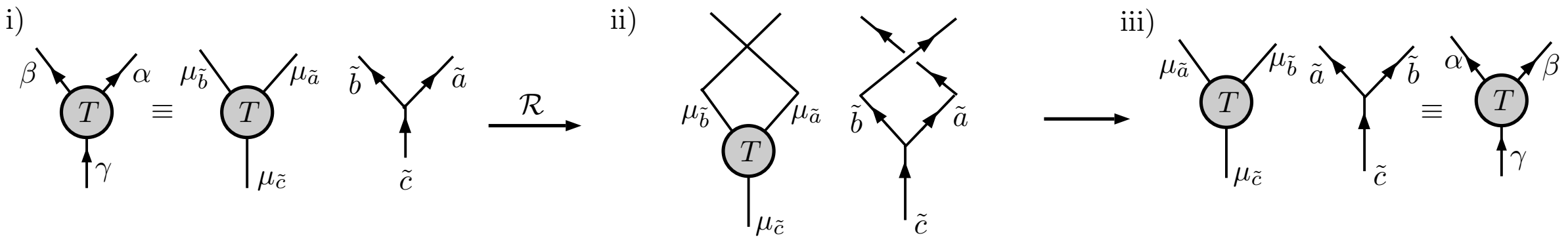

Figure 4.7: To braid the indices of the anyonic tensor $T^{\beta\alpha}_{\gamma}$: i) Decompose the anyonic tensor $T^{\beta\alpha}_{\gamma}$ into its degeneracy tensor and fusion network parts. ii) Apply the braid operation $\mathcal{R}$, by applying the permutation operation (reviewed in Sec. 2.2) on the degeneracy tensor and the braid operation on the splitting tree. iii) Recombine the degeneracy tensor and fusion part to give the desired braided anyonic tensor $T^{\alpha\beta}_{\gamma}$.

### Fusion tensor and loop factors

A trivalent tensor can be used to define a linear map from the tensor product of two Hilbert spaces $\mathbb{V}^{(\mathrm{A})}$ and $\mathbb{V}^{(\mathrm{B})}$ (which can possibly be degenerate) to a new composite Hilbert space $\mathbb{V}^{(\mathrm{C})}$. Let $\{|0\rangle, |1\rangle, \ldots, |d_{\mathrm{A}} - 1\rangle\}$ be the set of basis for $\mathbb{V}^{(\mathrm{A})}$ of dimension $d_{\mathrm{A}}$, and $\{|0\rangle, |1\rangle, \ldots, |d_{\mathrm{B}} - 1\rangle\}$ as the set of basis for $\mathbb{V}^{(\mathrm{B})}$ of dimension $d_{\mathrm{B}}$. Since $\mathbb{V}^{(\mathrm{C})} \cong \mathbb{V}^{(\mathrm{A})} \otimes \mathbb{V}^{(\mathrm{B})}$, the dimension of $\mathbb{V}^{(\mathrm{C})}$ is $d_{\mathrm{A}} d_{\mathrm{B}}$. We can let the basis of $\mathbb{V}^{(\mathrm{C})}$ be $\{|0\rangle, |1\rangle, \ldots, |d_{\mathrm{A}} d_{\mathrm{B}} - 1\rangle\}$. A linear map $T : \mathbb{V}^{(\mathrm{A})} \otimes \mathbb{V}^{(\mathrm{B})} \to \mathbb{V}^{(\mathrm{C})}$ can be written as

$$T = \sum_{a,b,c} T^{c}_{ab} \; |c\rangle \langle a| \otimes \langle b|, \tag{4.20}$$

which sends product basis $|a\rangle \otimes |b\rangle \in \mathbb{V}^{(\mathrm{A})} \otimes \mathbb{V}^{(\mathrm{B})}$ to basis $|c\rangle \in \mathbb{V}^{(\mathrm{C})}$ in a unique fashion. Technically, $T$ is a bijection which implements a one-to-one correspondence between $\{|a\rangle \otimes |b\rangle\}$ and $\{|c\rangle\}$. If $\mathcal{A} = \{0, 1, \ldots, d_{\mathrm{A}} - 1\}$ and $\mathcal{B} = \{0, 1, \ldots, d_{\mathrm{B}} - 1\}$ are the sets of labels for the basis vectors of $\mathbb{V}^{(\mathrm{A})}$ and $\mathbb{V}^{(\mathrm{B})}$, we bijectively map the elements of the (Cartesian) product set $\mathcal{A} \times \mathcal{B} = \{(a, b) | a \in \mathcal{A}, b \in \mathcal{B}\}$ to elements of the set $C = \{0, 1, \ldots, d_{\mathrm{A}} d_{\mathrm{B}} - 1\}$ using

$$(a, b) \mapsto c = (a d_B + b), \tag{4.21}$$

| a | $\mu_a$ | b | $\mu_b$ | c | $\mu_c$ |
|---|---|---|---|---|---|
| 0 | 1 | 0 | 1 | 0 | 1 |
| 0 | 1 | 1 | 1 | 1 | 1 |
| 1 | 1 | 0 | 1 | 1 | 2 |
| 1 | 1 | 1 | 1 | 0 | 2 |

Table 4.1: The mapping from tensor product state $|a, \mu_a\rangle \otimes |b, \mu_b\rangle$ to a new basis $|c, \mu_c\rangle$ using the $\mathbb{Z}_2$ fusion rule. Degeneracy basis labels $\mu_x$ for each charge $x \in \{a, b, c\}$ have been included to count distinct fusion into a particular charge.

where $c \in C$. In terms of basis vectors, this map is $|a\rangle \otimes |b\rangle \mapsto |c = ad_B + b\rangle$. The coefficient $T^c_{ab}$ is 1 for a valid map ($ab \to c$) and zero otherwise. As a simple example, let us consider qubits, i.e. let $\mathbb{V}^{(\mathrm{A})} = \mathbb{V}^{(\mathrm{B})} = \{|0\rangle, |1\rangle\}$ with the basis labels as elements in $\mathbb{Z}_2 = \{0, 1\}$. The map in Eq. (4.21) gives

$$\left.\begin{array}{ll}(0,0) \mapsto 0, & (0,1) \mapsto 1 \\ (1,0) \mapsto 2, & (1,1) \mapsto 3\end{array}\right\}, \tag{4.22}$$

whence $\mathbb{V}^{(\mathrm{C})} = \{|0\rangle, |1\rangle, |2\rangle, |3\rangle\}$ indexed by the set of labels in $\mathbb{Z}_4 = \{0, 1, 2, 3\}$.

Now, assume we add some structure to the map, then $a$ and $b$ will be related via a specified 'fusion rule'. The ordered pair $(a, b)$ will now have a defined relationship $(a, b) = a \times b$, where the '×' represents fusion. As a simple example, let us consider the $\mathbb{Z}_2$ fusion rule, which is just the ordinary addition of integers modulo two: $a \times b = (a + b) \mod 2$. The outcomes $a \times b \to c$ are

$$0 \times a \to a \quad \forall \mathrm{a} \in \mathbb{Z}_2 \qquad \text{and} \qquad 1 \times 1 \to 0. \tag{4.23}$$

The outcome $c = 1$ resulting from the fusion $0 \times 1$ and $1 \times 0$ is degenerate, so also is $c = 0$, which results from $0 \times 0$ and $1 \times 1$. The degenerate outcomes will now be indexed by an additional label, namely, the degeneracy index $\mu_c$ associated with label $c$, in order to maintain the uniqueness of the mapping. The linear map using the $\mathbb{Z}_2$ fusion rule is given in Table 4.1. In essence, because fusion of (charge) labels can induce degeneracy of (charge) labels, we therefore add a degeneracy index to each charge label.

Therefore, as previously discussed in Ref. [94], the linear map tensor for systems with Abelian symmetry can be written in general as

$$T = \bigoplus_{a,b,c} N^c_{ab} \sum_{\mu_a \mu_b \mu_c} \left(T^c_{ab}\right)^{\mu_c}_{\mu_a \mu_b} |c, \mu_c\rangle \langle a, \mu_a| \langle b, \mu_b|, \tag{4.24}$$

where the tensor $T$ is constructed as a block of tensors $T^c_{ab}$, with each block identified by the charge triple $(a, b, c)$, and the element within each block is indexed by the degeneracy indices $(\mu_a, \mu_b, \mu_c)$ associated with the charge triple $(a, b, c)$ respectively.

We generalize this to anyonic systems admitting Anyon × U(1) symmetries as follows: Let two sites of an anyonic system be described by a degenerate Hilbert space $\mathbb{V}^{(A)}$ and $\mathbb{V}^{(B)}$ with their basis vectors labelled by $\{\alpha = (\tilde{a}, \mu_{\tilde{a}})\}$ and $\{\beta = (\tilde{b}, \mu_{\tilde{b}})\}$. Let the anyonic fusion product define a "fusion map" $\tilde{N}^{\gamma}_{\alpha,\beta}$ from multi-indices $\alpha$ and $\beta$ to a new multi-index $\gamma$. The anyonic fusion map creates a new vertex, and we normalize it according to diagrammatic isotopy convention. The fusion tensor is represented as in Fig.4.8(a). Unlike the case of abelian symmetry, when working in the diagrammatic isotopy convention for non-Abelian anyons, the coefficient of the fusion tensor for a valid fusion map $\alpha \times \beta \to \gamma$ is the vertex normalization factor $\left(\frac{d_{\tilde{c}}}{d_{\tilde{a}} d_{\tilde{b}}}\right)^{1/4}$, and zero for an invalid map.

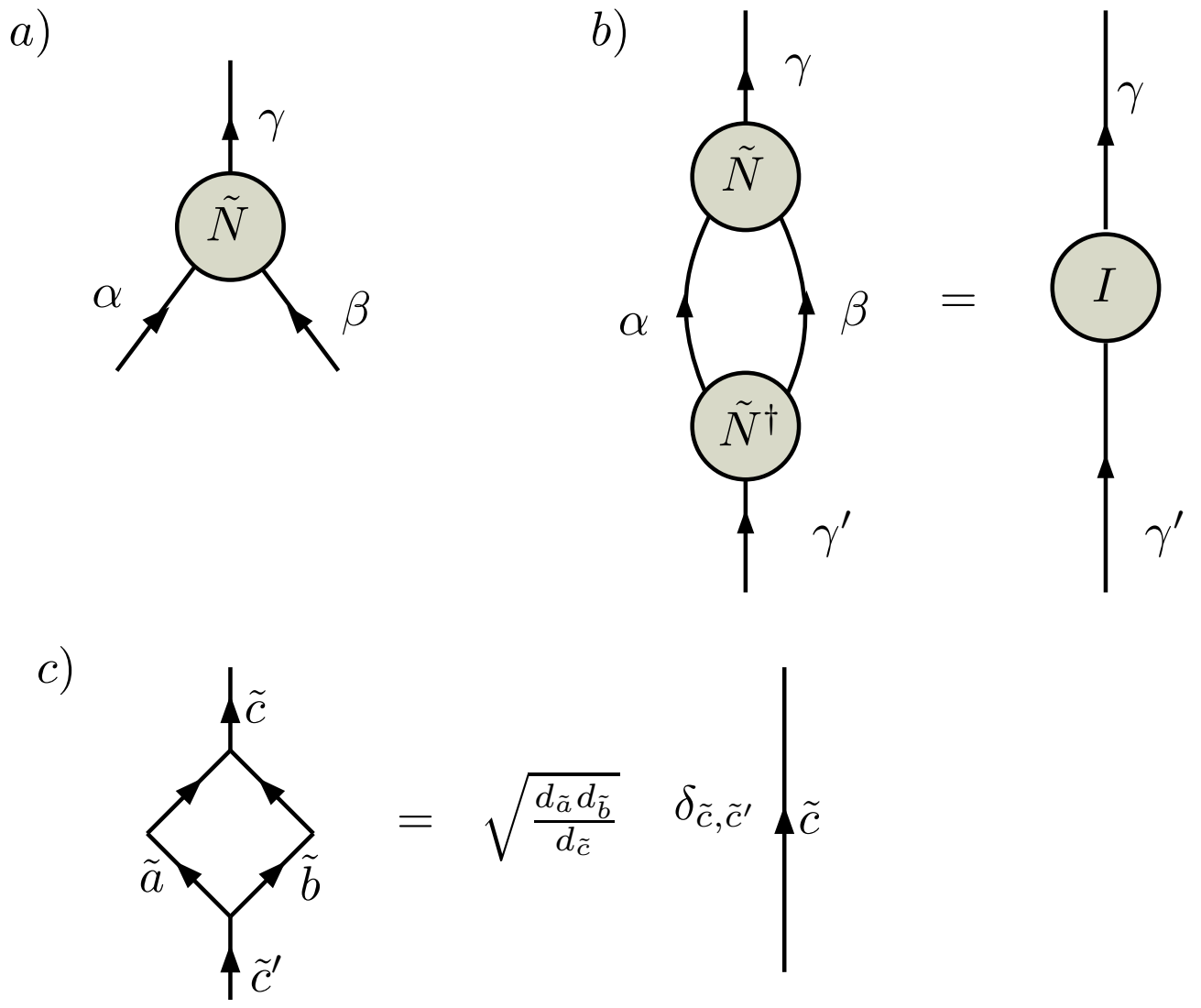

Figure 4.8: a) The diagrammatic representation of the anyonic fusion tensor $\tilde{N}$. This can be expressed in a block structure form, where the elements of $\tilde{N}$ are $\tilde{N}^{\gamma}_{\alpha\beta} = \left(\tilde{N}^{\tilde{c}}_{\tilde{a}\tilde{b}}\right)^{\mu_{\tilde{c}}}_{\mu_{\tilde{a}}\mu_{\tilde{b}}} = \left(\frac{d_{\tilde{c}}}{d_{\tilde{a}} d_{\tilde{b}}}\right)^{1/4}$ for a valid map $\alpha \times \beta \to \gamma$, and zero otherwise. b) The fusion tensor $\tilde{N}^{\gamma}_{\alpha\beta}$ and its Hermitian conjugate $\tilde{N}^{\alpha\beta}_{\gamma'}$ contracts to the identity operator defined on the new multi-index $\gamma$. c) Loop factor as given in Ref. [1]. This factor has been accounted for in the contraction diagram of b) to give the identity.

The anyonic fusion tensor $\tilde{N}^{\gamma}_{\alpha\beta}$ and its Hermitian conjugate, the splitting tensor $\tilde{N}^{\alpha\beta}_{\gamma}$, are linear maps and fulfill the condition that $\tilde{N}^{\alpha\beta}_{\gamma'} \tilde{N}^{\gamma}_{\alpha\beta} = I^{\gamma}_{\gamma'}$ (Einstein summation convention assumed), which is an identity operator on the new (degenerate) space $\mathbb{V}^{(C)}$ as shown in Fig.4.8(b). The loop present in the contraction in Fig.4.8(b) is eliminated using the relation in Fig.4.8(c). It should be noted that the vertex factor $\sqrt{\frac{d_{\tilde{c}}}{d_{\tilde{a}} d_{\tilde{b}}}}$ in the definition of the fusion tensor $\tilde{N}$ and splitting tensor $\tilde{N}^\dagger$ cancels with the loop factor $\sqrt{\frac{d_{\tilde{a}} d_{\tilde{b}}}{d_{\tilde{c}}}}$, and hence the identity matrix operator in Fig.4.8(b) does not contain any factor of the quantum dimension. Also, it should be noted that in Anyon×U(1) symmetry, the quantum dimension $d_{\tilde{a}}$ decomposes into the product $d_{\tilde{a}} = d_a d_{n_a}$, where $d_a$ is the quantum dimension of the anyonic charge $a$ and $d_{n_a}$ is the dimension of the U(1) charge, which always have a trivial value of one.

### Vertical Bends

Bending a charge line horizontally is trivial, as timelike (i.e. horizontal) slices of the fusion tree are invariant under topology-preserving deformations. However, vertically bending an anyon charge line is non-trivial and involves reversing the orientation of the anyon worldline. The details of how to resolve vertical bends in terms of F-moves were worked out in Chapter 3. We do not repeat the derivations here but only mention the minor changes in the presence of U(1) charges. Adapted to systems with Anyon × U(1) symmetry, the equation for the left and right bends respectively are

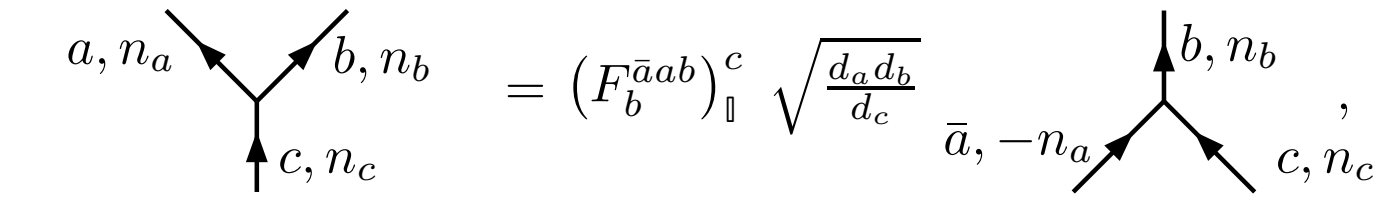

(4.25)

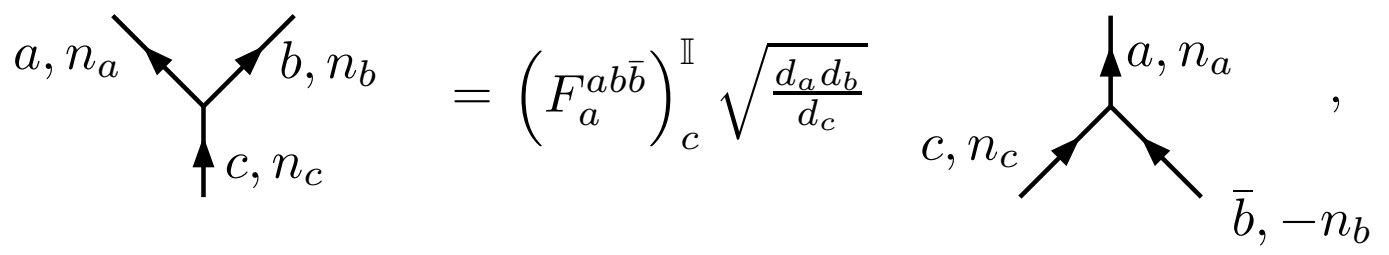

(4.26)

where in addition to the anyons satisfying their fusion rules, the U(1) charges around the vertex have to satisfy the condition $n_a + n_b = n_c$. The dual of the anyonic charge $a$ and the U(1) charge $n$ are denoted as $\bar{a}$ and $-n$ respectively, which are the new charge labels on the bent legs. We have only shown the F-factors of the anyons in the equations above, as these may be nontrivial, but the F-factors for U(1) charges are always trivial as long as the charges satisfy the modular addition fusion rule.

The vertical bends can be applied to anyonic tensors to transform an order-three tensor as,

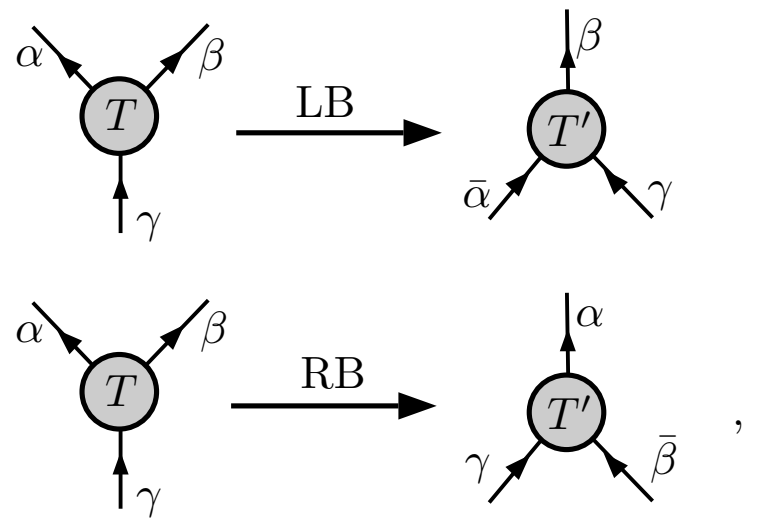

(4.27)

where the first equation is the left bend (LB) and the second is the right bend (RB). The bend factors in Eq. 4.25 and 4.26 have been absorbed into the transformed tensor $T'$. In terms of explicit symbolic expression, the left and right bends of order-three anyonic tensors are,

$$T'^{(\tilde{b},\mu_{\tilde{b}})}_{(\tilde{a},\mu_{\tilde{a}})(\tilde{c},\mu_{\tilde{c}})} = \left(F^{\tilde{a}^*\tilde{a}\tilde{b}}_{\tilde{b}}\right)^{\tilde{c}}_{\mathbb{I}} \sqrt{\frac{d_{\tilde{a}} d_{\tilde{b}}}{d_{\tilde{c}}}}\, T^{(\tilde{a},\mu_{\tilde{a}})(\tilde{b},\mu_{\tilde{b}})}_{(\tilde{c},\mu_{\tilde{c}})}, \tag{4.28}$$

$$T'^{(\tilde{a},\mu_{\tilde{a}})}_{(\tilde{c},\mu_{\tilde{c}})(\tilde{b},\mu_{\tilde{b}})} = \left(F^{\tilde{a}\tilde{b}\tilde{b}^*}_{\tilde{a}}\right)^{\mathbb{I}}_{\tilde{c}} \sqrt{\frac{d_{\tilde{a}} d_{\tilde{b}}}{d_{\tilde{c}}}}\, T^{(\tilde{a},\mu_{\tilde{a}})(\tilde{b},\mu_{\tilde{b}})}_{(\tilde{c},\mu_{\tilde{c}})}, \tag{4.29}$$

where $\tilde{a} = (a, n_a)$ is the composite anyonic-and-U(1) charge. The conjugation of the composite charge $\tilde{a}$ is now denoted as $\tilde{a}^*$.

## 4.6 Conclusion

In conclusion, by constructing the appropriate tensor objects having Anyon × U(1) symmetry (e.g. two index and three index tensors, fusion tensors, etc.), one can construct a MPS tensor network for anyonic systems admitting both anyonic and U(1) symmetry. This TN ansatz can be variationally optimised to ground state by means of anyonic TN algorithms such as anyonic TEBD algorithm [41] or anyonic DMRG [43].

Properties of the ground state such as ground state energy, entropic scaling and correlation functions can be computed using approaches similar to those used for non-symmetric tensor networks, but now modified for anyonic systems using the anyonic tensor objects and concepts introduced in this chapter and the preceding one. All the manipulations necessary to do the calculations basically involve contraction of tensors, where the details of the contraction comprises of F-moves, bending of charge lines, fusion and splitting of tensors, and elimination of loops until the tensor network is fully contracted.

In the next chapter, I present a practical implementation of the time-evolving block decimation (TEBD) algorithm for anyonic systems applied to symmetric MPS having either or both of anyonic and U(1) symmetries. I benchmarked my implementation by using the imaginary time evolution to compute the ground state of lattice models having both Abelian and non-Abelian anyonic symmetries in the thermodynamic limit.

# 5

# Implementation of Time Evolving Block Decimation algorithm for anyonic systems

## 5.1 Introduction

The anyonic-TEBD algorithm was presented in Ref. [41]. However, my impression is that the presentation might be a bit challenging for someone who is new to incorporating non-Abelian symmetries into tensor networks. Therefore, my aim in this chapter is to provide an introduction to anyonic-TEBD algorithm, often offering some programming tips (and pseudocodes) where necessary. Moreover, in this chapter in addition to previous chapters (most especially, Chapters 3 and 4), I provide some refinements to the anyonic-TEBD algorithm which were not mentioned in Ref. [41]. In particular, I will show that by using pairwise contraction of tensors (or at most contraction of three tensors a time), the contraction cost can be reduced below that in Ref. [41]—where the "whole network approach" was adopted. Also, the use of vertex normalisation was not consistent in Ref. [41], which I now use consistently here.

The anyonic-TEBD tensor network is very similar to the conventional non-symmetric TEBD, except that the tensors making up the network are *symmetric tensors*: these are tensors that incorporate structure arising from the (quantum) symmetry group of the anyon model of interest. They are usually sparse, and having a means to directly access the non-zero elements is what offers the computational gain achieved with symmetric tensor network simulations.

The TEBD algorithm, in general, is based on finding an efficient decomposition of both the low-energy state $|\psi\rangle$ of a quantum system and the imaginary-time gate $\hat{G}(\tau) = e^{-\hat{H}\tau}$ of a local Hamiltonian $\hat{H}$ into a network of local low-order tensors. The MPS offers an efficient decomposition for a one-dimensional non-critical quantum state—based on bipartite Schmidt decomposition, while Trotter-Suzuki method offers an efficient decomposition for the imaginary-time gate. These decompositions turn a one-dimensional quantum system into a "two-dimensional gate array," as shown in Fig. 5.1. The tensors in this diagram are charge-conserving, respecting the quantum symmetry of the anyon model. Arrows are used to indicate that charges on the legs and bonds of the TN could be non-self dual. In our diagrammatic representation, we adopt the convention that diagrammatic objects (e.g.

tensors, legs of tensors, etc.) are arranged from left-to-right and bottom-to-top of the page. Adopting this convention, all the program functions (or routines) I suggest in this chapter can be automated.

Similar to conventional TEBD, it can be seen from Fig. 5.1 that the contraction of the (imaginary-time) gate at each time interval $\delta\tau$ can be reduced to a sequence of (local) contractions over two sites, achieved by the exploiting the local structure of the tensor network. Apart from other programming details which pertain to anyons, Fig. 5.2 is the only contraction scheme needed to be implemented to contract the entire anyonic-TEBD tensor network.

To simulate any class of anyon model (e.g. Fibonacci, Ising, $\mathbb{Z}_q$ series for $q \in \mathbb{Z}_+$, etc.), the specific anyon data (e.g. fusion rules, quantum dimensions, braid-moves coefficients, etc.) need to be built directly into the tensors making up the network. In order words, no one single tensor network can simulate all anyon models, but each network is adapted to work for a specific anyon model. The data are used to construct the tensors that make up the network, and also the tensors which are used to manipulate the network. In the next section, I give an exhaustive list of routines that can be used to perform the anyonic-TEBD algorithm.

The remaining parts of this chapter are structured as: In Section 5.2, I present a list of routines that are sufficient to implement the anyonic-TEBD. In Section 5.3, I present the two-site contraction of anyonic tensors. In Section 5.4, I review the convergence measure used to check for convergence of the MPS in TEBD simulations. In Section 5.5, I present a list of test models which I used to benchmark my anyonic-U(1) MPS ansatz and my implementation of the anyonic-TEBD algorithm. Finally, I end the chapter with discussion and some concluding remarks.

## 5.2 List of anyonic-TEBD accessory functions

In a practical implementation of anyonic-TEBD, the functions needed to be implemented can be broadly divided into two different categories: (i) those that specify the anyon model data, for example, the anyon model fusion rules, F-move and R-move matrix elements, and the quantum dimensions $d$ of the charges in each anyon model. A more complex anyonic tensor network may need more data, but these suffice for MPS and TEBD simulations. (ii) Those that are used to create and manipulate the anyonic tensors which make up the tensor network.

In my presentation, I first review the set of functions in the first category, using Fibonacci or Ising anyons to give concrete example where necessary. Then I review the set of functions in the second category. Let me remind the reader again that the aim of this chapter is on the practical implementation of anyonic TEBD, and not on the mathematical framework which were already presented in Chapters 2, 3, and 4. As such, I will assume that the reader is already comfortable with ideas of tensor networks, non-Abelian anyons, and perhaps some anyonic tensor networks formalism.

### 5.2.1 Anyon data routines

In this subsection, I give a list of routines that can be used to programme the anyonic data which are necessary to manipulate an anyonic fusion network. For concreteness, I will make examples using Ising and/or Fibonacci anyons. We recall that the set of charges in the Ising anyon theory is $\mathcal{A}_{\text{Ising}} = \{0, 0.5, 1\}$ (or in symbolic form $\{\mathbb{I}, \sigma, \psi\}$), and that of Fibonacci anyon is $\mathcal{A}_{\text{Fib}} = \{0, 1\}$ (or $\{\mathbb{I}, \tau\}$).

Specifying the data for any anyon model is straightforward, and can easily be programmed.

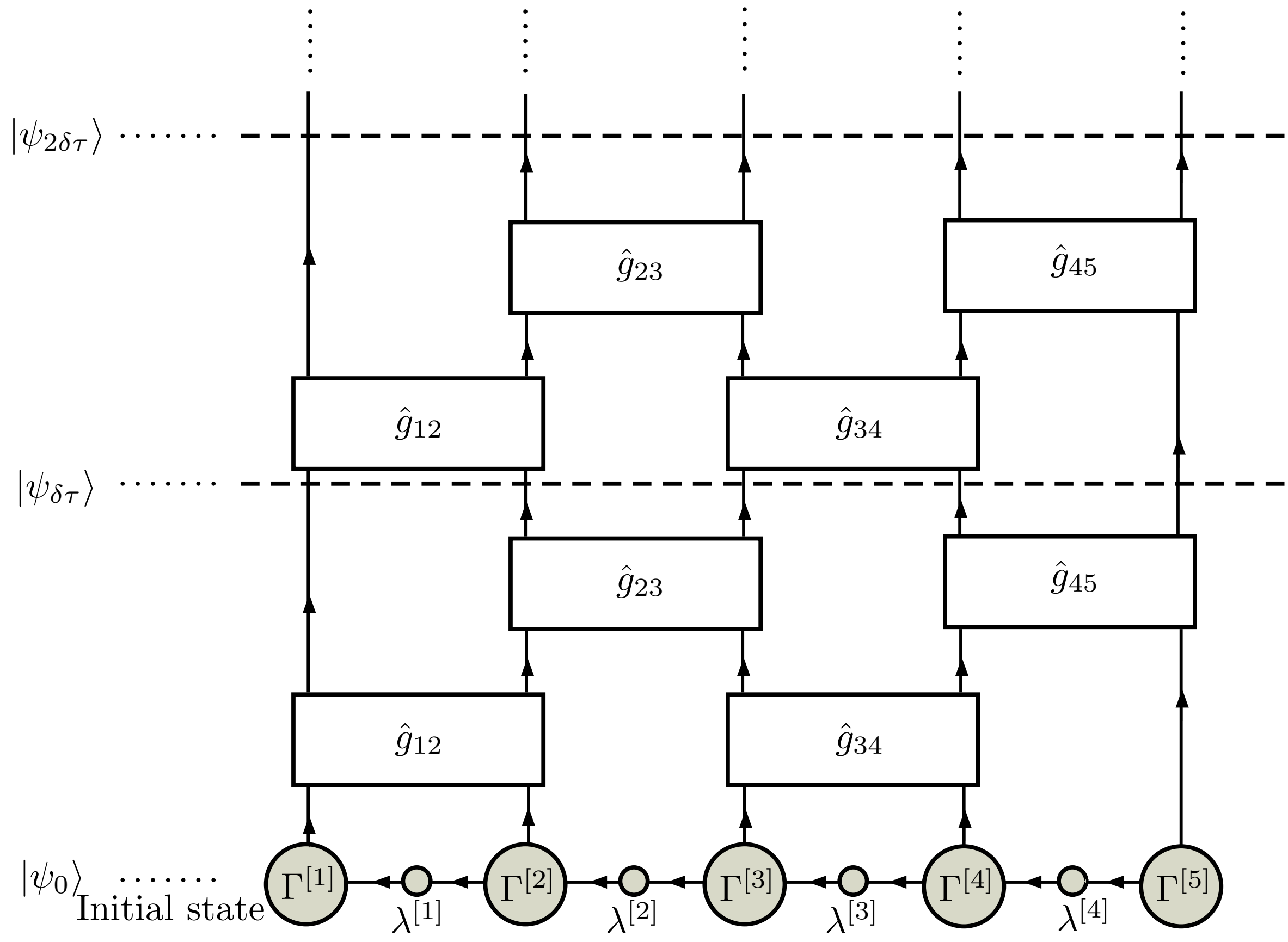

Figure 5.1: The anyonic MPS and TEBD tensor network. The horizontal axis is the lattice sites $n$, here n=5, and the vertical axis is the (imaginary) time. The initial MPS state is denoted $|\psi_0\rangle$. The (imaginary-time) gate $\hat{G}(\tau)$ is decomposed into a sequence of 'even" and "odd" sites gate for a local Hamiltonian $\hat{H}$ defined on two sites. The optimisation of the MPS from iteration to iteration is demarcated by dashed lines. In practice, one continues to perform this iteration in (imaginary) time until a desired convergence is achieved. Some comments are in order. First, I note that this network has been turned up-side-down as opposed to the conventional diagrammatic scheme adopted for MPS and TEBD tensor networks. This is done in order to have the leaves of the TN point upwards the page to make it similar to the popular diagrammatic convention used in specifying fusion tree basis of a system of anyons. Second, arrows have been used to indicate that anyonic charges can be non-self dual, whereas these would not be necessary in ordinary tensor network (i.e. TN without symmetries). The charge labels have been suppressed for clarity.

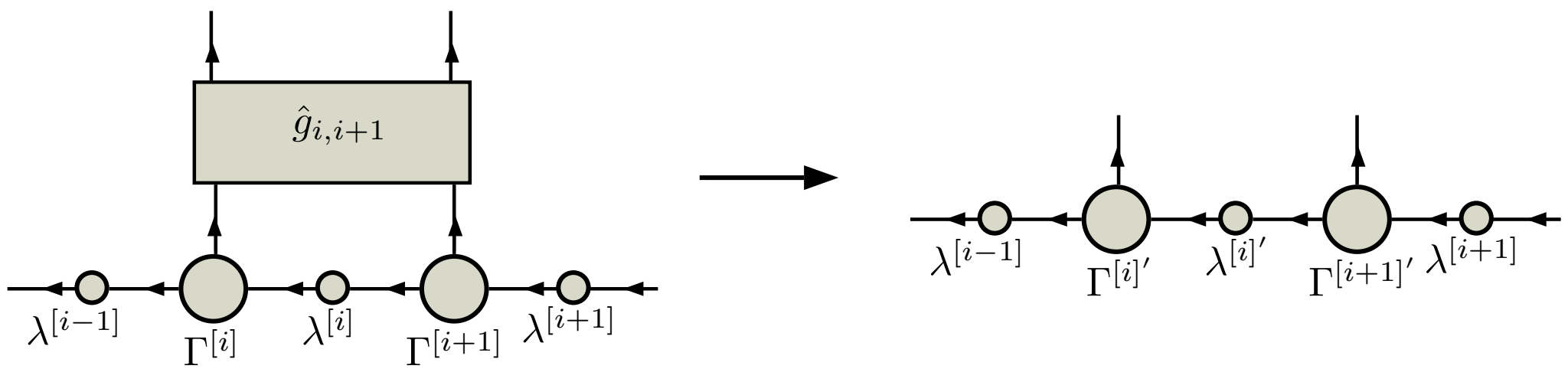

Figure 5.2: A two-site section of the network in Fig. 5.1. The network on the left is contracted to update the tensors on the two sites as shown in the right figure.

Some of the routines I give below depend recursively on others, and for the ease of referencing of a particular routine within another, I introduce non-standard mnemonics for the routines. Below is an example of how each routine will be stated subsequently:

*Routine "`routine name`"—a short descriptive title for the routine: a longer description of routine, and a possible implementation.*

1. Routine "`FuseAnyons`"—fusion of anyonic charges: Given two anyonic charges $a$ and $b$, we can determine the set of possible charge outcomes $c$ using the fusion rules of the anyon model. For this, a table of the fusion rules (or outcomes) can be created at the beginning of any simulation. Using Ising anyons as an example, its fusion rules are given in the table:

$\mathrm{X}_{\mathrm{Ising}} =$

| x | 0 | 0.5 | 1 |
|---|---|---|---|
| 0 | 0 | 0.5 | 1 |
| 0.5 | 0.5 | [0 1] | 1 |
| 1 | 1 | 0.5 | 0 |

, (5.1)

where "x" denotes fusion operation. This table can be saved using a data structure that allows for multiple types like scalars and vectors—since fusion of anyons can give more than one charge outcome which I conveniently group together as a vector of charge outcomes. In a software like MATLAB, a cell data structure can be used. In that case, the table may be saved as $X_{\text{Ising}} = \backslash\{0\,,0.5\,,1\,;0.5\,,[0\quad 1]\,,1\,;1\,,0.5\,,0\}$. To read the fusion outcome of charges $a$ and $b$, you can use $\mathtt{c} = X(2a+1, 2b+1)$. So, for instance, if $a = 0.5$ and $b = 0.5$, then c=$X\{2,2\}$, which gives $c = [0 \quad 1]$.

2. Routine "`isCompatible`"—check for the compatibility of charges around a vertex: Anyonic tensors are arranged into blocks of degeneracy tensors which are indexed by the charges of the anyon model. For example, each sector of an order-three anyonic tensor is indexed by a triplet of charges, which either forms a fusion vertex $(a, b, c)$, or a splitting vertex $(c, a, b)$, denoted respectively as:

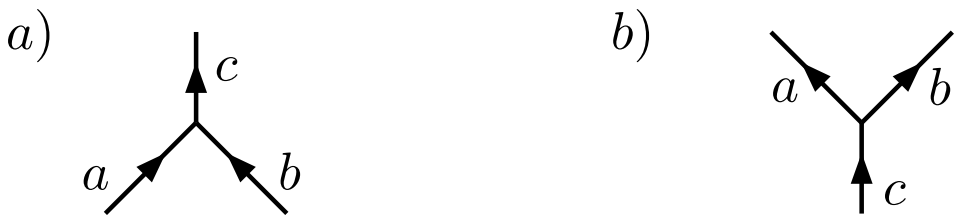

. (5.2)

To create these tensors, it becomes necessary to check for the "compatibility" of these fusion and splitting vertices using the fusion rules of the anyon model. The implementation of this routine recursively depends on Routine `FuseAnyons`. To determine if charges $(a, b, c)$ is a compatible fusion vertex, first fuse charges $(a, b)$ using Routine `FuseAnyons` to give an array of charges $d$, then check if any of the charge outcomes in $d$ corresponds with the charge $c$. If true, assign a *truth* value of 1, otherwise assign 0. The implementation for checking for a valid splitting vertex is similar. This function and other functions proposed here can be automated if adopting my diagrammatic convention which was earlier mentioned. So for instance, if the function's header is `flag=isFusionVertex(a,b,c)`, this ordering of the arguments of the function will mean that the program checks that charges $a$ and $b$ is "fusion-compatible" with charge $c$, while a different function with another header such as `flag=isSplittingVertex(c,a,b)` will check that charge $c$ splits into charges $a$ and $b$. Adopting this convention is not strictly necessary, unless that it allows one to pass high-level commands without worrying about the details, and therefore further enhances the automation of anyonic network algorithms.

3. Routine "`BraidFactors`"—R-move (braid) factors: The braiding of two charges $a$ and $b$ having a direct fusion channel with charge $c$, acquires a complex phase factor $R_c^{a,b}$. The implementation of this function depends on Routine `isCompatible`—which itself depends on Routine `FuseAnyons`, as previously stated. In implementation, when a call to the function is made by passing a triplet of charges $(a, b, c)$, the function should first check that the charges are fusion-compatible, then it returns the braid factor. The braid factors are saved to disk for each anyon model.

4. Routine "`FMoveFactors`"—F-Matrix factors: F-move is used to change fusion basis involving three anyonic charges $(a, b, c)$ fusing into charge $d$, with intermediate charges $e \in \{a \times b\}$ and $f \in \{b \times c\}$. The F-move diagram is:

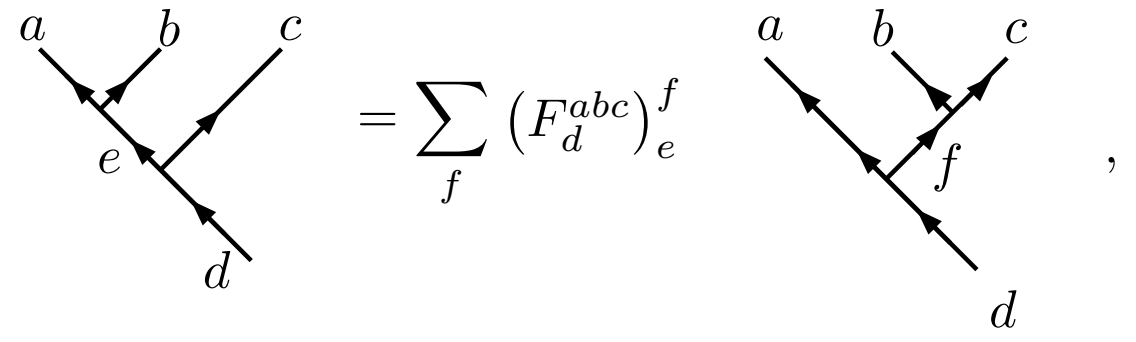

(5.3)

where the matrix elements are saved for each anyon model. As in the case of the R-move braid factors, these F-move matrix elements are entered manually, and saved to the computer's disk, and retrieved when needed.

5. Routine "`QDim`"—Quantum dimension: The last of the anyon model data we will need in anyonic TEBD is the quantum dimension $d$ of the charges in the model. These values are saved to disk.

The data of several different classes of anyons are tabulated in Chapter five of Ref. [1].

### 5.2.2 Anyonic tensor network routines

In this subsection I give a list and detailed explanation of the set of accessory functions which are used to create the anyonic tensor objects which form the anyonic MPS tensor network, and the tensors used to manipulate the network during the TEBD optimisation algorithm. As the MPS with TEBD is a trivalent tensor network, the tensors comprising the network are at most order-three.

Anyonic tensor elements are indexed by both charge $a$ and degeneracy index $\mu_a$. Let the maximum degeneracy in each charge sector $a$ be $\nu_a$, therefore, $\mu_a = 1, 2, \ldots, \nu_a$. In practice, the tensors are arranged as blocks of degeneracy tensors, with each block indexed by the charge $a$, and each element within a block by the degeneracy index $\mu_a$.

The charge-degeneracy (greek) index $\alpha$ should be understood as an array of the charges and their degeneracies that are supported on that tensor. Therefore, except where stated otherwise, the free legs (and shared legs) of the tensors are indexed by the (greek) charge-degeneracy indices (or simply referred to as indices for brevity). The summary of the plan of my exposition in the following is, first, I will illustrate how to "fuse" charge-degeneracy indices, which I later use to show how anyonic tensors are created and manipulated. I then introduce some primitives used for contracting anyonic tensors.

1. Routine "`FuseIndices`"—fuse charge-degeneracy indices: Let the charge-degeneracy indices to be fused be $\alpha = [a, \nu_a]$ and $\beta = [b, \nu_b]$, and their outcome be $\gamma = [c, \nu_c]$, i.e. $\alpha \times \beta \to \gamma$. The degeneracy of charge $c$ is obtained from the relation

$$\nu_c = \sum_{a,b} \nu_a \nu_b \delta_{a\times b,c}, \tag{5.4}$$

where $\delta_{a\times b,c}$ checks that charges $a$ and $b$ is fusion-compatible with charge $c$, which can be checked with Routine `isCompatible`. Note that $\delta_{a\times b,c} = 1$ if compatible, 0 if not. The basic steps of a naive implementation of this routine are:

(a) Create an initial array of charge-degeneracy index $\gamma = [c, \nu_c]$, where the initial degeneracy counts, $\nu_c$, for each charge $c$ is set to zero, $\nu_c = 0$.

— Using a "for-loop" for each of steps $(b)$, $(c)$, and $(d)$ —

(b) Select a new value of the charge $c$ in the array $\gamma$

(c) Select a new value of the charge $a$ in the array $\alpha$

(d) Select a new value of the charge $b$ in the array $\beta$

---

(e) Determine if the selected charges $(a, b, c)$ are fusion-compatible. If yes, proceed to next step, otherwise return to either steps (d), (c), or (a), depending on the number of charges remaining at each iteration.

(f) Get the degeneracy $\nu_a$ and $\nu_b$ of charges $a$ and $b$, respectively. Increase the degeneracy $\nu_c$ of charge $c$ through $\nu_c = \nu_c + (\nu_a \times \nu_b)$.

(g) Return back to step (d) if there are still charges in the array $\beta$, otherwise go to the next step.

(h) Return back to step (c) if there are still charges in the array $\alpha$, otherwise go to the next step.

(i) Finally, update the degeneracy count $n_c$ of charge $c$ in the array $\gamma$, and proceed to the next step.

(j) Return back to step (b) if there are still charges in the array $\gamma$, otherwise end program.

2. Routine "`AnyonicMatrix`"—create anyonic matrices: An anyonic matrix (or two-index tensor) has two index labels on the legs of the matrix. In order to create an anyonic matrix, first determine the charge-degeneracy indices $\alpha$ and $\beta$ on the two legs of the matrix $M$, which may be obtained from the fusion of all the anyonic charges on the leaves of the fusion and splitting trees of the diagram, as shown below:

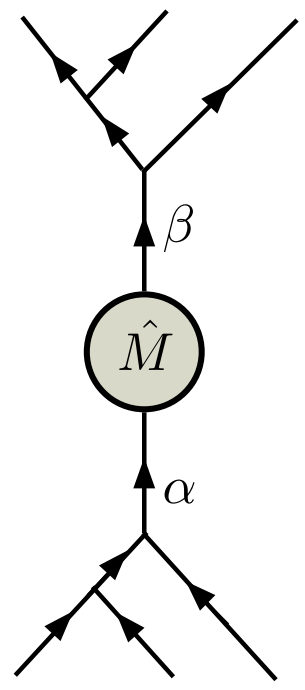

where, matrix $M$ is charge-conserving and constrained to have non-zero elements only when the charges $a \in \alpha$ and $b \in \beta$ on its legs are equal. In implementation, iterate through the set of charges in $\alpha$ and $\beta$, check for equality of charges $a$ and $b$, if true, select the sector's degeneracy matrix and populate it with the correct matrix elements. The matrix elements may either be computed—in the case where the matrix is a representation of a certain operation performed on the anyons on sites, or the matrix elements may be inserted "by hand" (e.g. zeros, ones, or random values), which can be used to initialise the matrix.

3. Routine "`Anyonic3Tensor`"—create an order-three tensor: To satisfy anyonic fusion rules unambiguously, we do not permit a tensor with more than three legs, with each leg labelled by a charge-degeneracy index. Depending on whether the anyons form a fusion or splitting vertex, we can have two types of anyonic order-three tensor $T$ which satisfy the anyonic fusion rules: $(a)$ a tensor with two incoming legs indexed by $\alpha$ and $\beta$ and fusing into one outgoing leg indexed by $\gamma$, or $(b)$ a tensor with one incoming leg indexed by $\gamma$ and splitting into two outgoing legs indexed by $\alpha$ and $\beta$, as shown below:

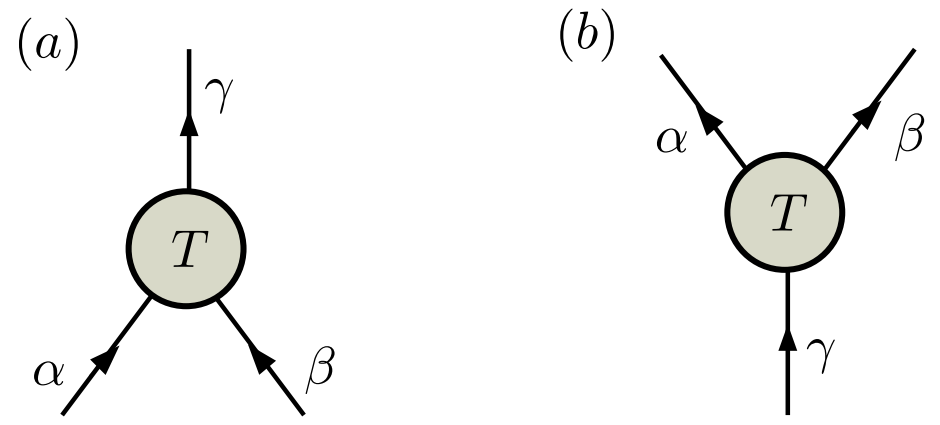

It is noted that if the elements of the tensor in $(b)$ are complex-conjugated, the tensor $T$ in $(b)$ can be considered a conjugate of the tensor in $(a)$, and therefore, the distinction between the two tensors might be unnecessary. However, the benefit of the distinction comes from the automation of the creation of the tensors, as the ordering of the legs and the direction of fusion or splitting, matters. The practical implementation of this routine is similar to the steps given in Routine `FuseIndices`. As an example, the basic steps needed to create the order-three tensor T in Eq. (3)(b) are:

— Using a "for-loop" for each of steps $(a)$, $(b)$, and $(c)$ —

(a) Select a new value of the charge $c$ in the array $\gamma$

(b) Select a new value of the charge $a$ in the array $\alpha$

(c) Select a new value of the charge $b$ in $\beta$

---

(d) Determine if the selected charges $(c, a, b)$ forms a compatible splitting vertex. If yes, proceed to next step, otherwise return to either steps (c), (b), or (a), depending on the number of charges remaining at each iteration.

(e) Get the degeneracy $n_c$, $n_a$, and $n_b$ of charges $c$, $a$, and $b$, respectively.

(f) Create the specific order-three tensor $T_c^{a,b}$ of size $n_c \times n_a \times n_b$. When working in diagrammatic isotopy, we multiply the tensor $T$ by the vertex normalisation factor $\left(\frac{d_c}{d_a d_b}\right)^{1/4}$.

(g) Save the tensor $T_c^{a,b}$ into its sector which is indexed by charge index $a$, $b$, and $c$.

(h) Return back to step (c) if there are still charges in the array $\beta$, otherwise go to the next step.

(i) Return back to step (b) if there are still charges in the array $\alpha$, otherwise go to the next step.

(j) Return back to step (a) if there are still charges in the array $\gamma$, otherwise end all iterations.

(k) Return the created anynonic tensor $T$, and end the program.

The anyonic-MPS ansatz is constructed using an order-three tensor having a charge-compatible splitting vertex, and an order-two charge-conserving tensor (or matrix) having one ingoing and one outgoing leg.

4. Routine "`FusionTensor`"—create fusion and splitting tensors: These are special cases of the anyonic three-index tensors. The fusion tensor, which I denote as $X^{\gamma}_{\alpha\beta}$, maps the Hilbert spaces of two anyonic sites—whose basis states are indexed by $\alpha$ and $\beta$—to a new coarse-grained site indexed by $\gamma$, while the splitting tensor, denoted as $X^{\dagger\,\alpha\beta}_{\gamma}$, which is the Hermitian conjugate of the fusion tensor, is used to split an Hilbert space into two Hilbert spaces. The elements of $X^{\gamma}_{\alpha\beta}$ and $X^{\dagger\,\alpha\beta}_{\gamma}$ are 1s or 0s. The fusion and splitting tensors are used to reshape anyonic tensors, for example, it can convert a three-index tensor into a two-index tensor (or matrix) using a fusion tensor, and vice versa. The fusion and splitting tensors are diagrammatically denoted as:

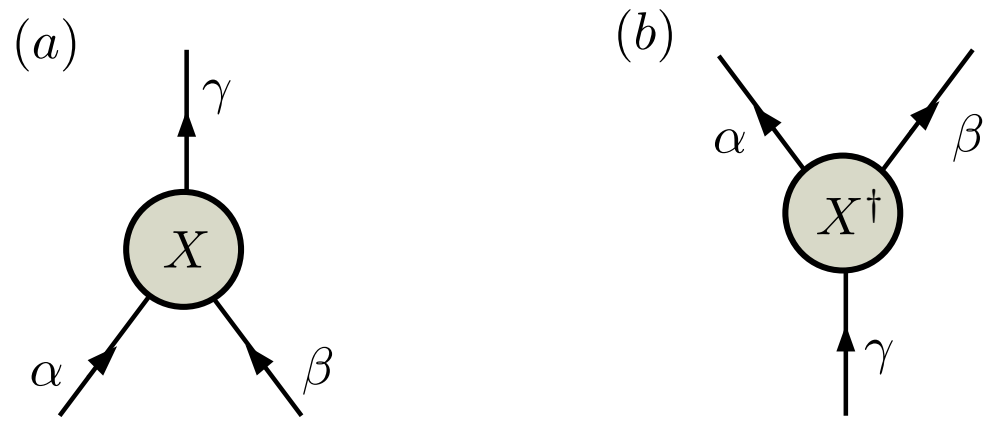

,

where $X$ represents the fusion map, and $X^{\dagger}$ represents the splitting map. Wherever any of these tensors appear in a network, their meaning and purpose should be understood.

Implementing the routines that create these tensors is similar to the implementation of the creation of three-index tensors above, except for a minor difference. After selecting a compatible fusion $(a, b, c)$ or splitting vertex, proceeding from steps $(e)$ of the basic steps above, get their degeneracy $\nu_a$, $\nu_b$, and $\nu_c$ of charges $a$, $b$ and $c$. Map the degeneracy indices $\mu_a = [1 : \nu_a]$ and $\mu_b = [1 : \nu_b]$ to "new" unique values of degeneracy index $\mu_c$ of charge $c$ by setting the element to 1 (and multiply by $\left(\frac{d_c}{d_a d_b}\right)^{1/4}$ when working in diagrammatic isotopy), otherwise all other values are set to 0. The values of $\mu_c$ are indexed starting from 1 to $\nu_c$, where $\nu_c = \sum_{a,b} \nu_a \nu_b \delta_{a\times b,c}$.

5. Routine "`ContractTensor`"—contraction of tensors: Two tensors are contracted when they have shared indices. In the first instance, we will consider the contraction of two tensors that does not involve any loop, then we later consider contractions of tensors that involve loops. Some examples of different possible contractions that does not involve any loop are:

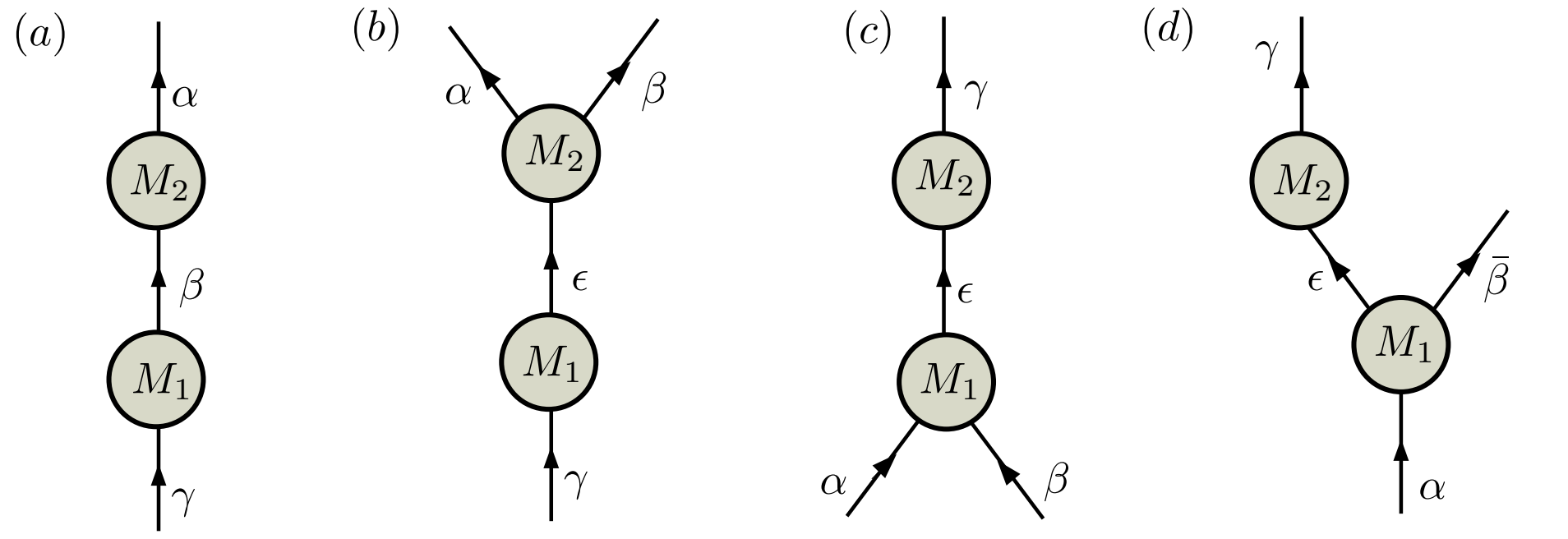

(5.5)

It should be noted that the contraction between two order-three tensors along a single shared leg is not permitted, as the resulting tensor will be order-4 (i.e. having four legs) which does not have a unique fusion tree decomposition. I now provide a short description of how to implement these contractions using Eq. (5.5)(b). First, recognise that the contraction of tensors $M1$ and $M_2$ will yield a order-3 tensor, say named $T_{12}$, with indices $\gamma$, $\alpha$, and $\beta$. Create an initial "placeholder" tensor for $T_{12}$ using Routine `Anyonic3Tensor`, populated with zeros. Then, start by iterating over the *free* indices $(\gamma, \alpha, \beta)$ and *shared* index $\epsilon$. Get their charges $c \in \gamma$, $a \in \alpha$, $b \in \beta$, and $e \in \epsilon$. Perform the following checks: (i) that charges $c$ and $e$ of tensor $M_1$ are conserved, i.e. $c = e$, (ii) that charges $(e, a, b)$ of tensor $M_2$ forms a compatible splitting vertex, and (iii) that charges $(c, a, b)$ of the new tensor $T_{12}$ forms a compatible splitting vertex. After all these checks, select the appropriate degeneracy tensors to be contracted. Contract them, and save the resulting tensor into the appropriate sector in $T_{12}$. Repeat this procedure until all compatible sectors have been identified.

6. Routine "`VerticalBends`"—vertical bends of an anyonic tensor leg: During the manipulation of anyonic tensors, it might prove convenient to bend the leg of an anyonic tensor vertically up or down, and later undo the bend. For instance, bends may be used to change the orientation (or structure) of a tensor in order to put the network into a form suitable for contraction. Mathematically, bends correspond to "raising" or "lowering" the indices of a tensor. A vertical bend only introduces a factor into the tensor. This factor is computed from the splitting and fusion vertices, which are given as:

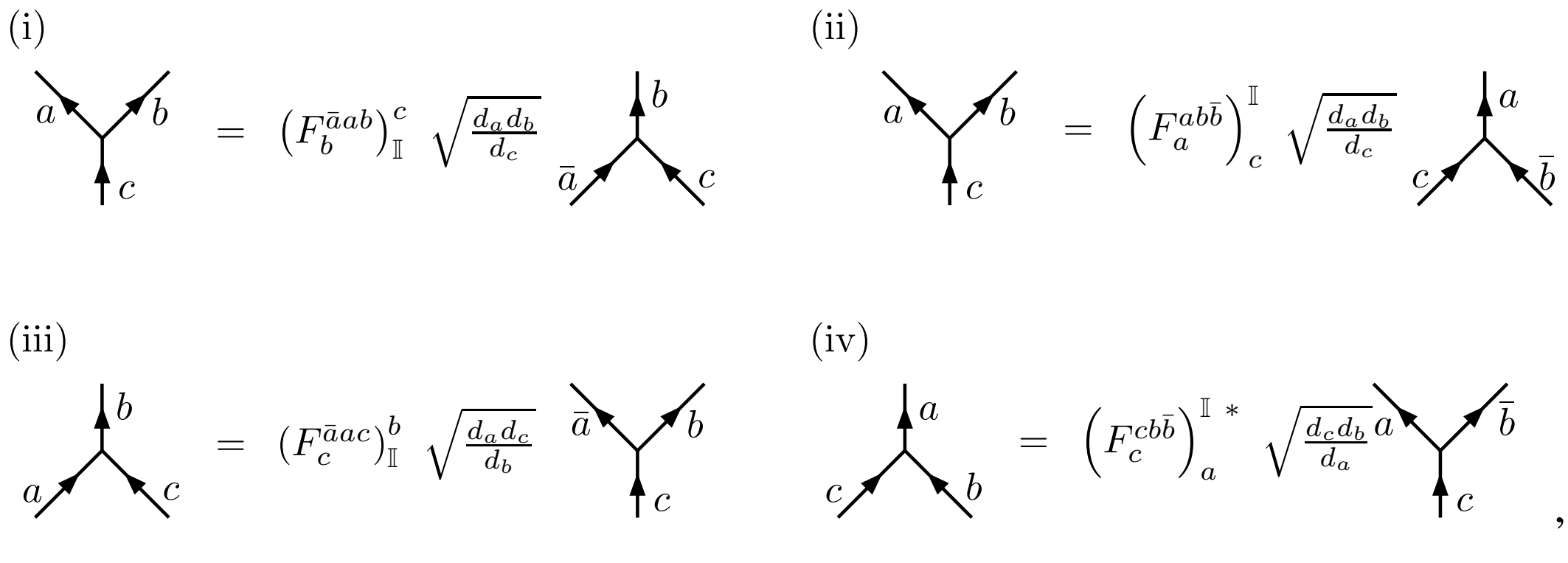

,

where in (i), the outgoing leg with charge $a$ is bent down to become incoming, in (ii) the outgoing leg with charge $b$ bends down, in (iii) the incoming leg with charge $a$ bends up to be outgoing, and in (iv) the incoming leg with charge $b$ bends up. In all, the charge of the bent leg is conjugated after the bend. It should be noted that these bend factors are only valid for unitary anyon models. Therefore, to bend any leg of a tensor $T$, create a charge-conserving order-2 tensor which has the (appropriate) bend factors as its elements. Then, contract the "bending" tensor with the tensor $T$. Examples are shown in the diagrammatic equations below:

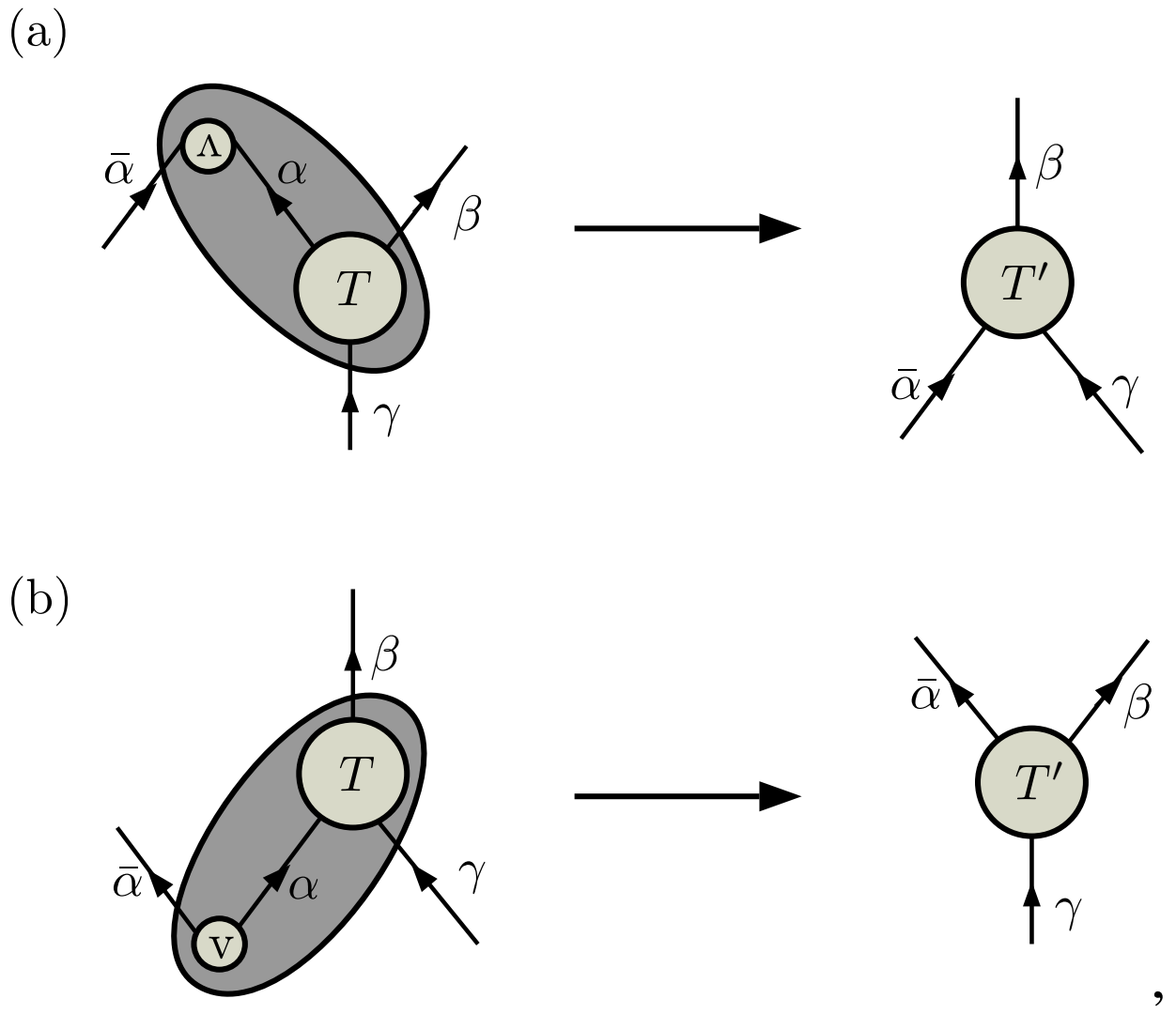

, (5.6)

where in (a), the outgoing indices $\alpha$ is bent down using the tensor $\Lambda$ to give the new tensor $T'$, while in (b), the incoming leg $\alpha$ is bent up using the tensor v. Similarly, the right leg can be bend up or down. The shaded part are to be contracted. Previously, I mentioned that bends can be used to change the structure of an anyonic tensor, for instance, to put it into a suitable form. As a concrete example, Eq. (5.5)($d$) can be converted into Eq. (5.5)($c$), by bending the right outgoing leg having the greek index $\bar{\beta}$.

7. Routine "`RemoveLoop`"—contraction of tensors with a loop: The contraction of two order-three tensors along two shared legs involves a loop as below:

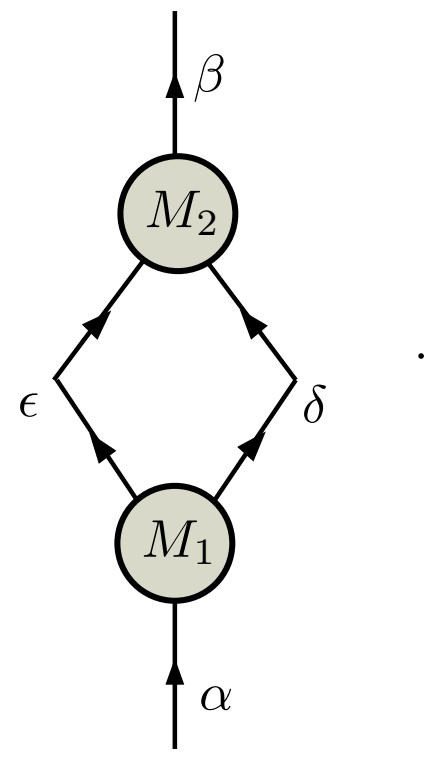

.

Similar to the contraction above in Routine `ContractTensor`, first recognise that this contraction will yield a two-index tensor, say named $M_{12}$, with indices $\alpha$ and $\beta$ on the bottom and top leg, respectively. The indices $\epsilon$ and $\delta$ on the shared legs have been matched, i.e. we have implicitly assumed that the charge-degeneracy indices on the left and right legs of the tensors $M_1$ and $M_2$ are the same. To contract this network, iterate through all the charges, check for compatibility of splitting and fusion vertices on tensors $M_1$ and $M_2$ respectively, and also the conservation of charges in the resulting two-index tensor $M_{12}$. Then contract the degeneracy tensors and multiply the resulting tensor by the appropriate loop factor $\sqrt{\frac{d_a d_b}{d_c}}$. Save the resulting degeneracy tensor into the appropriate sector of $M_{12}$.

8. Routine "`StarBubble`"—"Star-bubble" relation for anyonic tensors: After considering the contraction of two order-three tensors with two shared indices (i.e with a loop), we now consider the contraction of three order-three anyonic tensors with three shared indices. An example is shown below:

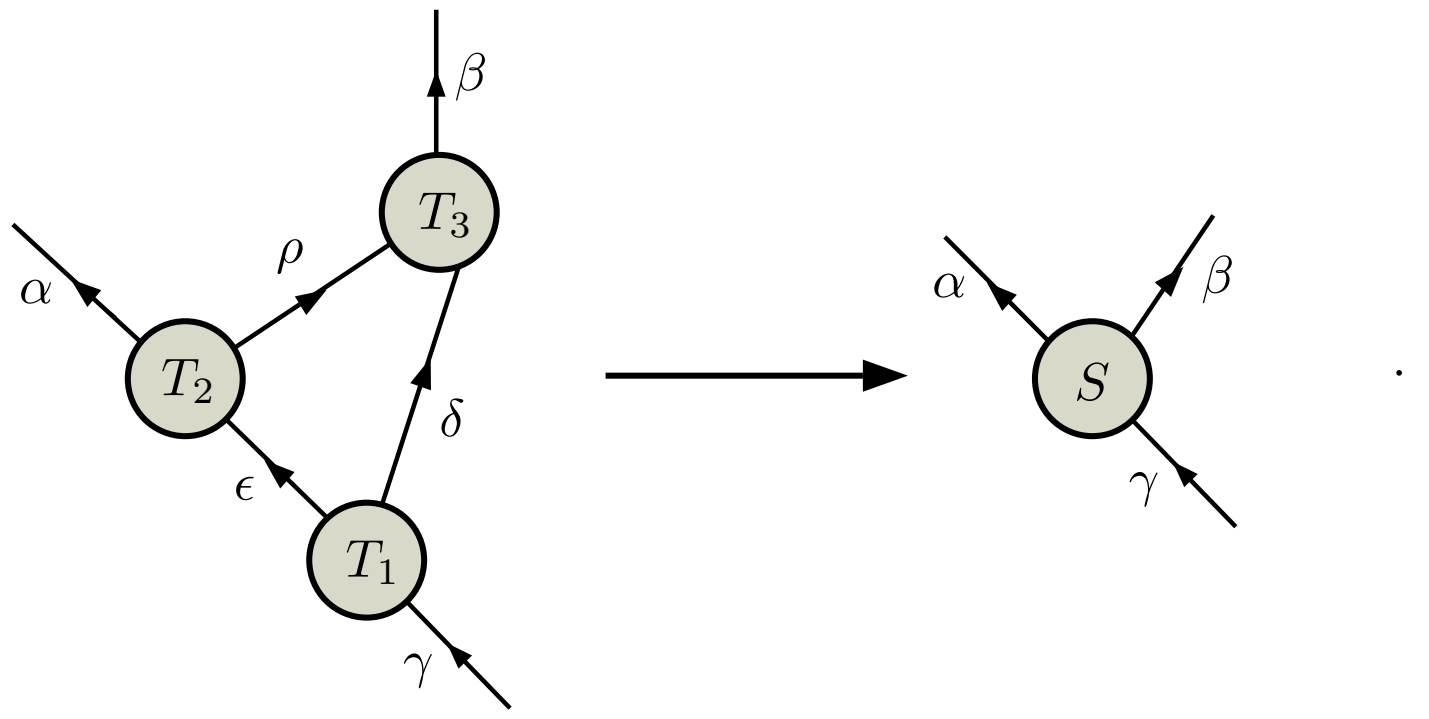

(5.7)

This diagrammatic equation has three shared indices $\epsilon$, $\delta$, $\rho$ and three free indices $\alpha$, $\beta$, and $\gamma$. The contraction of the network on the left yields tensor $S$ on the right, having the free indices $\alpha$, $\beta$, and $\gamma$. The implementation of this contraction is similar to previous ones but with a minor difference coming from the loop factor associated with the tensor network. The summary of its implementation is: iterate through all the charges $c \in \gamma$, $a \in \alpha$, $b \in \beta$, determine if the triplet $(c, a, b)$ forms a compatible splitting vertex of the new tensor $S$. If true, iterate through the remaining charges $e \in \epsilon$, $p \in \rho$ and $d \in \delta$, check for valid fusion and splitting vertices on all the nodes of the tensor network. If all these conditions are satisfied, contract their degeneracy tensors and multiply by the loop factor—which is given below. Save the resulting degeneracy tensor into the appropriate sector of tensor $S$. Repeat this process until all the charges have been iterated through.

Meanwhile, the loop factor mentioned above is computed from the anyon fusion network of the tensor network on the left of Eq. (5.7) as:

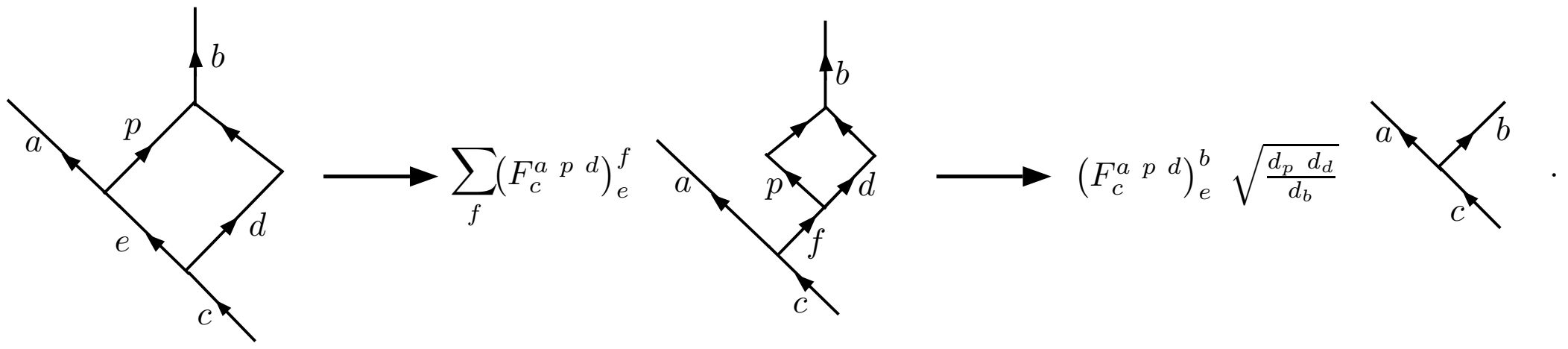

(5.8)

The loop factor used to multiply the degeneracy tensors of the tensor $S$ is factor $= \left(F_c^{apd}\right)_e^b \sqrt{\frac{d_p\ d_d}{d_b}}$. The factor $\sqrt{\frac{d_p\ d_d}{d_b}}$ involving the quantum dimensions normalises the vertex of tensor $S$ if adopting the diagrammatic isotopy convention, otherwise no factor should be associated with the contraction of anyons loops apart from the coefficient of the F-move.

The implementation of the star-bubble relation for anyonic tensors will prove versatile in anyonic TEBD algorithm, since it can be reused to contract other variants of the "star-bubble" network. If the network to be contracted differs from the above, it can be diagrammatically adjusted to look similar to the one above, and can therefore be contracted using the same routine. If however the contraction network already looks structurally similar to the one above, but has a different orientation of the arrows, the arrows should be reversed, and the charge conjugated. Then, proceed with the contraction as supposed.

9. Routine "`SVD`"—singular value decomposition (SVD) and truncation: SVD is used to decompose a matrix $M$ into its factors as $M = U\lambda V^\dagger$, where $U$ and $V$ are unitary, $U^\dagger U = UU^\dagger = \mathbb{I}$, and $\lambda$ is a diagonal matrix of singular (or Schmidt) values $\lambda_i$, which are non-negative real numbers. In a software like MATLAB, the singular values come ordered in decreasing order.

   An anyonic matrix $M$ is charge-conserving, which implies that its elements are arranged into diagonal blocks of degeneracy matrix, $M = \bigoplus_a M_a$, where $M_a$ is a degeneracy matrix in charge sector $a$. The anyonic SVD is represented diagrammatically as:

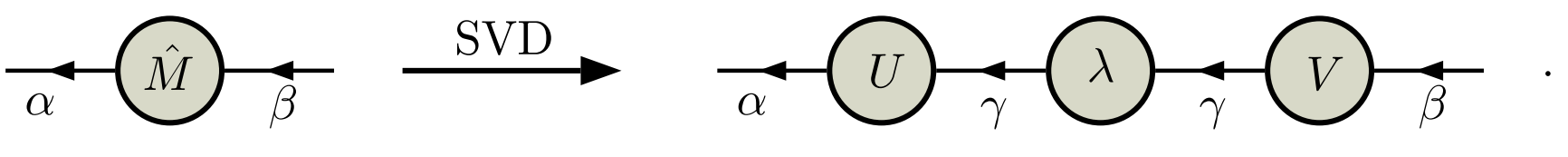

(5.9)

   In practical implementation, the anyonic SVD is performed block-wise: iterate through the charges $a \in \alpha$ and $b \in \beta$, check for conservation (i.e. equality) of charges $a$ and $b$. If true, select its degeneracy matrix $M_a$, decompose it into its subfactors $U_a$, $\lambda_a$, and $V_a$. Save all those factors also in blocks like the original matrix $M$. It should be noted in Eq. (5.9) that, though I have used different charge-degeneracy labels on the bonds of the subfactors $U$, $\lambda$ and $V$, the charges $a \in \alpha$, $c \in \gamma$, and $b \in \beta$ must be conserved, i.e. all equal, albeit their degeneracies may differ.

   Truncation: In a practical simulation on a computer with modest resources, the Hilbert space on the bonds on the MPS needs to be truncated after SVD. I will represent the truncation diagrammatically as,

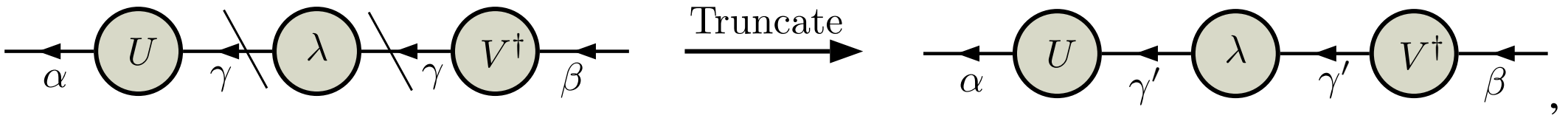

(5.10)

   where the "slash" represent a cut-down on the Hilbert space on the bonds connecting matrices $U$, $\lambda$, and $V$. The basis of the reduced space on the bonds is now denoted by $\gamma'$. The need for a truncation is determined by two control values, namely, the desired *precision* $\epsilon$ which controls the smallest value of the Schmidt values to be kept, and the bond dimension $\chi$ which controls the maximum number of states kept on the bonds of the MPS. The truncation should be performed with respect to any of the two control values, or both of them.

   Truncation in non-symmetric MPS is simple. The Schmidt values of $\lambda$ are ordered in decreasing order as $\lambda_1 > \lambda_2 > \ldots > \lambda_{k-1} > \lambda_k$. If $k > \chi$, truncate all values $\lambda_i$ for $i > \chi$, i.e. keep only $\{\lambda_i \mid 1 \leq i \leq \chi\}$. In addition, also truncate those values $\lambda_i$ of the diagonal matrix $\lambda$ that are less than the precision $\epsilon$, if there are. Also, truncate the corresponding basis in matrices $U$ and $V$.

   For truncation in anyonic MPS, it should be realised that the matrices $U$, $\lambda$, and $V$ are arranged in blocks of degneracy matrices, and the singular values are not a priori ordered. Moreover, the different charge sectors of the Schmidt matrix are weighted by the vertex normalisation factor when working in diagrammatic isotopy formalism. To realise this, note that the bipartite Schmidt decomposition of an anyonic state $|\psi\rangle$ is [41]

$$|\psi^{[A:B]}\rangle = \bigoplus_{a} \sum_{\mu_a}^{\chi_a} (\lambda_a)_{\mu_a} \, |\phi_{a,\mu_a}\rangle \otimes |\phi_{\bar{a},\mu_a}\rangle \tag{5.11}$$

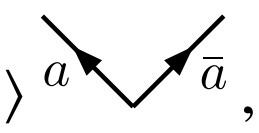

where $a$ is the charge label of each sector, $\mu_a$ is the degeneracy index of charge sector $a$, and $\chi_a$ is the maximum number of states in that sector. When working in diagrammatic isotopy, the vertex normalisation factor $1/\sqrt{d_a}$ of each charge sector $a$ has been absorbed into the Schmidt values $(\lambda_a)_{\mu_a}$. Therefore, to order the singular values accurately, multiply each sector—indexed by charge $a$, by $\sqrt{d_a}$, then order the spectrum in descending order, and proceed with the truncation as in the non-symmetric case.

## 5.3 Anyonic two-site contraction

In the Introduction 5.1 of this chapter, we alluded to the fact that the entire anyonic MPS-TEBD tensor network can be contracted once we know how to perform the two-site contraction (see Fig. 5.2). The sequence of steps needed to perform the two-site contraction using the routines given in Section 5.2 are shown in Fig. 5.3. Any shaded part of the network is to be contracted. I explain these steps here: (i) Absorb $\lambda_1$ into $T_1$, possibly bending leg labelled 1 down before contraction. Also, absorb tensors $\lambda_2$ and $\lambda_3$ into tensor $T_2$ as shown in the diagram, applying bends where necessary. (ii) The contractions in (i) give the resulting network of two order-three tensors $T_1$ and $T_2$. (iii) Apply the fusion map $X_1$ to the two physical legs of the tensors $T_1$ and $T_2$ in (ii). Apply the two-site gate $\hat{g}$, and also a splitting map $X_1^\dagger$ to restore the two initially contracted legs of $T_1$ and $T_2$. Contract the shaded part of this diagram using the Routine "`star-bubble`" relation to give the tensor $S$. (iv) Contract the obtained tensor $S$ with the gate $\hat{g}$ to give a new tensor—which I again named $S$ for simplicity, resulting in (v). (vi) Fuse legs $1'$ of $X_1^\dagger$ and 3 of $S$ using the fusion map $X_2$. Let the splitting map used to split the legs later be $X_2^\dagger$, which is shown adjacent to the shaded region, and separated by a dashed line (to signify that it is not part of this contraction but to be used later). (vii) The bent leg is straightened. (viii) Reverse the arrow on bond $1'$, and conjugate its charge. Contract the network using the "star-bubble" routine which yields the tensor $S$ in (ix). (x) Now "fuse" legs $2'$ and 8 using the splitting map $X_3^\dagger$. The fusion map $X_3$ used to "split" the fused leg $(8, 2')$ is shown beyond the dashed line. Contract this network of two tensors which form a closed loop to give a matrix $T$ in (xi). (xii) Decompose the tensor $T$ into its factors $U$, $\lambda$, and $V$ using the SVD routine. (xiii) Then, apply the splitting map $X_2^\dagger$ [shown in (vi)], and the fusion map $X_3$ [shown in (x)]. Contract the shaded part to give the resulting network (xiv). Finally, restore the canonical form of the MPS by introducing the identity matrices $\mathbb{I} = \lambda_1 \lambda_1^{-1}$ and $\mathbb{I} = \lambda_3^{-1}\lambda_3$ on the left and right hand side respectively as shown in the diagram. (xv) Contract the shaded part of the network to give the updated tensors $T_1'$ and $T_2'$. The two-site update is done and we get the resulting two-site network in (xvi) consisting of the updated tensors $T_1'$, $\lambda_2'$, and $T_2'$, and the non-updated tensors $\lambda_1$ and $\lambda_3$ restoring the original structure of the MPS network.

The two-site contraction scheme is used iteratively to perform the TEBD for both finite and infinite anyonic-MPS. After every iteration, the MPS is checked for convergence. The iteration stops when the MPS has converged to the desired precision, in the case where we wish to compute the ground state.

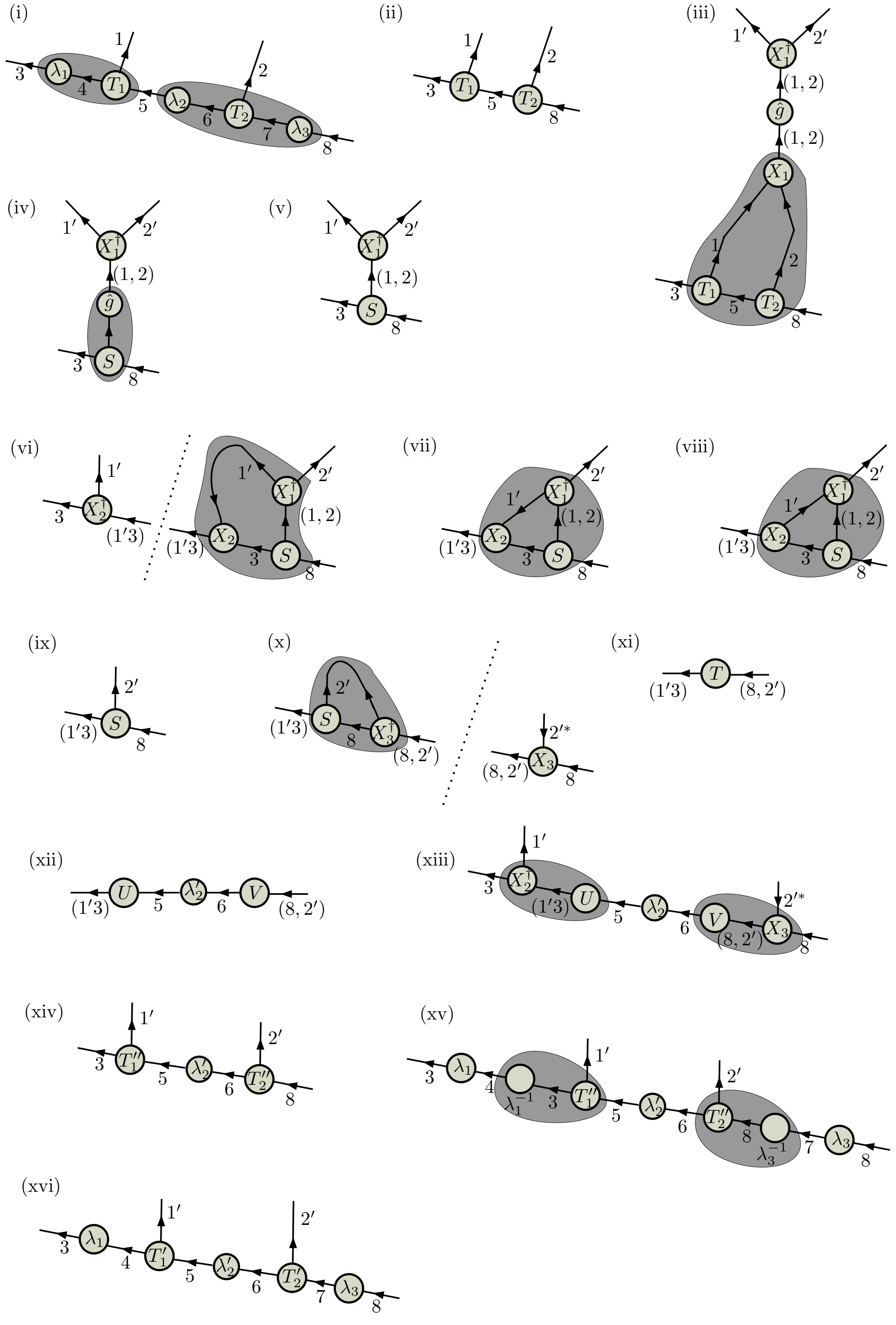

Figure 5.3: Graphical illustration of the anyonic two-site contractions. The explanation of the sequence of contractions is given in the text. The legs and bonds of the tensor network have been labelled by means of integers for the ease of identifying the legs of the tensors being manipulated.

## 5.4 Measure of convergence of simulation

We now consider how to check for the convergence of the MPS in the TEBD algorithm. The convergence measure I use is the *temporal deviation* of the state $|\psi\rangle$ from iteration $\tau$ to $\tau+1$. Let $\xi = \| |\psi_{\tau+1}\rangle - |\psi_\tau\rangle \|$ be the norm of the deviation of the state $|\psi\rangle$ from one iteration to the next. If the Schmidt decomposition of the system into two subsystems is,

$$|\psi\rangle = \sum_i \lambda_i \, |\phi_i^{(A)}\rangle \otimes |\phi_i^{(B)}\rangle, \tag{5.12}$$

where $|\phi_i^{(x)}\rangle$ is an orthonormal basis of both partitions ($x \equiv$ A or B). Then,

$$\begin{aligned}
\xi &= \Big\| \, |\psi_{\tau+1}\rangle - |\psi_\tau\rangle \, \Big\|, &(5.13)\\
&= \Bigg\| \sum_i \Big( \lambda_i(\tau+1) - \lambda_i(\tau) \Big) \, |\phi_i^{(A)}\rangle \, |\phi_i^{(B)}\rangle \Bigg\|, &(5.14)\\
&= \sqrt{\sum_i \Big( \lambda_i(\tau+1) - \lambda_i(\tau) \Big)^2}, &(5.15)
\end{aligned}$$

where I made the assumption that the Schmidt bases no longer change with iteration. (This can be regarded as a reasonable assumption in the long time limit).

For anyonic systems, let $i \equiv (a, \mu_a)$ be the basis index, where $a$ is the anyonic charge index and $\mu_a$ is the degeneracy index. The expression of temporal deviation for anyonic systems becomes,

$$\xi^2 = \sum_a d_a \sum_{\mu_a} \Big( \lambda_{a,\mu_a}(\tau+1) - \lambda_{a,\mu_a}(\tau) \Big)^2, \tag{5.16}$$

where $d_a$ is the quantum dimension of charge $a$ attributed to the loop factor resulting from the inner product.

The quantity $\xi$ can be easily computed from the Schmidt coefficients of the MPS. We make a cut along any bond of the MPS (see the MPS diagram in Section 4.3 of Chapter 4) to separate the network into parts: a left subnetwork, a central bond (Schmidt) matrix, and a right subnetwork. We then compute the temporal deviation from the entries of the Schmidt matrix. To determine the accuracy of our simulation, we check to see if $\xi$ is less than the precision required for a good approximation to the ground state. If true, we halt the simulation and proceed to compute properties of the ground state, otherwise we continue the simulation until convergence, or perhaps halt it if not converging.

## 5.5 Test models and results

We provide some examples to demonstrate that our Anyon×U(1)-symmetric MPS ansatz along with the anyonic-TEBD algorithm may be used to simulate systems of itinerant (non-Abelian) anyons at any rational filling fraction. These examples also help to illustrate instances where more than one symmetry is exploited in a tensor network, with one of these symmetries being anyonic. The anyonic lattice models we consider are itinerant and interacting hardcore particles on a one-dimensional chain and a two-leg ladder, with the types of particles being either bosons, fermions, or Fibonacci anyons.

In one dimension, all models of free itinerant particles (e.g. bosons, fermions, and anyons) with hardcore constraint have the same ground state properties, since they do not experience their own particle statistics; because they are not able to exchange positions. However, in higher dimensions (e.g. two-leg ladder or two dimensions), there are several paths by which particles may exchange positions. Therefore, beyond 1D, ground state properties of hardcore bosons, spinless fermions and hardcore anyons should reflect the influence of their exchange statistics. We test our method using itinerant Fibonacci anyons on a chain (the Golden Chain [88]) and itinerant-and-braiding (henceforth, simply "braiding") Fibonacci anyons on a two-leg ladder (or the Golden ladder), and show how the ground state energies differ from those of hardcore bosons and spinless fermions. We also present results of the entanglement entropies (EE) in the ground state of the Golden Ladder model—the model is a system of Fibonacci anyons on a pair of coupled chains, where particles interact by means of ferromagnetic (FM) or antiferromagnetic (AFM) Heisenberg interactions. We compute how entangled a block of sites is with the rest of the system using the Von-Neumann formula for EE. In our computation of the EE for Fibonacci anyons on both the chain and ladder system, we have ensured consistency of the ordering of the anyons on the systems with their ordering on the fusion tree according to the prescription given for how to compute measures of entanglement in systems of particles with non-Abelian statistics as in Ref. [96]. From the block scaling of the EE, we extract the central charges of the conformal field theories associated with the infra-red limits of these models. Analytical solutions for these models are generally not known, but we establish the validity of our method by using it to compute known results for spinless fermions and hardcore bosons, and also by comparing results for selected anyonic systems with those obtained using anyonic DMRG [43]. In general, our results are found to be accurate to 4 or 5 decimal places for the ground state energy.

## 5.5.1 Itinerant hardcore particles on a one-dimensional chain

I give some diagnostic test results for hopping and interacting anyons on a chain using an anyonic $t$-$J$ Hamiltonian which is analogous to the $t$-$J$ model for electrons. To make the analogy more apparent, I briefly review the electronic $t$-$J$ model.

### Electronic t-J Model

The electronic t-J model is a popular model which has been widely used to study some of the cooperative phenomena occurring in electronic systems. The model is defined on a lattice $\mathcal{L}$ of itinerant and interacting electrons. On a lattice with hardcore constraint, each site can either be empty or host a single electron which can be a "spin up" or a "spin down." The Hilbert space of each site has a set of basis $\{|0\rangle, |\uparrow\rangle, |\downarrow\rangle\}$ for "empty," "spin-up," and "spin-down" state respectively. The total Hilbert space of the system is $\mathbb{V} = \left(\mathbb{C}^3\right)^{\otimes L}$, where $L$ is the number of sites on lattice $\mathcal{L}$.

The electronic t-J Hamiltonian consists of two competing terms: a term corresponding to the kinetic energy of the electrons, and an interaction between their spin degrees of freedom. The t-J Hamiltonian is

$$\hat{H} = -t \sum_{\langle ij \rangle} \hat{c}_i^\dagger \hat{c}_j + J \sum_{\langle ij \rangle} \hat{S}_i \cdot \hat{S}_j, \tag{5.17}$$

where the first term is the kinetic energy with hopping strength $t$ and $\hat{c}_i^\dagger$ $(\hat{c}_i)$ is the creation (annihilation) operator which satisfies fermionic anticommutation relations. The second term is the Heisenberg spin-spin interaction which can be rewritten in terms of projector of nearest spins to the singlet state using the fact that

$$\hat{S}_i \cdot \hat{S}_j = \frac{1}{2}\left[(\hat{S}_i + \hat{S}_j)^2 - \hat{S}_i^2 - \hat{S}_j^2\right]. \tag{5.18}$$

The addition of two spin-$1/2$ charges is given by the rule,

$$\frac{1}{2} \otimes \frac{1}{2} = 0 \oplus 1. \tag{5.19}$$

Let $\hat{S} = \hat{S}_i + \hat{S}_j$ and choose units such that $\hbar = 1$. Then the relation

$$\hat{S}^2 \left|s, m\right\rangle = s(s+1)\left|s, m\right\rangle, \tag{5.20}$$

means $\hat{S}^2$ has two eigenvalues, 0 (when $s = 0$) and 2 (when $s = 1$). Therefore $\hat{S}^2$ can be written in terms of projectors to the singlet and triplet subspaces as $(\hat{S}_i + \hat{S}_j)^2 = 0\hat{\pi}_{ij}^{(0)} + 2\hat{\pi}_{ij}^{(1)}$, where $\hat{\pi}^{(0)}$ and $\hat{\pi}^{(1)}$ are the projectors to singlet and triplet subspaces. Therefore,

$$\hat{S}_i \cdot \hat{S}_j = -\hat{\pi}_{ij}^{(0)} + \frac{1}{4}, \tag{5.21}$$

where the identity, $\mathbb{I} = \hat{\pi}^{(0)} + \hat{\pi}^{(1)}$ has been used in the last step. Therefore, the $t$-$J$ Hamiltonian simplifies to

$$\hat{H} = -t\sum_{\langle ij\rangle} \hat{c}_i^\dagger \hat{c}_j - J\sum_{\langle ij\rangle} \pi_{ij}^0 + \text{const.} \tag{5.22}$$

For $J > 0$ the Hamiltonian favours neigbouring spins forming singlets (antiferromagnetic), and for $J < 0$ it favours triplet formation (ferromagnetic). I adapt the electronic t-J model to anyons.

### Anyonic t-J Hamiltonian on a chain

Patterned after the form of the electronic t-J Hamiltonian, its anyonic analogue consists of a hopping term and an anyon-anyon interaction term. The hopping term moves an anyon from an occupied site to a vacant site, while the interaction term is Heisenberg-like in nature, favouring the projection of two anyons into a particular fusion channel. The Hamiltonian is a sum of local terms $\hat{H} = \sum_i H^{[i,i+1]}$, and is expressed diagrammatically as,

$$\hat{H}^{[i,i+1]} = \sum_{\substack{a_i, a_{i+1}, \\ a_i', a_{i+1}', c}} H(a_i, a_{i+1}, a_i', a_{i+1}', c)$$

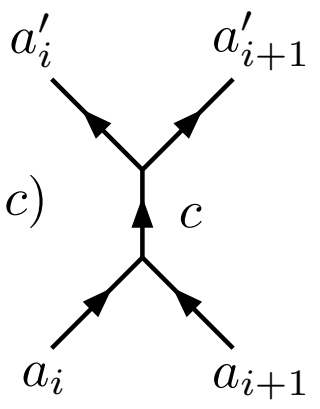

, (5.23)

where the values of the function $H(a_i, a_{i+1}, a_i', a_{i+1}', c)$ are determined by the physics of the model being constructed. Similar to how we represent anyonic operators as matrices, the Hamiltonian is represented in matrix form with basis chosen from the fusion space of the participating anyons. The conservation of charge $c$ implies that the Hamiltonian has a block structure as $\hat{H}^{[i,i+1]} = \bigoplus_c \hat{H}_c^{[i,i+1]}$.

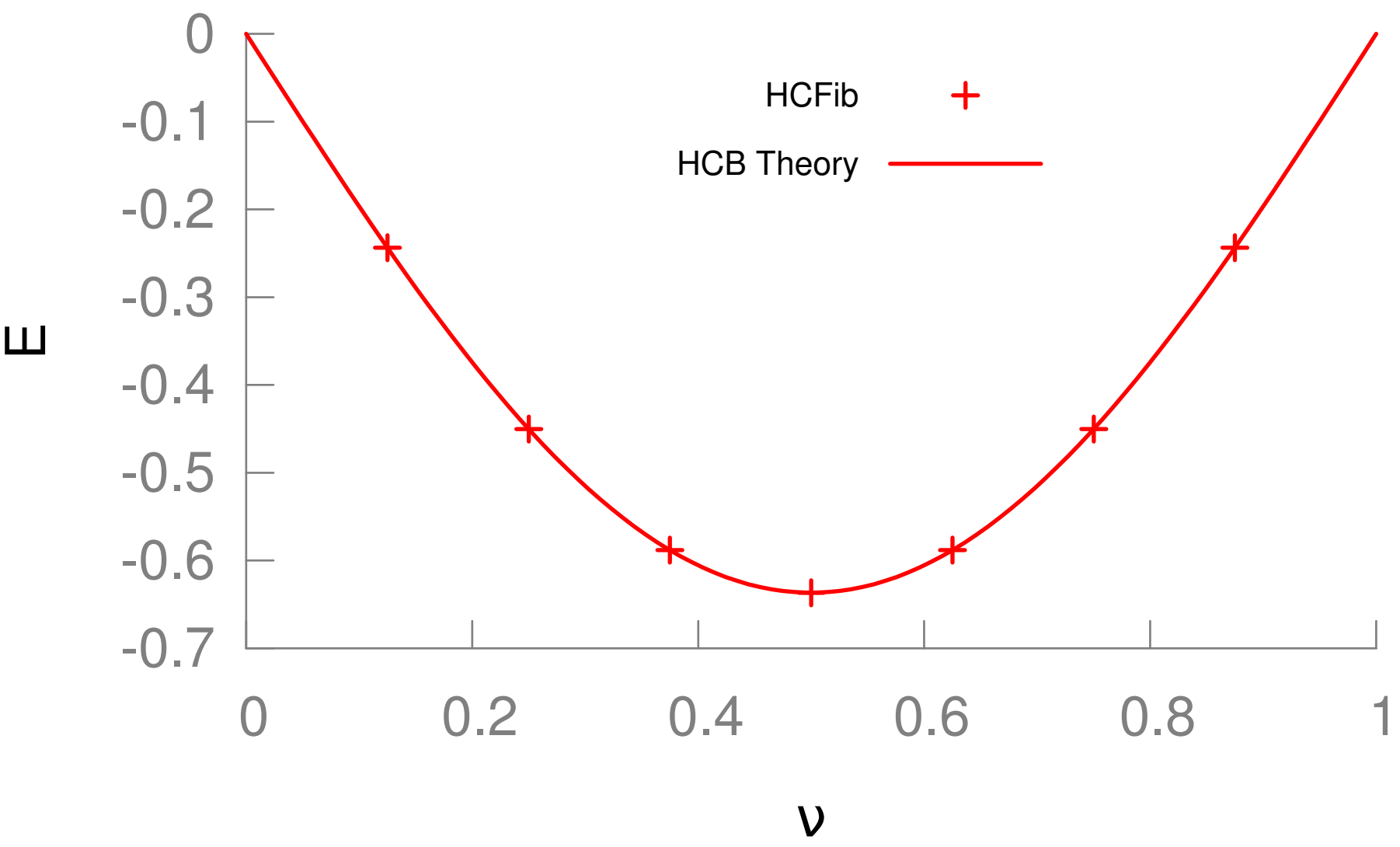

Figure 5.4: Ground state energy ($E$) of itinerant hard-core Fibonacci anyons on an infinite chain at different filling fractions ($\nu$). The datapoints come from numerical simulations, while the smooth curve is the plot of the ground state energy for an equivalent system of spinless fermions (or hardcore boson) at the thermodynamic limit.

To give a systematic and concrete treatment of both terms of the Hamiltonian, I give an explicit construction for Fibonacci anyons. In what follows, to start, I discuss the case when the interaction strength $J = 0$, i.e. when particles are free to hop. Afterwards, i discuss the case when particles interact with $J \neq 0$.

***Hopping term***: The hopping of a Fibonacci anyon in 1D means the neigbouring site has to be vacant—corresponding to the vacuum charge $\mathbb{I}$. The kinetic operator can thus be represented as

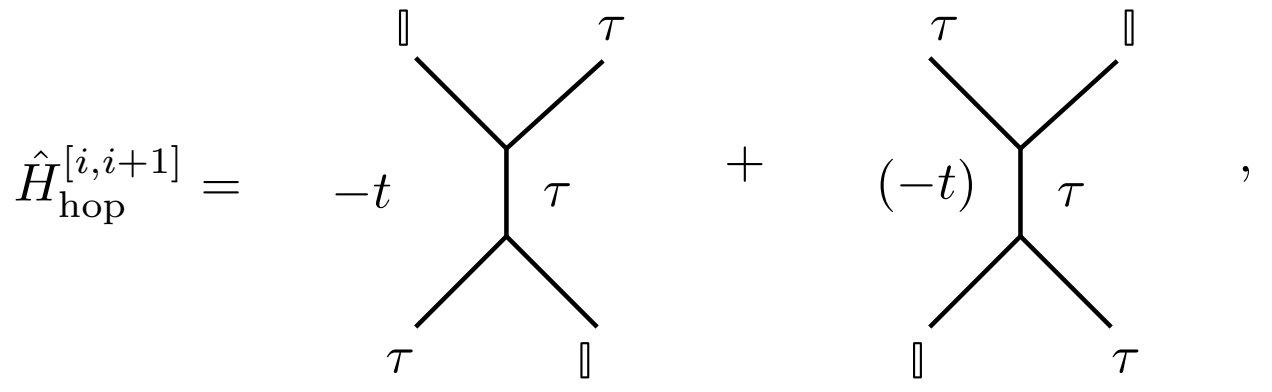

(5.24)

being analogous to the fermionic term ($\hat{c}_{i\sigma}^{\dagger}\hat{c}_{i+1,\sigma}$ + h.c) that translate a fermion between sites $i$ and $i+1$. Since the anyonic hopping term requires that there is a vacant site with vacuum charge $\mathbb{I}$, this implies that the hopping term is nonzero only when ($a_i = \mathbb{I}, a_{i+1} = \tau, a'_i = \tau, a'_{i+1} = \mathbb{I}$) and when ($a_i = \tau, a_{i+1} = \mathbb{I}, a'_i = \mathbb{I}, a'_{i+1} = \tau$). For dynamics in one dimension with hardcore constraints, the underlying exchange statistics of the particle do not affect the ground state properties (though the degeneracy of the ground states may differ for different particle species). Therefore, free itinerant hardcore Fibonacci anyons, spinless fermions and hardcore bosons all have the same ground state energies at any rational filling on the 1D lattice.

We use our Anyon×U(1) symmetric TEBD algorithm to compute the ground state energies of itinerant Fibonacci anyons, spinless fermions and hardcore bosons on a 1D lattice. We obtained the same ground state energies for all three cases up to 4 or 5 decimal places. This owes to the fact that particles are not allowed to exchange positions on the lattice, and thus particle statistics do not affect the ground state properties.

In the absence of any interaction, the Hamiltonian reduces to the well known tight-binding Hamiltonian for a system of electrons. The model has an exact analytical expression for ground state. I recall this solution for spinless fermions, which is equally valid for hardcore boson (HCB) and hardcore anyons in 1D. When $J = 0$ in Eq. (5.17), the Hamiltonian

$$\hat{H} = -t \sum_{j=1}^{L} \hat{c}_j^{\dagger} \hat{c}_{j+1} + \text{h.c.}, \tag{5.25}$$

describes free itinerant spinless fermions on a one dimensional lattice. This Hamiltonian can be diagonalised using Fourier transform from real to reciprocal space. We expand the fermion operators on each site as

$$\hat{c}_j = \frac{1}{\sqrt{L}} \sum_k \hat{c}_k e^{ikj}, \tag{5.26}$$

where $k$ is the quasimomentum. The lattice is embedded on a ring with periodic boundary condition, and therefore,

$$\hat{c}_{L+1} = \hat{c}_1,$$

from which the allowed values of the momentum are determined to be,

$$k = \frac{2\pi m}{L},$$

where $m = 0, 1, \ldots, L-1$. Therefore,

$$\hat{H} = -2t \sum_k \hat{c}_k^{\dagger} \hat{c}_k \cos(k). \tag{5.27}$$

Since particles are spinless fermions, or can be treated as hardcore bosons, each quasimomentum state can only accommodate a single particle. Integrating this equation in the continuum up to the Fermi level gives the ground state energy $E$,

$$E = -2tL \frac{\sin(\pi N/L)}{\pi}. \tag{5.28}$$

Define $\nu = N/L$ as the particle density. The mean energy per site is,

$$\epsilon = E/L = -2t \frac{\sin(\pi \nu)}{\pi}. \tag{5.29}$$

In Fig.5.4, I plot the numerical values of the energy of the ground state of itinerant Fibonacci anyons (derived from my simulations) against the values computed using the analytical formula of an equivalent system of spinless fermions given by Eq. (5.29). Note that the numerical ground state energy of spinless fermions and hardcore bosons are similar to that of Fibonacci anyons, and are not reproduced.

***Heisenberg interaction term***: Next, we include an anyonic Heisenberg interaction term in addition to the hopping term. The anyonic Heisenberg interaction is constructed in analogy to the Heisenberg spin-spin interaction. For 100% filling this model was first proposed and studied by Feiguin et al. [88], and is known as the Golden Chain. The anyonic Heisenberg interaction takes the form

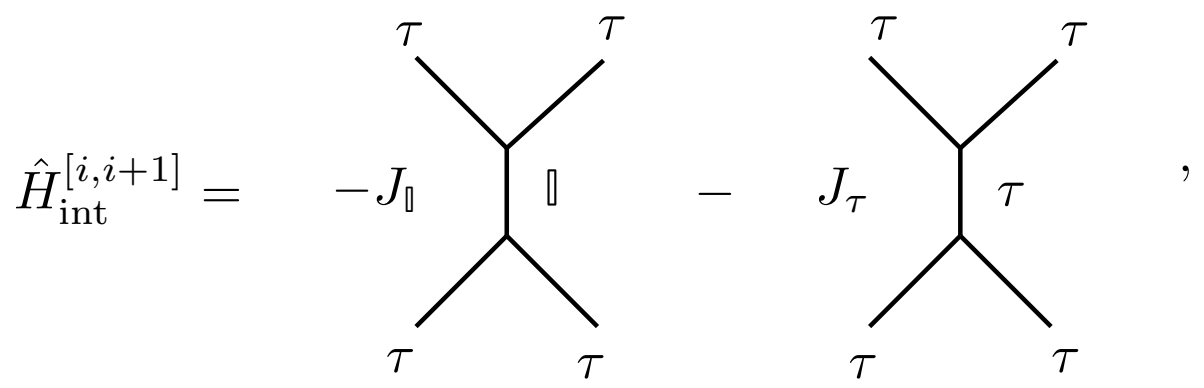

(5.30)

where $J_{\mathbb{I}} > J_{\tau}$ corresponds to an antiferromagnetic interaction favouring fusion of the two Fibonacci anyons to the vacuum charge $\mathbb{I}$, and $J_{\mathbb{I}} < J_{\tau}$ corresponds to a ferromagnetic interaction favouring projection to the Fibonacci charge $\tau$.

When a Heisenberg interaction is introduced into a system of itinerant Fibonacci anyons, the extensive degeneracy of the free anyon system is lifted. The Hilbert space of the interacting itinerant anyon system admits the decomposition

$$\mathcal{H} = \mathcal{H}_{\text{config}} \otimes \mathcal{H}_{\text{fusion}} \tag{5.31}$$

where $\mathcal{H}_{\text{config}}$ is the space of particle configurations, and $\mathcal{H}_{\text{fusion}}$ is the space of valid labellings of the fusion tree. The Hamiltonian admits an equivalent decomposition, and the Hamiltonian for a system of free particles acting on $\mathcal{H}_{\text{config}}$ is associated with a central charge of 1. When a Heisenberg-type interaction is added, this acts on $\mathcal{H}_{\text{fusion}}$, lifting the degeneracy of the states in this subspace. For a critical interaction, the total central charge is additive, and may be written as $1 + c$, where 1 is the contribution from the itinerant anyon model acting on $\mathcal{H}_{\text{config}}$ and $c$ is the contribution from the interactions on the fusion portion of the Hilbert space [82]. This is alluded to as spin charge separation in anyonic systems. From our numerical simulations, when $J_{\mathbb{I}} > J_{\tau}$ (antiferromagnetic), we obtained $c = 0.708$ and when $J_{\mathbb{I}} < J_{\tau}$ (ferromagnetic), we obtained $c = 0.84$, making total central charges of 1.708 and 1.84 respectively. These are very close to the expected central charges of $1 + 7/10$ for antiferromagnetic interaction and $1 + 4/5$ for ferromagnetic interaction for 1D systems exhibiting spin-charge separation.

### 5.5.2 Anyonic $t$-$J$ model on ladder

Nonabelian anyons have nontrivial braid factors, making their simulation difficult. For such systems, numerical approaches based on Monte Carlo schemes are plagued by a form of the sign problem. I offer numerical evidence that anyonic tensor networks (such as the anyonic MPS) are able to simulate anyonic systems on geometries beyond one dimension, in situations where the anyons experience braiding. To model how braiding statistics affect the ground state of anyons, I introduce the anyonic $t$-$J$ model on the ladder, as a generalisation of the model already considered on a chain. Each site on the ladder supports only two types of charges, namely, either a vacuum charge $\mathbb{I}$ or a single Fibonacci anyon $\tau$. Unlike in one dimension, anyons on the ladder can exchange positions and consequently braid.

To model the braiding of anyons on the ladder in a consistent manner, we impose a linear ordering to the anyons by attaching ficticious "strings" to the anyons and oriented them leftward of their on-site position (see Fig. 5.5). When an anyon hops from one site to another on either the top or bottom chain of the ladder, it braids with any adjacent anyonic charge along its trajectory, with the strings acting as a convenient mnemonic to visualise the orientation of the braid.

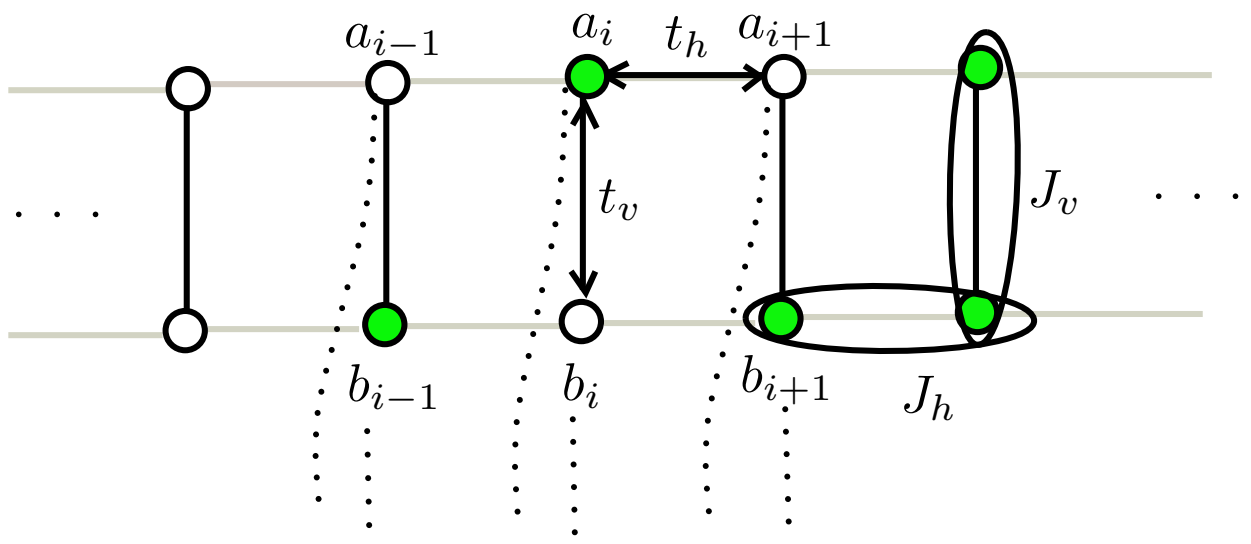

Figure 5.5: A ladder of itinerant anyons. Ficticious strings are attached to each nontrivial charge to indicate that they can participate in nontrivial braids as they exchange positions with neigbouring anyons. For example, the nontrivial anyonic charge $a_i$ braids with the nontrivial anyonic charge $b_{i-1}$ as it hops horizontally to a new site with $a_{i-1} = 0$. The labels $J_v$ and $J_h$ represent the amplitudes for projecting the corresponding circled pairs into the vacuum sector.

With reference to Fig.5.5, the anyonic $t$-$J$ Hamiltonian can be written as

$$\begin{aligned}\hat{H} = -t_h \sum_{i=1}^{N-1} \left(\hat{b}_{a_i \to a_{i+1}=\mathbb{I}} + \hat{b}_{b_i \to b_{i+1}=\mathbb{I}} + \text{h·c}\right) \\ - J_h \left(\hat{\Pi}^{\mathbb{I}}_{a_i,a_{i+1}} + \hat{\Pi}^{\mathbb{I}}_{b_i,b_{i+1}}\right) \\ - \frac{t_v}{2} \sum_{i=1}^{N} \left(\hat{b}_{a_i \to b_i=\mathbb{I}} + \hat{b}_{a_{i+1} \to b_{i+1}=\mathbb{I}} + \text{h·c}\right) \\ - \frac{J_v}{2} \left(\hat{\Pi}^{\mathbb{I}}_{a_i,b_i} + \hat{\Pi}^{\mathbb{I}}_{a_{i+1},b_{i+1}}\right),\end{aligned} \tag{5.32}$$

where $(t_h, t_v)$ and $(J_h, J_v)$ are the hopping and interaction amplitudes for anyons on the legs and rungs of the ladder. The vacuum charge is denoted by $\mathbb{I}$. The operator $\hat{b}_{x \to y=\mathbb{I}}$ moves a nontrivial charge $x$ into a new site having trivial vacuum charge $y = \mathbb{I}$ while it braids the charge $x$ with any other charge along its path. The projector $\hat{\Pi}^{\mathbb{I}}_{x,y}$ projects the nontrivial anyonic charges $x$ and $y$ into a vacuum charge $\mathbb{I}$. The anyonic interaction is antiferromagnetic when $J > 0$ but becomes ferromagnetic when $J < 0$. The Hamiltonian along the rung has been symmetrized with half a contribution from each of the rungs $i$ and $i + 1$.

Below, I show an explicit derivation of the Hamiltonian terms which can be arranged as a charge-conserving matrix operator. The local Hamiltonian $\hat{h}$ is derived on a plaquette whose vertices are labeled $(a, b, c, d)$ for brevity as shown in Fig.5.6. The local Hamiltonian is written as

$$\begin{aligned}\hat{h} = -t_h \left(\hat{b}_{a \to c=\mathbb{I}} + \hat{b}_{b \to d=\mathbb{I}} + \text{h·c}\right) - J_h \left(\hat{\Pi}^{\mathbb{I}}_{a,c} + \hat{\Pi}^{\mathbb{I}}_{b,d}\right) \\ - \frac{t_v}{2} \left(\hat{b}_{a \to b=\mathbb{I}} + \hat{b}_{c \to d=\mathbb{I}} + \text{h·c}\right) - \frac{J_v}{2} \left(\hat{\Pi}^{\mathbb{I}}_{a,b} + \hat{\Pi}^{\mathbb{I}}_{c,d}\right).\end{aligned} \tag{5.33}$$

Depending on the imposed fusion order, some of the operators will be diagonal in the fusion basis. The two most convenient fusion order are shown in Fig.5.6. Let the first basis be denoted as $|\text{I}\rangle = |(ab;\alpha)(cd;\beta)(\alpha\beta;\gamma)\rangle$ with the fusion order $((a,b)(c,d))$ where the anyons $(a, b)$ and $(c, d)$ are first fused independently, then fuse their outcomes, and let the second basis be $|\text{II}\rangle = |(ac;\kappa)(bd;\lambda)(\kappa\lambda;\gamma)\rangle$ with fusion order $((a,c)(b,d))$. Using a series

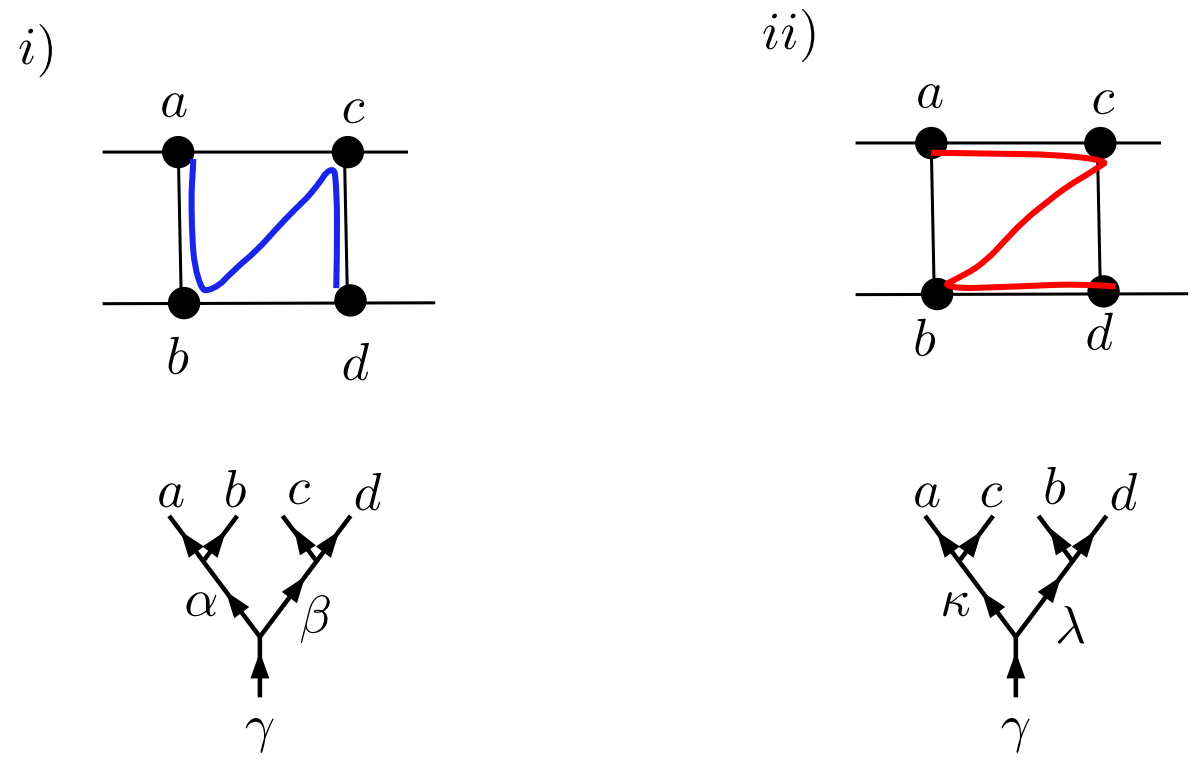

Figure 5.6: The two convenient fusion orderings along with their respective fusion trees shown underneath. The first fusion order couples charges $(a, b)$ and $(c, d)$ while the second couples charges $(a, c)$ and $(b, d)$.

of $F$-moves and $R$-moves, the first basis transforms into the second basis as

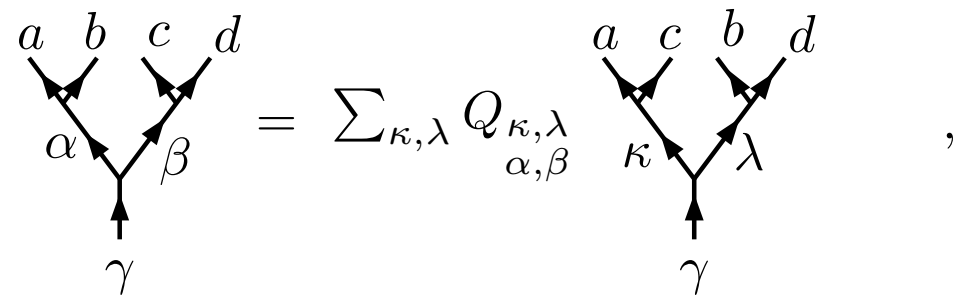

(5.34)

where the tensor $Q^{\kappa\lambda}_{\alpha\beta}$ is given by,

$$Q^{\kappa\lambda}_{\alpha\beta} = \sum_{\eta,\theta} \left[\left(F^{\alpha cd}_{\gamma}\right)^{-1}\right]^{\eta}_{\beta} \left(F^{abc}_{\gamma}\right)^{\theta}_{\alpha} R^{bc}_{\theta} \left[\left(F^{acb}_{\gamma}\right)^{-1}\right]^{\kappa}_{\theta} \left(F^{\kappa bd}_{\gamma}\right)^{\lambda}_{\eta} . \tag{5.35}$$

Its derivation and the rest of the derivation of the Hamiltonian $\hat{h}$ are shown as an illustrative example in the next section.

### 5.5.3 Derivation of the anyonic $t$-$J$ Hamiltonian on a ladder

Before giving an explicit derivation of the Hamiltonian $\hat{h}$ on a plaquette, I show how the two bases in Fig. 5.6 transform to one another.

### Fusion Tree Basis Transformation

The transformation between the two chosen bases of Fig.5.6 are obtained as follows:

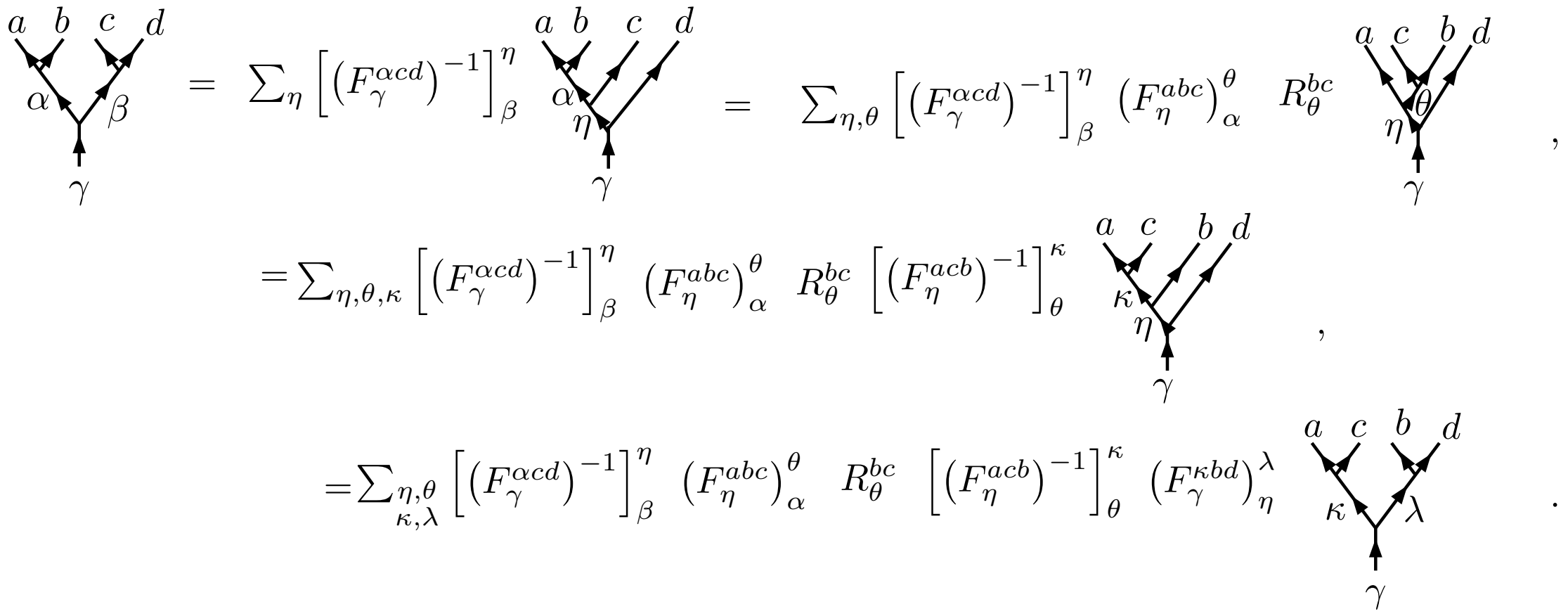

$$= \sum_{\eta} \left[\left(F_{\gamma}^{\alpha c d}\right)^{-1}\right]_{\beta}^{\eta} = \sum_{\eta,\theta} \left[\left(F_{\gamma}^{\alpha c d}\right)^{-1}\right]_{\beta}^{\eta} \left(F_{\eta}^{abc}\right)_{\alpha}^{\theta} R_{\theta}^{bc} ,$$

$$= \sum_{\eta,\theta,\kappa} \left[\left(F_{\gamma}^{\alpha c d}\right)^{-1}\right]_{\beta}^{\eta} \left(F_{\eta}^{abc}\right)_{\alpha}^{\theta} R_{\theta}^{bc} \left[\left(F_{\eta}^{acb}\right)^{-1}\right]_{\theta}^{\kappa} ,$$

$$= \sum_{\substack{\eta,\theta \\ \kappa,\lambda}} \left[\left(F_{\gamma}^{\alpha c d}\right)^{-1}\right]_{\beta}^{\eta} \left(F_{\eta}^{abc}\right)_{\alpha}^{\theta} R_{\theta}^{bc} \left[\left(F_{\eta}^{acb}\right)^{-1}\right]_{\theta}^{\kappa} \left(F_{\gamma}^{\kappa b d}\right)_{\eta}^{\lambda} . \tag{5.36}$$

The above equation can be written more succinctly as

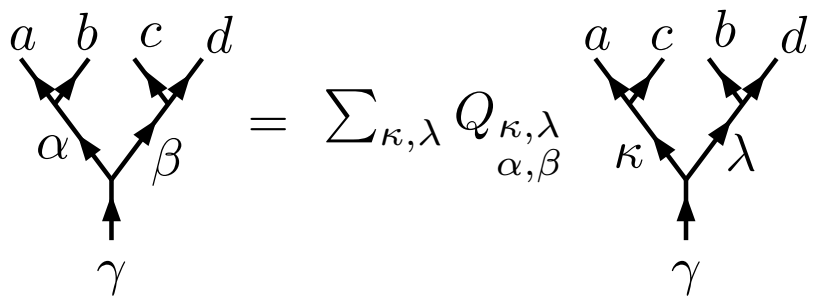

$$= \sum_{\kappa,\lambda} Q_{\alpha,\beta}^{\kappa,\lambda} , \tag{5.37}$$

where the tensor $Q_{\alpha\beta}^{\kappa\lambda}$ is defined according to

$$Q_{\alpha\beta}^{\kappa\lambda} = \sum_{\substack{\eta,\theta \\ \kappa,\lambda}} \left[\left(F_{\gamma}^{\alpha c d}\right)^{-1}\right]_{\beta}^{\eta} \left(F_{\eta}^{abc}\right)_{\alpha}^{\theta} R_{\theta}^{bc} \left[\left(F_{\eta}^{acb}\right)^{-1}\right]_{\theta}^{\kappa} \left(F_{\gamma}^{\kappa b d}\right)_{\eta}^{\lambda} , \tag{5.38}$$

which is the relation in Eq. (5.35). Using Dirac bra-ket notation, Eq. (5.36) can alternatively be written as

$$|(ab;\alpha)(cd;\beta)(\alpha\beta;\gamma)\rangle = \sum_{\kappa,\lambda} Q_{\alpha\beta}^{\kappa\lambda} |(ac;\kappa)(bd;\lambda)(\kappa\lambda;\gamma)\rangle . \tag{5.39}$$

### Anyonic $t$-$J$ Hamiltonian on a plaquette

The anyonic local Hamiltonian $\hat{h}$ on a plaquette is given by

$$\begin{aligned} \hat{h} = & -t_h \left(\hat{b}_{a\to c=\mathbb{I}} + \hat{b}_{b\to d=\mathbb{I}} + \text{h·c}\right) - J_h \left(\hat{\Pi}_{a,c}^{\mathbb{I}} + \hat{\Pi}_{b,d}^{\mathbb{I}}\right) \\ & - \frac{t_v}{2} \left(\hat{b}_{a\to b=\mathbb{I}} + \hat{b}_{c\to d=\mathbb{I}} + \text{h·c}\right) - \frac{J_v}{2} \left(\hat{\Pi}_{a,b}^{\mathbb{I}} + \hat{\Pi}_{c,d}^{\mathbb{I}}\right) . \end{aligned} \tag{5.40}$$

Whereas charge label $a$ may take any value from the particle spectrum, the nontrivial anyonic charge will be denoted by $a_0$. For example, in the Fibonacci anyon theory, $a_0 = \tau$. The derivation is quite general and can be used with any other anyon model. Note that numerical factors such as vertex normalization factors and loop factors are not accounted for here. We account for them during the implementation of the anyonic TEBD algorithm.

We proceed by first deriving all the kinetic energy terms and then derive all the interaction terms similarly. All the operators in the Hamiltonian are applied to the fusion tree on the left hand of Eq. (5.36) which is represented in Dirac notation in Eq. (5.39).

Kinetic terms: The terms contributing to the kinetic energy are the braid operators whose matrix elements are derived below.

(i) The matrix element of the braid operator $\hat{b}_{a\to b=\mathbb{I}}$ is given by

$$\langle \hat{b}_{a\to b=\mathbb{I}}\rangle = \delta_{a,a_0}\delta_{b,\mathbb{I}},\delta_{a',b}\delta_{b',a}\delta_{c',c}\delta_{d',d}\delta_{\alpha',\alpha}\delta_{\beta',\beta}, \tag{5.41}$$

where we have used, for the sake of conciseness, the notation $\langle \hat{b}_{a\to b=\mathbb{I}}\rangle$ as a shorthand for

$$\langle (a'b';\alpha')(c'd';\beta')(\alpha'\beta';\gamma)|\hat{b}_{a\to b=\mathbb{I}}|(ab;\alpha)(cd;\beta)(\alpha\beta;\gamma)\rangle. \tag{5.42}$$

(ii) The matrix element of the braid operator $\hat{b}_{c\to d=\mathbb{I}}$ is given by

$$\langle \hat{b}_{c\to d=\mathbb{I}}\rangle = \delta_{c,c_0}\delta_{d,\mathbb{I}},\delta_{a',a}\delta_{b',b}\delta_{c',d}\delta_{d',c}\delta_{\alpha',\alpha}\delta_{\beta',\beta}. \tag{5.43}$$

(iii) The matrix element of the operator $\hat{b}_{a\to c=\mathbb{I}}$ involves braiding of anyonic charge $a$ with $b$. The charge $c$ has to be vacuum for the process to have a nonzero amplitude. Its action on the basis $|(ab;\alpha)(cd;\beta)(\alpha\beta;\gamma)\rangle$ is given by

$$\begin{aligned}
&\hat{b}_{a\to c=\mathbb{I}}\,|(ab;\alpha)(cd;\beta)(\alpha\beta;\gamma)\rangle\\
&= \textstyle\sum_{\kappa,\lambda} Q^{\kappa\lambda}_{\alpha\beta}\hat{b}_{a\to c=\mathbb{I}}\,|(ac;\kappa)(bd;\lambda)(\kappa\lambda;\gamma)\rangle,\\
&= \textstyle\sum_{\kappa,\lambda} Q^{\kappa\lambda}_{\alpha\beta}\delta_{a,a_0}\delta_{c,\mathbb{I}}\,|(ca;\kappa)(bd;\lambda)(\kappa\lambda;\gamma)\rangle\,.
\end{aligned}$$

The matrix element $\langle \hat{b}_{a\to c=\mathbb{I}}\rangle$ is

$$\begin{aligned}
\langle \hat{b}_{a\to c=\mathbb{I}}\rangle \;=\; &\textstyle\sum_{\substack{\kappa',\lambda'\\ \kappa,\lambda}} \langle (a'c';\kappa')(b'd';\lambda')(\kappa'\lambda';\gamma)|\, Q^{*\kappa'\lambda'}_{\alpha'\beta'}Q^{\kappa\lambda}_{\alpha\beta}\\
&\times\delta_{a,a_0}\delta_{c,\mathbb{I}}\,|(ca;\kappa)(bd;\lambda)(\kappa\lambda;\gamma)\rangle,
\end{aligned} \tag{5.44}$$

which simplifies to

$$\langle \hat{b}_{a\to c=\mathbb{I}}\rangle = \sum_{\kappa,\lambda} Q^{\kappa\lambda}_{\alpha\beta}(Q^{\dagger})^{\alpha'\beta'}_{\kappa\lambda}\,\delta_{a,a_0}\delta_{c,\mathbb{I}}\delta_{a',c}\delta_{c',a}\delta_{b',b}\delta_{d',d}. \tag{5.45}$$

(iv) The matrix element $\langle \hat{b}_{b\to d=\mathbb{I}}\rangle$ is similarly given by

$$\langle \hat{b}_{b\to d=\mathbb{I}}\rangle = \sum_{\kappa,\lambda} Q^{\kappa\lambda}_{\alpha\beta}(Q^{\dagger})^{\alpha'\beta'}_{\kappa\lambda}\,\delta_{b,b_0}\delta_{d,\mathbb{I}}\delta_{a',a}\delta_{c',c}\delta_{b',d}\delta_{d',b}. \tag{5.46}$$

Interaction terms: The interaction terms consist of projectors whose matrix elements are derived similarly to the kinetic terms. The projection favours fusion of nontrivial anyons to the vacuum charge.

i) The action of the projector $\hat{\Pi}^{\mathbb{I}}_{ab}$ on the fusion basis is given as

$$\hat{\Pi}^{\mathbb{I}}_{ab}\,|(ab;\alpha)(cd;\beta)(\alpha\beta;\gamma)\rangle = \Pi^{\alpha}_{ab}\,|(ab;\alpha)(cd;\beta)(\alpha\beta;\gamma)\rangle \tag{5.47}$$

where the element $\Pi^{\alpha}_{ab} = 1$ if $\alpha = \mathbb{I}$ (vacuum) and $a = a_0$, $b = b_0$ (i.e. nontrivial charges). The matrix element of the projector $\hat{\Pi}^{\mathbb{I}}_{ab}$ is thus

$$\langle \hat{\Pi}^{\alpha=\mathbb{I}}_{ab}\rangle = \delta_{a',a_0}\delta_{b',b_0}\delta_{a',a}\delta_{b',b}\delta_{\alpha',\alpha}\delta_{\alpha,\mathbb{I}}\delta_{c',c}\delta_{d',d}\delta_{\beta,\beta'}. \tag{5.48}$$

(ii) The matrix element $\hat{\Pi}^{\mathbb{I}}_{cd}$ of the projector is similarly given as

$$\langle \hat{\Pi}^{\beta=\mathbb{I}}_{cd}\rangle = \delta_{a',a}\delta_{b',b}\delta_{\alpha',\alpha}\delta_{c,c_0}\delta_{d,d_0}\delta_{\beta,0}\delta_{c',c}\delta_{d',d}\delta_{\beta,\beta'}. \tag{5.49}$$

(iii) The action of the projector $\hat{\Pi}^{\mathbb{I}}_{ac}$ is

$$\begin{aligned}\hat{\Pi}^{\mathbb{I}}_{ac} \quad & |(ab;\alpha)(cd;\beta)(\alpha\beta;\gamma)\rangle \\ &= \sum_{\kappa,\lambda} Q^{\kappa\lambda}_{\alpha\beta}\Pi^{\kappa}_{ac}\,|(ac;\kappa)(bd;\lambda)(\kappa\lambda;\gamma)\rangle \\ &= \textstyle\sum_{\kappa,\lambda} Q^{\kappa\lambda}_{\alpha\beta}\delta_{a,a_0}\delta_{c,c_0}\delta_{\kappa,\mathbb{I}}\,|(ac;\kappa)(bd;\lambda)(\kappa\lambda;\gamma)\rangle\,.\end{aligned}$$

The matrix element $\langle\hat{\Pi}^{\mathbb{I}}_{ac}\rangle$ is

$$\langle\hat{\Pi}^{\mathbb{I}}_{ac}\rangle = \sum_{\kappa,\lambda} Q^{\kappa\lambda}_{\alpha\beta}(Q^{\dagger})^{\alpha'\beta'}_{\kappa\lambda}\,\delta_{\kappa,\mathbb{I}}\delta_{a,a_0}\delta_{c,c_0}\delta_{a',a}\delta_{c',c}\delta_{b',b}\delta_{d',d}. \tag{5.50}$$

(iv) The matrix element $\langle\hat{\Pi}^{\mathbb{I}}_{bd}\rangle$ is similarly given by

$$\langle\hat{\Pi}^{\mathbb{I}}_{bd}\rangle = \sum_{\kappa,\lambda} Q^{\kappa\lambda}_{\alpha\beta}(Q^{\dagger})^{\alpha'\beta'}_{\kappa\lambda}\,\delta_{b,b_0}\delta_{d,d_0}\delta_{\lambda,\mathbb{I}}\delta_{a',a}\delta_{b',b}\delta_{c',c}\delta_{d',d}. \tag{5.51}$$

### 5.5.4 Results of the anyonic $t$-$J$ model on ladder

We exploit the anyonic and U(1) symmetries of the model both in the MPS ansatz and in the Hamiltonian $\hat{H}$, and use the TEBD algorithm to compute the ground state energies of itinerant Fibonacci anyons on the ladder at different filling fractions. Since the MPS has a one-dimensional structure, we map the ladder to a chain by fusing the anyonic charges on each rung to make a new single site. The vertical and the horizontal hopping rates are set equal to one, $t_h = 1$ and $t_v = 1$ while the vertical and horizontal Heisenberg interactions $J_v$ and $J_h$ are set to zero.

There are no known analytical results for the ground state of itinerant Fibonacci anyons on a ladder, but we test the validity of our method against the ground state energies of itinerant hardcore bosons and spinless fermions on the ladder shown in Fig. 5.7. The phase diagram of this model for unit filling fraction was studied in Ref. [2].

It can be seen from the figure that incorporating the capacity for anyons to braid around one another results in an increase in the ground state energy per particle. This fact is reminiscent of the property that a system of identical fermions have a higher energy than bosons due to Pauli exclusion principle in real space. This also implies that there might exist a Pauli-like exclusion principle for anyons too, at least in some regimes [97]. We also see from the figure that while the bosons and fermions have a paricle-hole symmetry which is reflected in the symmetric ground state energy around half-filling $\nu = \frac{1}{2}$, the system of Fibonacci anyons on the ladder does not display this symmetry. One of the consequences of particle-hole symmetry is that the ground state energies $E$ at filling fractions $\nu$ and $1-\nu$ should be equal. While this is known for fermions and bosons, and reproduced by our numerical results as shown in Table.5.1, we see from our numerical results that this no longer holds for some non-abelian anyon model such as Fibonacci anyons, though in this instance the breakdown of particle-hole duality is weak in the sense that it has only a very small impact on ground state energies. Interference of braiding particles raises the ground state energies, and thus the higher filling fractions $\nu > 1/2$, e.g. $\nu = 5/8$, have slightly higher energies than the $1-\nu$ states, e.g. $\nu = 3/8$.

The origin of the breakdown in the particle-hole duality is in the difference of the fusion degrees of freedom of the particle types. For systems of bosons or fermions, the fusion space is one-dimensional, independent of the number of particles. While for non-Abelian anyon models, such as the Fibonacci anyons, the fusion space grows exponentially with the number of anyons and hence is not symmetric under particle hole exchange. Braiding acts non-trivially on the fusion degrees of freedom and changes the ground state energy in way that is not particle hole symmetric. In contrast, if particles are confined to one dimension then braiding cannot take place, and in the absence of other interactions such as Heisenberg-like interactions, the expanded Hilbert space has no effect on ground state energies. Thus the ground state energies for Fibonacci anyons on the ladder do not exhibit particle/hole duality, as braiding is possible (Fig. 5.7 and Table 5.1), whereas in 1D, the ground state energy is symmetric and independent of particle statistics (Fig. 5.4).

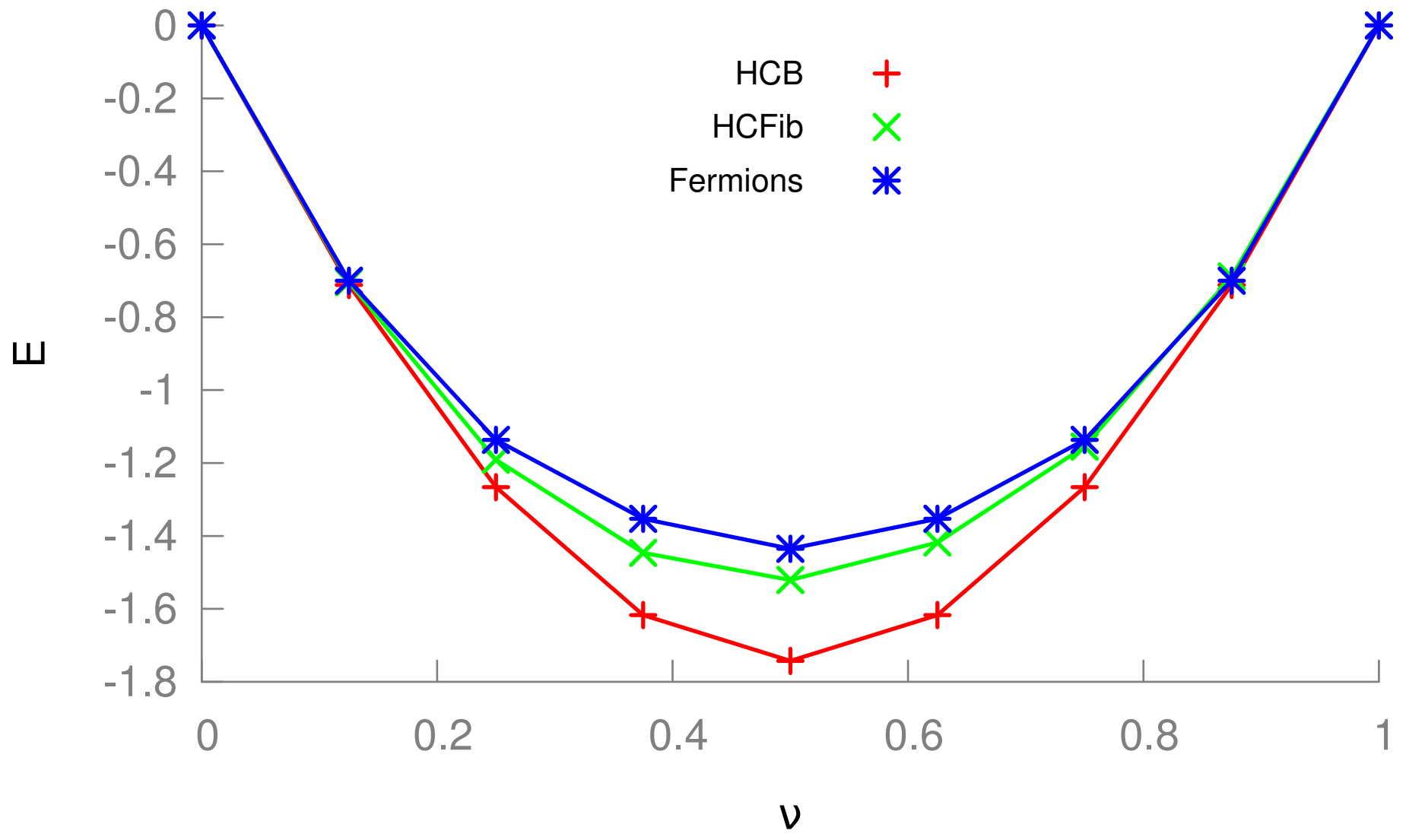

Figure 5.7: The ground state energies of hardcore bosons (HCB), spinless fermions and itinerant Fibonacci anyons (HCFib) on a two-leg ladder when only hopping is turned on. The line is a guide to the eye.

### 5.5.5 Phase diagram of the Golden Ladder

We further test our ansatz by studying the entanglement structures of ground states of interacting Fibonacci anyons on the ladder at unit filling. This model has been studied in Ref. [2], and we verify our ansatz by reproducing known phases of the model at specific values of the tunable parameters. At unit filling, there is a single localized Fibonacci anyon per site of the ladder and therefore hopping rates are everywhere zero. This is a quasi-1D generalisation of the Golden Chain [88], which might be called the Golden Ladder. The relative interaction strengths of the legs and rungs of the ladder, including both antiferromagnetic and ferromagnetic couplings may be parametrized on a circle (see Fig. 5.8, where the ferromagnetic or antiferromagnetic nature of the interactions in each sector is indicated).

We evolve this model to ground state using TEBD, and compute the scaling of the block entanglement entropy from von Neumann's relation,

$$S(r) = -\mathrm{Tr}(\hat{\rho}_r \log \hat{\rho}_r), \tag{5.52}$$

where $\hat{\rho}_r$ is in general the reduced density matrix of a block of $r$ sites, here $r$ rungs. From conformal field theory, the scaling of entanglement entropy on a system with an open

| $\nu$ | $E_{\mathrm{HCB}}$ | $E_{\mathrm{HCFib}}$ | $E_{\mathrm{SF}}$ |
|---|---|---|---|
| 0 | 0 | 0 | 0 |
| 1/8 | -0.71162 | -0.70397 | -0.70015 |
| 2/8 | -1.26597 | -1.19102 | -1.13658 |
| 3/8 | -1.61707 | -1.44620 | -1.35273 |
| 4/8 | -1.74300 | -1.52085 | -1.43534 |
| 5/8 | -1.61707 | -1.41803 | -1.35271 |
| 6/8 | -1.26597 | -1.15486 | -1.13660 |
| 7/8 | -0.71162 | -0.68857 | -0.70015 |
| 1 | 0 | 0 | 0 |

Table 5.1: The values of the ground state energy $E$ at various filling fractions $\nu$ corresponding to figure Fig. 5.7. The subscripts in $E_{(\bullet)}$ are "HCB" for hardcore bosons, "HCFib" for hardcore Fibonacci anyons, and "SF" for spinless fermions. The values are given to five decimal places. The ground state energies of bosons and fermions are symmetric around half-filling, but not so for Fibonacci anyons.

boundary [98] is

$$S(r) = \frac{c}{3} \log r, \tag{5.53}$$

where $c$ is the central charge of the system at criticality. This relation means that, for a critical model, the entanglement block scaling—computed from the MPS ground state representation—should display a logarithmic relation with the block size. The central charge $c$ can then be extracted from the relationship

$$c = 3\,\frac{S(r_2) - S(r_1)}{\log r_2 - \log r_1}. \tag{5.54}$$

The block entanglement entropy for various parameter regimes are shown in Fig. 5.9, and their central charges are indicated in Fig. 5.10. As seen in Fig. 5.9 the finite bond dimension of the MPS causes entanglement to artificially plateau over larger distances $r = |r_2 - r_1|$, but calculation of $c$ using Eq. (5.54) may be performed for any separation $r$ prior to this plateau, where an appropriate linear correlation is obtained between $S(r_2) - S(r_1)$ and $\log r_2 - \log r_1$.

One can interpret this Fig. 5.10 by considering how the physics of the interacting Fibonacci anyon changes as the parameterization angle $\theta$ is varied. When $\theta = 0$, there are no couplings along the rungs and we have 2 chains of Fibonacci anyons with antiferromagnetic interactions. The system in this parameter regime is gapless and has a central charge which is twice that of a single chain, i.e. $2 \times 7/10$. Even though the MPS most naturally yields exponentially decaying correlators, we are nevertheless able to extract an approximate value for the central charge $c = 1.405$ from the linear part of the curve. When $\theta = \pi/4$ and $\theta = 3\pi/4$, the vertical couplings are antiferromagnetic favouring pairs of Fibonacci anyon fusing into the vacuum charge. This phase is gapped with central charge $c = 0$. When $\theta = \pi/2$, the Hamiltonian favours fusion of pair of $\tau$ charges on each rung to the vacuum charge, and is hence a product state which is unique and gapped. The phase is not critical and has a central charge $c = 0$. At the $\theta = \pi$ point, the horizontal coupling $J_h = -1$ is ferromagnetic, while the vertical coupling $J_v$ is zero, and the ladder reduces to two copies of a ferromagnetic Golden Chain. From our numerical simulation, we computed a central charge of $c = 1.629$ which is close to the expected theoretical value of $c = 2 \times 4/5$. At the point $\theta = 5\pi/4$, the horizontal and vertical couplings are ferromagnetic. Fusion of the $\tau$ charges on the rungs and legs favours projection to the $\tau$ channel (triplet state). This can easily be pictured by considering a linearized or snaked version of the ladder. Nearest neighbour $\tau$ charges on the rung becomes nearest neighbour on the chain and nearest neighbours on the legs becomes either nearest or next-to-nearest neighbour on the chain [89]. Heuristically, fusion to the $\tau$ fusion channel makes the ladder effectively like a single Fibonacci chain and therefore has the same central charge as a single chain. We obtain a central charge of $c = 0.801$ which is close to the expected $c = 4/5$. When $\theta = \frac{3\pi}{2}$, the vertical coupling $J_v = -1$ is ferromagnetic while the horizontal coupling is zero. This favours projection of neighbouring $\tau$ charges on the rungs into the $\tau$ channel. The ladder reduces to a chain of decoupled $\tau$ charges which has an exponentially large degeneracy in intermediate fusion degrees of freedom. Hence a generic ground state at this point obeys a volume law rather than area law. This system is gapped and not described by conformal field theory. The parameter point $\theta = 7\pi/4$ correspond to horizontal antiferromagnetic coupling on the leg and vertical ferromagnetic coupling which is effectively an antiferromagnetic interacting chain. The obtained central charge is $c = 0.704$, being close to the expected value of $c = 7/10$. Our findings are in agreement with known results showing that the entire upper semicircle is gapped while the lower semicircle is gapless with the exception of the indicated point at $\theta = 3\pi/2$. Table 5.2 compares the extracted central charges with their expected theoretical values.

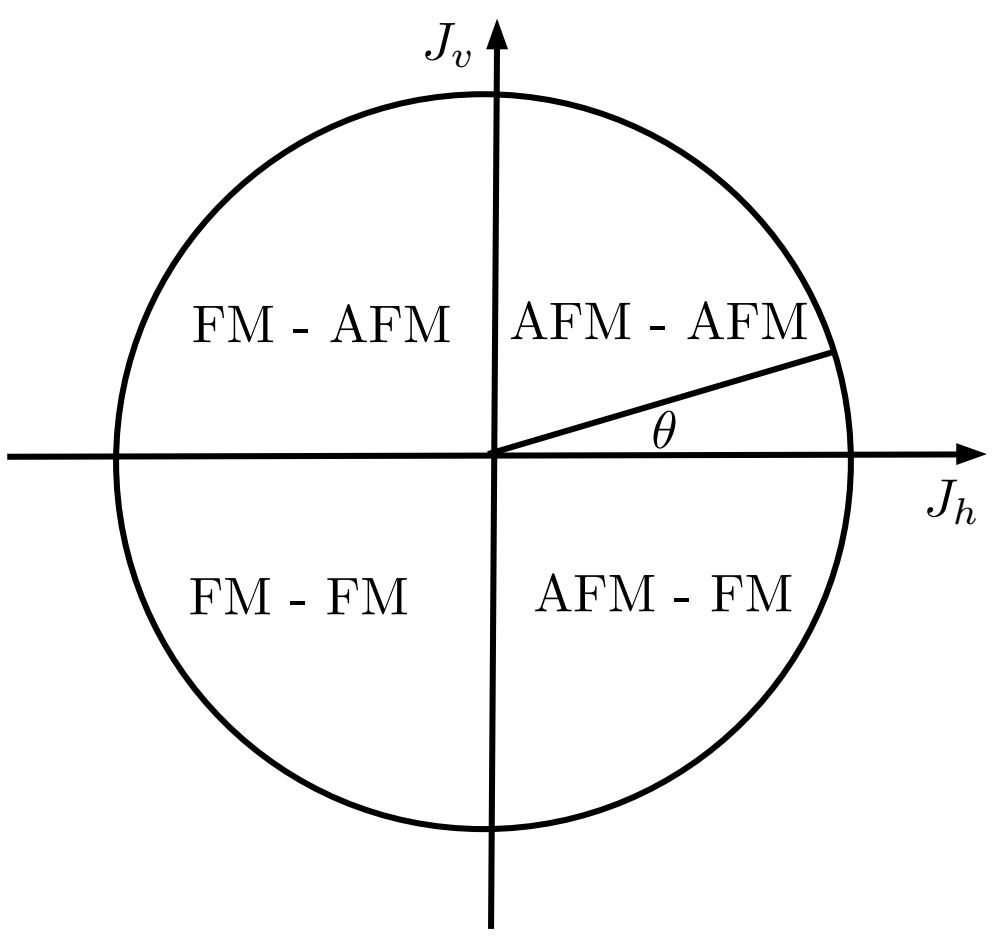

Figure 5.8: The horizontal and vertical interaction strengths $(J_h, J_v)$ on the legs and rungs of the ladder are parameterized by $(\cos\theta, \sin\theta)$ along the legs and rungs respectively. The labels within each quadrant indicate the nature of the interaction, whether antiferromagnetic or ferromagnetic.

| $\theta$ | $c_{\text{sim}}$ | $c_{\text{Theo.}}$ |
|---:|---:|---:|
| 0 | 1.405 | $2\times 7/10$ |
| $\pi/4$ | 0 | 0 |
| $\pi/2$ | 0 | 0 |
| $3\pi/4$ | 0 | 0 |
| $\pi$ | 1.629 | $2\times 4/5$ |
| $5\pi/4$ | 0.801 | $4/5$ |
| $3\pi/2$ | 0 | 0 |
| $7\pi/4$ | 0.704 | $7/10$ |

Table 5.2: The table shows the obtained central charges $c_{\text{Sim.}}$ from our numerical simulations, compared against their theoretical values $c_{\text{Theo.}}$ known from conformal field theory, at the interaction strengths parameterized by $\theta$ according to Fig. 5.8. The values correspond to the points shown in Fig. 5.10. Where a model is not critical, and hence not described by CFT, we have substituted their central charge $c$ with zero.

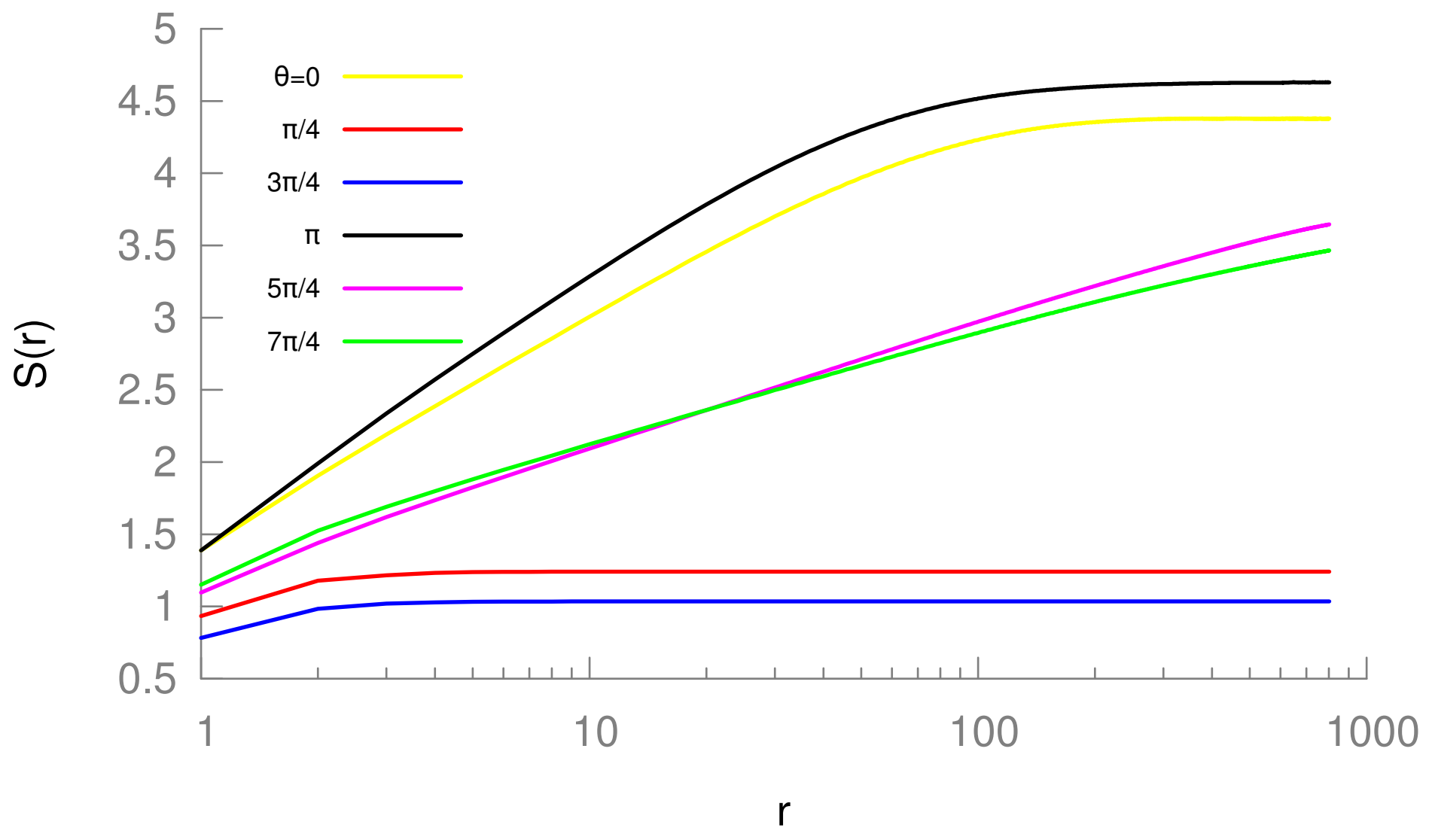

Figure 5.9: Scaling of entanglement entropy $S$ as a function of block size $r$, for different angles on the circle ($\theta$), which correspond to different ratios of coupling strength between the legs and rungs of the ladder.

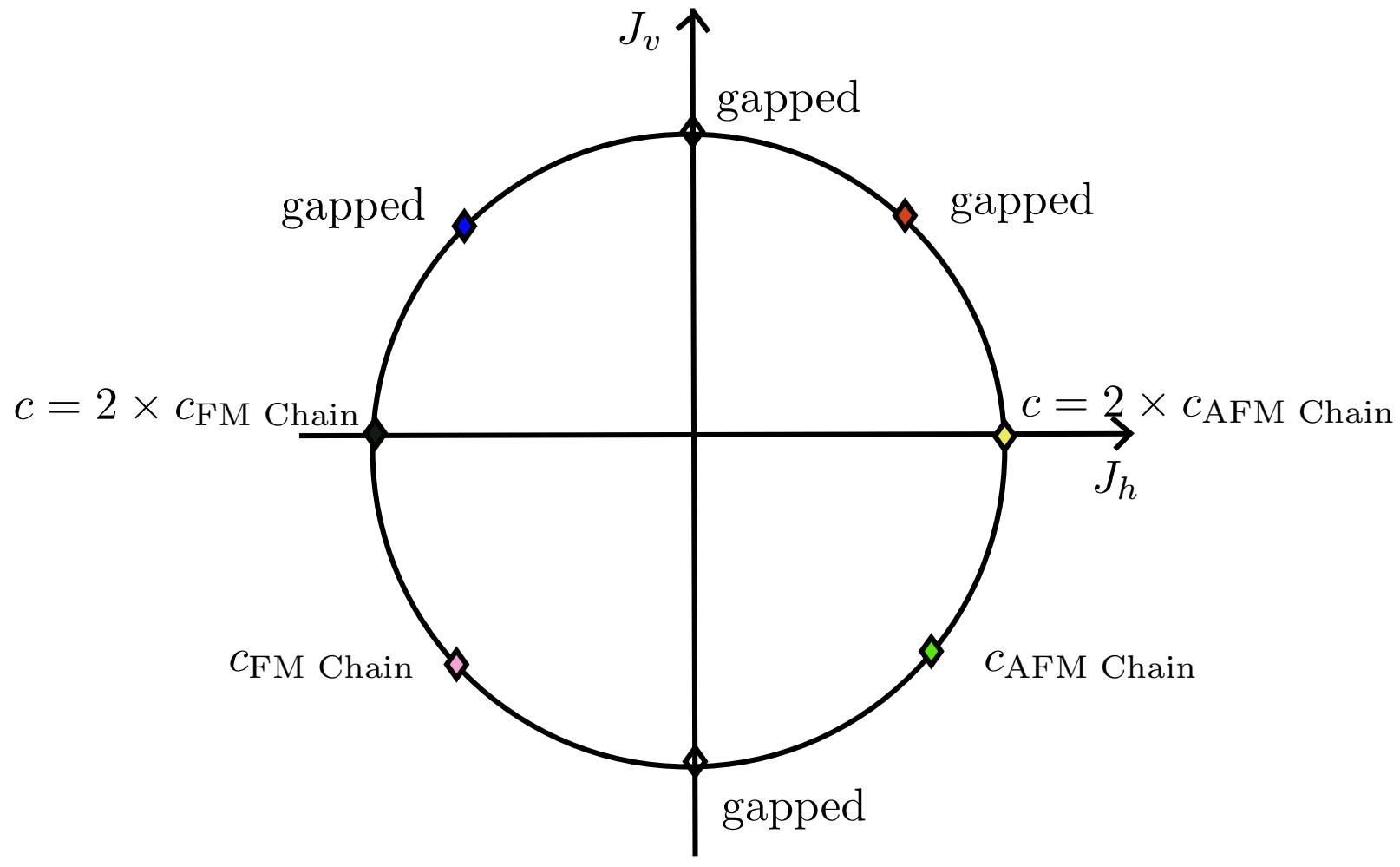

Figure 5.10: The central charge of the underlying CFT extracted from the scaling of the entanglement entropy of Fig. 5.9 are shown at the paremeter points we considered. When vertical coupling is set to zero and $J_h$ range from $-1$ to $+1$, we obtain central charge which doubles that of single critical FM or AFM chains which lies on the equator. The theoretical values of $c_{\text{AFM Chain}}$ and $c_{\text{FM Chain}}$ are $c_{\text{AFM Chain}} = 7/10$ and $c_{\text{FM Chain}} = 4/5$. Phase boundaries for this model may be found in Fig. 11(a) of Ref. [2]. A summary of the central charges of this model is given in Table 5.2.

## 5.6 Conclusion

In this chapter we showed how the anyonic tensor network formalism of Refs. [39, 41, 43] may be applied in the context of particles admitting multiple charge labels, specifically an anyonic charge and a U(1) charge, here corresponding to particle number. We constructed test models involving both hopping and interaction terms, with this construction being explicitly elaborated in Section 5.5.3. Application of the anyonic infinite TEBD algorithm [41] permitted calculation of the ground states of these systems, their entanglement entropies, and central charges. In doing so, we successfully reproduced elements of the phase diagrams for those systems which have previously been obtained using exact diagonalisation [2, 81, 82, 88, 89].

This chapter consequently demonstrates the feasibility of applying anyonic TEBD to systems of particles admitting both anyonic and U(1) conserved charges. The method presented here can be used to probe new regimes of the physics of anyons such as equilibrium phases of quasi-1D systems of braiding anyons at arbitrary density as well as non-equilibrium dynamics of anyons in two dimensional systems at low density.

# 6

# Phase transitions in braided non-Abelian anyonic system

## 6.1 Introduction

As described in Chapter 5, the formation of different possible phases in a collective system of particles are dictated by various competing terms in the Hamiltonian of the system, including the local interaction between particles, hopping of particles between sites, the dimensionality and topology of the system, geometric frustrations, and particle statistics. For a review of these concepts, see these Refs. [7–9].

While a lot of work have been done on interacting and itinerant anyons on one dimensional lattice chains [2, 81, 82, 88, 89] where the particles are restrained from braiding if hardcore constraint is imposed, not much have been done on higher dimensional systems where particles have braiding degrees of freedom. Of a particular interest would be to know how braid statistics affect the ground state properties of a collective of anyons. The wave function of a many-body system of non-Abelian anyons can acquire nontrivial phase factors when the anyons braid. Braiding is a topological interaction which may induce a phase transition. In Ref. [99], the authors studied “finite-legged” ladders of Fibonacci anyons, where the anyons can interact and braid with one another. In Ref. [100], we developed an Anyon$\times$U(1) MPS ansatz, and used it to study a specific case of braided Fibonacci anyons on a two-leg ladder. Among other results, we numerically showed that particle-hole duality breaks down for non-Abelian anyons as would be expected. In Ref. [97], the author studied a ladder of $\mathbb{Z}_3$ anyons, and found phases with normal and superfluid behaviours. In all these works, braid statistics is shown to affect, and perhaps even induce the phases formed.

In this Chapter, we study a model of anyons on a two-legged ladder system. A two-leg ladder is a minimal geometry that allows for particle exchanges, and therefore provides intuition into the properties of infinite two dimensional systems of anyons. Each site of the ladder can either be empty or has a single non-trivial anyon. The anyons are allowed to interact and braid around each other. In analogy with fermionic and bosonic Hubbard model, we dub this *anyonic Hubbard model* on a ladder. In the numerical results presented, we affirmatively establish that phases of anyonic systems depend nontrivially on the braid statistics. In other words, different anyon species under the same model microscopic Hamiltonian will have different phase diagrams in the presence of braiding operation. This conjecture is reinforced by a doing a comparative analyses using Fibonacci and Ising anyons—which are two prominent anyon species with different braid statistics.

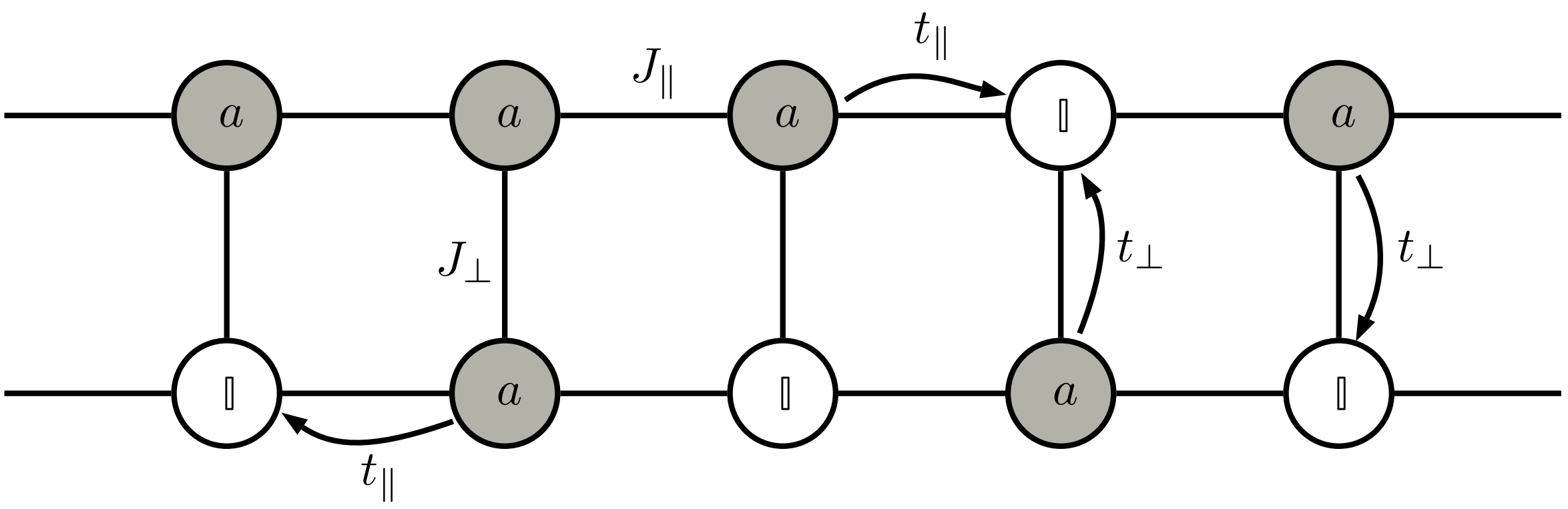

Figure 6.1: A two-leg ladder of interacting and itinerant non-Abelian anyons. An empty site is equivalent to a vacuum charge. When a site is occupied with a charge $a$, the charge can be either a Fibonacci anyon $\tau$ or an Ising anyon $\sigma$. (The fermion $\psi$ in $\mathcal{A}^{\text{Ising}}$ will not be physically supported on sites).

The rest of this Chapter are organised as follows: In Section 6.2, we introduce and describe our model. In Section 6.3, we present the phase diagrams of the model, and in Section 6.4, we present the numerical results of the order parameters for the phases. We end with a conclusion in Section 6.5.

## 6.2 Anyonic Hubbard model on a ladder

### 6.2.1 Model and its Hilbert space

The system we study is a two-leg ladder of interacting and braiding non-Abelian anyons. The ladder has a finite vertical width of two and an infinite horizontal dimension. See Fig. 6.1. There is an imposed "hardcore" constraint on each site to penalise the occupation of more than one particle. The specific particles supported are chosen from the Fibonacci and Ising anyons particle spectrum: $\mathcal{A}^{\text{Fib}} = \{\mathbb{I}, \tau\}$, and $\mathcal{A}^{\text{Ising}} = \{\mathbb{I}, \sigma, \psi\}$, where $\mathbb{I}$ represent vacuum charge (or empty site), $\tau$ is one Fibonacci anyon, $\sigma$ is one Ising anyon, and $\psi$ is a fermion obtained when two Ising anyons fuse. All the particles except the fermion $\psi$ are allowed on site. The fusion algebra and other anyonic data of these particles are recalled in Appendix A. We study the model with each anyons species independently, and not an eclectic mix of both Ising and Fibonacci anyons.

We use a number of conventions to write the charges of the system: the particles on each site may be enumerated explicitly in terms of the charge $a$ of the anyon, or if we use an additional U(1) symmetry [i.e. particle (number) density conservation] present in the system, the particles can be enumerated in terms of the composite charge label $c = (a, n)$, where $a$ is the anyonic charge and $n$ is the number charge associated to the anyon charge, which can also be re-written in subscript style as $a_n$. We associate a number $n = 0$ to the vacuum charge and $n = 1$ to one Fibonacci or Ising anyon. Table 6.1 summarises our charge enumeration convention.

As the ladder has a finite vertical width, it might sometimes prove convenient to view it as a one dimensional system by coarse-graining the rungs to a single site, thereby collapsing the ladder into a chain. The charges on each sites of the chain would be the fusion outcomes of the charges on the rungs of the original ladder. Table 6.2 lists the combined charges on the rungs. In addition, we represent the anyonic charges on sites and their fusion outcomes on the rungs using a set of diagrammatic representations shown in Fig. 6.2.

| Fibonacci anyons | Ising anyons |
|---|---|
| $a = \begin{cases} \mathbb{I} \equiv (\mathbb{I}, 0) \equiv \mathbb{I}_0 \\ \tau \equiv (\tau, 1) \equiv \tau_1 \end{cases}$ | $a = \begin{cases} \mathbb{I} \equiv (\mathbb{I}, 0) = \mathbb{I}_0 \\ \sigma \equiv (\sigma, 1) \equiv \sigma_1 \end{cases}$ |

Table 6.1: Table shows onsite charges and our different enumeration convention. The vacuum charge is written as $\mathbb{I} \equiv (\mathbb{I}, 0) \equiv \mathbb{I}_0$. One Fibonacci and Ising anyon is written as $\tau \equiv (\tau, 1) \equiv \tau_1$ and $\sigma \equiv (\sigma, 1) \equiv \sigma_1$.

| Fibonacci anyons | Ising anyons |
|---|---|
| $(\mathbb{I}, 0) \times (\mathbb{I}, 0) = (\mathbb{I}, 0) \equiv \mathbb{I}_0$ | $(\mathbb{I}, 0) \times (\mathbb{I}, 0) = (\mathbb{I}, 0) \equiv \mathbb{I}_0$ |
| $(\mathbb{I}, 0) \times (\tau, 1) = (\tau, 1) \equiv \tau_1$ | $(\mathbb{I}, 0) \times (\sigma, 1) = (\sigma, 1) \equiv \sigma_1$ |
| $(\tau, 1) \times (\mathbb{I}, 0) = (\tau, 1) \equiv \tau_1$ | $(\sigma, 1) \times (\mathbb{I}, 0) = (\sigma, 1) \equiv \sigma_1$ |
| $(\tau, 1) \times (\tau, 1) = \begin{cases} (\mathbb{I}, 2) \equiv \mathbb{I}_2 \\ (\tau, 2) \equiv \tau_2 \end{cases}$ | $(\sigma, 1) \times (\sigma, 1) = \begin{cases} (\mathbb{I}, 2) \equiv \mathbb{I}_2 \\ (\psi, 2) \equiv \psi_2 \end{cases}$ |

Table 6.2: Table shows the combined total charges on the rungs, derived from fusing the onsite charges of the two sites on the rung. In the fusion $(a_1, n_1) \times (a_2, n_2)$, the left ordered pair represent the charge on the top site and the right one represent the charge on the bottom site of the rung. Fusion of the anyonic charges follow the fusion rules of the specific anyon model (see Appendix A), and the number charges fuse using ordinary addition. As in Table 6.1, the subscript $n$ in $a_n$ is the number of non-trivial charges (e.g. $\tau$ or $\sigma$) fusing into the anyon charge $a$.

The Hilbert space of our model admits both anyonic (quantum) symmetry and U(1) symmetry. The U(1) symmetry is defined in terms of conserved global particle density. At any fixed global particle number $N$, particles can have different spatial configurations, which are specified by a configuration set $\mathcal{I}_N$. As the particles are non-Abelian anyons, they also have a fusion space when $N > 0$. The Hilbert space structure is therefore a direct sum,

$$\bigoplus_{i \in \mathcal{I}_N} \mathbb{V}_i^{(N)}, \tag{6.1}$$

where $i$ is the index of the spatial configurations of the particles and $\mathbb{V}_i^{(N)}$ is the anyonic fusion space of that spatial configuration. For all possible particle numbers $N$ of the system, the total Hilbert space is therefore,

$$V = \bigoplus_{N=0}^{\infty} \bigoplus_{i \in \mathcal{I}_N} \mathbb{V}_i^{(N)}. \tag{6.2}$$

### 6.2.2 Hamiltonian

We construct an anyonic Hubbard-like Hamiltonian $\hat{H}$ for the dynamics of particles on the ladder (see Fig. 6.1), which consists of a kinetic and a chemical potential term, jointly

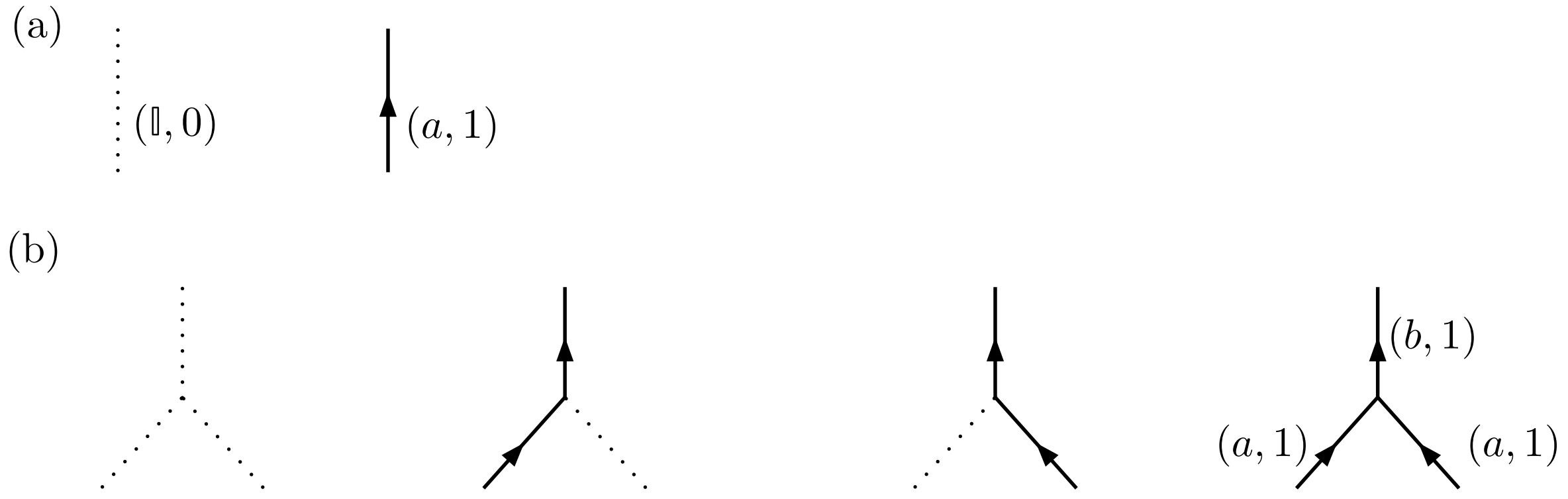

Figure 6.2: Diagrammatic representations of anyonic charges. Sub-figure (a) shows two types of lines representing the two types of charges permitted on each site: a vacuum charge $(\mathbb{I}, 0)$ and a non-trivial charge $(a, 1)$, where $a = \tau$ or $\sigma$. Sub-figure (b) shows the charge fusion outcomes on each rung, where the lines carry the same meaning as in (a). The charge label $b$ in the last vertex diagram can be $b = \mathbb{I}$ if $a$ is either a $\tau$ or $\sigma$—as both can annihilate to vacuum, or $b = \tau$ if $a = \tau$, or $b = \psi$ if $a = \sigma$—all based on the fusion rules of the specific anyon model. Note that as Fibonacci and Ising anyons are self dual charges, lines would not need arrowheads, but in the presence of a U(1) number charge—which is not self-dual, the lines would need arrowheads.

represented by $\hat{H}_{\text{t-}\mu}$, and an interaction term $\hat{H}_{\text{J}}$,

$$\hat{H} = \hat{H}_{\text{t-}\mu} + \hat{H}_{\text{J}}. \tag{6.3}$$

As with fermions and bosons, the kinetic term accounts for the hopping of particles between sites, but now in addition for being anyons, it also braids them when they hop passed one another. The chemical potential term tunes the rational filling of the ladder. The interaction term acts locally between nearest-neighbouring anyons. The fusion outcome of two Fibonacci anyons is $\tau \times \tau \longrightarrow \mathbb{I} + \tau$, and that of two Ising anyons is $\sigma \times \sigma \longrightarrow \mathbb{I} + \psi$. The interaction term is an Heisenberg-type one, which favours projection of either two $\tau$ particles or two $\sigma$ particles to the vacuum. If the projection to vacuum charge sector is assigned the lowest energy, it is termed "antiferromagnetic" otherwise it is termed "ferromagnetic." Note that for the case of Fibonacci anyons, ferrromagnetic interactions favour fusing to a non-Abelian charge, whereas for Ising anyons, both interactions fuse to a an Abelian charge.

To capture the physics of braiding, it is necessary to map the sites of the two dimensional ladder to a canonical ordering in one dimension which can then be used to order charges on the leaves of a fusion diagram. Using Fig. 6.3 as a reference, the ladder should be viewed from bottom to top, where the particles on the top leg of the ladder are assumed to be "behind" those on the bottom leg, and we adopt a zig-zag linear ordering of the sites, which minimizes the interaction length the most. The anyons are fused in the "standard" convention: from left to right.

We derive below the terms of the Hamiltonian with respect to this orientation and ordering.

### Kinetic and chemical potential term

We derive the contributions to the kinetic energy and chemical potential term using Fig. 6.3(i). The kinetic term of the Hamiltonian receives contributions from a possible combination of the following three scenarios: (1) the hopping of a non-trivial charge between sites $2i - 1$ and $2i + 1$ on the top leg, while passing behind site $2i$ on the bottom leg, and braiding with any non-trivial charge there, otherwise, the passage does not introduce any braid factor into the state of the system. (2) Similarly, the hopping of a non-trivial charge between sites $2i$ and $2i + 2$ on the bottom leg, while passing in front of site $2i + 1$ on the top leg, and also braiding with any non-trivial charge there, otherwise that passage also does not introduce any braid factor into the state of the system. (3) And lastly, the hopping of a charge between sites $2i - 1$ and $2i$ on the rung, which does not introduce any braid factor—as they are nearest neighbouring sites based on our choice of linear ordering. Note that, because of the hardcore constraint, there needs to be a vacant site for hopping to take place. The chemical potential is the energy needed to add or remove particles from the sites of the ladder.

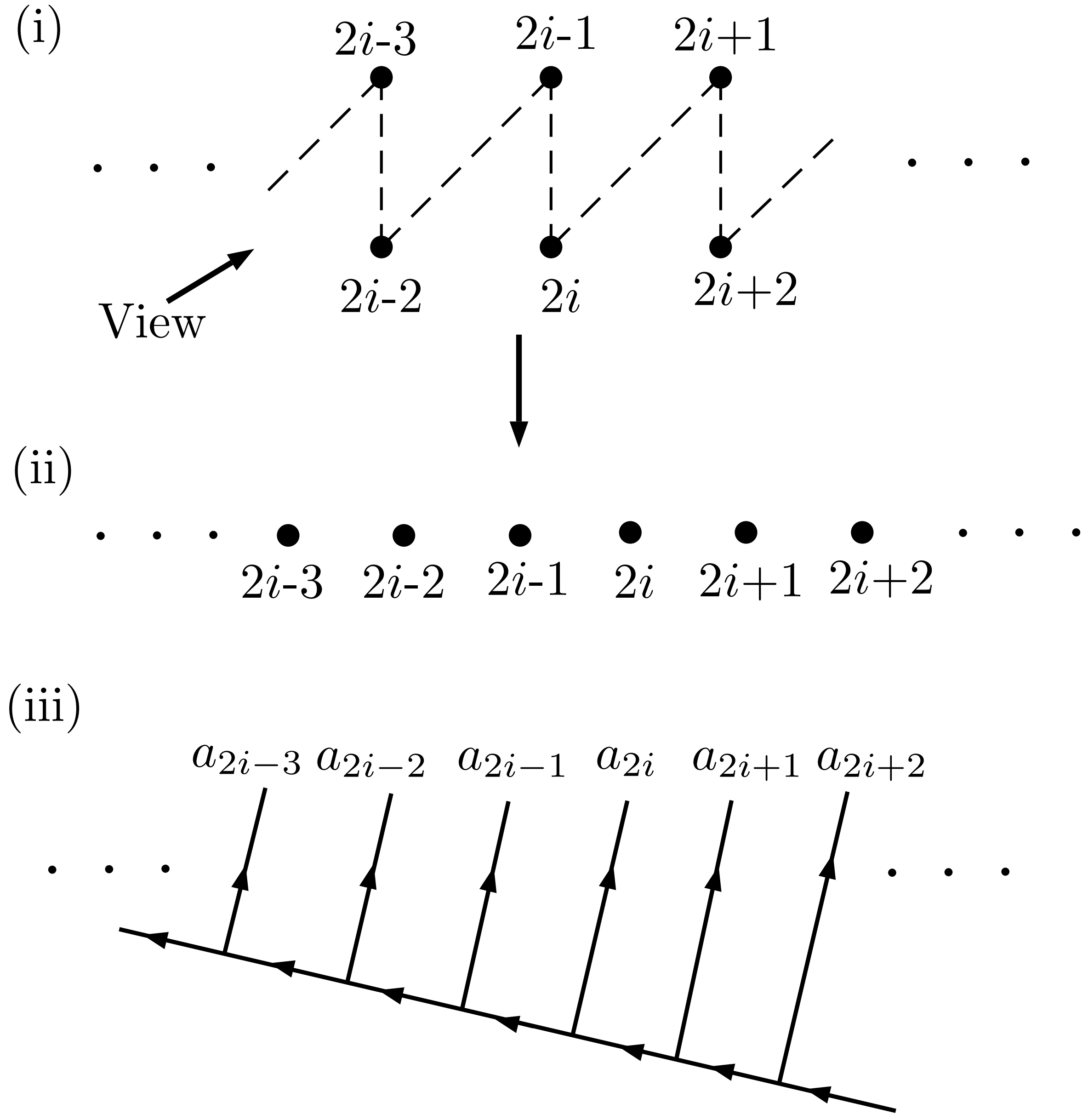

Figure 6.3: (i) Orientation of ladder: The ladder is viewed at an angled position from the bottom to top. Particles on the top leg are regarded to be "behind" those on the bottom leg. (ii) Linear ordering: The ladder is projected to one dimension; the ladder is "stretched out" into a chain. (iii) Fusion ordering: The particles on the sites of the ladder are fused in the standard way—from left to right. The charge label of site $i$ is denoted as $a_i$ (which can be a vacuum charge or a non-trivial charge). The intermediate fusion outcomes on the links of the fusion tree have been suppressed for clarity.

In the convenient diagrammatic formalism similar to those used in Refs. [1, 101], the kinetic and chemical potential term, $\hat{H}_{\text{t-}\mu}$, is,

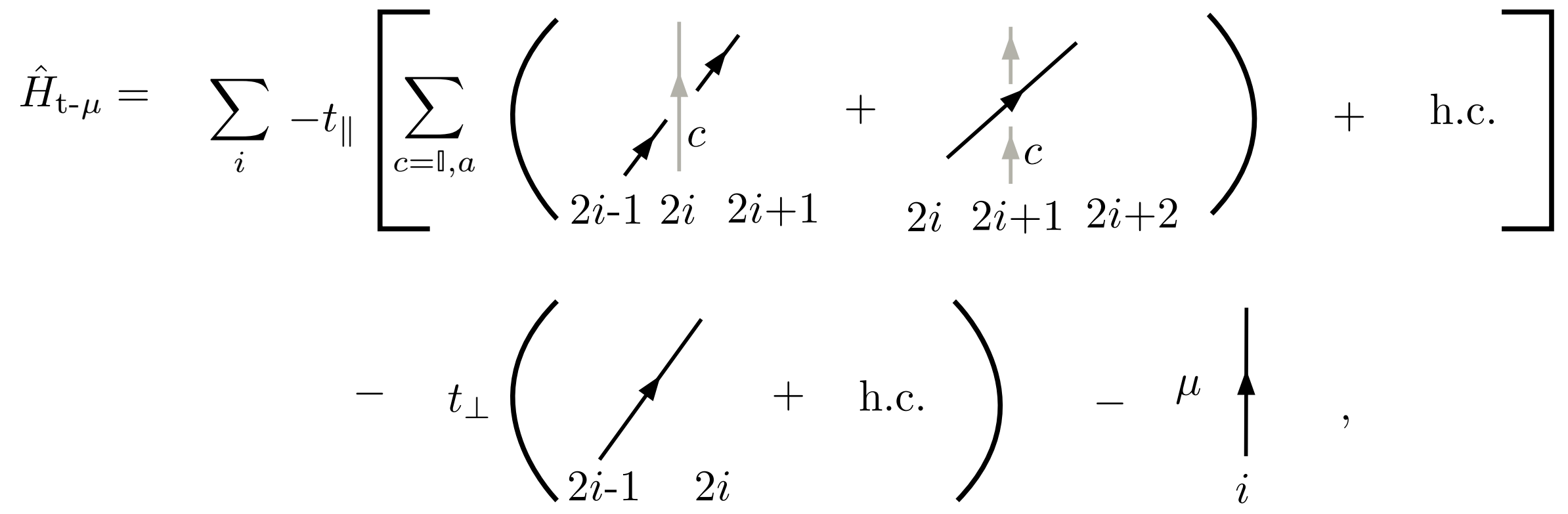

(6.4)

where $t_{\|}$ is the horizontal hopping strength on the legs, $t_\perp$ is the transverse hopping strength on the rungs, and $\mu$ is the chemical potential. The term “h.c.” denotes the Hermitian conjugate of all the diagrammatic terms preceding it. In our diagrammatic expression, a solid black line represents a non-trivial charge $a$, which can be $\tau$ or $\sigma$ depending on the model, while a grey line represents either the trivial vacuum charge $\mathbb{I}$ or a nontrivial charge $a$.

Note that, when $c = \mathbb{I}$, hopping does not introduce any braid factor, but when $c = a$, the hopping term acquires a braid factor $R^{aa}$ for “under-crossing,” and $(R^{aa})^\dagger$ for “over-crossing.” The matrix representation of these braid factors, for both Fibonacci and Ising anyons, are given in Appendix A.

### Heisenberg interaction term

The interaction term is a sum of Heisenberg-like couplings applied to anyons. For two non-trivial anyonic charges with label $a$, and fusion outcomes $a \times a \to \mathbb{I} + \cdots$, an antiferromagnetic coupling energetically favours projection of the two charges $a$ into the vacuum channel $\mathbb{I}$, while ferromagnetic coupling favours the other charge sectors.

With reference to Fig. 6.3(i), we project the non-trivial charges $a$ on sites $(2i-1, 2i+1)$ on the top leg and sites $(2i, 2i + 2)$ on the bottom leg to the vacuum charge $\mathbb{I}$ with an interaction strength $J_{\|}$, and also project the charges $a$ on sites $(2i - 1, 2i)$ on the rung to the vacuum charge $\mathbb{I}$ with an interaction strength $J_\perp$.

Diagramatically, the interaction Hamiltonian is,

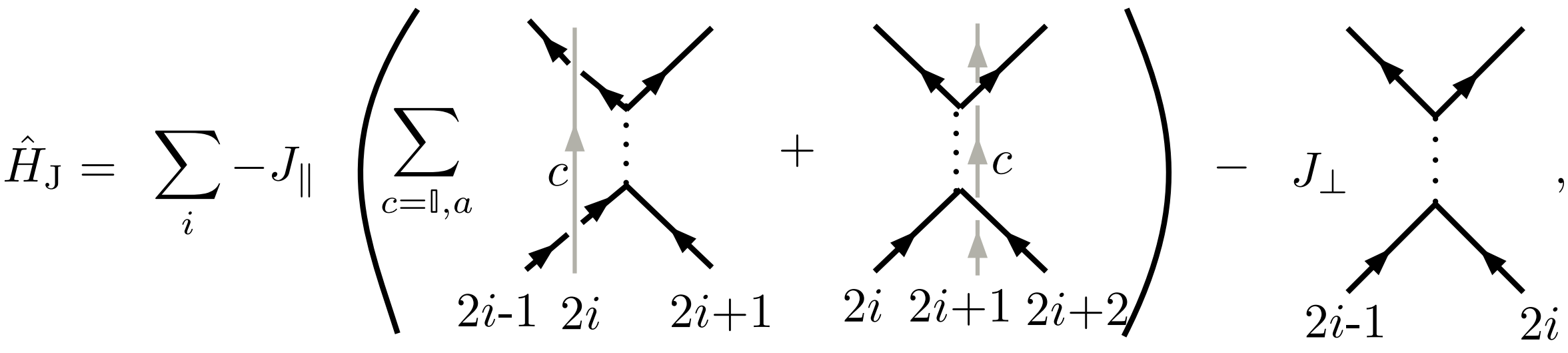

(6.5)

where the dotted line represents the vacuum charge $\mathbb{I}$. For this work we focus on the fully antiferromagnetic interaction regime, i.e. $J_{||}, J_{\perp} > 0$.

## 6.3 Results

In this section we present the main results explaining the various phases present in ground states of braided non-Abelian anyons on the ladder. We performed numerical simulations for both Fibonacci and Ising anyons using three different types of anyonic tensor networks algorithms: anyonic×U(1)-MPS [100], anyonic TEBD [41], and anyonic-DMRG [43]. Except where otherwise indicated, the results presented in this work are based on numerical simulations with a bond dimension of $\chi = 200$. For clarity, a detailed presentation of numerical results used to identify the phases is deferred to Sec. 6.4.

The Hamiltonian defined in Eq. (6.4) and 6.5 has several parameters: hopping $t_{\perp}$ of particles on the rungs (or simply "rung hopping"), hopping on legs $t_{||}$ (or "leg hopping"), a chemical potential term $\mu$ that controls the filling of the ladder, an (Heisenberg) interaction on rungs $J_{\perp}$ (or "rung interaction"), and interaction on the legs $J_{||}$ (or "leg interaction"), and therefore, the parameter space is very large. To recapitulate, we are mainly interested in studying the effect of braid statistics on the ground states of anyonic systems, and therefore the leg hopping $t_{||}$ and the chemical potential $\mu$ are the two parameters that we make tunable. In order to maintain a ladder model, we set $t_{\perp} = 1$—i.e. we do not consider an instance of having two copies of a single chain (when $t_{\perp} = 0$). The ground space of *free* non-Abelian anyons is exponentially degenerate in the number of anyons in the system, which become lifted in the presence of inter-particle interaction. Henceforth, we set the interaction strengths equal $J_{||} = J_{\perp} \equiv J = 1$. Furthermore, at any non-zero filling of the ladder, and at any non-zero value of $t_{||}$, the anyons braid with strength $\sim (t_{||} R^{aa})$ which also acts as an interaction on the fusion degrees of freedom.

### 6.3.1 Order parameters

The relevant phases in our system are identified using both local order parameters and non-local entropic calculations. Average particle density per rung, $\nu = \langle \hat{n} \rangle$, and pair correlation function, $C(r)$, are the calculated local order parameters, while the block scaling of the entanglement entropy (EE), $S(r)$, is non-local as it is computed from the reduced density matrix of many blocks of varying size.

The pair correlation function $C(r)$ is the expectation value of a pair particle propagator, $\hat{C}(r)$. We define the anyonic pair particle propagator as an operator that fuses a pair of particles into a particular charge outcome at rung $i$ and splits off at a distant rung $j$, as defined

in the diagram below:

$(i)$

$\hat{C}_{\mathbb{I}}(r) =$ 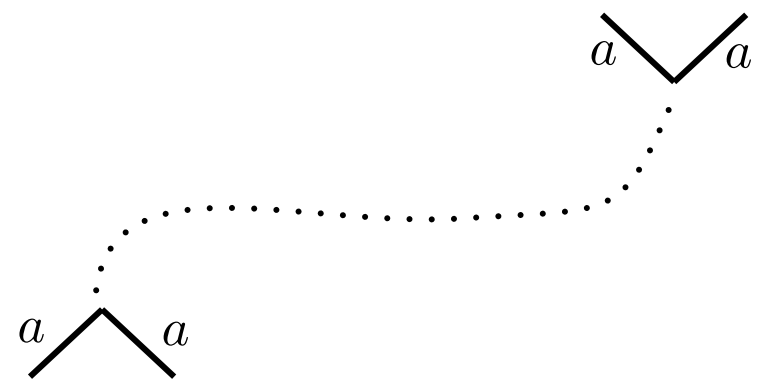

,

$(ii)$

$\hat{C}_b(r) =$ 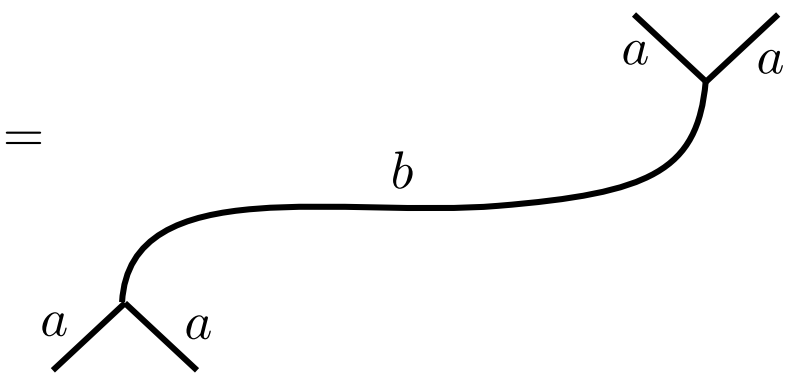

, (6.6)

where $a = \tau$ or $\sigma$. In $(i)$, two anyons of charge $a$ fuse into the vacuum charge $\mathbb{I}$ for both anyon models, and in $(ii)$, the two charges $a$ can also fuse into a non-vacuum charge $b$, where $b = \tau$ for Fibonacci anyons and $b = \psi$ for Ising anyons.

Some insight into the expected behaviour of this correlation function $C(r)$ for anyons can be gleaned from the behaviour of the pair correlation function for spinless fermions on a ladder as investigated in Ref. [102]. In the infinite ladder limit, the correlation of the (quasi) ordered superfluid phase decays polynomially as,

$$C(r) \sim r^{-\frac{1}{K}}, \tag{6.7}$$

where $r$ is the inter-particle separation, and $K$ is the Luttinger parameter. In the (quasi) disordered (non-superfluid) phase, $C(r)$ decays exponentially,

$$C(r) \sim e^{-r/\xi}, \tag{6.8}$$

where $\xi$ is the correlation length.

With regards to the nonlocal order parameter, we compute the von Neumann EE,

$$S(l) = -\mathrm{Tr}\left(\hat{\rho}_l \log \hat{\rho}_l\right). \tag{6.9}$$

where $\hat{\rho}_l$ is the reduced density matrix of a contiguous block of $l$ sites on the lattice. This quantifies how entangled the block is with the rest of the lattice. We coarse-grain rungs on the ladder to sites of a 1-D chain in order to compute this quantity. For critical systems with open boundaries and in the thermodynamic limit, the entanglement entropy has a simple relation from conformal field theory (CFT) [98, 103, 104],

$$S(l) = \frac{c}{3} \log l, \tag{6.10}$$

where $c$ is the central charge of the CFT.

## 6.3.2 Phase diagrams

The phase diagrams for the ground states of the model Hamiltonian $\hat{H}$ for both Fibonacci and Ising anyons are presented in Fig. 6.4 (planar view) and Fig. 6.5 (three-dimensional view). As mentioned above the only varying parameters in our simulations are the chemical potential $\mu$ which populates the ladder with particles and the horizontal hopping $t_\parallel$ which braids the particles. As the microscopic Hamiltonians for both anyon models are the same, we attribute the differences in the phase diagrams to the differing fusion outcomes and braid statistics of the anyon models as described below.

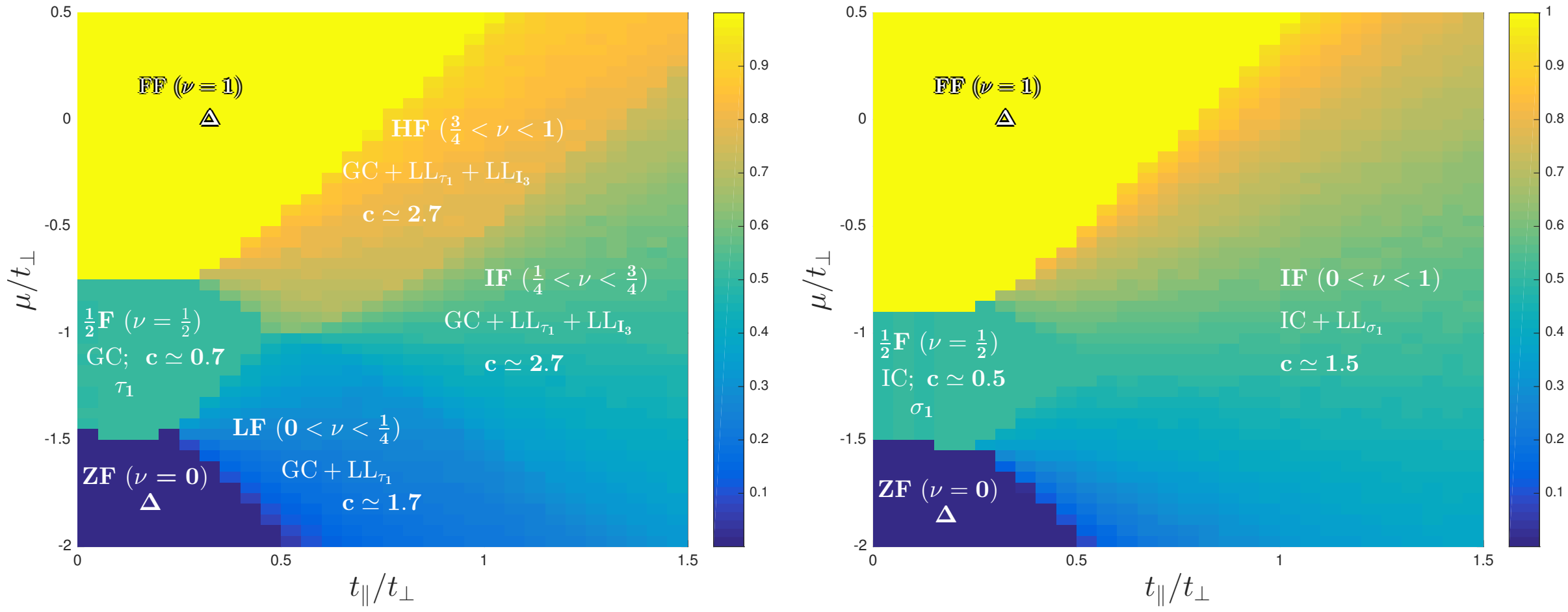

(a) Fibonacci anyons (b) Ising anyons

Figure 6.4: Phase diagrams for ground states of interacting and braiding ($a$) Fibonacci and ($b$) Ising anyons on a ladder. The horizontal axis is the leg hopping ratio $t_\parallel/t_\perp$, the vertical axis is the chemical potential $\mu/t_\perp$, and the plotted quantity is the filling fraction $\nu = \langle\hat{n}\rangle$. Six distinct phases are apparent for Fibonacci anyons and four for Ising anyons. The phases are described according to: the nature of filling and filling fraction (in round brackets), the central charge for the gapless phases, and the types of mobile excitations of the Luttinger liquid. The abbreviations are: ZF—zero filling ($\nu = 0$), $\frac{1}{2}$F—half filling ($\nu = 1/2$), FF—fully filled ($\nu = 1$), LF—low filling ($0 < \nu < \frac{1}{4}$), IF—intermediate filling ($\frac{1}{4} < \nu < \frac{3}{4}$), HF—high filling ($\frac{3}{4} < \nu < 1$), Δ—gapped, $c$—central charge, GC—Golden chain, IC—Ising chain, $LL_{\tau_1}$—Luttinger liquid of hopping $\tau_1$ charges, $LL_{\sigma_1}$—Luttinger liquid of hopping $\sigma_1$ charges, and $LL_{\mathbb{I}_3}$—Luttinger liquid of hopping "neutral Fibonacci triplets".

While the phase diagrams are very different at high leg hopping ratio $t_\parallel/t_\perp$, they look similar at low ratios. Hence, we divide the explanation of the phase diagrams into two subsections: one for low leg hopping ratio and the other for high hopping ratio.

### Low hopping ratio

At low leg hopping ratio ($0 \leq t_\parallel/t_\perp \lesssim 0.5$) the two phase diagrams in Fig. 6.4 are similar, and share a common description. Briefly stated, as the chemical potential $\mu/t_\perp$ varies at low $t_\parallel/t_\perp$, there is a first order phase transition from a phase with zero filling to a phase which is half-filled, and finally to a phase where the ladder is fully filled. These transitions are sharply revealed by jumps in the filling fraction as a function of the chemical potential as shown in Fig. 6.5.

The explanation for this behaviour is as follows. The chemical potential term is the energy required to add or remove particles from the system. At very low values of $\mu$, the system has negligible number of particles, and averages to zero particle density which explains the appearance of the "deep blue" region on the bottom left in both phase diagrams (with the abbreviation ZF—for zero filling). This phase is trivially gapped ($\Delta$) with the dominant occupation being the trivial vacuum charge $\mathbb{I}_0$ on all sites of the ladder.

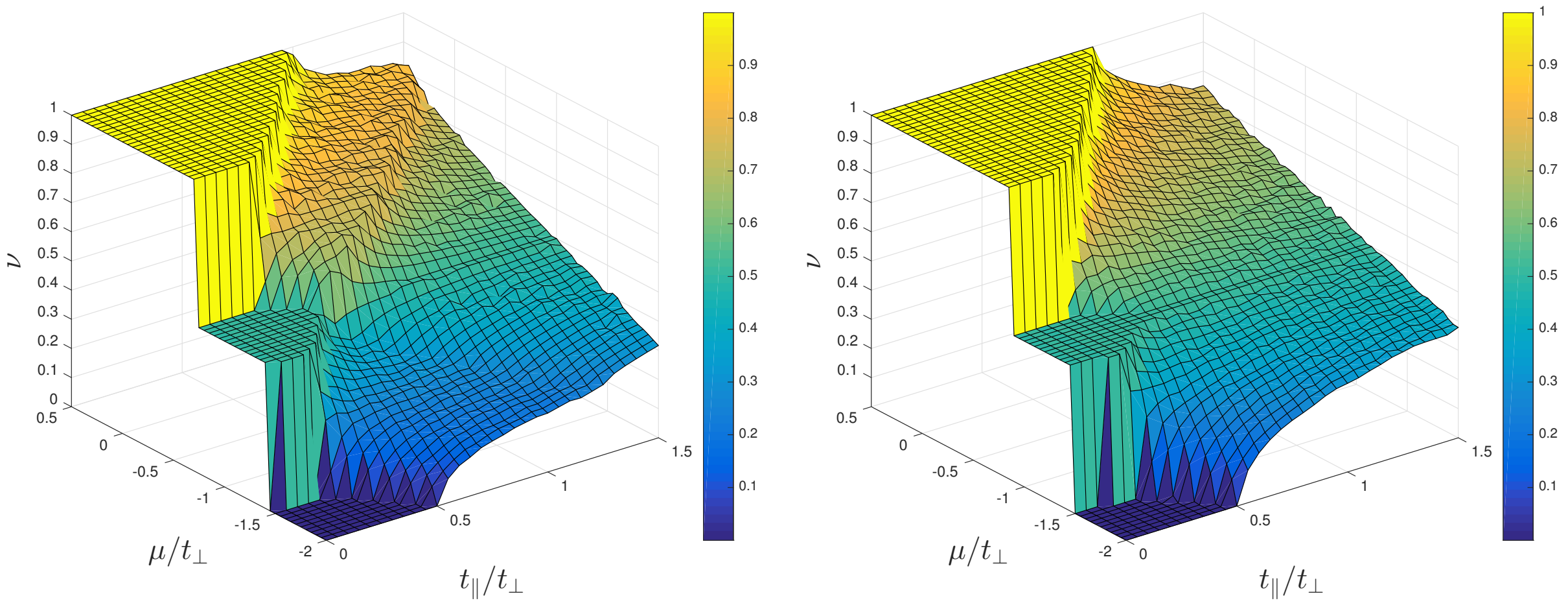

(a) Fibonacci anyons

(b) Ising anyons

Figure 6.5: Three-dimensional versions of Fig. 6.4. The vertical axis is the filling fraction—which was collapsed in Fig. 6.4. First order phase transitions can be seen as the chemical potential $\mu/t_\perp$ varies at low leg hopping ratio $t_\parallel/t_\perp$. Second order phase transitions are also evident at higher leg hopping $t_\parallel/t_\perp$ values for varying values of the chemical potential $\mu/t_\perp$.

As the chemical potential increases, we observe a first order phase transition from the zero-filling phase into a phase where the ladder is half-filled ($\frac{1}{2}$F) and has an average particle density of $\nu = 1/2$ per rung (i.e. one particle per rung). We note that, there is no phase transition into an $\epsilon$-filling phase, where $\epsilon$ is a small number. On each local rung, the chemical potential term is $-\mu\hat{n}_i$, where $\langle\hat{n}_i\rangle = \{0, 1, 2\}$, i.e. particles are added locally in discrete amounts, from no particle to a maximum of two on every rung. In the case of single particle occupancy per rung, due to the non-zero value of $t_\perp = 1$, the particle will be in a superposition $|+\rangle_i = \frac{1}{\sqrt{2}}(|1, a\rangle + |2, a\rangle)$ of the top and bottom site of the rung, where "1" and "2" represent the top and bottom site respectively, and $a = \tau$ or $\sigma$.

At $t_\parallel/t_\perp = 0$, when particles do not hop across horizontally on legs, the configuration on the ladder acts effectively like a single chain of interacting anyons. It has been shown that an antiferromagnetic interacting anyon chain is critical and described by CFT, with known central charges $c = 0.7$ for Fibonacci anyons (known as the Golden chain) [88], and $c = 0.5$ for the Ising anyon chain (equivalent to the Ising model CFT). In the phase diagram Fig. 6.4, we found the half-filled phase to be critical and computed the central charges of their CFT to be, $c_{\text{Fib}} = 0.7029$ for Fibonacci anyons, and $c_{\text{Ising}} = 0.499$ for Ising anyons, which are very close to their theoretical values.

As the leg hopping ratio increases, particles use available vacancies to braid past one another. From our simulations, we found that the particles still arrange into an effective chain-like configurations with a central charge near their theoretical values. This may be explained by the fact that when non-Abelian anyons move past each other they experience braiding, which in the confined geometry of a ladder mimics a local interaction, and like the Heisenberg interaction, depends on the local fusion outcomes (see Refs. [105, 106] for more details). Thus for small enough hopping ratio, the anyons still behave as a single antiferromagnetically interacting chain.

Increasing the leg hopping $t_{\|}/t_{\perp}$ beyond a critical ratio $t_c$, we find evidence for a second-order phase transition into another phase. This critical ratio differs for Fibonacci and Ising anyons, and even more, the physics of the phases differ as explained below.

As the chemical potential increases further, at low hopping ratio $t_{\|}/t_{\perp}$, there is another first-order phase transition from the half-filled phase into a phase where the ladder is fully filled (FF). It has been shown in previous works ( Refs. [2, 42]) that two-leg ladder with Fibonacci or Ising anyons is gapped. This can be simply understood: there is a strong vertical AFM Heisenberg interaction $J_{\perp} = 1$ that favours fusing pair of charges on every rung into the vacuum charge, which is a product state under renormalisation. We give another qualitative reason for the gapped ground state behaviour at higher values of the chemical potential. At $\mu/t_{\perp} = 0$, for example there remains only two competing interactions in the Hamiltonian, namely hopping and interaction strengths on the rungs and legs of the ladder, where $t_{\perp} = J = 1$, and $t_{\|} < t_c < 1$. The energy for one free particle to hop on a rung is $t_{\perp} = 1$, which is the same as the interaction energy ($J = 1$) for two particles on a rung. However, since $J > t_{\|}$, it is energetically favourable to place two neighbouring particles on the legs of the ladder rather than placing one per rung with a vacant site that allows for hopping, and therefore the ground state of the ladder becomes fully filled.

### High hopping ratio

When the leg hopping ratio $t_{\|}/t_{\perp}$ increases further than a certain critical value $t_c$, there is a second-order phase transition into another phase, as shown in Figs. 6.4 and 6.5. From the calculations of the pair correlation function using Eq. (6.6) (i), we found these phases to be superfluid-like forming anyonic Luttinger liquid. In addition, from the calculations of the block scaling of the entanglement entropy using Eq. (6.9), we found that the anyons in these phases "fractionalize" into interacting anyonic degrees of freedom and itinerant hardcore particles. In other words, the interaction and kinetic terms in the Hamiltonian decouple into two "weakly" non-commuting terms, and as such can be treated separately. The same effect has been shown in Refs. [82, 83] to happen in a one-dimensional chain of interacting and itinerant non-Abelian anyons. The interacting and itinerant anyons separate into distinct interacting and itinerant parts analogous to spin charge separation in a chain of spinful electrons. The CFT has a central charge of $1 + c$, where $c$ is the central charge of (Heisenberg) interacting anyons and the number 1 is the central charge of itinerant bosons.

From our results, we found distinct behaviour between Fibonacci and Ising anyons:

*Ising anyons*—The filling fraction ($0 < \nu < 1$) varies continuously between 0 and 1, which means there is no phase transition as the chemical potential varies at high leg hopping ratio $t_{\|}/t_{\perp}$. We refer to this phase as the "Intermediate Filling" (IF) phase. The ground state is superfluid. The Ising anyons ladder behave effectively like a chain, having a central charge of $1 + c$, where we computed $c \simeq 0.499$ for the interacting anyons, which is very close to the theoretical value of the Ising CFT central charge $c_{\text{theo}} = 0.5$, and the number 1 is the central charge of the hopping hardcore $\sigma_1$ charge, which is the only mobile particle in this phase.

*Fibonacci anyons*—The phase diagram of Fibonacci anyons at high leg hopping ratio is more elaborate. There are three distinct superfluid and spin-charge separation phases.

At low chemical potential, the ladder is sparsely filled with particles and has a "low filling" (LF) fraction, $0 < \nu < 1/4$ per rung, meaning there is an average of one $\tau_1$ charge per every four sites of a plaquette. The central charge of the CFT of this phase was computed to be $c \simeq 1.762$. The nearest theoretical value to our value is $c_{\text{theo}} = 1.7 \equiv (0.7 + 1)$ according to the phenomenology of spin-charge separation. The Fibonacci anyon ladder therefore behaves

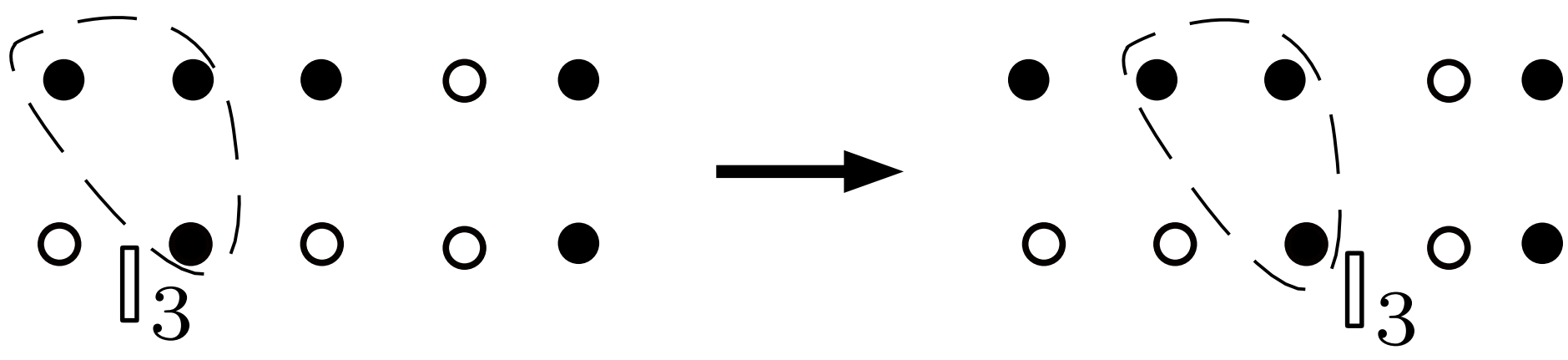

Figure 6.6: Illustration of the hopping of a neutral Fibonacci triplet $\mathbb{I}_3$. Let a filled dot represent a single Fibonacci $\tau$ charge, a vacant site as a vacuum charge $\mathbb{I}$, and an ellipse round a group of three $\tau$'s fusing into a total vacuum charge as a neutral Fibonacci triplet $\mathbb{I}_3$. When a single $\tau$ hops out of the triplet into a vacant site, due to the high density packing, another triplet can be formed, as shown. Thus, it appear as if the triplet hops on the ladder and a hole hopped in the reverse direction.

like a one dimensional chain in this regime.

As the chemical potential $\mu/t_\perp$ increases, more particles are added to the ladder. We found a phase transition into another phase, where the filling fraction is bounded as $\frac{1}{4} < \nu < \frac{3}{4}$, which we call the "intermediate filling" (IF) phase. Increasing $\mu/t_\perp$ further, there is another phase transition into another phase where the filling fraction is $\frac{3}{4} < \nu < 1$. We dub this phase as the "high filling" (HF) phase. Apart from their filling fraction, these two phases share a number of same properties including, they are critical and described by CFT, they are superfluid, and exhibit spin-charge separation—described by Luttinger liquid theory. We computed the central charge of the CFT of the IF phase to be $c \simeq 2.743$, and of the HF phase to be $c \simeq 2.8574$. These two values are in the vicinity of the theoretical value of $c_{\text{theo}} = 2.7$, though the offset of these numerical values can be attributed to the much entanglement created from braiding and the finite value of the bond dimension of our MPS simulations. Similar to the LF phase, this central charge value can also be explained using the phenomenology of anyonic spin-charge separation. The Heisenberg interaction term contributes a central charge value of $c = 0.7$, where the ladder behaves effectively like the Golden chain (GC) [88]. We attribute the remaining value of $c = 2.0$ to the kinetic energy term, which comes from the two types of distinct particle types with mobile degrees of freedom on the ladder. Though, $\tau_1$ is the only particle type with mobile degree of freedom in our Hamiltonian, giving a Luttinger liquid (labelled $\text{LL}_{\tau_1}$) and accounts for a central charge value of $c = 1.0$, we conjecture that, due to the ladder geometry, particles can group into "neutral Fibonacci triplets" $\mathbb{I}_3$ which have mobile degree of freedom, forming another Luttinger liquid $\text{LL}_{\mathbb{I}_3}$ with a central charge of $c = 1.0$. Three Fibonacci anyons have three fusion channels, which includes an $\mathbb{I}_3$, where the three $\tau$ charges can fuse into the vacuum charge $\mathbb{I}_3$—the subscript 3 is the number of $\tau$ charges. Due to a high packing density of the $\tau_1$ on the ladder, the neutral triplet $\mathbb{I}_3$ can hop when a single $\tau_1$ hops as illustrated in Fig. 6.6. The neutral triplet can move freely on the ladder and fusion tree and hence is different from the hopping of a single $\tau_1$ anyon.

Summing all the central charges together explains why the IF and HF phases have a central charge value of $c \sim 2.7$. We labelled these two phases as "$\text{GC} + \text{LL}_{\tau_1} + \text{LL}_{\mathbb{I}_3}$" for one Golden chain and Luttinger liquid of two different mobile excitations.

## 6.4 Numerical results

In this section we present the results of the calculations of the central charges and the superfluid correlation functions within the critical phases of the phase diagrams Fig. 6.4.

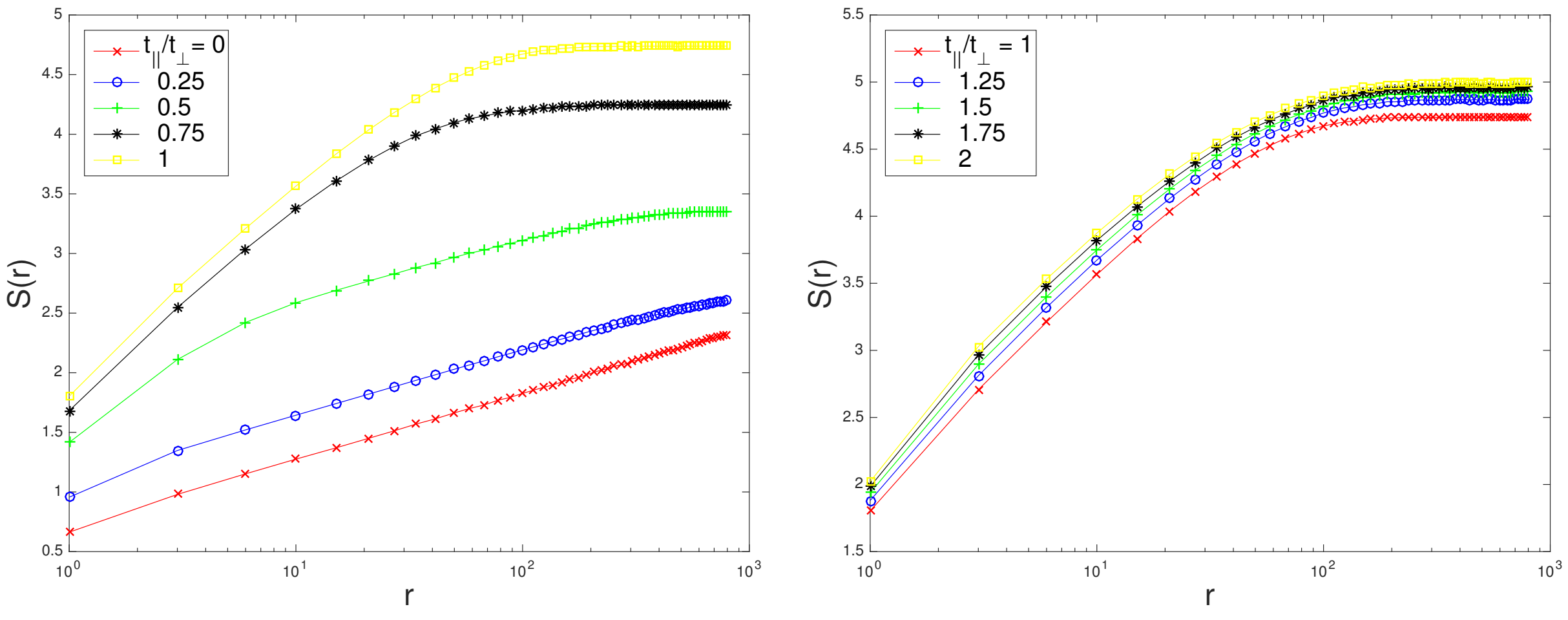

(a) (b)

Figure 6.7: Block scaling of the entanglement entropy for Ising anyons on the ladder at half filling ($\nu = 1/2$) when (a) the longitudinal hopping $t_{\parallel}/t_{\perp} = [0, 1]$ with increments of 0.25, and (b) $t_{\parallel}/t_{\perp} = [1, 2]$ with increments of 0.25.

The central charges are computed from the block scaling of the entanglement entropy $S(r)$ in Eqs. 6.9 and 6.10.

## 6.4.1 Scaling of the entanglement entropy

There are four distinct critical phases in the Fibonacci anyons phase diagram and two in the Ising anyons phase diagram. We first present the entanglement entropy plots of the Ising anyons, then followed by plots for Fibonacci anyons.

### Ising anyons

At low values of $t_{\parallel}/t_{\perp}$ and for $\mu/t_{\perp}$ between $-1.5$ and $-1$, the filling fraction is $\nu = 0.5$ per rung. We therefore use an Anyon×U(1)-MPS ansatz [42], where we set the filling fraction to $\nu = \frac{1}{2}$ to stay exactly in the half-filling sector.

The plots of $S(r)$ for the leg hopping ratio $t_{\parallel}/t_{\perp} = [0 : 0.25 : 1]$ and $t_{\parallel}/t_{\perp} = [1 : 0.25 : 2]$ are shown in Fig. 6.7(a) and (b), respectively. The expected linear scaling of the $S(r)$ in logarithmic scale (from Eq. (6.10)) can be seen in the plots of $t_{\parallel}/t_{\perp} = \{0, 0.25, 0.5\}$, but this linear scaling is not sharply evident for values of $t_{\parallel}/t_{\perp} \geq 0.75$ due to the entanglement created from braiding of the anyons and the finite bond dimension of the MPS. In such cases, a finite entanglement analysis [107, 108] (or known as finite-$\chi$ scaling) can be employed to fit a straight line to the entanglement entropy scaling. In the infinite limit of $\chi$, the scaling of the entanglement entropy would be linear, and has a central charge closer to the supposed theoretical value. For example Fig. 6.8 shows the finite-$\chi$ scaling of the EE, $S(r)$, at $t_{\parallel}/t_{\perp} = 1$. The linearity of the EE, $S(r)$, improves as the bond dimension $\chi$ increases.

The central charge can be computed from the slope of $S(r)$,

$$c = 3 \times \frac{S(r_2) - S(r_1)}{\log r_2 - \log r_1}, \tag{6.11}$$

where $r_1$ and $r_2$ are two points on the line. Some of the central charges computed for Ising anyons are shown in Table 6.3. It can be seen that at low values of $t\|/t_{\perp}$, particles are weakly delocalised and the ladder is an effective interacting Ising anyon chain. The central charge of the CFT is $c \sim 0.5$. At high hopping ratio $t_{\|}/t_{\perp}$, for example at $t_{\|}/t_{\perp} = 1$, there is spin-charge separation, and the central charge is $c \sim 1.5$, as previously explained.

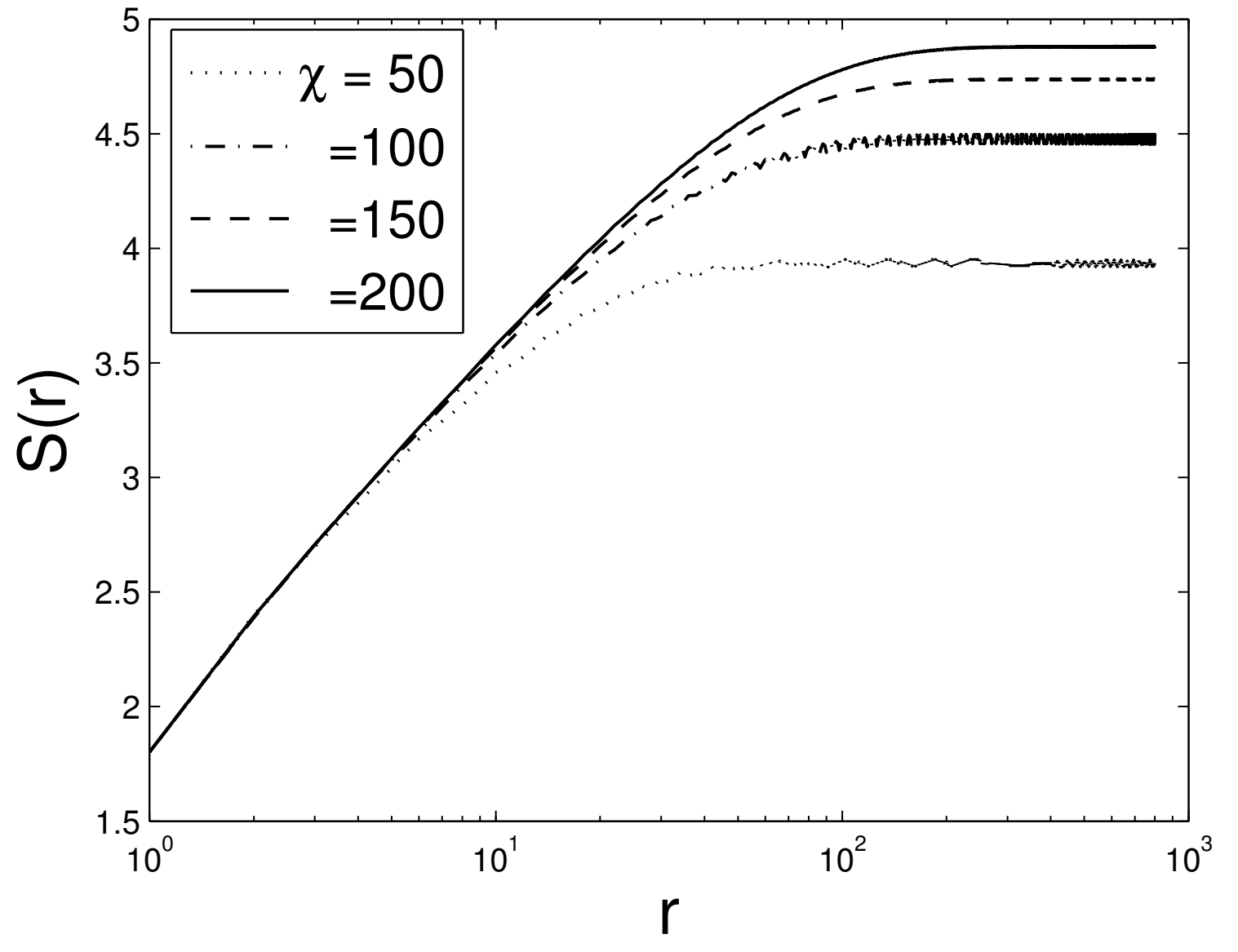

Figure 6.8: Finite entanglement analysis of block scaling of EE for Ising anyons at the filling fraction $\nu = 1/2$ and at $t_{\|}/t_{\perp} = 1$. The "bond dimension," $\chi$ in the legend controls the amount of entanglement allowable in the ground state. The computed central charge at these parameters is $c_{\text{num.}} = 1.499$.

| $t_{\|}/t_{\perp}$ | 0 | 0.25 | 0.5 | 1 | 2 |
|---|---|---|---|---|---|
| $c_{\text{num.}}$ | 0.505 | 0.499 | 0.502 | 1.499 | 1.534 |

Table 6.3: Numerical values of the central charges for Ising anyons extracted from the linear scaling of the entanglement entropy $S(r)$ at half filling.

To confirm that there is no other phase transition as the chemical potential $\mu/t_{\perp}$ varies at high leg hopping ratio $t_{\|}/t_{\perp}$, we computed the central charges at two other points in the phase diagram: $(\mu/t_{\perp}, t_{\|}/t_{\perp}) = \{(-2.0, 1.1), (0, 1.2)\}$. The plots of the EE for these parameters and their central charges are given in Fig. 6.9. Therefore, there is just a single Luttinger liquid and a single interacting anyon chain in the Ising anyons spin-charge separated phase.

### Fibonacci anyons

Like in the Ising anyons, at low values of $t_{\|}/t_{\perp}$ and for $\mu/t_{\perp}$ between $-1.5$ and $-1$, the filling fraction is $\nu = 0.5$ per rung, i.e. there is an average of one particle per rung. The ground state of this phase is critical and described by CFT with a central charge of $c = 0.7$ for AFM Heisenberg interaction. We therefore constrained the ground state ansatz to the half-filled sector using Anyon×U(1) MPS, and vary $t_{\|}/t_{\perp} \in [0 : 0.25 : 2]$. The plots of the scaling of the EE are shown in Fig. 6.10 (a) and (b). Some values of the central charges extracted from the block EE scaling (where finite-$\chi$ was done if necessary) are presented in Table. 6.4.

We computed the central charges of the three gapless (spin-charge separated) phases: at $(\mu/t_{\perp}, t_{\|}/t_{\perp}) = (-1.5, 0.75)$ in the "low filling" (LF) phase, the block scaling of the EE and

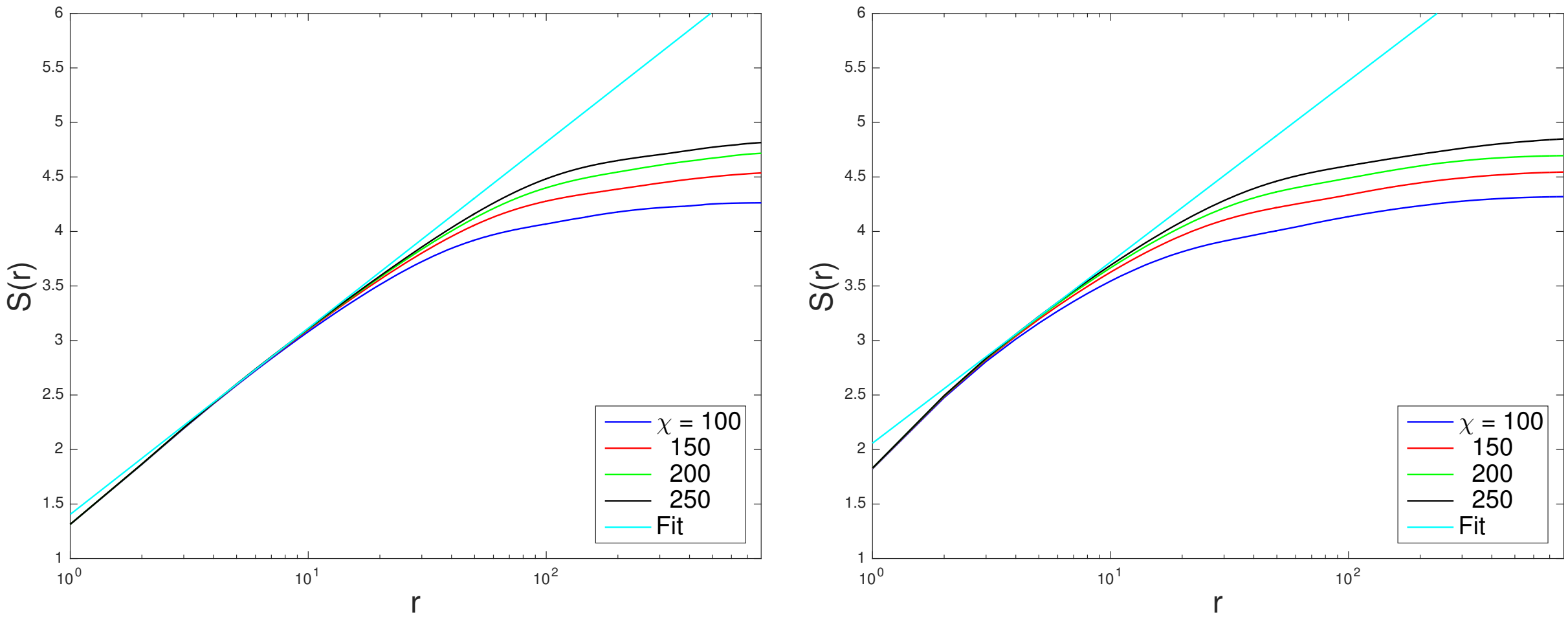

(a) (b)

Figure 6.9: Finite-$\chi$ scaling of the block entanglement entropy when (a) $(\mu/t_\perp, t_\parallel/t_\perp) = (-2, 1.1)$, and (b) $(\mu/t_\perp, t_\parallel/t_\perp) = (0, 1.2)$ [in the Ising phase diagrams Fig. 6.4(b)]. The central charge computed from the linear fit in (a) is $c = 1.542$, and the central charge from the linear fit in (b) is $c = 1.501$, which are both near the theoretical value of the central charge $c = 1.5$ for a spin-charge separation phase.

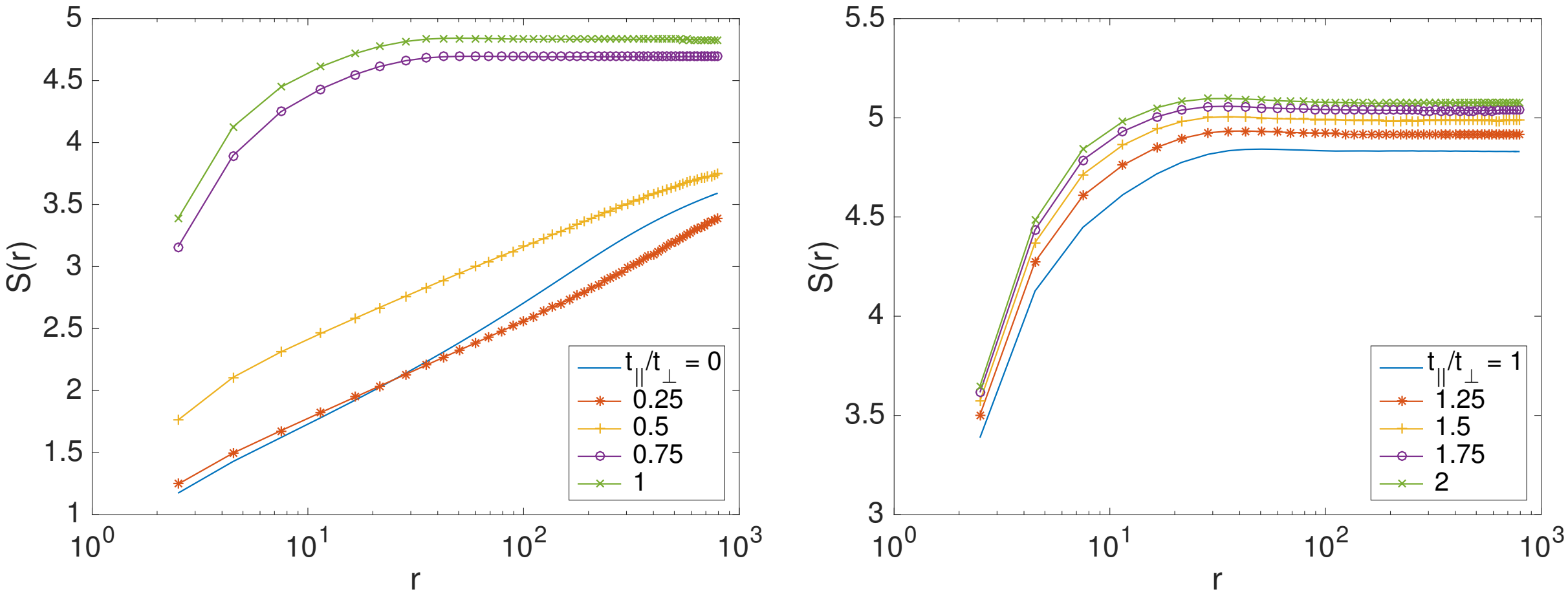

(a) (b)

Figure 6.10: Block scaling of the entanglement entropy $S(r)$ with block size $r$ of Fibonacci anyons at half filling ($\nu = 0.5$) for (a) $t_\parallel/t_\perp = [0, 1]$ with increments of 0.25, and (b) for $t_\parallel/t_\perp = [1, 2]$ with increments of 0.25. The entropy quickly saturates at high values of $t_\parallel/t_\perp$ due to entanglement from braiding.

| $t_\parallel/t_\perp$ | 0 | 0.25 | 0.5 | 1 | 2 |
|---|---|---|---|---|---|
| $c_{\text{num.}}$ | 0.754 | 0.703 | 0.701 | 2.819 | 2.921 |

Table 6.4: Numerical values of the central charge of Fibonacci anyons at half filling.

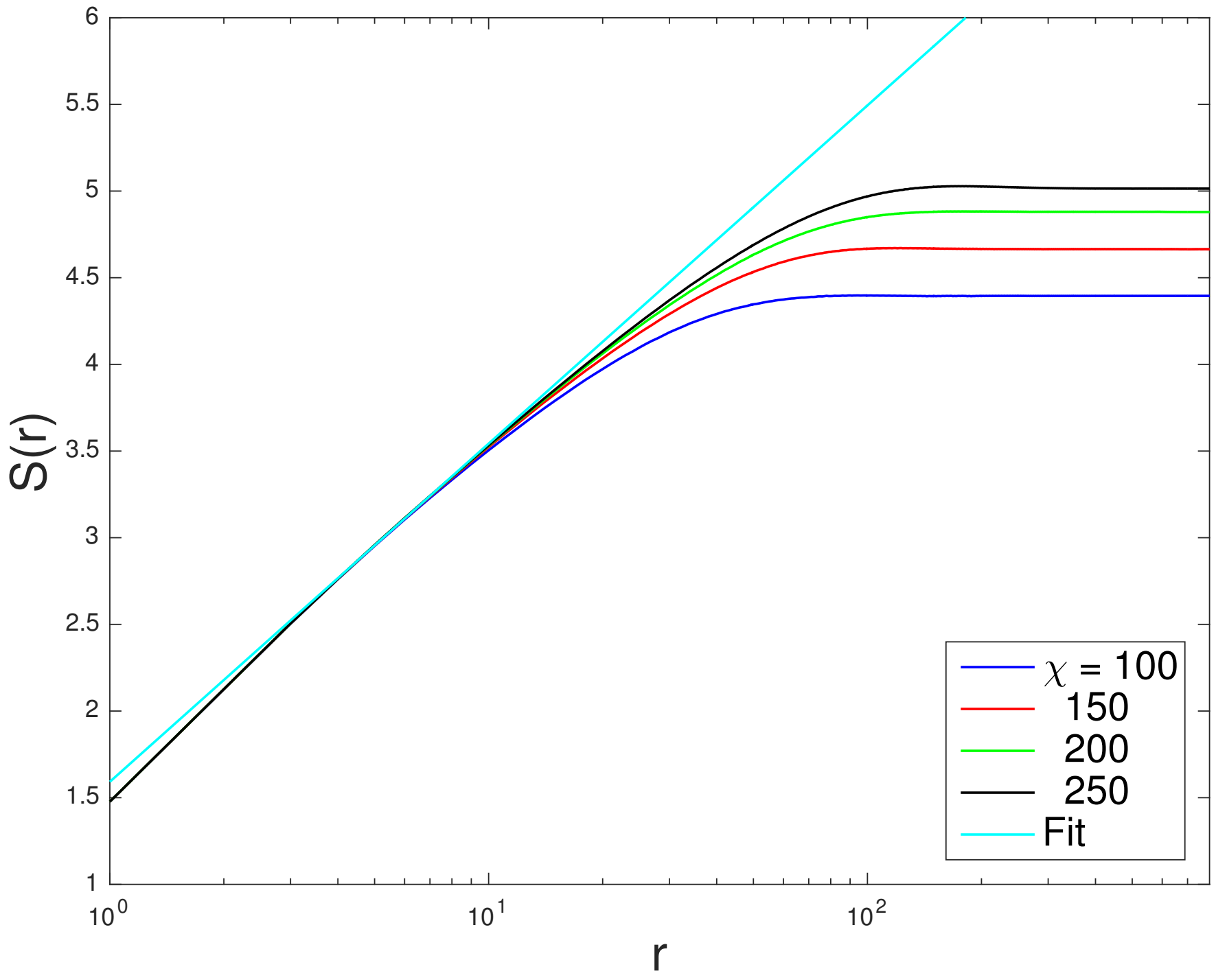

Figure 6.11: The block of the EE of Fibonacci anyons at $(\mu/t_\perp, t_\parallel/t_\perp) = (-1.5, 0.75)$ in the "low filling" phase. The central charge extracted from the linear fit is $c \simeq 1.762$, which is in close range of $c_{\text{theo}} = 1.7 \equiv (0.7 + 1)$ as explained in the text.

the central charge are given in Fig. 6.11; at $(\mu/t_\perp, t_\parallel/t_\perp) = (0, 1.25)$ in the "high filling" (HF) phase, the block scaling of the EE and the central charge are given in Fig. 6.12 (a); and finally, at $(\mu/t_\perp, t_\parallel/t_\perp) = (-1, 1.5)$ in the "intermediate filling" (IF) phase, the block scaling of the EE and the central charge are given in Fig. 6.12 (b).

### 6.4.2 Scaling of superfluid correlation function

We computed the scaling of the pair particle propagator defined in Eq. (6.6)(i) for Ising and Fibonacci anyons only at half-filling as the leg hopping ratio $t_\parallel/t_\perp$ varies from 0 to 2 in steps of 0.25. This calculation will reinforce the conclusion that the phases at high $t_\parallel/t_\perp$ are superfluid, i.e. quasi-ordered. Fig. 6.13 shows the plots of superfluid correlation for Ising and Fibonacci anyons. It can be deduced that, at half filling, the phase transition for Ising anyons occurs at $t_c = (t_\parallel/t_\perp)_c \sim 1$, while for $t_c \in [0.5, 0.75]$ for Fibonacci anyons. A better value of $t_c$ for Fibonacci anyons is obtained from the Fig. 6.14 which is a plot of the superfluid correlation for values of $t_\parallel/t_\perp = \{0.5, 0.75\}$. At half filling, the critical ratio for Fibonacci anyons is $t_c \sim 0.62$.

The Luttinger parameter can be obtained from a fit to the plot of the superfluid correlation. Though the value of Luttinger parameter $K$ for non-Abelian anyons is not known, we got the following numbers for points within the superfluid phase: for Ising anyons at $t_\parallel/t_\perp = 1$, we obtained $K_{\text{Ising}} \simeq 0.954$, and for Fibonacci anyons at $t_\parallel/t_\perp = 0.75$, we obtained $K_{\text{Fib}} \simeq 1.014$ (see Fig. 6.15).

In summary, at half filling there is a phase transition from a phase where the anyons form effective interacting chain to a different phase, which is superfluid and the anyons fractionalise into interacting and itinerant part. The critical points for both anyon models are different, which we attribute to differing braid statistics since the microscopic Hamiltonian of both models is the same.

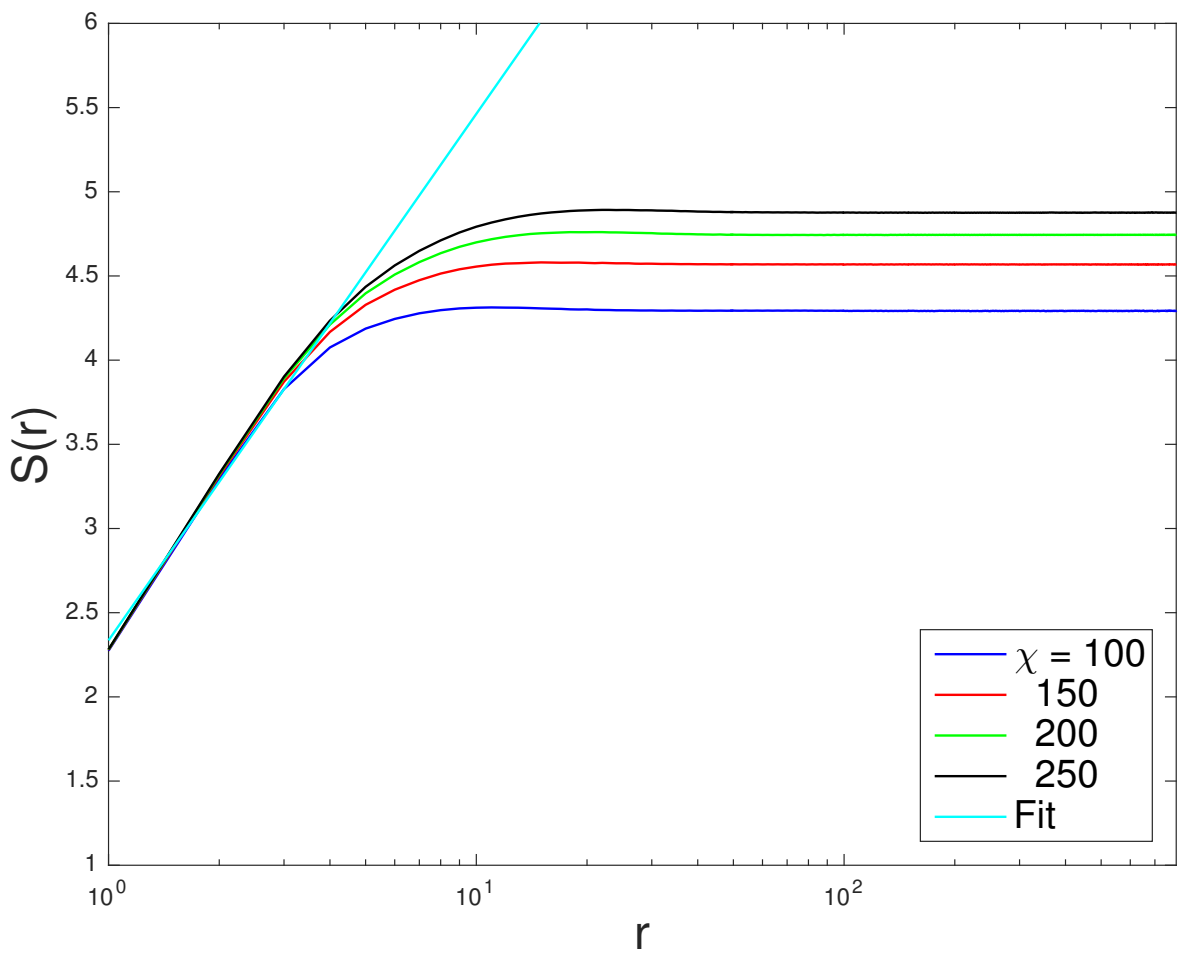

(a)

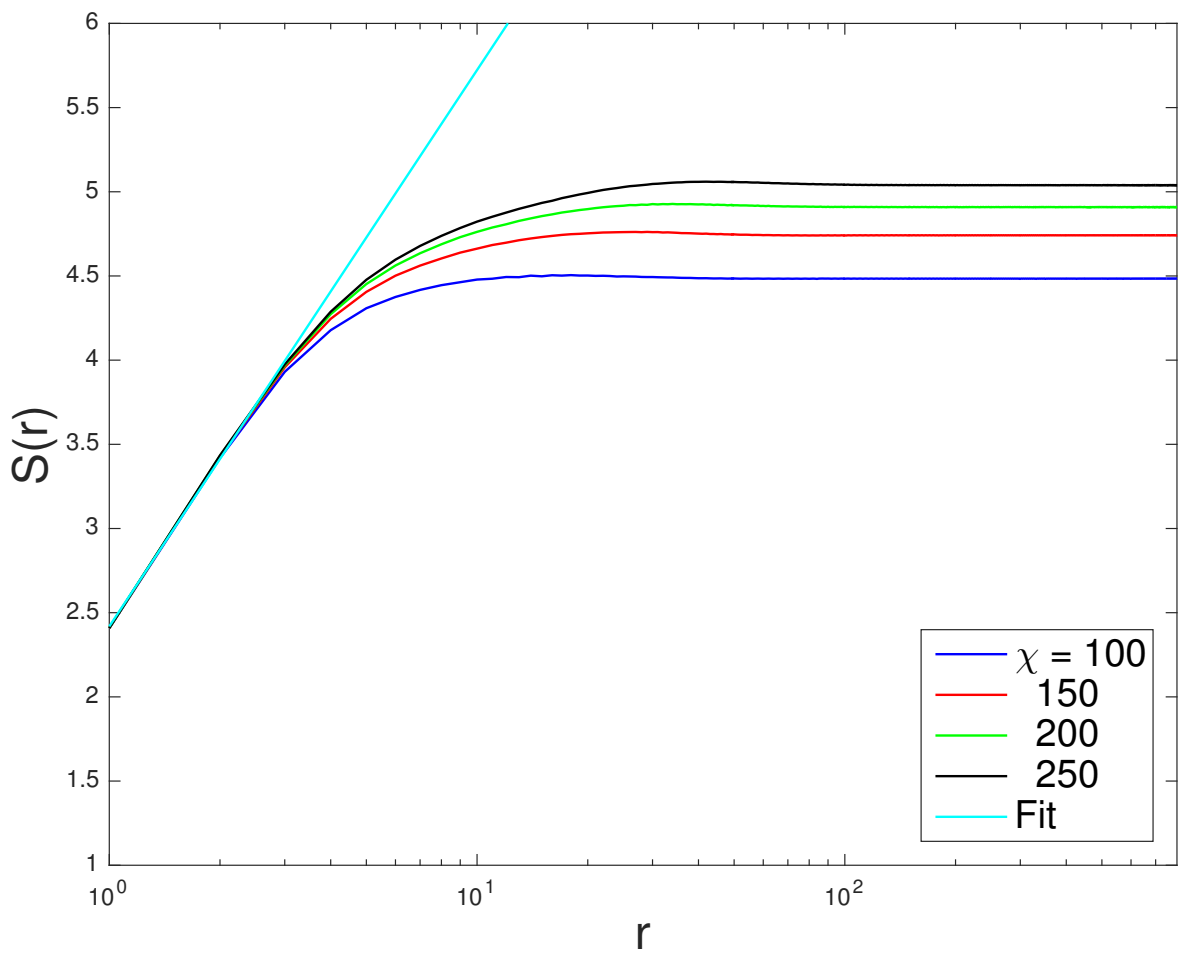

(b)

Figure 6.12: The block scaling of the EE of Fibonacci anyons at $(\mu/t_\perp, t_\parallel/t_\perp) = (0, 1.25)$—the "high filling" phase and at $(\mu/t_\perp, t_\parallel/t_\perp) = (-1, 1.5)$—the "intermediate filling" phase, respectively. The central charges from the linear fits are: (a) $c \simeq 2.743$ and (b) $c \simeq 2.846$. Both values are close to theoretical value of $c_{\text{theo}} = 2.7 \equiv (0.7 + 1 + 1)$ which is the aggregate sum of the central charges for one AFM Golden chain and one Luttinger liquid of two distinct particle types as explained in the text. The offset in the values of the central charges can be attributed to the small bond dimension of the MPS and much braid-induced entanglement in the ground state at high filling fraction.

## 6.5 Conclusion

In this work, we have investigated the role of two different braid statistics—Ising and Fibonacci anyons—using an anyonic Hubbard model on a two-leg ladder. We found disparate phase diagrams for the two anyon models, where their differences arise from their braid and fusion statistics. (The braid statistics is intimately related to the fusion statistics). We found that braiding has a "localising" effect, and at half-filling and low longitudinal hopping ratio, the ladder form an effect interacting chain, which is gapless and critical like the truly one-dimensional interacting chain model. At high hopping ratio, there is a phase transition to an anyonic superfluid phase which exhibit spin-charge separation. While the Ising anyon ladder showed properties like the one dimensional chain, the Fibonacci anyon ladder showed a behaviour unlike its 1-D chain equivalent, even showing support for the formation of a new composite particle—the neutral Fibonacci triplet $\mathbb{I}_3$.

There are many important questions yet to be answered, such as how the physics change as the dimensionality of the system is increased from two-leg to $W$-leg ladder, and ultimately reaching the limit $W \to \infty$, i.e. infinite two-dimensional limit. Does fractionalisation still survives at these higher dimensions? and what types of mobile excitations are possible in the ground state? Is there a general theory for any anyon model?

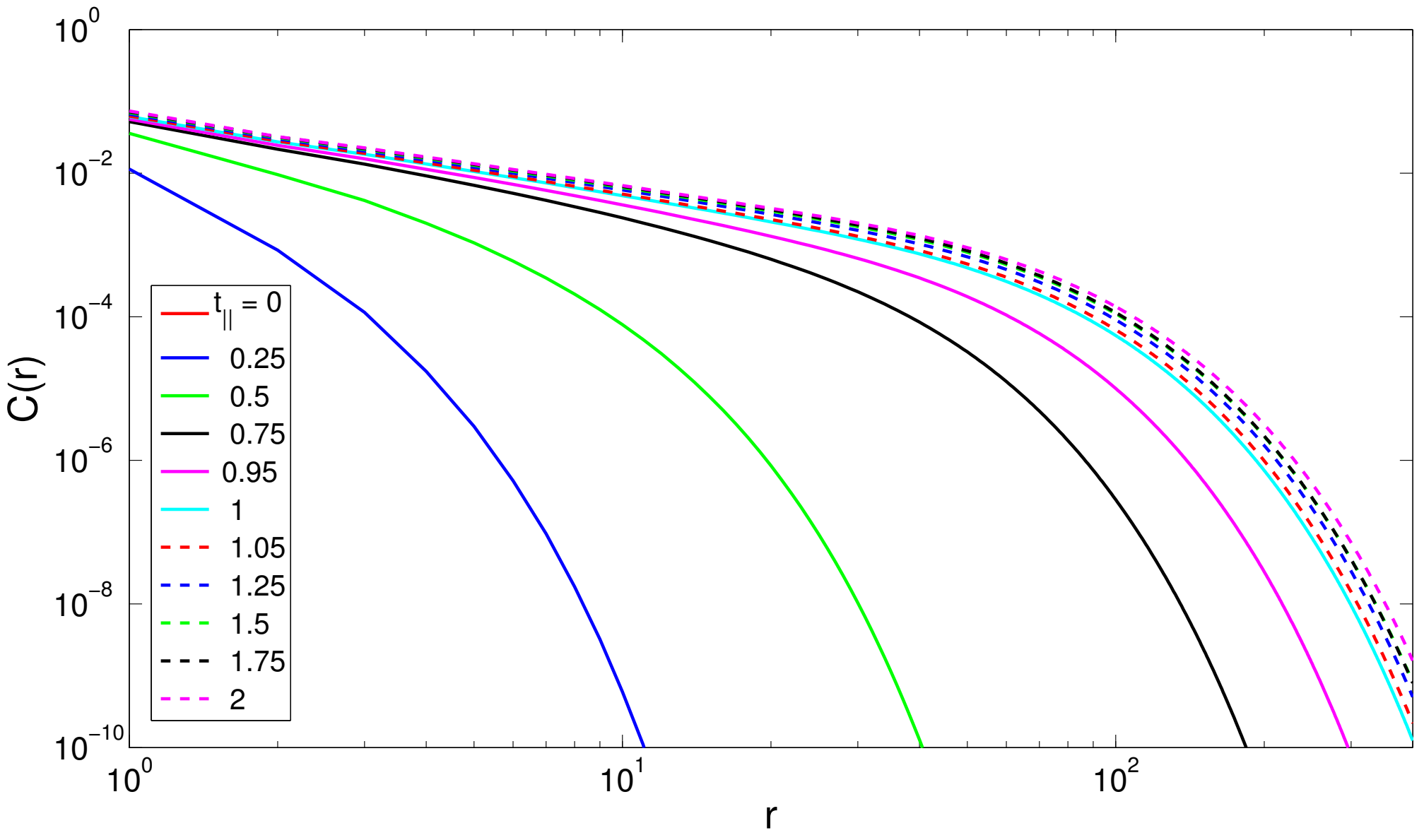

(a)

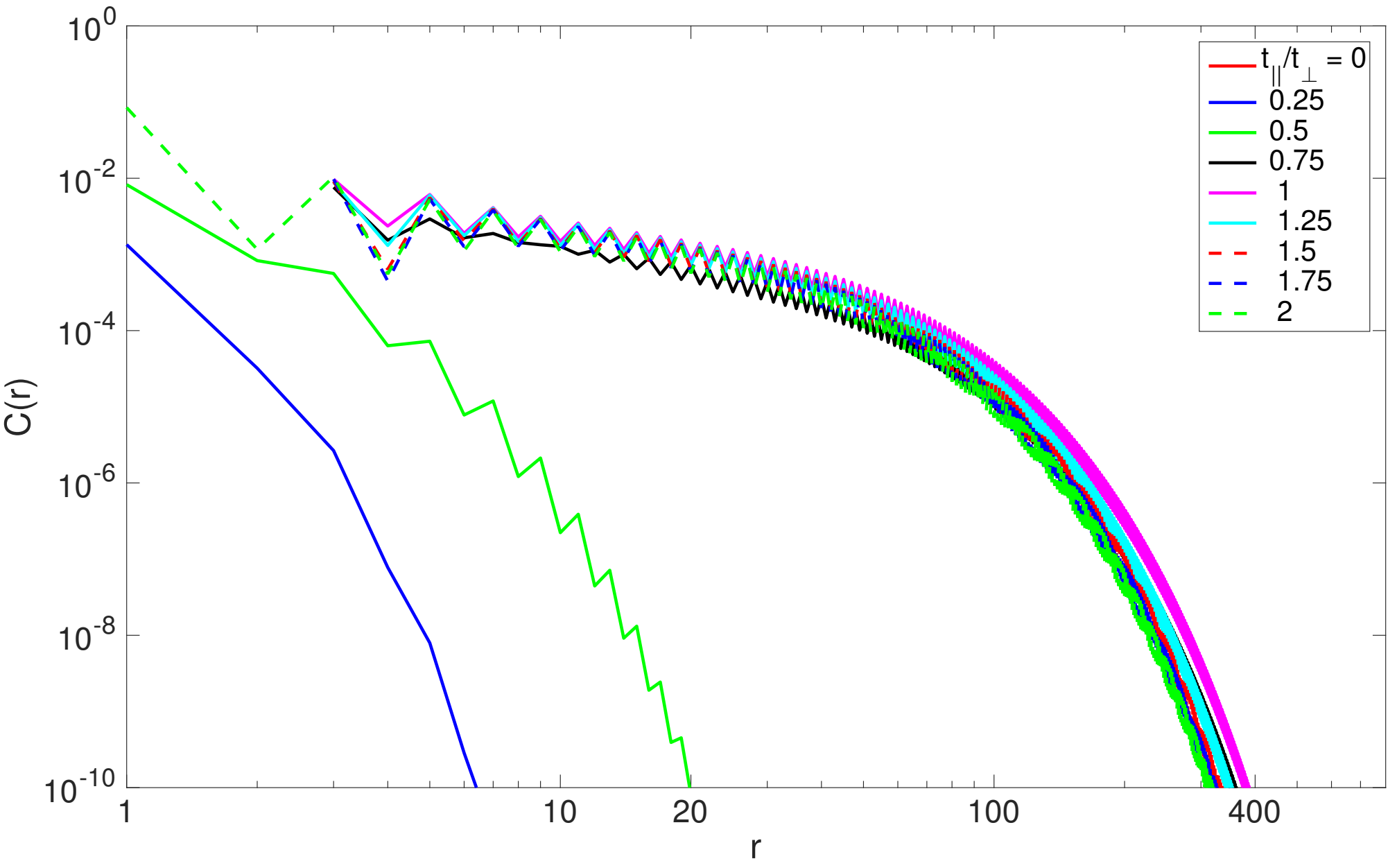

(b)

Figure 6.13: Superfluid correlation function for (a) Ising anyons and (b) Fibonacci anyons at half filling ($\nu = 1/2$) for horizontal hopping ratio $t_{\|}/t_{\perp} = [0, 2]$ with increments of 0.25. The critical point for Ising anyons is around $(t_{\|}/t_{\perp})_c \simeq 1$, while for Fibonacci anyons, it is $(t_{\|}/t_{\perp})_c \in [0.5, 0.75]$.

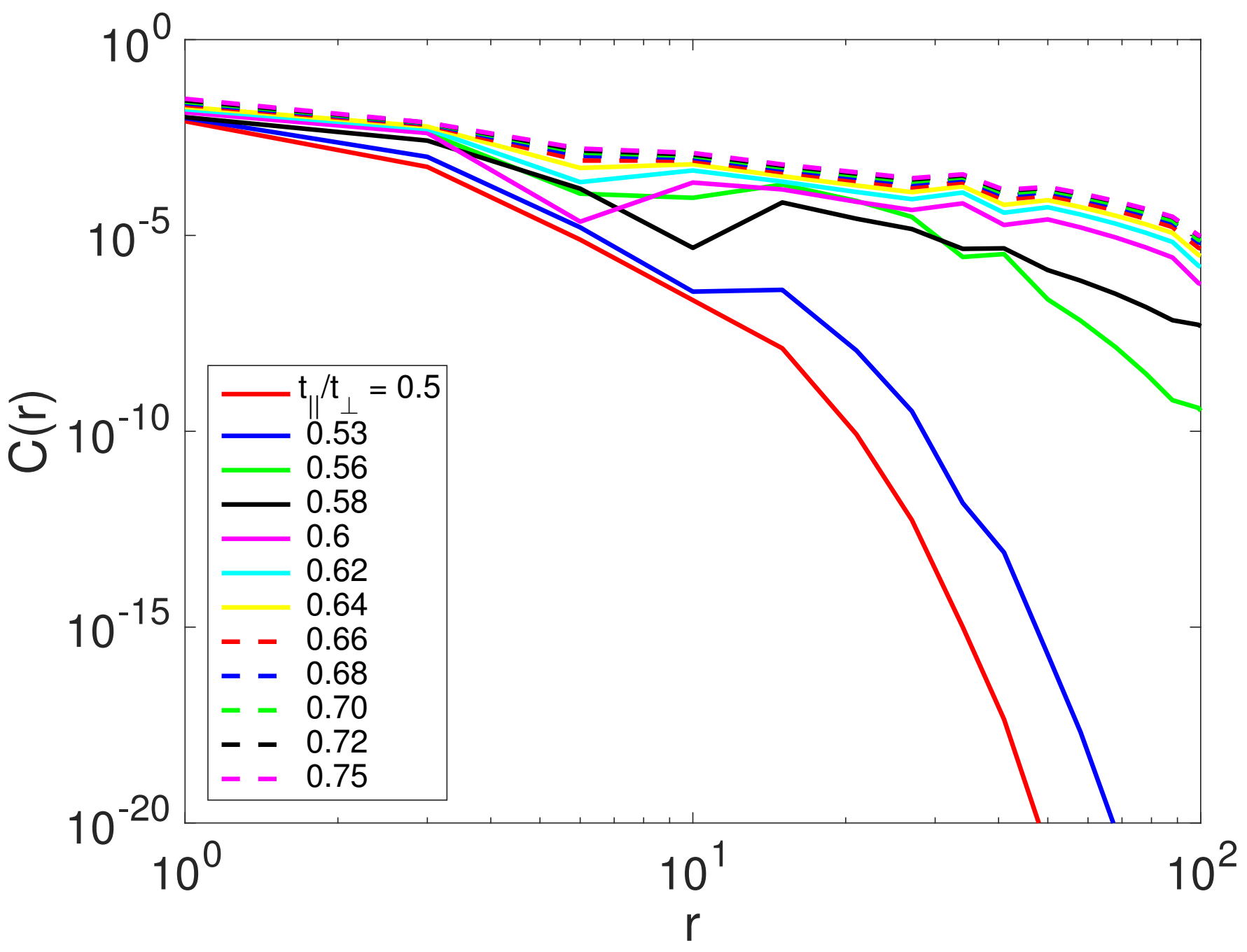

Figure 6.14: A "zoomed-in" version of the superfluid correlation function Fig. 6.13(b) for Fibonacci anyons at half filling for $t_{\|}/t_{\perp}$ between 0.5 and 0.75. The critical ratio seems to be $t_c \sim 0.62$.

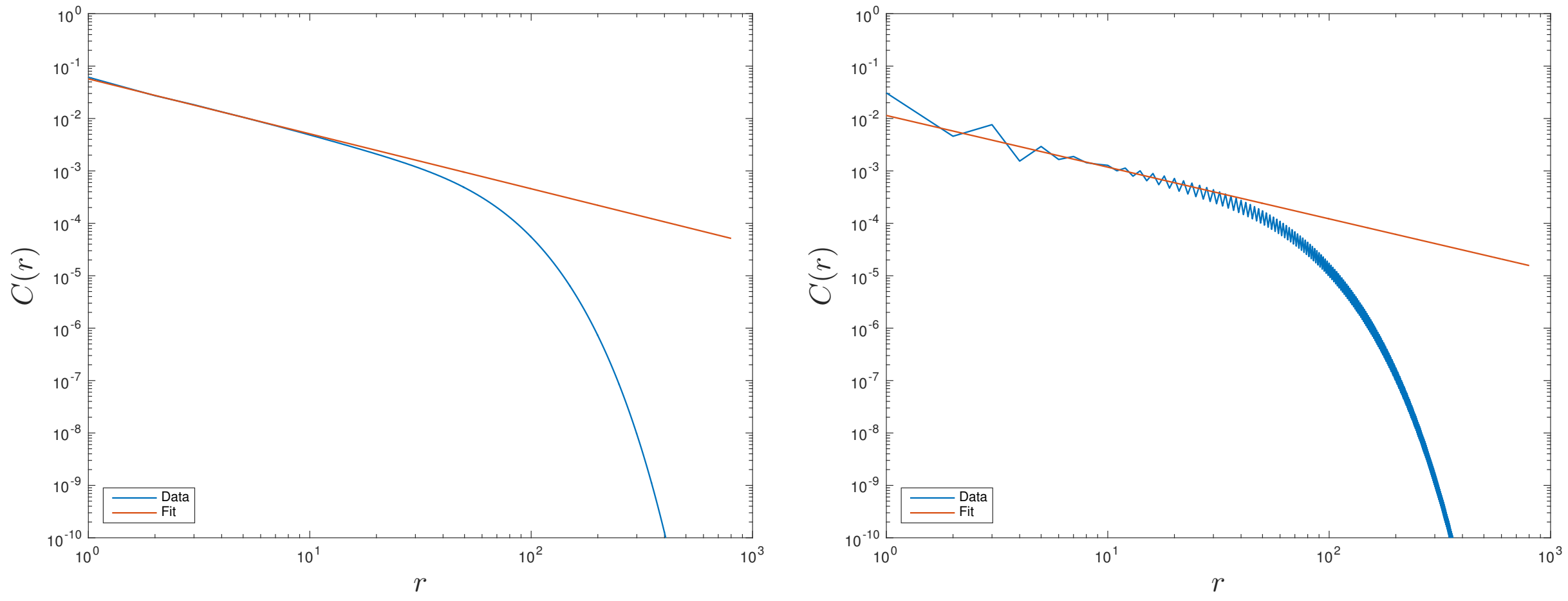

(a) (b)

Figure 6.15: Plots of the correlation function within the superfluid phase at half filling and a linear fit to determine the Luttinger parameter $K$. Plots are in log-log scale. (a) For Ising anyons at $t_{\|}/t_{\perp} = 1$, the Luttinger parameter obtained is $K_{\text{Ising}} \simeq 0.954$, and (b) for Fibonacci anyons at $t_{\|}/t_{\perp} = 0.75$, the Luttinger parameter is $K_{\text{Fib}} \simeq 1.014$.

# 7

# Single Particle Dynamics using Quantum Walks

## 7.1 Introduction

Discrete quantum systems can be used to study novel quantum phenomena such as quantum walks (QW). Quantum walks have received much attention over the past two decades, initially motivated as being the quantum analogue of the well known classical random walk (RW). Initial interest in QWs focused on the behaviour of the spatial evolution of the walks. In the classical case the spatial probability distribution of the walker after a time $\tau$ is Gaussian with a width $\sigma(\tau) \sim \tau^{1/2}$ while the typical coined quantum walk has a bimodal spatial distribution with a width that scales linearly with time $\sigma(\tau) \sim \tau$ [109–111].

Beyond fundamental interest, researchers have also shown that quantum walks can be the basis for constructing quantum algorithms, such as, graph searching [112–114], graph isomorphism testing [115, 116], and towards full blown quantum computing [117], which in part, is due to the presence of quantum entanglement in quantum walks [117, 118]. A number of interesting physical implementations of quantum walks have been proposed including trapped ions [119], Bose-Einstein condensates [120], linear optics [121], neutral atoms in optical lattices [122] and circuit quantum electrodynamics [123] and a number of experimental demonstrations including Nuclear Magnetic Resonance (NMR) [124] and photonic system [125]. Finally, an emerging direction of study for quantum walks is their use in quantum simulation, for example, the simulation of bosonic or fermionic quantum walks using integrated photonics [126].

In this work we look at the possibility of using a biased 1D quantum walk as an element within a larger quantum device to route quantum information either in one direction or another by appropriately biasing the quantum walk. By considering the entire quantum walk as a single "lumped element" whose routing action depends on the coin bias (see Fig 7.1), we investigate this idea further and find that we must generalize the concept of a quantum probability current density which is typically defined for a continuous time & space setup, to the case of a coined quantum walk with discrete time & space. Although there are many ways to perform such a generalization, we argue that to be physically relevant the generalized discrete probability current density must satisfy a discretized continuity equation which essentially encodes the microscopic detailed balance of the probability density with time. We show that the "lumped element" current , obtained by summing over the entire quantum walk lattice, reaches a steady state and for specific initial states we derive an analytic form for this steady state current. We show that by altering a phase factor within the biased coin we can engineer the magnitude and the direction(routing) of the steady state current. The control phase and steady state total current exhibits a sinusoidal current-phase relationship indicating some similarity to the behaviour of a Josephson junctions. Finally we illustrate that conservative 1D Hamiltonian systems can also exhibit steady state dynamics similar to the lumped element quantum walk router.

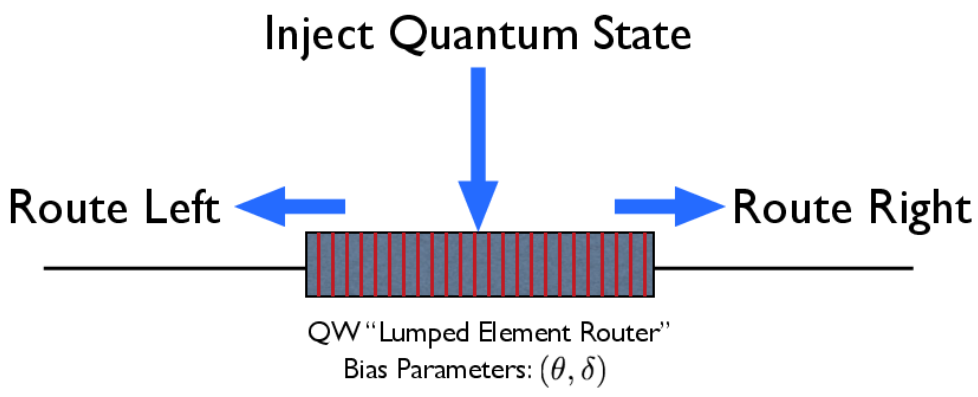

Figure 7.1: Using the "lumped element" biased quantum walk as a router. We examine whether one can control the routing of an initial quantum state injected at the middle of a quantum walk to either end of the lumped element. We consider a 1D biased quantum walk with a large number of lattice sites as a "lumped element", which can be used to redirect quantum probability to either end and into attached quantum wires (not described here). The bias parameters $(\theta, \delta)$ parametrize the type of coin used in the quantum walk.

As mentioned above, central to our study is the existence of a probability current density which satisfies an appropriately physical continuity relation. We note that although quantum walks have received much attention and that their evolution can be studied analytically we are aware of only two previous studies that make use of a QW probability current density [127, 128]. Moreover in both cases, neither of their proposed currents satisfies a continuity equation and thus are physically of little relevance. Below we will derive a form for the local probability current density which manifestly satisfies the local discrete continuity equation for probability. We derive an analytical expression for the *total current* when summed over the spatial lattice of the QW - this corresponds to the current through the entire lumped element of the QW router - and we show that this total current reaches a steady state whose value can be controlled by considering a two-parameter family of $SU(2)$ biased coined QWs. We find that the total stationary current is an oscillatory function of this one parameter and we discuss the similarity in behaviour to the current-phase relation of Josephson junctions. We then show that one can find a more symmetric expression for the local probability current density through a central difference approach to the continuity equation and finally we explore what types of continuous conservative dynamical systems possess similar stationary currents.

## 7.2 Discrete-time coined quantum walk

We quickly review the model of the typical coined quantum walk. For a detailed overview of quantum walks, see Ref. [110]. In this model, a particle with an internal chirality (or two-state coin) resides on a discrete 1D lattice and is displaced conditioned on the state of the quantum coin, where the latter periodically experiences an operation similar to coin tossing in a classical random walk. Let the position eigenkets of the particle at position $n$ be $|n\rangle \in \mathcal{H}_p$ and with an internal chirality state $|c\rangle \in \mathcal{H}_c$, where $\mathcal{H}_p$ and $\mathcal{H}_c$ are the position and coin Hilbert space respectively. $\mathcal{H}_p = \{|n\rangle : n \in \mathbb{Z}\}$, $\mathcal{H}_c = \{|\uparrow\rangle, |\downarrow\rangle\}$, $|\uparrow\rangle$ and $|\downarrow\rangle$ are the coin states that specifies the direction of motion. The combined Hilbert space $\mathcal{H} = \mathcal{H}_c \otimes \mathcal{H}_p$ and the coin flip operation is performed via a unitary coin operator, $\hat{C}$, which operates only on the coin Hilbert space. The most general form of this operator is

$$\hat{C}(\xi, \theta, \eta) = \begin{pmatrix} e^{i\xi}\cos\theta & e^{i\eta}\sin\theta \\ e^{-i\eta}\sin\theta & -e^{-i\xi}\cos\theta \end{pmatrix}, \tag{7.1}$$

which is a three parameter unitary operator. The unitary conditional translation (or step) operator $\hat{S}$ is,

$$\hat{S} = |\uparrow\rangle\langle\uparrow| \otimes |n+1\rangle\langle n| + |\downarrow\rangle\langle\downarrow| \otimes |n-1\rangle\langle n|, \tag{7.2}$$

which implies that, if the particle is at position $n$ with an internal coin state $|\uparrow\rangle$, $\hat{S}$ shifts the particle to position $n+1$, while if its chirality state is $|\downarrow\rangle$, $\hat{S}$ shifts it to to position $n-1$. If we put $\theta = \frac{\pi}{4}$, $\xi = \eta = 0$, we realize the Hadamard coin that has been used extensively to study quantum walks [109, 129].

In our study of the transport properties of the quantum walk we use the two-parameter unitary coin operator $\hat{C}(\theta, \delta)$

$$\hat{C}(\theta, \delta) = \begin{pmatrix} \cos\theta & e^{i\delta}\sin\theta \\ e^{-i\delta}\sin\theta & -\cos\theta \end{pmatrix}, \tag{7.3}$$

which we find sufficient to study the properties of quantum transport in our model namely: the spread of the quantum walk through $\theta$ and also any bias of the walk in the "lumped element" through $\delta$. Moreover, we later show that any other phase factor in a three-parameter coin appear only in combination with the phase factors of the initial coin state of the walker, a fact that has been noted too by previous authors [130]. Hence, we mainly use a two-parameter coin operator.

We represent the state of the quantum particle at any time $t$ by a state vector

$$\begin{aligned} |\psi(t)\rangle &= \sum_{n=-\infty}^{\infty} \begin{pmatrix} \alpha_{n,t} \\ \beta_{n,t} \end{pmatrix} \otimes |n\rangle \\ &= \sum_{n=-\infty}^{\infty} \alpha_{n,t}|\uparrow, n\rangle + \beta_{n,t}|\downarrow, n\rangle, \end{aligned} \tag{7.4}$$

where $\alpha_{n,t}$ and $\beta_{n,t}$ are the amplitudes associated with walker which has a given chirality. We have chosen the representation $|\uparrow\rangle = \begin{pmatrix} 1 \\ 0 \end{pmatrix}$ and $|\downarrow\rangle = \begin{pmatrix} 0 \\ 1 \end{pmatrix}$. The update rule for our quantum walk is given by

$$|\psi(t+1)\rangle = \hat{S}(\hat{C} \otimes \mathbb{I})|\psi(t)\rangle, \tag{7.5}$$

from which we obtain the (recurrence) relations satisfied by the amplitudes $\alpha_{n,t}$ and $\beta_{n,t}$,

$$\begin{aligned}\alpha_{n,t} &= \cos\theta\ \alpha_{n-1,t-1} + e^{i\delta}\sin\theta\ \beta_{n-1,t-1},\\ \beta_{n,t} &= e^{-i\delta}\sin\theta\ \alpha_{n+1,t-1} - \cos\theta\ \beta_{n+1,t-1}.\end{aligned} \tag{7.6}$$

These are the basic recurrence relations defining the evolution of the quantum walk on the 1D lattice which we will use to derive the probability current density. Denoting the probability $\rho(n,t)$ for the particle to be found at position $n$ at time $t$ as, $\rho(n,t) = |\alpha_{n,t}|^2 + |\beta_{n,t}|^2$, we have the conservation of probability with time, i.e. $\sum_{n=-\infty}^{n=\infty}\rho(n,t) = 1,\ \forall t$.

## 7.3 Probability current density of the coined quantum walk

Some authors have previously found an expression for the probability current density for the QW of the form [127, 128]

$$j(n,t) = |\alpha_{n,t}|^2 - |\beta_{n,t}|^2\ . \tag{7.7}$$

which however does not satisfy the continuity relation, and hence is not suitable for our purpose. We now outline how to derive an expression for the current density in our quantum walk from the recurrence relations (7.6) starting from the continuity equation. The continuity equation (in its continuous form), can be written as $\partial_x j = -\partial_t\rho$, and implies that the net flow of probability $\partial_x j$, into/out from a region is equal to the rate of change of the overall probability, $-\partial_t\rho$, in that region. We consider the discrete version of the continuity equation to be

$$-\vartriangle_t \rho(n,t) = \vartriangle_n j(n,t) \tag{7.8}$$

where $\vartriangle_{n,t}$ is a forward difference operator in space and time given as $\vartriangle_t \rho(n,t) \equiv \rho(n,t+1) - \rho(n,t)$, and $\vartriangle_n j(n,t) \equiv j(n+1,t) - j(n,t)$.

The left-hand-side of (7.8) can be computed easily if we consider that the local probability $\rho(n,t) = |\alpha_{n,t}|^2 + |\beta_{n,t}|^2$ and then use the recurrence equations (7.6) to get $\rho(n,t+1)$,

$$\begin{aligned}-\vartriangle_t \rho(n,t) &= |\alpha_{n,t}|^2 + |\beta_{n,t}|^2 - \cos^2\theta|\alpha_{n-1,t}|^2\\ &\sin^2\theta|\beta_{n-1,t}|^2 - \sin^2\theta|\alpha_{n+1,t}|^2 + \cos^2\theta|\beta_{n+1,t}|^2\\ &+ \sin 2\theta\,\mathrm{Re}\left\{e^{i\delta}\left(\alpha^*_{n+1,t}\beta_{n+1,t} - \alpha^*_{n-1,t}\beta_{n-1,t}\right)\right\}.\end{aligned} \tag{7.9}$$

This equation can be arranged in a more suggestive form as

$$\begin{aligned}-\vartriangle_t \rho(n,t) &= \left(\cos^2\theta\ |\alpha_{n,t}|^2 - \sin^2\theta\ |\alpha_{n+1,t}|^2\right)\\ &- \left(\cos^2\theta\ |\alpha_{n-1,t}|^2 - \sin^2\theta\ |\alpha_{n,t}|^2\right)\\ &+ \left(\sin^2\theta\ |\beta_{n,t}|^2 - \cos^2\theta\ |\beta_{n+1,t}|^2\right)\\ &- \left(\sin^2\theta\ |\beta_{n-1,t}|^2 - \cos^2\theta\ |\beta_{n,t}|^2\right)\\ &+ \sin 2\theta\,\mathrm{Re}\left[e^{i\delta}\left(\alpha^*_{n+1,t}\beta_{n+1,t} + \alpha^*_{n,t}\beta_{n,t}\right)\right]\\ &- \sin 2\theta\,\mathrm{Re}\left[e^{i\delta}\left(\alpha^*_{n,t}\beta_{n,t} + \alpha^*_{n-1,t}\beta_{n-1,t}\right)\right],\end{aligned} \tag{7.10}$$

and from this the following expression for the probability current density can be read off,

$$\begin{aligned}j(n,t) &= \cos^2\theta\left(|\alpha_{n-1,t}|^2 - |\beta_{n,t}|^2\right)\\ &- \sin^2\theta\left(|\alpha_{n,t}|^2 - |\beta_{n-1,t}|^2\right)\\ &+ \sin 2\theta\,\mathrm{Re}\left[e^{i\delta}\left(\alpha^*_{n-1,t}\beta_{n-1,t} + \alpha^*_{n,t}\beta_{n,t}\right)\right]\end{aligned} \tag{7.11}$$

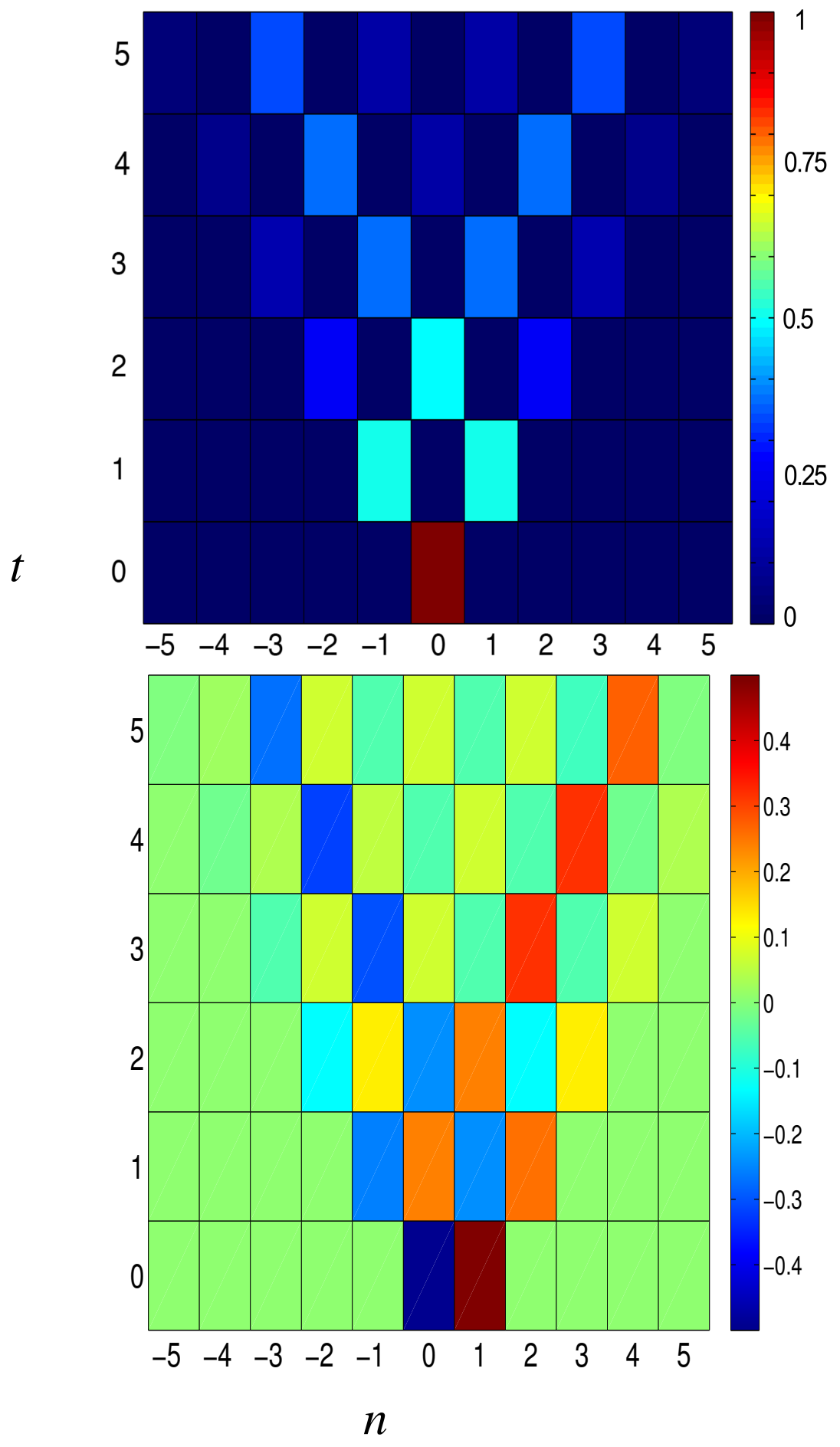

Figure 7.2: For a QW which is initialised to yield a spatially symmetric walk, where $(\alpha_{n,0}, \beta_{n,0}) = \frac{1}{\sqrt{2}}(1, i)\delta_{n,0}$, i.e. the walker is initially located at the origin, and using $\theta = \pi/4, \delta = 0$ (a Hadamard coin); we plot (up) the probability for the walker to be found at spatial location $n$ at time $t$, i.e $\rho(n, t)$; (down) the probability current density $j(n, t)$. We see that although the probability evolves in a spatially symmetric fashion the forward-difference defined probability current density breaks this symmetry.

It is interesting to note that the current density $j(n, t)$ in (7.11), is more involved than had been initially supposed (7.7). Interestingly we find a dependence on the interferences between $\alpha_{n,t}$ and $\beta_{n,t}$ but since the amplitudes $\alpha_{n,t}$ and $\beta_{n,t}$ oscillates forever, $j(n, t)$ does not achieve a steady state. In the next section we instead consider the total cumulative current over the entire 1D lattice and find that this indeed does achieve a steady state. As an illustration we show in Fig 7.2 the probability current density for the Hadamard QW ($\delta = 0$), as a function of time for a localized symmetric initial coin state $(\alpha_{n,0}, \beta_{n,0}) = \frac{1}{\sqrt{2}}(1, i)\delta_{n,0}$. We observe that the probability current density as defined using the forward differences has some peculiarities, i.e. although the probability distribution for the evolving QW is spatially symmetric about the origin the introduction of the forward difference has introduced an apparent symmetry breaking into the associated probability current density. Despite this, the forward difference probability current density still satisfies the associated continuity equation (7.8). This apparent asymmetry will be remedied later on when we describe how to use a central difference form of (7.8).

## 7.4 Steady state current

We now examine the asymptotic behaviour of the "total" current on the entire 1D lattice. Let the total cummulative current $J(t)$ be defined as

$$J(t) = \sum_{n=-\infty}^{\infty} j(n,t) \tag{7.12}$$

By using the dynamical recurrence equations we can derive a very compact analytical expression for the steady state value of this current e.g. $J_\infty \equiv \lim_{t\to+\infty} J(t)$. We also investigate the steady state value $J_\infty$ by numerical simulation using (7.6) and find perfect agreement with our analytic formula. We now define some useful quantities, the global probability amplitudes and global interference terms as, $\rho_+(t) \equiv \sum_{n=-\infty}^{\infty} |\alpha_{n,t}|^2$, $\rho_-(t) \equiv \sum_{n=-\infty}^{\infty} |\beta_{n,t}|^2$, and $Q(t) \equiv \sum_{n=-\infty}^{\infty} \alpha^*_{n,t}\beta_{n,t}$. From (7.6), we can find

$$\begin{aligned} |\alpha_{n,t+1}|^2 &= \cos^2\theta\, |\alpha_{n-1,t}|^2 + \sin^2\theta\, |\beta_{n-1,t}|^2 \\ &\quad + \sin 2\theta\, \mathrm{Re}\left(e^{i\delta}\alpha^*_{n-1,t}\beta_{n-1,t}\right) , \\ |\beta_{n,t+1}|^2 &= \sin^2\theta\, |\alpha_{n+1,t}|^2 + \cos^2\theta\, |\beta_{n+1,t}|^2 \\ &\quad - \sin 2\theta\, \mathrm{Re}\left(e^{i\delta}\alpha^*_{n+1,t}\beta_{n+1,t}\right) , \end{aligned} \tag{7.13}$$

and summing over space in (7.13) gives,

$$\begin{aligned} \rho_+(t+1) &= \cos^2\theta\, \rho_+(t) + \sin^2\theta\, \rho_-(t) + \sin 2\theta\, \mathrm{Re}\left(e^{i\delta}Q(t)\right) , \\ \rho_-(t+1) &= \sin^2\theta\, \rho_+(t) + \cos^2\theta\, \rho_-(t) - \sin 2\theta\, \mathrm{Re}\left(e^{i\delta}Q(t)\right) . \end{aligned} \tag{7.14}$$

In the long-time limit, we define the following $\rho_+(t\to\infty) = \Omega_+$, $\rho_-(t\to\infty) = \Omega_-$, $Q(t\to\infty) = Q_0$, and $J(t\to\infty) = J_\infty$.

In this limit (7.14), in matrix form becomes,

$$\begin{pmatrix} 1 & -1 \\ -1 & 1 \end{pmatrix} \begin{pmatrix} \Omega_+ \\ \Omega_- \end{pmatrix} = 2\cot\theta\, \mathrm{Re}\left(e^{i\delta}Q_0\right) \begin{pmatrix} 1 \\ -1 \end{pmatrix} \tag{7.15}$$

which, with $\Omega_+ + \Omega_- = 1$, has the solution,

$$\begin{pmatrix} \Omega_+ \\ \Omega_- \end{pmatrix} = \frac{1}{2} \begin{pmatrix} 1 + 2\cot\theta\, \mathrm{Re}(e^{i\delta}Q_0) \\ 1 - 2\cot\theta\, \mathrm{Re}(e^{i\delta}Q_0) \end{pmatrix} . \tag{7.16}$$

This remarkable result has previously been derived by Romanelli [131]. The steady state current $J_\infty$ then easily proceeds from (7.11), (7.12) and (7.16) as

$$J_\infty = 2\cot\theta\, \mathrm{Re}\left(e^{i\delta}Q_0\right) . \tag{7.17}$$

The above equation clearly indicates that the total current in our quantum walk depends on a number of factors including a) the interferences through $Q_0$ which ultimately depends on the initial state $|\psi(t=0)\rangle$ and on b) the bias parameter $\delta$.

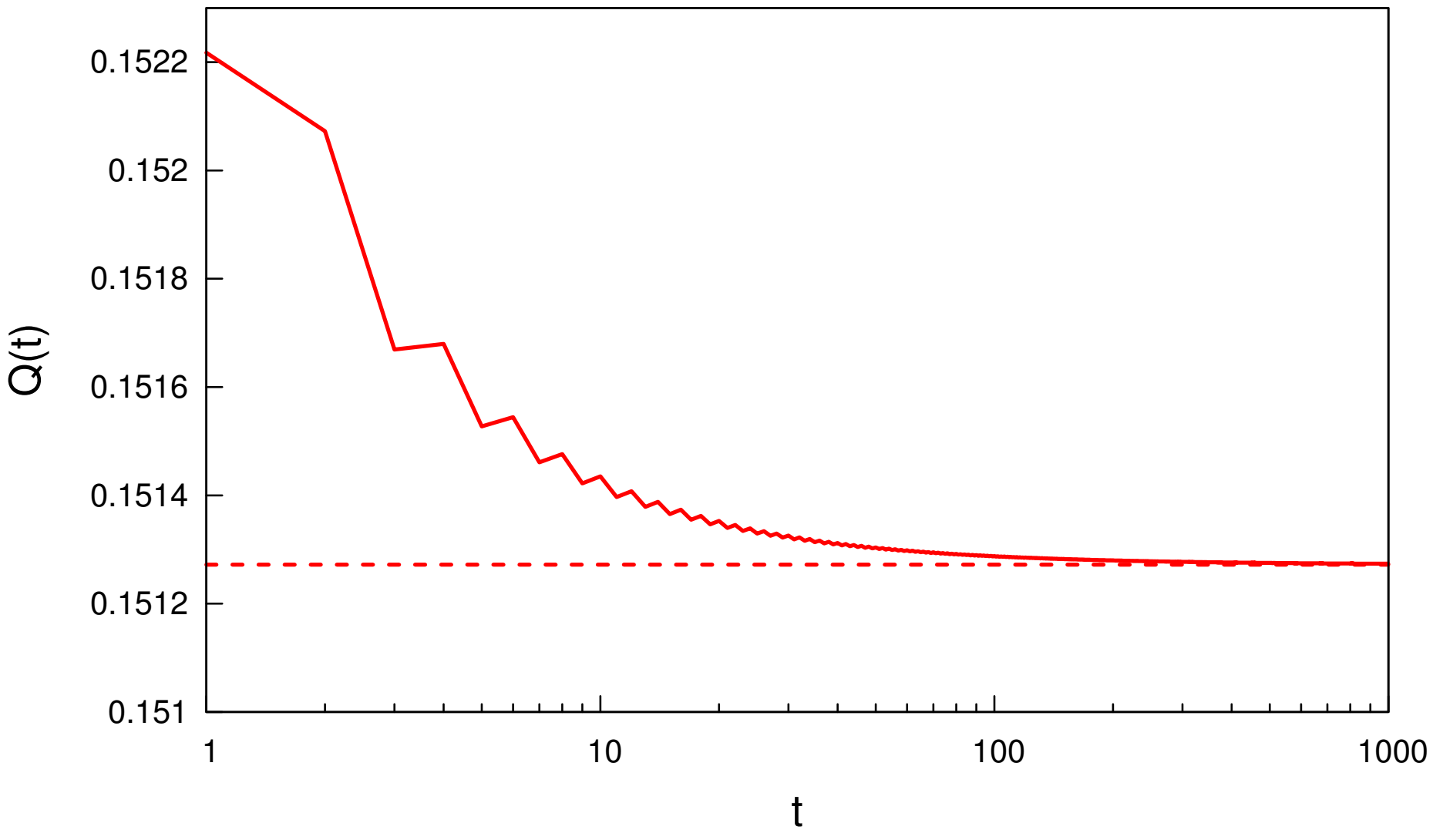

Figure 7.3: Global interference term $Q(t)$ for a symmetric localized initial coin state $(\alpha_{n,0}, \beta_{n,0}) = \frac{1}{\sqrt{2}}(1, i)\delta_{n,0}$ and with coin paramters $\theta = \pi/4$, $\delta = 5\pi/12$, smoothed in time using a moving window average of width 10. The solid line is the numerical result while the dashed line is the asymptotic value $Q_0$.

The global interference term $Q_0$ can be computed using a Fourier series method and this same method has been quite successful in analyzing the asymptotics of quantum walks [109, 132, 133]. Using this method and assuming a sharply localized initial state $|\psi(t=0)\rangle$ at $n = 0$ as

$$|\psi(t=0)\rangle = \begin{pmatrix} \cos\frac{\phi}{2} \\ e^{i\gamma}\sin\frac{\phi}{2} \end{pmatrix} \otimes |n=0\rangle \tag{7.18}$$

where $\phi \in [0, \pi]$ and $\gamma \in [0, 2\pi]$, with some effort (see Appendix B), one can derive $Q_0$ to be,

$$Q_0 = \frac{(1-\sin\theta)e^{-i\delta}\tan\theta}{2}\Big[\cos\phi + \sin\phi\Big(e^{-i(\delta+\gamma)}\tan\theta + i\sin(\delta+\gamma)\frac{\cos\theta}{1-\sin\theta}\Big)\Big] \quad . \tag{7.19}$$

Through numerical simulation one can observe the asymptotic approach of $Q(t)$ to $Q_0$ (see Fig(7.3)), in the long time limit.

From this we can express the steady state total current $J_\infty$, for the particular case of the initial state (7.18), as

$$J_\infty = (1-\sin\theta)\left[\cos\phi + \sin\phi\cos(\delta+\gamma)\tan\theta\right] \quad , \tag{7.20}$$

which depends not only on the initial state through $\phi, \gamma$, but also on the nature of the dynamics through $\theta$ and $\delta$ as shown in Fig 7.4. The term $\cos(\delta+\gamma)$ indicates that $J_\infty$ can be controlled in an identical fashion either by the coin bias factor $\delta$ or the phase of the initial state. One has the freedom to adjust $J_\infty$ dynamically irrespective of the initial state.

Finally it is curious to note that the sinusoidal dependence of the current on the phase $\delta$ (or $\gamma$), is very reminiscent of the sinusoidal current-phase relationship (CPR), found in Josephson junctions [134]. The highly nonlinear dependence of the Josephson current on the phase difference across the junction has led to numerous quantum devices, the most prominent being the Superconducting Quantum Interference Device (SQUID). Indeed, based on the Josephson CPR the total current in a SQUID varies sinusoidally with the magnitude of the trapped flux.

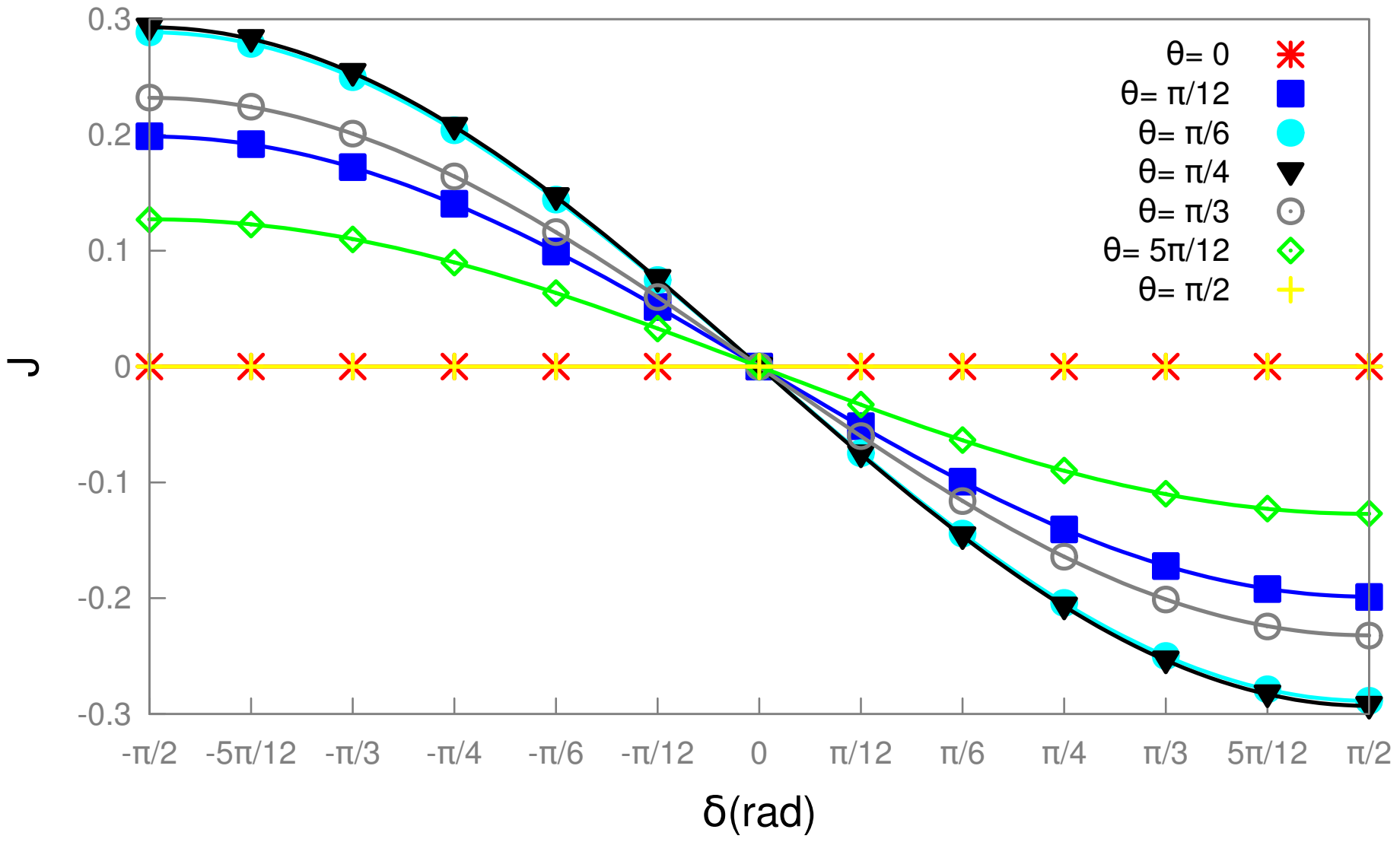

Figure 7.4: Total "steady" current $J_\infty$ against $\delta$ for various coin values $\theta$ for a symmetric localized initial coin state $(\alpha_{n,0}, \beta_{n,0}) = \frac{1}{\sqrt{2}}(1, i)\delta_{n,0}$. The points are the results of numerical simulation using (7.11) while the lines are the analytical result (7.20). We note that $J_\infty(-\delta) = -J_\infty(\delta)$, i.e. $J_\infty(\delta)$ is an odd function of $\delta$.

## 7.5 Symmetric probability current density

Above we derived a probability current density according to a forward difference approximation to the continuity equation (7.8), and we noted that this lack of symmetry in this discretization led to peculiar asymmetric behaviours in the associated probability current density. We now show how this can be remedied by choosing a central difference version of the continuity equation, where we now choose

$$-\,\Delta_t^C\ \rho(n,t) = \Delta_n^C\ j^C(n,t) \tag{7.21}$$

where $\Delta_{n,t}^C$ are the central difference operators in space and time given as $\Delta_t^C\ \rho(n,t) \equiv (\rho(n,t+1) - \rho(n,t-1))/2$, and $\Delta_n^C\ j(n,t) \equiv (j^C(n+1,t) - j^C(n-1,t))/2$. We will find that

$$j^C(n,t) = \cos^2\theta\left(|\alpha_{n,t}|^2 - |\beta_{n,t}|^2\right) + \sin 2\theta\,\mathrm{Re}\left[e^{i\delta}\alpha_{n,t}^*\,\beta_{n,t}\right] \ . \tag{7.22}$$

To show this we use the recurrence relations (7.6), to express $\rho(n, t+1)$ in terms of the amplitudes $\alpha_{k,t}$ and $\beta_{l,t}$, at time step $t$. We then note that these same recurrence relations can be recast in the matrix format

$$\begin{pmatrix} \alpha_{n+1,t} \\ \beta_{n-1,t} \end{pmatrix} = \begin{pmatrix} \cos\theta & e^{i\delta}\sin\theta \\ e^{-i\delta}\sin\theta & -\cos\theta \end{pmatrix} \begin{pmatrix} \alpha_{n,t-1} \\ \beta_{n,t-1} \end{pmatrix}. \tag{7.23}$$

and as the transformation matrix in (7.23), is unitary we straight away can re-express $\rho(n, t-1)$ in terms of quantites at time $t$,

$$\rho(n,t-1) = |\alpha_{n,t-1}|^2 + |\beta_{n,t-1}|^2 = |\alpha_{n+1,t}|^2 + |\beta_{n-1,t}|^2 \ . \tag{7.24}$$

With (7.24), one can write $\rho(n,t+1)-\rho(n,t-1)$ involving quantities expressed only at time step $t$, to be

$$\triangle_t\ \rho = \rho(n,t+1) - \rho(n,t-1)$$

$$\begin{aligned}\triangle_t \rho = &\cos^2\theta|\alpha_{n-1,t}|^2 + (\sin^2\theta - 1)|\beta_{n-1,t}|^2 \\ &+ \sin 2\theta\ \mathrm{Re}[e^{i\delta}\alpha^*_{n-1,t}\beta_{n-1,t}] \\ &+ (\sin^2\theta - 1)|\alpha_{n+1,t}|^2 + \cos^2\theta|\beta_{n+1,t}|^2 \\ &- \sin 2\theta\ \mathrm{Re}\left[e^{i\delta}\alpha^*_{n+1,t}\ \beta_{n+1,t}\right] \\ \\ = &\cos^2\theta\left(|\alpha_{n-1,t}|^2 - |\alpha_{n+1,t}|^2\right) + \cos^2\theta\left(|\beta_{n+1,t}|^2\right. \\ &\left.-|\beta_{n-1,t}|^2\right) + \sin 2\theta\ \mathrm{Re}\left[e^{i\delta}\left(\alpha^*_{n-1,t}\beta_{n-1,t} - \alpha^*_{n+1,t}\beta_{n+1,t}\right)\right] \\ \\ = &\cos^2\theta|\alpha_{n-1,t}|^2 - \cos^2\theta|\beta_{n-1,t}|^2 \\ &+ \sin 2\theta\ \mathrm{Re}\left[e^{i\delta}\alpha^*_{n-1,t}\ \beta_{n-1,t}\right] \\ &- \left[\cos^2\theta|\alpha_{n+1,t}|^2 - \cos^2\theta|\beta_{n+1,t}|^2\right. \\ &\left.\qquad + \sin 2\theta\ \mathrm{Re}\left[e^{i\delta}\alpha^*_{n+1,t}\ \beta_{n+1,t}\right]\right]. \end{aligned} \tag{7.25}$$

Now using the central difference form for the continuity equation as $\rho(n,t+1)-\rho(n,t-1) = -\left(j^C(n+1,t) - j^C(n-1,t)\right)$, by inspection we obtain the central difference probability current density (7.22).

In Fig. 7.5 we plot out the behaviour of $j^C(n,t)$, for the same symmetric initial state as in Fig 7.2, and now we observe that the spatial symmetry of the QW's evolution is maintained by $j^C$. We also can define the total symmetric probability current $J^C(t) \equiv \sum_{n=-\infty}^{\infty} j^C(n,t)$, and we find that $J^C(t)$, reaches a steady-state value which is identical to that found in the forward difference case.

## 7.6 Do we expect the total probability current to have a steady-state?

We have seen that for the coined QW the total current attains a steady state. We now ask the question whether this is typical or not? In classical mechanics, we associate a steady state current or momentum with terminal velocity, i.e acceleration in a dissipative medium. Our biased-coined QW is completely unitary and thus one may pose the question: in a purely Hamiltonian system (classical or quantum), what type of dynamics will result in steady state momenta or probability current? We show that such conservative dynamics are possible and may yield a continuous space-time analogue of our biased/directed quantum walk.

Looking now within classical mechanics, a steady-state current typically corresponds to a terminal velocity or momentum. Considering a massive particle moving in one dimension, and assuming that it's momentum attains a terminal value of $p_f$ as $t \to \infty$, and that we have no interest in the detailed dynamics before steady-state, we can phenomenologically model the long-time dynamics as,

$$\dot{p} = \lambda(p_f - p) \tag{7.26}$$

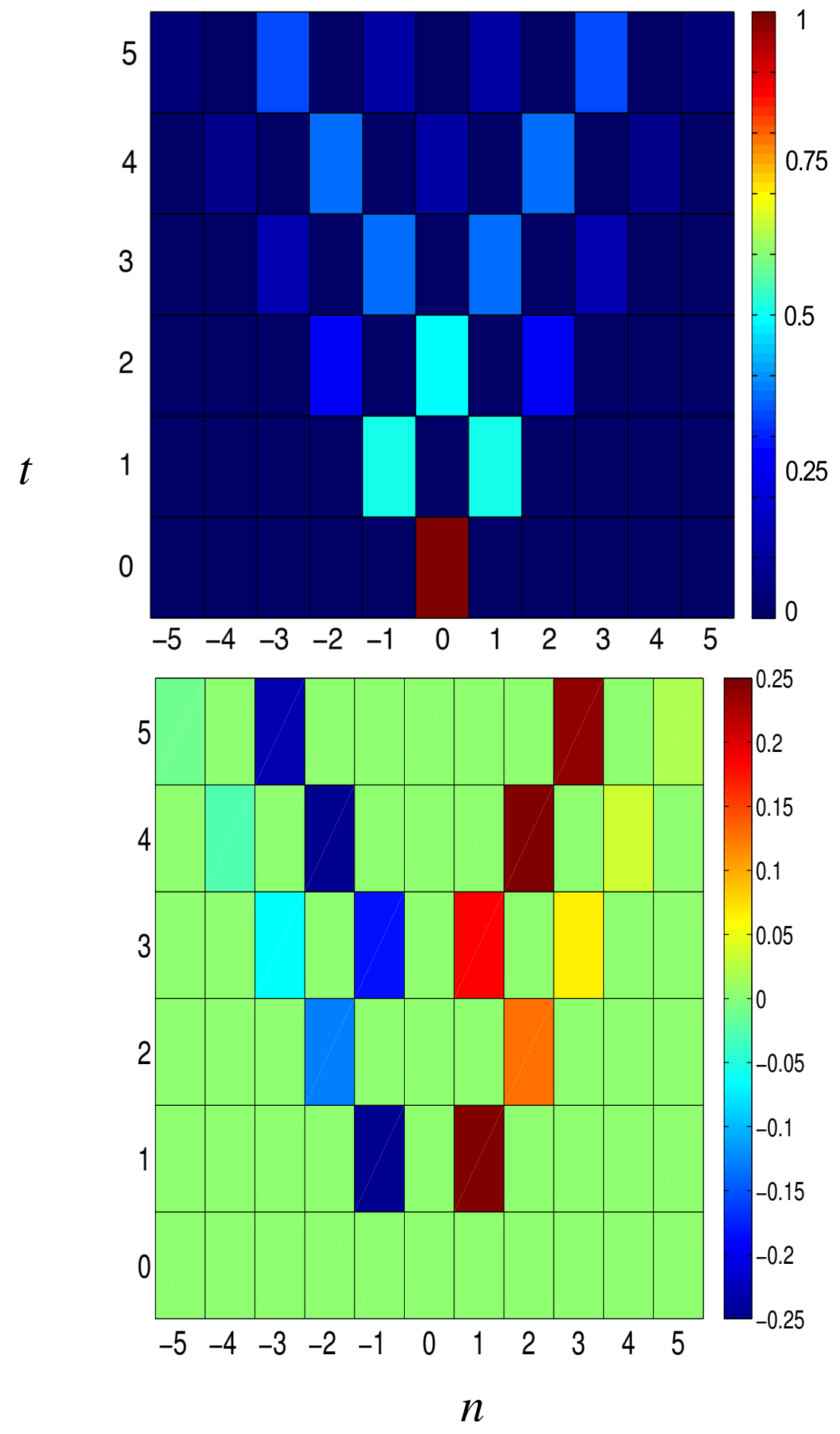

Figure 7.5: For a QW which is initialised in an identical fashion to Fig. 7.2, we plot (up) the probability density $\rho(n,t)$; (down) the symmetric probability current density $j^C(n,t)$. We see that now the current density is symmetric in space.

where $p_f$ is the terminal momentum, i.e at $t = \infty$, $p_\infty = p_f$ and $\lambda > 0$. Now taking the classical Hamiltonian to be

$$H = \frac{p^2}{2m} + \lambda W(x,p) \tag{7.27}$$

where $W(x,p)$ depends on *both* $x$ and $p$, using Hamilton's equations of motion, $\dot{x} = \frac{\partial H}{\partial p}$, $\dot{p} = -\frac{\partial H}{\partial x}$, we can construct a sample $H$ as

$$H = \frac{p^2}{2m} - \lambda x(p_f - p) \ . \tag{7.28}$$

from which the equation of motion are: $\dot{x} = \frac{p}{m} + \lambda x$, $\dot{p} = \lambda(p_f - p)$. The solution to this dynamics is:

$$p(t) = p_f(1 - e^{-\lambda t}) \quad , \quad x(t) = \frac{p_f e^{\lambda t}}{2m\lambda}(1 - e^{-\lambda t})^2 \ , \tag{7.29}$$

with the initial conditions: $p(0) = 0$, $x(0) = 0$. With this sample Hamiltonian the particle reaches a steady state momentum $p_f$ as it travels to $x \to \text{sgn}(p_f) \times \infty$.

The quantization of this sample classical Hamiltonian turns out to be somewhat ambiguous. If we shift $p$ by $p_f$, i.e $p \rightarrow p_f + p$, the classical Hamiltonian becomes,

$$H = \frac{(p + p_f)^2}{2m} + \lambda xp \tag{7.30}$$

To quantize this equation, we can let $p \rightarrow \hat{p}$, and $\hat{x}\hat{p} \rightarrow \frac{1}{2}(\hat{x}\hat{p} + \hat{p}\hat{x})$ to have the Hamiltonian hermitian. However there are many ways of promoting the classical quantity $xp$ to an hermitian operator and many issues relating to these difficulties have been addressed before in the literature under the so-called "H = xp model" ([135, 136]).

## 7.7 Conclusion

Both the theory pertaining to, and the experimental implementation of, quantum walks continues to attract widespread interest. Despite this little research has been done to understand their transport properties. In this work we derived expressions for the probability current density based on a discretisation of the probability continuity equation. For the specific case of an initial localised state we were able to derive an analytical expression for the total spatial current and showed that it reached a steady state. Curiously this steady state total current satisfied a type of sinusoidal current phase relationship akin to the current behaviour in Josephson junctions. With some effort we were able to derive an expression for the current density when we used central difference approximations and found that this symmetric probability current density behaved more intuitively. Finally we asked the question whether one can find conservative classical/quantum continuous systems whose dynamics leads to steady-state currents/momenta and found a wide class of such Hamiltonians.

# 8
# Summary and Outlook

In this thesis, I have studied what can be broadly categorised as *topological phases of matter* using *anyonic tensor network techniques*. In particular, I have studied how to use anyonic tensor network algorithms to simulate the physics of braided non-Abelian anyons, using two prominent species of anyons: Fibonacci and Ising anyons, and I investigated how braid statistics affect the ground state of a system of anyons, whose dynamics naturally allow for braiding. I gave the construction of an Anyonic×U(1)-MPS, which is an ansatz that can be used to encode the state of an anyonic system at any rational filling fraction. Then I studied a particular class of an anyonic Hubbard model on a ladder, where braid statistics can influence the phases. I found that different anyon models have different critical points, even when the microscopic Hamiltonians are similar, thereby indicating that the observed differences come from braiding. Therefore, it can be said the results in this thesis have pushed forward the boundaries of the physics of anyonic systems, anyonic tensor networks, topological phases, and the study of the influence of braid statistics in the formation of new phases of matter. These results also serve as indication of more exciting new physics yet to be discovered.

There are a number of different problems where the methods and results in this thesis could become useful. The anyonic tensor network algorithms presented here can be used to perform many further investigations into the equilibrium physics of anyons, such as quasi-1D models with other species of anyons, different lattice configurations, and so forth. In addition, the same ansatz in conjunction with the time-evolving block-decimation (TEBD) algorithm can be used to study non-equilibrium physics, such as the time evolution of an anyonic state. From elementary quantum mechanics we have a good understanding of how fermionic and bosonic statistics affect the state of a physical system in the presence of other degrees of freedom, e.g. position degrees of freedom. Bosons prefer to “stay closer together”, while fermions prefer to “stay apart”. But what do anyons do? The exchange statistics of Abelian anyons interpolate between those of fermions and bosons, and hence the expected influence of anyonic statistics might in some way be expected to likewise interpolate between the extreme ends of fermions and bosons. This kind of physics can be studied using anyonic-TEBD, though with caution, because tensor network techniques are classical algorithms, and braiding, during time evolution, can generate a large amount of entanglement. This may render tensor networks not useful, but certain regimes and limits can still be studied.

One area of interest in quantum information science is determining the error thresholds in topological quantum memories. The fusion space of anyons is non-local. States of anyons can be used as memories, which should naturally have a high tolerance to certain types of error, as information is encoded across many anyons; the information is carried collectively, rather than by individual anyons. It is however possible that unwanted anyon pair may be created from vacuum and which braid with the logical anyons, thereby destroying the logical information encoded into the system. In principle, these errors can be corrected if we can locate the erroneous anyons, undo the unintentional braids, fuse those anyons back into the vacuum. Studies of this kind are known as *topological quantum error correction*. Our methods can be used to simulate error models, to determine the efficiency of the error-correction protocols on, for example, a finite two-dimensional lattice of anyons [85–87].

As topological phases of matter is still currently a hot topic, there are still more questions that are yet to be answered beyond what I have done in my thesis. Topological phases are phases which have a topological order that may be largely "hidden" from being detected by local order parameters. In my thesis, I have used the scaling of the entanglement entropy and scaling of superfluid correlators as order parameters. In addition to these order parameters, the use of *fidelity* has also been proposed as a measure of phase transition [137]. Some others have proposed studying topological phase transitions through *Bose condensations* [138] and through the use of *topological S-matrix* [139] as an order parameter. One of my future goals is to try out these methods using the formalism introduced and exploited in this thesis.

In addition, the field of tensor networks (TN) is still actively evolving. More so, anyonic tensor networks represent a particular class of TN, and will therefore evolve as well. In addition, I have recently become aware that the geometric approach to topological quantum field theories bears much resemblance to tensor networks.[1] Some of these connections have also been explored here [140]. Anyons are topological charges which can also be thought of as punctures in TQFT, and hence the geometric approach of TQFT may be related to anyonic tensor networks in some ways. I am excited at the prospect of further exploring the intimate connections between these two fields.

Finally, the theoretical and mathematical appeal of anyons makes them really exciting to study. Even more, the technological possibility of using anyons to do something useful, like ushering in the next phase of (quantum) computing, makes them precious gems. It is an exciting time to be working on anyons!

[1] I learnt this from Robert Dijkgraaf lectures on TQFT at the 2015 Prospects in Theoretical Physics (PiTP) school at the Institute of Advanced Study, Princeton.

# A

# Data of some anyon models

The data for the anyon models I studied in this thesis are given here.

An anyon model is minimally specified by the following data: a set of charges $\mathcal{A}$, the fusion rules of the charges, the braid matrix $R$, and the F-tensor $F$. All other topological quantities can be derived from these data.

Some of the anyon models I used or studied in this thesis include: Fibonacci and Ising anyons—which are two prominent examples of non-Abelian anyons, $\mathbb{Z}_2$ (spinless) fermions, and $\mathbb{Z}_\infty$ bosons—which can be considered as examples of Abelian anyons in a general sense. Although in the case of bosons, we impose an additional hardcore constraint to limit the charges permitted on physical sites to either 0 (for no charge) or 1 (for a single boson).

## A.1 Fibonacci anyon data

The Fibonacci anyon model consists of two charges: vacuum ($\mathbb{I}$) and Fibonacci anyon ($\tau$), hence $\mathcal{A} = \{\mathbb{I}, \tau\}$. The charges have quantum dimensions: $d_\mathbb{I} = 1$ and $d_\tau = \frac{1+\sqrt{5}}{2}$. The fusion rules obeyed by the charges are

$$\mathbb{I} \times \mathbb{I} = \mathbb{I}, \qquad \mathbb{I} \times \tau = \tau \times \mathbb{I} = \tau, \qquad \tau \times \tau = \mathbb{I} + \tau. \tag{A.1}$$

The fusion tensor $N$ has components $N^c_{ab} = 0$ when $a \times b \nrightarrow c$ for all $a, b, c \in \mathcal{A}$. The nonzero components are given by

$$N^{\mathbb{I}}_{\mathbb{I}\mathbb{I}} = N^{\tau}_{\tau\mathbb{I}} = N^{\tau}_{\mathbb{I}\tau} = N^{\mathbb{I}}_{\tau\tau} = N^{\tau}_{\tau\tau} = 1. \tag{A.2}$$

The $R$-matrix has nonzero components

$$R^{\tau\tau}_{\mathbb{I}} = e^{-4\pi i/5},\ R^{\tau\tau}_{\tau} = e^{3\pi i/5},\ R^{\mathbb{I}\tau}_{\tau} = R^{\tau\mathbb{I}}_{\tau} = R^{\mathbb{I}\mathbb{I}}_{\mathbb{I}} = 1, \tag{A.3}$$

for compatible charges, and zero otherwise. The nontrivial $F$-move coefficients are

$$(F^{\tau\tau\tau}_{\tau})^f_e = \begin{pmatrix} \phi^{-1} & \phi^{-\frac{1}{2}} \\ \phi^{-\frac{1}{2}} & -\phi^{-1} \end{pmatrix}, \tag{A.4}$$

where $\phi = \frac{1+\sqrt{5}}{2}$, and $e, f \in \{\mathbb{I}, \tau\}$. The remaining $F$-move coefficients are given by

$$\left(F_d^{abc}\right)_e^f = N_{ab}^e N_{bc}^f N_{ec}^d N_{af}^d, \tag{A.5}$$

which is equal one for compatible charges.

## A.2 Ising anyon data

The Ising anyon model consists of three charges: vacuum ($\mathbb{I}$), Ising anyon ($\sigma$), and a fermion ($\psi$), hence $\mathcal{A} = \{\mathbb{I}, \sigma, \psi\}$. The charges have the following quantum dimensions: $d_{\mathbb{I}} = 1$, $d_\sigma = \sqrt{2}$, and $d_\psi = 1$. The fusion rules obeyed by these charges are

$$\mathbb{I} \times a = a, \qquad \sigma \times \sigma = \mathbb{I} + \psi, \qquad \sigma \times \psi = \sigma, \qquad \psi \times \psi = \mathbb{I}, \tag{A.6}$$

where $a \in \mathcal{A}$. The fusion tensor $N$ has components $N_{ab}^c = 1$ when $a \times b \to c$ for all $a, b, c \in \mathcal{A}$, otherwise $N_{ab}^c = 0$.

The $R$-matrix has these nontrivial braid factors:

$$R_{\mathbb{I}}^{\sigma\sigma} = e^{-i\pi/8}, \; R_{\psi}^{\sigma\sigma} = e^{i3\pi/8}, \; R_{\sigma}^{\sigma\psi} = R_{\sigma}^{\psi\sigma} = e^{-i\pi/2}, R_{\mathbb{I}}^{\psi\psi} = -1, \tag{A.7}$$

and these trivial braid factors:

$$R_{\mathbb{I}}^{\mathbb{I}\mathbb{I}} = R_{\sigma}^{\mathbb{I}\sigma} = R_{\sigma}^{\sigma\mathbb{I}} = R_{\psi}^{\mathbb{I}\psi} = R_{\psi}^{\psi\mathbb{I}} = 1, \tag{A.8}$$

for compatible charges, and zero otherwise. The nontrivial $F$-move coefficients are

$$(F_\sigma^{\sigma\sigma\sigma})_e^f = \begin{pmatrix} \frac{1}{\sqrt{2}} & \frac{1}{\sqrt{2}} \\ \frac{1}{\sqrt{2}} & -\frac{1}{\sqrt{2}} \end{pmatrix}, \tag{A.9}$$

where $e, f \in \{\mathbb{I}, \psi\}$, and

$$\left(F_\psi^{\sigma\psi\sigma}\right)_\sigma^\sigma = \left(F_\sigma^{\psi\sigma\psi}\right)_\sigma^\sigma = -1. \tag{A.10}$$

The remaining $F$-move coefficients are given by

$$\left(F_d^{abc}\right)_e^f = N_{ab}^e N_{bc}^f N_{ec}^d N_{af}^d, \tag{A.11}$$

which equals one for compatible charges.

## A.3 Fermions and Bosons Data

Fermions and bosons can generally be studied within the framework of anyons, and consequently using anyonic tensor networks such as the anyonic MPS.

As in the case of Fibonacci and Ising anyons, we give the (almost trivial) data for these particle types.

The particle spectrum for bosons is the $\mathbb{Z}_\infty = \{0, 1, 2, \ldots, \infty\}$, i.e. the set of positive integers, and the particle set for (spinless) fermions is $\mathbb{Z}_2 = \{0, 1\}$. As mentioned above, it should be noted that when we impose an hardcore constraint on bosons, the particle set reduces to $\mathbb{Z}_{\text{HCB}} = \{0, 1\}$, albeit the charges on the bonds of the MPS can still, in principle, grow to infinity.

The fusion rule for bosons is ordinary addition of integers, i.e.,

$$a \times b \to c \equiv a + b, \tag{A.12}$$

where $a, b \in \mathbb{Z}_\infty$, and the fusion rule for fermions is addition modulo 2, and therefore the fusion rules modifies to

$$a \times b \to c \equiv (a + b) \mod 2, \tag{A.13}$$

for $a, b \in \mathbb{Z}_2$. As a simple example of the distinction between the fusion rules of fermions and bosons, $1 \times 1 = 2$ if the charges are considered as bosons, but $1 \times 1 = 0$ if the charges are regarded as fermions.

As it is known from elementary quantum mechanics, the wave function of fermions and bosons acquire a phase factor of either $-1$ and $+1$ respectively, when a pair of those particles are exchanged. These exchange factors are the permutation factors of bosons and fermions, which written in the language of anyons are given by:

$$R_0^{00} = R_1^{10} = R_1^{01} = 1, \quad R_0^{11} = -1, \tag{A.14}$$

for fermions, while for bosons

$$R_c^{ab} = 1, \tag{A.15}$$

for all valid charges $a, b$, and $c$ in $\mathbb{Z}_\infty$. These exchange factors are considered trivial when compared to the exotic exchange (or braid) factors of non-Abelian anyons.

The $F$-matrix for both particle types is given as

$$\left(F_d^{abc}\right)_e^f = N_{ab}^e N_{bc}^f N_{ec}^d N_{af}^d, \tag{A.16}$$

for all charges satisfying the fusion rules of the two particle types.

# B

# Derivation of the global interference term $Q_0$

We show here the derivation of the global interference term $Q_0$ which we defined as $Q_0 = \lim_{t\to\infty}\sum_{n=-\infty}^{\infty}\alpha_{n,t}^{*}\beta_{n,t}$ in real space. We will derive the expression for $Q_0$ directly in Fourier space as it is more convenient and has been reportedly successful in analyzing the properties of quantum walks.

Let the amplitudes $\alpha_{n,t}$ and $\beta_{n,t}$ be grouped together as $\psi_n(t)$ written as a column vector,

$$\psi_n(t) = \begin{pmatrix} \alpha_{n,t} \\ \beta_{n,t} \end{pmatrix} = \alpha_{n,t}\,|0\rangle + \beta_{n,t}\,|1\rangle\,. \tag{B.1}$$

where vectors $|0\rangle = \begin{pmatrix} 1 \\ 0 \end{pmatrix}$ and $|1\rangle = \begin{pmatrix} 0 \\ 1 \end{pmatrix}$ are introduced for notational convenience in the derivation of $Q_0$. Note that the vectors $|0\rangle$ and $|1\rangle$ are not necessarily related to the coin states $\{|\uparrow\rangle, |\downarrow\rangle\}$.

Let the Fourier transform of $\psi_n(t)$ be $\tilde{\psi}_k(t)$ given as

$$\tilde{\psi}_k(t) = \sum_{n=-\infty}^{\infty}\psi_n(t)e^{ikn} \tag{B.2}$$

with $k \in [-\pi, \pi]$. The recurrence relation Eq.7.6 in Fourier space becomes

$$\begin{pmatrix} \tilde{\alpha}_k(t) \\ \tilde{\beta}_k(t) \end{pmatrix} = \begin{pmatrix} e^{ik}\cos\theta & e^{i(k+\delta)}\sin\theta \\ e^{-i(k+\delta)}\sin\theta & -e^{-ik}\cos\theta \end{pmatrix}\begin{pmatrix} \tilde{\alpha}_k(t-1) \\ \tilde{\beta}_k(t-1) \end{pmatrix}, \tag{B.3}$$

i.e.

$$\tilde{\psi}_k(t) = \hat{U}_k\tilde{\psi}_k(t-1), \tag{B.4}$$

iterating the "Markovian" equation recursively gives,

$$\tilde{\psi}_k(t) = \hat{U}_k^t\tilde{\psi}_k(t=0), \tag{B.5}$$

where $\tilde{\psi}_k(0)$ is the Fourier transform of the localized initial state $\psi_n(0)$. We assume here that our initial state $\psi_n(0)$ is localized at the origin of our lattice and is given as $\psi_n(0) =$

$\begin{pmatrix} \cos\frac{\phi}{2} \\ e^{i\gamma}\sin\frac{\phi}{2} \end{pmatrix}\delta_{n,0}$ as in Eq.(7.18), situated on a Bloch sphere, with $\phi \in [0,\pi]$ and $\gamma \in [0,2\pi]$.

Its Fourier transform is $\tilde{\psi}_k(0) = \begin{pmatrix} \cos\frac{\phi}{2} \\ e^{i\gamma}\sin\frac{\phi}{2} \end{pmatrix}$.

The Fourier transform of $Q_0$ becomes

$$Q_0 = \lim_{t\to\infty}\int_{-\pi}^{\pi}\frac{dk}{2\pi}\tilde{\alpha}_k^*(t)\tilde{\beta}_k(t). \tag{B.6}$$

As $U$ is unitary, it can be diagonalized with complex eigenvalues and eigenvectors and written in *spectral representaion* as,

$$U_k = (\lambda_k^1)\,|\phi_1(k)\rangle\langle\phi_1(k)| + (\lambda_k^2)\,|\phi_2(k)\rangle\langle\phi_2(k)| \tag{B.7}$$

from which the expression for $U_k^t$ follows as

$$U_k^t = (\lambda_k^1)^t\,|\phi_1(k)\rangle\langle\phi_1(k)| + (\lambda_k^2)^t\,|\phi_2(k)\rangle\langle\phi_2(k)|, \tag{B.8}$$

where $\lambda_k^1 = e^{i\omega_k}$ and $\lambda_k^2 = -e^{-i\omega_k}$ are the eigenvalues with their respective eigenvectors $|\phi_1(k)\rangle$ and $|\phi_2(k)\rangle$ as

$$\begin{aligned} |\phi_1(k)\rangle &= N(k)\begin{pmatrix} e^{i(k+\delta)}\sin\theta \\ e^{i\omega_k} - e^{-ik}\cos\theta \end{pmatrix} \\ |\phi_2(k)\rangle &= N(\pi-k)\begin{pmatrix} e^{i(k+\delta)}\sin\theta \\ -e^{-i\omega_k} - e^{ik}\cos\theta \end{pmatrix}, \end{aligned} \tag{B.9}$$

where $\omega_k$ is determined from $\sin\omega_k = \cos\theta\sin k$. $N(k)$ given as

$$N(k) = \frac{1}{\sqrt{2-2\cos\theta\cos(\omega_k - k)}}, \tag{B.10}$$

is a normalization factor that ensures the orthonormality of the eigenvectors $|\phi_1(k)\rangle$ and $|\phi_2(k)\rangle$.

From Eq.(B.5), we have,

$$|\tilde{\psi}_k(t)\rangle = \tilde{\alpha}_k(t)\,|0\rangle + \tilde{\beta}_k(t)\,|1\rangle = \cos\frac{\phi}{2}\hat{U}_k^t\,|0\rangle + e^{i\gamma}\sin\frac{\phi}{2}\hat{U}_k^t\,|1\rangle, \tag{B.11}$$

from which we could easily identify the amplitudes $\tilde{\alpha}_k(t)$ and $\tilde{\beta}_k(t)$ to be given as

$$\begin{aligned} \tilde{\alpha}_k(t) &= \cos\frac{\phi}{2}\langle 0|\,U_k^t\,|0\rangle + e^{i\gamma}\sin\frac{\phi}{2}\langle 0|\,\hat{U}_k^t\,|1\rangle, \\ \tilde{\beta}_k(t) &= \cos\frac{\phi}{2}\langle 1|\,U_k^t\,|0\rangle + e^{i\gamma}\sin\frac{\phi}{2}\langle 1|\,\hat{U}_k^t\,|1\rangle, \end{aligned} \tag{B.12}$$

where $\tilde{\alpha}_k(t) = \langle 0\,|\tilde{\psi}_k(t)\rangle$ and $\tilde{\beta}_k(t) = \langle 1\,|\tilde{\psi}_k(t)\rangle$.

With the expression we have for $U_k^t$, we could compute the expressions for $\tilde{\alpha}_k(t)$ and $\tilde{\beta}_k(t)$ as

$$\begin{aligned} \tilde{\alpha}_k(t) &= \cos\frac{\phi}{2}a_k(t) + e^{i\gamma}\sin\frac{\phi}{2}b_k(t), \\ \tilde{\beta}_k(t) &= \cos\frac{\phi}{2}c_k(t) + e^{i\gamma}\sin\frac{\phi}{2}d_k(t), \end{aligned} \tag{B.13}$$

where $a_k(t), b_k(t), c_k(t)$ and $d_k(t)$ are certain oscillatory functions which we list below,

$$a_k(t) = e^{i\omega_k t}\sin^2\theta N^2(k) + (-1)^t e^{-i\omega_k t}\sin^2\theta N^2(\pi - k)$$

$$\begin{aligned} b_k(t) = {} & e^{i\omega_k t}e^{i(k+d)}(e^{-i\omega_k} - e^{-ik}\cos\theta)\sin\theta\, N^2(k) \\ & - (-1)^t e^{-i\omega_k t}e^{i(k+d)}(e^{i\omega_k} \\ & + e^{-ik}\cos\theta)\sin\theta\, N^2(\pi - k), \end{aligned}$$

$$\begin{aligned} c_k(t) = {} & e^{i\omega_k t}e^{-i(k+d)}(e^{i\omega_k} - e^{ik}\cos\theta)\sin\theta\, N^2(k) \\ & - (-1)^t e^{-i\omega_k t}e^{-i(k+d)}(e^{-i\omega_k} \\ & + e^{ik}\cos\theta)\sin\theta\, N^2(\pi - k), \end{aligned}$$

$$\begin{aligned} d_k(t) = {} & e^{i\omega_k t}(1 - 2\cos\theta\cos(\omega_k - k) + \cos^2\theta)\, N^2(k) \\ & + (-1)^t e^{-i\omega_k t}(1 + 2\cos\theta\cos(\omega_k + k) \\ & + \cos^2\theta)\, N^2(\pi - k). \end{aligned}$$

With all the above equations, we can evaluate $Q_0$ from (B.6 as

$$Q_0 = \cos^2(\frac{\gamma}{2})E_0 + \sin^2(\frac{\gamma}{2})F_0 + \frac{e^{i\varphi}\sin\gamma}{2}G_0 + \frac{e^{-i\varphi}\sin\gamma}{2}H_0 \tag{B.14}$$

where

$$\begin{aligned} E_0 &= \lim_{t\to\infty}\int_{-\pi}^{\pi}\frac{dk}{2\pi}a_k^*(t)c_k(t)\,, \\ F_0 &= \lim_{t\to\infty}\int_{-\pi}^{\pi}\frac{dk}{2\pi}b_k^*(t)d_k(t)\,, \\ G_0 &= \lim_{t\to\infty}\int_{-\pi}^{\pi}\frac{dk}{2\pi}a_k^*(t)d_k(t)\,, \\ H_0 &= \lim\int_{-\pi}^{\pi}\frac{dk}{2\pi}b_k^*(t)c_k(t)\,. \end{aligned}$$

The asymptotics of the integrals above are easy to calculate but lengthy. Using, Stationary Phase Approximation(SPA), integral of the form $\int f(k)e^{i\varphi(k)t}dk$ vanishes as $t^{-\frac{1}{2}}$ for a non-vanishing $\varphi''(k)$ as shown in ([109]) and hence every time-dependent part of the integrals $E_0$, $F_0$, $G_0$ and $H_0$ gives negligible contribution and we can drop them in the the evaluation of the integrals to obtain,

$$E_0 = \frac{e^{-i\delta}\tan\theta}{2}(1 - \sin\theta) \tag{B.15}$$

$$F_0 = -\frac{e^{-i\delta}\tan\theta}{2}(1 - \sin\theta) \tag{B.16}$$

$$G_0 = \frac{\sin\theta}{2} \tag{B.17}$$

$$H_0 = e^{-2i\delta}\left((1 - \sin\theta)\tan^2\theta - \frac{\sin\theta}{2}\right) \tag{B.18}$$

Hence, $Q_0$ is given as,

$$Q_0 = \frac{(1-\sin\theta)e^{-i\delta}\tan\theta}{2}\Big[\cos\phi + \sin\phi\Big(e^{-i(\delta+\gamma)}\tan\theta + i\sin(\delta+\gamma)\frac{\cos\theta}{1-\sin\theta}\Big)\Big] \tag{B.19}$$

This result can also be checked if we compare it with numerical results of $Q(t)$ in the long time limit as shown in Fig.(7.3).